\documentclass[backref=page,cornellheadings,smallerheadings]{fiu_FV}
\pdfoutput=1

\renewcommand{\bibname}{LIST OF REFERENCES}
\usepackage{color}   
\usepackage{url}
\usepackage{appendix}
\usepackage{capt-of}
\usepackage{siunitx}
\usepackage{subfig}
\usepackage{caption}
\usepackage{dcolumn}   
\usepackage{xcolor} 
\usepackage[numberlinked=true]{footnotebackref}
\usepackage{cite}
\usepackage{empheq}
\usepackage{notoccite}

\usepackage{epsfig}
\usepackage[]{graphicx} 
\graphicspath{ {figures/} }

\usepackage{physics}
\usepackage{slashed}

\newcommand{\sh}[1]{#1\hskip -6pt  / }
\newcommand{\shl}[1]{#1\hskip -5pt  / }

\usepackage{amsmath}
\usepackage{amssymb}
\usepackage{bm}
\usepackage{float}

 \usepackage[utf8x]{inputenc}
\usepackage{fontenc}

\usepackage{hyperref}
\hypersetup{
    colorlinks=false, 
    linktoc=all,     
    linkcolor=blue,  
}

\PassOptionsToPackage{hyperfootnotes=true}{hyperref}

\title{Probing the Structure of Deuteron at Very Short Distances}
\author{Frank Vera}
\conferraldate{July}{2021}
\defensedate{July 1, 2021}
\advisor{Misak Sargsian}
\memberone{Rajamani Narayanan}
\membertwo{Werner Boeglin}
\memberthree{Mirroslav Yotov}
\degreefield{PHYSICS}
\college{College of Arts, Sciences and Education}
\collegedean{Dean Michael R. Heithaus}
\gradschooldean{Andrés G. Gil}

\begin{document} 

\setcounter{page}{1}
\pagenumbering{roman}
\pagestyle{plain}


\maketitle

\makeapproval{3}

\makecopyright

\begin{dedication}
To my wife and my parents.
\end{dedication}

\begin{acknowledgments}

I am deeply indebted to the members of my dissertation committee, Dr. Rajamani Narayanan, Dr. Werner Boeglin, and Dr. Mirroslav Yotov, for their contributions and encouragement during the development of my research. 
My special gratitude to my adviser, Dr. Misak Sargsian, whose guidance enabled me to complete this work. I will never forget his lessons on the importance of understanding beyond the calculations, seeking transparent explanations, and the rewards of hard work.  

I am very grateful to the graduate program directors, Dr. Brian Raue and Dr. Jorge Rodriguez, for their valuable assistance in navigating the graduate school. In addition, I recognize the administrative staff of the FIU physics department. Especially Elizabeth Bergano-Smith, Omar Tolbert, and Maria Martinez for helping me with their technical expertise.

My thanks to Dr. Lei Guo, Dr. Brian Raue, Dr. Wim Cosyn, and my fellow Ph.D. students Oswaldo Artiles and Trevor Reed for many enlightening discussions. In this regard, I am especially in debt to Christopher Leon. My intellectual experience in the graduate program was considerably enriched and much more enjoyable because of so many helpful discussions with Chris.

I acknowledge the Department of Energy for providing financial support for my research.

My greatest gratefulness goes to my wife, whose support and love over these years are beyond reason.


\end{acknowledgments}

\begin{abstract}

We study the electro-disintegration of deuteron at quasi-elastic kinematics and high transferred momentum as a probe for the short distance structure in nuclei. 
In this reaction, an electron hits a nucleus of deuterium, which breaks up into a pair of nucleons (proton-neutron). 
We focus our attention on events where fast nucleons emerge,  corresponding to nuclear configurations where the bound nucleons have a high relative momentum (exceeding 700 MeV/c).
The present research is relevant to physical systems where high-density nuclear matter is present. This condition covers a wide range of physics, from neutron stars to nuclei stability and the repulsive nuclear core.

Our calculations differ from previous studies in two crucial features. 
One is that our definition of the deuteron wave function, as it is used in high-energy electro-disintegration, depends on terms (of a relativistic origin) that can be ordered based on their 
relative contribution to the deuteron's internal momentum distribution.
These terms, related to the off-shell properties of the nucleon-nucleon bound-state, do not occur in non-relativistic quantum mechanics.
However, they become increasingly important in describing configurations with a high nucleon-nucleon relative momentum. 
The second essential difference is that we account for the off-shell nature of the bound nucleon that enters  on the definition of the  (half-off-shell) electromagnetic current.
We avoid many of the difficulties inherent to the relativistic nature of the processes involved by adopting a theoretical framework known as Light Front dynamics.
Simultaneous simplifications in the definition of the relativistic wave function for the proton-neutron bound state and the treatment of the (half-off-shell) electromagnetic current for the bound nucleon are among the essential advantages resulting from the use of the Light Front dynamics.
Furthermore, the rescattering between the emerging nucleons in the final stage of the reaction is also simplified within the Light Front framework.

Our new theoretical calculations provide new ways for the exploration of the relativistic structure of nuclear matter. 
In particular, it can provide the explanation for the results of recent experiments conducted at the Jefferson Laboratory, which for the first time have probed very high (relativistic) internal momentum configurations in the deuteron ($\sim1$~GeV/c).
\end{abstract}

\contentspage


\figurelistpage

\normalspacing
\setcounter{page}{1}
\pagenumbering{arabic}
\pagestyle{cornell}

\chapter{INTRODUCTION} \label{Intro}

Despite the overwhelming experimental support pointing to quantum chromodynamics (QCD) as the fundamental theory of strong interaction, it is not fully known how the nuclear force originates from QCD. Therefore, knowledge about the mechanism through which this emergence occurs is essential to understand the nuclear dynamics at a fundamental level and the phase transition from quark-gluon to hadronic matter. Since the nuclear force is responsible for binding atomic nuclei together, this knowledge is also a prerequisite for solving many open problems in nuclear theory, such as the stability of strongly interacting matter, the existence of hidden color configurations (multi-quark states) in the nuclei, as well as the origin of the nuclear core.

The lack of a complete practical theory for the study of the strong force had pushed forward a program focused on developing models. An example is the hadronic model of nuclear physics, where it is assumed that nuclei can be described in terms of constituent baryons interacting through the exchange of mesons.
It is expected, however, that this picture breaks down once the distance between constituent baryons is less than their radius. In this case, the internal structure of the baryons will play an explicit role in the nuclear dynamics.
In this regard, the study of dense nuclear matter plays a crucial role, providing a transition from baryonic to quark-gluon degrees of freedom.

The main goal of our research is the study of nuclear dynamics at short distances. Considerable progress has been made in  understanding  the nuclear force at distances up to  {$r\approx 1/m_\pi \approx 1.8$ fm}\footnote{One fermi (fm) corresponds to $10^{-15}$ meters.} within effective theories  based on the chiral symmetry of QCD and formulated in terms of  hadronic degrees of freedom. 
In these theories, the short-range phenomena are parameterized through  contact interactions, which are then included in the renormalization scheme.  As a result, the dynamic aspect of the short-range interactions is  not resolved.

Another approach in describing the nuclear force is the one boson exchange model {(OBE)}, which  has succeeded in explaining several  properties of the nucleon-nucleon ($NN$) interaction at intermediate to short distances, such as the large tensor forces. However, OBE cannot   fully describe  the interaction strength without including phenomenological parameterizations~(see, e.g., \cite{Njm}). Also, it contains the concept of strongly virtual  exchanged mesons,  which is not well defined \cite{Feynman}.

Yet another successful approach to describe NN interactions is the phenomenological parameterization (potential models), in which the general spin-isospin structure of the NN interaction is reproduced by adding a finite number of Yukawa-type interactions with the long distance interaction given by the pion exchange term (see e.g. \cite{V18}).
In this approach, the ansatz parameters used to describe the {$NN$} potential are obtained by fitting experimental phase shifts.

While OBE and phenomenological potential models reproduce the $NN$ scattering phase shifts with high precision, they considerably diverge in their predictions for the nucleon momentum distributions in the deuteron above {$500$~MeV/c}. For example, the probability (based on OBE) for nucleon momentum of about {$600$~MeV/c} in the deuteron differ by a factor of two for AV18\cite{V18} and {CD-Bonn}\cite{CDBonn}  potentials. This indicates that there are limitations for OBE and phenomenological $NN$ potentials to gain reliable information about the short-range structure of $NN$ forces in particular and nuclear forces in general.
Moreover, the above approaches are non-relativistic and attempt to include relativistic effects turn out to be either inconsistent or make them tremendously complicated, resulting in a set of coupled channel equations. Another limitation of the above approaches is that there is no clear way of including the quark-gluon degrees of freedom in such a way that it is consistent with QCD.

Our main strategy for probing short-range nuclear forces is to study pairs of bound nucleons that occupy a small space-time region. In other words, we  search for bound nucleons that have relative momenta much larger than the average momentum found in the deuteron.
The most suitable theoretical description for describing high-energy processes is based on the light-front formalism, which allows avoiding or suppressing vacuum fluctuations.
In this case, the nucleons are treated relativistically, and the application of diagrammatic methods\cite{Sargsian:2001ax,Artiles-Sargsian2016} result in light-front wave functions described by Weinberg-type equations\cite{Weinberg1966}. There have been many extensive studies on the light-front nuclear dynamics (e.g., Refs.\cite{Frankfurt:1981mk,Frankfurt:1977vc,KP91,Miller00}). However, these studies did not focus on (semi) exclusive reactions\footnote{Where one or more produced particles are detected, in addition to the scattered electron.}, which will allow us to test many properties of the nuclear structure. 
In our research, we have focused primarily on deuteron electron-disintegration experiments at high momentum transfer\footnote{In the present work, the approach chosen to calculate the high momentum contribution to the nuclear structure is similar to the methods applied in high energy physics for the description of the parton distributions in the nucleon, where partons are treated fully relativistic.}, which are part of the research program of the Jefferson Laboratory, located in Virginia, USA.

\section{Why the Deuteron}

The program of nuclear physics is to provide a description of the strong interaction (QCD). From a modern perspective, it presents a roadmap similar to that of atomic physics, which led to the establishment of Quantum Electrodynamics (QED) as the (contemporary) fundamental theory of electromagnetic interaction\footnote{The quantum field theory of electrons and photons.}. For QCD and QED, the infinite-dimensional problem resulting from the field-theoretical picture can be 
reformulated
as a perturbative series\footnote{For each order in the perturbative expansion, a finite number of terms need to be calculated. Each term represents the interaction among a finite number of particles.}. However, while for the electromagnetic interaction, the perturbative series approximates the solution beyond the precision of each experiment carried out so far, the same does not happen with QCD. Finding non-perturbative solutions for QCD remains an open problem.

As an alternative of tackling the complicated nuclear problem directly from the fundamental QCD degrees of freedom (quarks and gluons)\footnote{Although important achievements have been made, the computation of solutions remains impractical.}, we rely on simplified models.
The first step is to consider the nucleon as a (quasi-)particle, and the internal structure is quantitatively parameterized by functions called form factors\footnote{Details about the form factors will be provided in subsequent chapters.}.
An example of a simplified version of the nuclear interaction  
is provided by the aforementioned phenomenological approach,
which reduces the many-body field theoretical problem\footnote{Where the interaction is carried by an exchanged particle.} to an interaction potential. 
The total potential incorporates contributions from two-body interactions, three-body interactions, etc., i.e.,  
\begin{equation}
V_\text{A} = V_{2} + V_{3} + \dots  
\end{equation}
where, $\text{A}$ is the number of nucleons, and $V_{i}$ corresponds to the $i$th-body  potential. It has been found that the two-body potential ($V_{2}$) reproduces about 90\%  of the binding energy of light nuclei ($\text{A} \lesssim 12$) and becomes even more dominant as the number of nucleons increases \cite{Atti:2015eda}.

The deuteron is the simplest nucleus, consisting (predominantly) of a proton-neutron-bound state. Hence, it is the simplest laboratory for studying the two-body nuclear problem. 
Furthermore, it was found that in nuclei, the configurations involving two nucleons (correlated) with a significant relative momentum are dominated by proton-neutron pairs, i.e., over proton-proton and neutron-neutron counterparts. 
Thus, a better understanding of the distribution of the high internal momentum in the deuteron will improve the extraction of the correlations of two nucleons in heavier nuclei. Therefore, our research is important to describe high-density nuclear matter and the short-range structure of the nuclear force.

%
%
%
%

\section{Modern view of the Deuteron }

Being the simplest nucleus, the deuteron represents a unique testing ground for the emergence of nuclear forces from the fundamental interaction of quarks and gluons. One expects that QCD degrees of freedom become relevant for situations where the bound proton and neutron in the deuteron are substantially overlapping. Such a situation can be achieved by probing deuteron at large relative momenta, comparable with the masses of its constituent nucleons.

Despite the apparent simplicity of the deuteron, the complex character of the nuclear forces and the hadronic spectrum creates a rather diverse picture for the composition of the deuteron. Because of the positive parity and total spin J=1 of the deuteron, it represents a pseudo-vector state. In addition, the deuteron has zero total isospin, $\text{T}=0$  (iso-singlet). As a result,  the modern view of the deuteron can be summarized as a decomposition into  Fock states\footnote{By Fock state, we adopt the definition of a state characterized by a well-defined (fixed) particle number.} with total spin equal to 1 and isospin equal to 0 \cite{Boeglin:2015cha}, 
\begin{equation}
\Psi_d = \Psi_{pn} + \Psi_{\Delta\Delta} + \Psi_{NN^*} + \Psi_{hc} +  \Psi_{NN\pi} \cdots
\label{Fstates}
\end{equation}
where  "$\cdots$" includes the contributions from higher Fock components with potentially higher mass constituents.

The relative contribution of the Fock components in the deuteron wave function, starting from the lowest number of constituents, is organized  as follows:

\noindent {\bf \boldmath $pn$ Component:} The empirical evidence suggests that the $pn$ component dominates in the deuteron wave function for large internal momenta ($\sim 600-700$~MeV/c)\cite{Frankfurt:2008zv}. 
Theoretical calculations based on the $pn$ component of the deuteron wave function (e.g., Refs.\cite{Sargsian:2009hf,Cosyn:2010ux,Arrington:2011xs}) provide a good  description for a wide variety  of the processes involved in probing deuteron's wave function for up to  $600-650$~MeV/c internal momenta.

\medskip

\noindent {\bf \boldmath $\Delta\Delta$  and  $NN^*$ Components:}  Energetically, the next closest two-particle (isosinglet, pseudovector) Fock components of  the deuteron  are the  $\Delta\Delta$  and  $NN^*$, the latter representing a radial excitation of one of the nucleons in the deuteron. 
Such excitations require energies in the order of $600$~MeV, corresponding to internal momentum of about $ 800$~MeV/c.
Due to the large-cross-section of the $\pi N\rightarrow \Delta$ transition, it is expected that the $\Delta\Delta$ component will dominate over the $NN^*$ component.
The overall contribution of the $\Delta\Delta$ component has been experimentally constrained to  $\le 1$\%\cite{Frankfurt:1988nt}, while there is no evidence yet for the possible mixture of the  $NN^*$  component.

\medskip

\noindent {\bf \boldmath Hidden Color Component:} One of the unique predictions of QCD is the possible existence of the hidden color component in the deuteron wave function.   
The color decomposition of the very same six quarks present in deuteron's $pn$ component allows for a 6-quark color-singlet configuration consisting of two colored (octet) baryons. The colored constituents configuration (hidden color component) has been estimated to be responsible for almost $80$\% of the high momentum  wave function strength \cite{Harvey:1980rva,Ji:1985ky}. 
Such component is expected to dominate at considerable high  excitation energies of the NN system, with the six-quark representation of deuteron describing the sum of all the possible two-baryonic states. 
Since large excitations are relevant to the nuclear core, there is an interesting possibility that the $NN$ repulsive core results from the cancellation in the overlap between hidden color-dominated configurations and the $NN$ component.

\medskip

\noindent {\bf \boldmath ${NN\pi}$ Component:} The most dominant three-particle  Fock component of  the deuteron  is  the $NN\pi$ component, which is expected to become relevant at excitation energies close to the pion threshold (corresponding to  internal momenta  $\sim 370$~MeV/c). 
There is evidence from reactions probing meson exchange currents that this component starts to dominate at missing momentum  $\sim 350$~MeV/c, which is consistent with the  above estimate of the pion threshold (see for example Refs.\cite{Hockert:1974qt,Fabian:1976ne}).
However, in experiments at high energy and momentum transfer (above the proton mass), aimed at probing short distance nuclei structure, no such evidence is observed for relative $NN$ momenta up to $\sim 650$~MeV/C (e.g., \cite{Ulmer:2002jn,Egiyan:2007qj,Boeglin:2011mt}).
The latter suppression can be explained by the fact that the transition   $N\rightarrow \pi N$ behaves as  $\sim \exp{\lambda t}$ with $\lambda \ge 3$~GeV$^{-2}$ (hard form factor) \cite{Frankfurt:1988nt}. 
Therefore, the processes in which  the high energy probe couples to the exchanged pion in the ${NN\pi}$ deuteron component are significantly suppressed at high momentum transfer (above the proton-pion system mass).

The above discussion   indicates that the dominant inelastic component  in the deuteron may be the $\Delta\Delta$ rather than the $NN\pi$ component. This will extend the $pn$ dominance for internal momenta up to $\sim 750$~MeV/c, which lays below  the expected threshold of the $\Delta\Delta$, and possibly hidden color  components in the wave function of the deuteron.

We conclude that for internal momenta up to $\sim 750$~MeV/c, 
the proton-neutron Fock component must be the dominant one in the deuteron wave function.
The conventional quantum mechanical description for the $pn$ component becomes questionable at such high momentum because of significant relativistic effects. 
Our calculation of the exclusive deuteron-electro-disintegration reaction consistently accounts for relativistic effects in the bound-proton-neutron component of the deuteron wave function and the rescattering of the outgoing (produced) nucleons.
The implementation of the light-front (effective) Feynman diagrammatic developed in this work  represents the light-front extension of the generalized eikonal approximation (GEA), which was developed earlier\cite{Sargsian:2009hf,Frankfurt:1996xx,Sargsian:2001ax}  to account for the relativistic motion of the nucleons in the final state interaction amplitude. 

Our research is motivated by the comprehensive program of electron scattering experiments with high transferred momentum, which aims to reveal the structure of nuclei. (e.g., \cite{arnps, Boeglin:2015cha, srcprog}).
With this experiments, a deeply bound nucleon in the nucleus can be probed by a photon with large virtuality\footnote{The definition of virtuality for the exchanged photon is provided in Eq.(\ref{Q2}).}, producing a final nucleon with very high 3-momentum\footnote{Roughly equal to that of the virtual photon.}.
The high-energy nature of these   processes allows for chief simplifications in the description of  the scattering process.  
One of the main characteristics of high energy scattering is that the process evolves along a light-like direction (e.g., \cite{Feynman, KS(1970), LB(1980), Frankfurt:1981mk, BPP(1997)}), making the light-front framework the most natural choice to describe the reaction. 
The crucial advantages of a  description are the suppression of  the negative energy contribution in the full propagator of the bound nucleon and the  possibility of identifying the ``good" component of the electromagnetic current, for which the off-shell effects are minimal
Off-shell effects are the sources of many ambiguities in the calculation of the scattering cross-section. Therefore, its minimization is the first step towards meaningful estimates.
All these features will be described in detail in the following chapters.

\chapter{DEUTERON ELECTRO-DISINTEGRATION}
\label{Chap 2 - Electro-disintegration}

Experiments performed at quasi-elastic kinematics, in which 
electrons are scattered off  a deeply bound nucleon producing a struck nucleon ($N_f$) in the final state, are the main tools to directly probe bound nucleons with large momenta.
Experiments of this kind have a very small differential cross-section. Therefore,  they require high-intensity electron beams, which became available only after the advent of the Continuous Electron Beam Facility (CEBAF) at Jefferson Laboratory (JLab), located in Virginia. The first of such experiments\cite{Ulmer:2002jn,Egiyan:2007qj,Boeglin:2011mt}  reaching large values of  transferred invariant momentum ($Q^2\ge 1$~GeV$^2$)  were performed at JLab,  and   for the first time succeeded  in a direct measurement of the deuteron momentum distribution up to $550$~MeV/c\cite{Boeglin:2011mt}. After the 12~GeV upgrade of CEBAF, a new era of long-awaited electro-disintegration experiments\cite{Yero:2020} are probing the deuteron structure at unprecedentedly large internal momenta, up to $1$~GeV/c.  
Such measurements are opening up an entirely new stage for investigating the QCD origin of nuclear forces.  

The advantage of exclusive electro-disintegration experiments is that, in the simplest model known as the plane wave impulse approximation, it directly probes the deuteron momentum distribution. 
However, because the outgoing proton and neutron interact strongly, there is a substantial contribution from the final state interaction effects, which must be properly accounted for. 
 With the birth of the experimental program on high momentum transfer electro-disintegration reactions, various theoretical groups have made intensive efforts to calculate the effects of FSI.
 (e.g. Refs.\cite{CiofidegliAtti:2000xj,Laget:2004sm,Jeschonnek:2008zg,Sargsian:2009hf}). 
All these efforts are based on the high-energy nature of the scattering process.  However, the previous works still use the non-relativistic description of the deuteron wave function.  As a result, their validity becomes increasingly questionable for the description of processes that probe deuteron's internal momentum at values comparable to the nucleon mass ($\sim 1$ GeV/c).
One of the main problems in the interpretation of the experimental data is to account for the relativistic dynamics of the deuteron adequately.

\section{Kinematic Definitions}
\label{kin_defs}

We investigate the reaction, 
\begin{equation}
e + d \rightarrow e^\prime + N_\text{f} + N_\text{r}
  \label{reaction}  
\end{equation}
where the deuteron nucleus is disintegrated after interacting with a high energy electron. The result of this event is the production of two nucleons (proton and neutron) in the final state together with the scattered electron. 
Two particles are detected in coincidence in the final state, the scattered electron and one of the nucleons. 
The measured momentum of the detected nucleon is labeled, $p_\text{f}=(E_\text{f},{\bf p}_\text{f})$. The second nucleon will be called the recoil nucleon, and labeled as $p_\text{r}=(E_\text{r},{\bf p}_\text{r})$, its momentum is  reconstructed from the energy-momentum conservation,
\begin{equation}
 p_d + p_e = p_e' + p_\text{f} +  p_\text{r}
 \end{equation}
 where, $p_d=(E_d,{\bf p}_d)$ is the deuteron's four-momentum, and $p_e=(E_e,{\bf p}_e)$ ($p_e'=(E_e',{\bf p}_e')$) is the initial (final) four-momenta of the electron. In the current kinematics, the initial and final energies of the electron are much larger than its mass, 
\begin{align}
 |{\bf p}_{e}|, \  |{\bf p}_{e}'| \ \gg \ m_e 
\label{HEP e}
\end{align}
from where it follows that can be treated as massless particles.

\begin{figure}[h]
	\centering
	\includegraphics[scale=0.5]{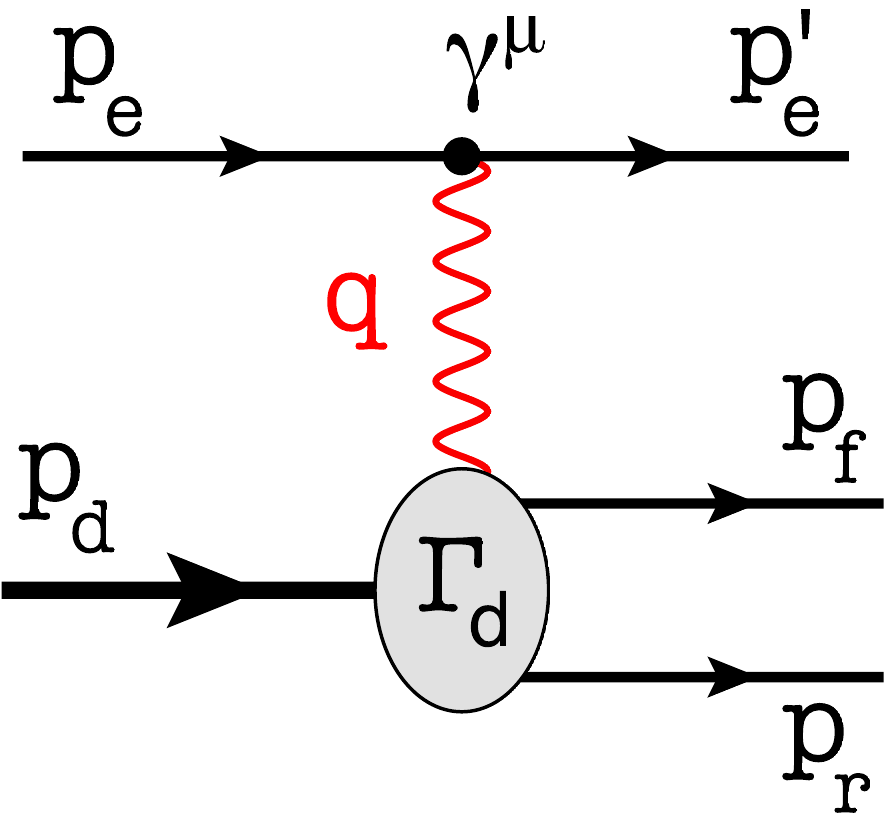}
	\caption{Deuteron electro-disintegration in the one-photon-exchange approximation.}
	\label{edeNN}
\end{figure}

In quantum field theory, the interaction between electron and deuteron is mediated by the exchange of virtual photons. 
The graph shown in Fig.(\ref{edeNN}) is known as the one photon exchanged approximation, which is a highly accurate approximation due to the small coupling constant of the electromagnetic interaction ($\alpha \approx 1/137$).
The four-momentum of the exchanged photon (momentum transfer, {\color{red}q})  is, 
\begin{equation}
q=(q^0, {\bf q})= p_e - p_e' = (E_e,{\bf p}_e) - (E_e',{\bf p}_e') 
\end{equation} 
It is very convenient to work with kinematic variables that are Lorentz invariant, since their values are the same for any inertial reference frame. A common choice are the (negative) invariant  momentum transfer squared,  
\begin{equation}\label{Q2}
 Q^2 = -q^2= {\bf q}^2 - q_0^2
 \end{equation} 
the fraction of the target's momentum longitudinal to the momentum transfer (Bjorken scaling variable)\footnote{Note that with this definition we have, $0 \le x \le 2$.},
\begin{equation}
x = {- q \cdot q \over p_d \cdot q}= {Q^2 \over p_d \cdot q}
\end{equation}
and, the 
center of mass energy of the photon-deuteron system,   
 \begin{equation}
s = (p_d+q)^2  
\end{equation}

\section{Reference Frames}
\label{reframe}

\begin{figure}[h]
	\centering
	\includegraphics[scale=0.5]{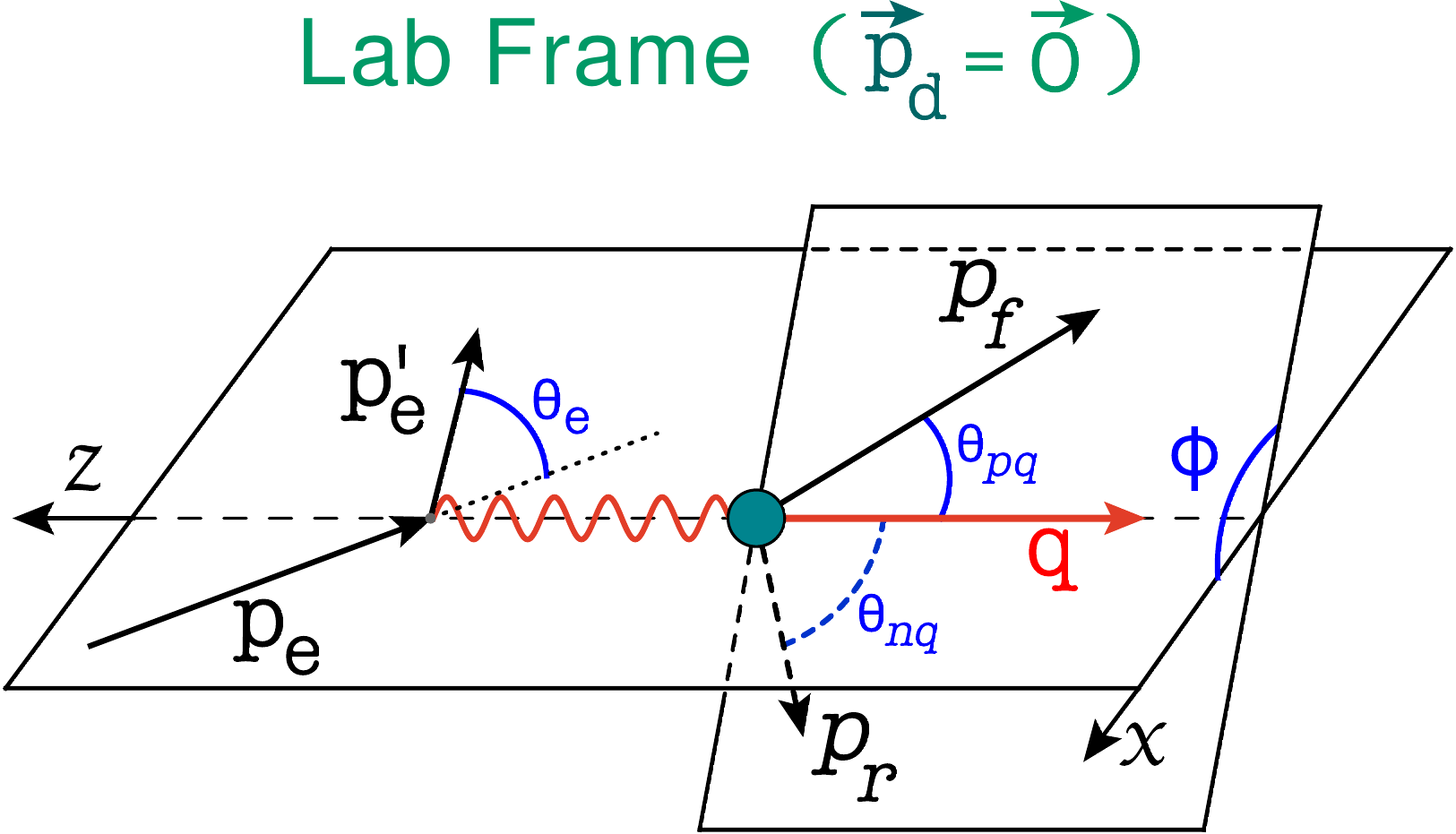}
	\caption{Deuteron Lab (rest) Frame.}
\label{Lab Frame}
\end{figure}

In our calculations, we will primarily consider two reference frames. One is the Lab frame of the deuteron sketched in Fig.(\ref{Lab Frame}) with coordinate axes defined by,  
\begin{equation}
(\hat x_\text{lab}, \hat y_\text{lab},\hat z_\text{lab}) = \left( { \vec q \times \hat y_{\text{lab}} \over | \vec q \times \hat y_{\text{lab}}|} ,  { \vec p_{e} \times  \vec p_{e}' \over |\vec p_{e} \times  \vec p_{e}'|} , -{\vec q \over |\vec q|} \right)
 = \left( {\hat y_\text{lab} \times \hat z_\text{lab} \over |\hat y_\text{lab} \times \hat z_\text{lab}|} , {\hat p_e \times \hat z_\text{lab} \over |\hat p_e \times \hat z_\text{lab}|} , - \hat q_\text{lab} \right)
\end{equation} 
where, $\vec p_e$ and $\vec p_e'$  are the (known) 3-momentum vectors of the incoming and outgoing electron, respectively.

\begin{figure}[h]
	\centering
	\includegraphics[scale=0.5]{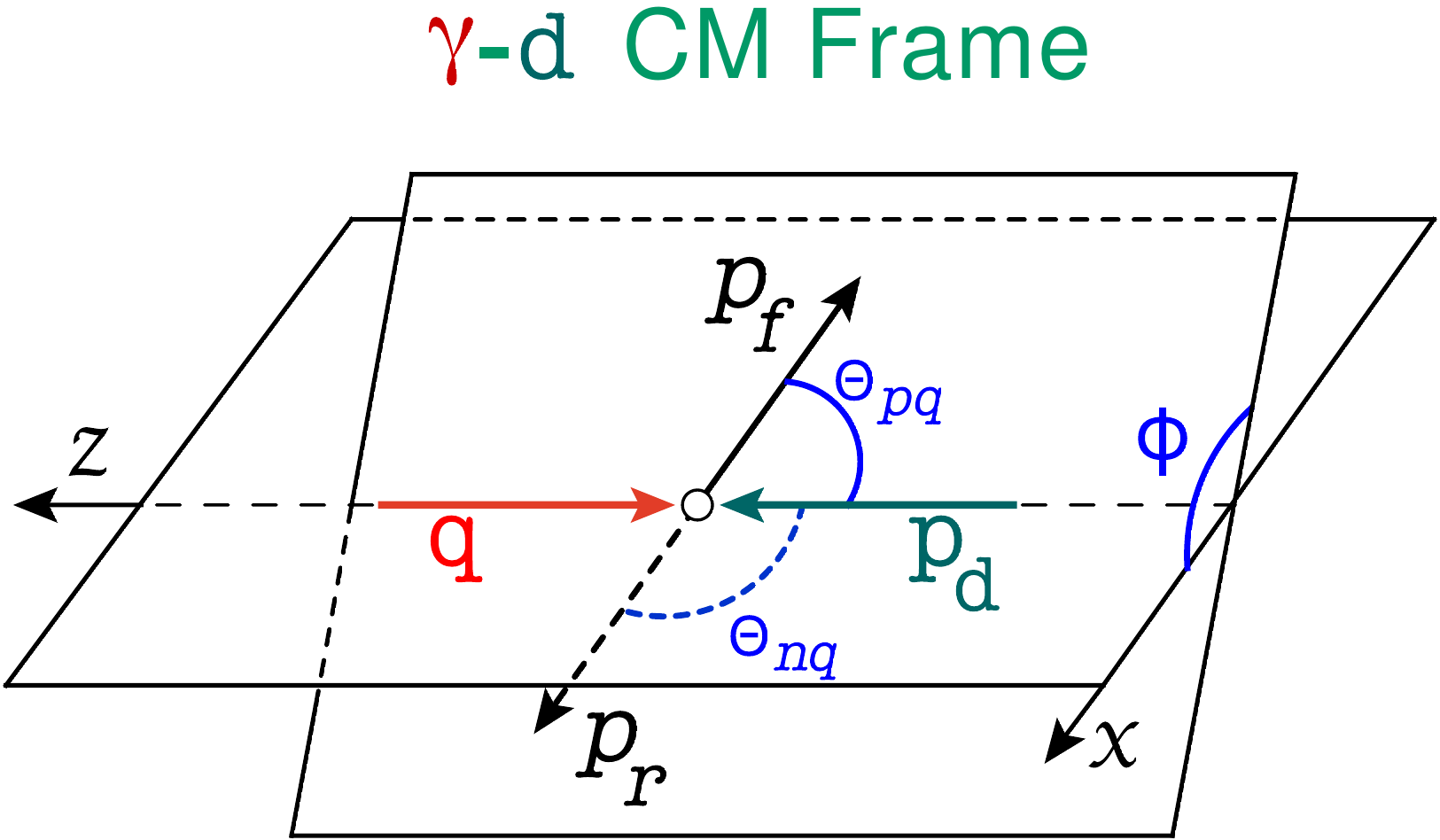}
	\caption{Center of momentum frame for the initial two-body system, deuteron and virtual-photon, and also  for the produced proton-neutron.}
	\label{CM Frame}
\end{figure}

The second is the center of momentum frame of the virtual photon and deuteron  (${\color{red}\gamma^*} d$-CM), Fig.(\ref{CM Frame}),
 given by the equivalent conditions ${\bf p}_{d}+{\bf q}=0={\bf p}_\text{r}+{\bf p}_\text{f}$, i.e. it is also the CM frame of the produced proton-neutron nucleons.  
 The coordinate axes  coincide with the Lab frame ones,
\begin{equation}
(\hat x_\text{CM}, \hat y_\text{CM},\hat z_\text{CM}) = ( \hat x_\text{lab}, \hat y_\text{lab},\hat z_\text{lab}) 
\end{equation}

In the CM reference frame the four-momenta of the deuteron and virtual photon are given in terms of invariant kinematical variables by,
\begin{eqnarray}
p_d^\mu & \equiv & (E_d,{p_{d}}_{z}, {{\bf p}_{d}}_\text{T}) =  
\left({{Q^2\over x} + m_d^2\over \sqrt{s}}, {Q^2\over x\sqrt{s} }\sqrt{1+ {m_d^2x^2\over Q^2}}, {\bf 0}_\text{T} \right)  \nonumber \\
q^\mu &\equiv & (q_0,q_z, {\bf q}_\text{T}) =  \left({{Q^2\over x} -Q^2\over \sqrt{s}}, -{Q^2\over x\sqrt{s} }\sqrt{1+ {m_d^2x^2\over Q^2}}, {\bf 0}_\text{T} \right) 
\label{refframeP}
\end{eqnarray}
where,  the invariant mass $s$ can be written as,
\begin{equation}
s = (p_d+q)^2 = Q^2 {2-x\over x} + m_d^2 
\end{equation}
and the Bjorken variable $x$ is given in the Lab frame by,
\begin{equation}
  x = {Q^2 \over m_d q^0_\text{lab}}
  \label{xb}
\end{equation}
 with $q^{0}_\text{lab}$ being the energy transferred in the Lab frame and $m_d$ the deuteron mass.
The mass of the electron is negligible compared with their energies (Eq. \ref{HEP e}), in which case we have, $Q^2 = 4 E_e E_e ^\prime \sin^2{\theta_e\over 2}$, where
 $E_e$ ($E_e^\prime$)  is the energy of the incoming (scattered) electron, and $\theta_e$ the scattering angle, all their values as measured  in the Lab frame.

\section{Reaction Dynamics }

\begin{figure}[h]
\hspace{-0.3in}
\includegraphics[scale=1]{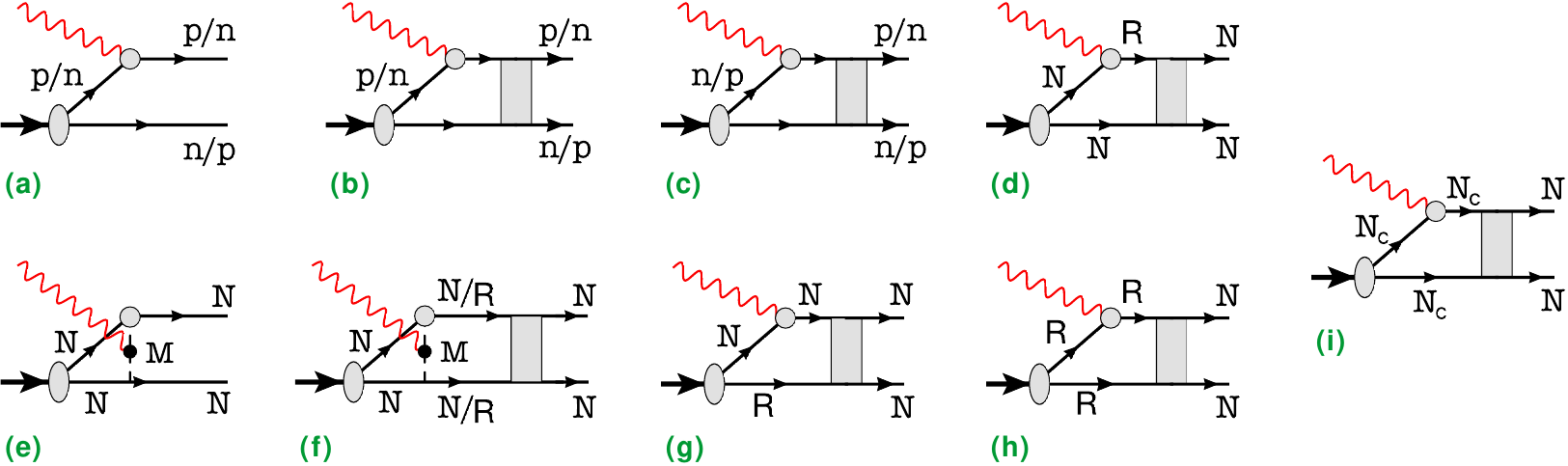}
\caption{Diagrams contributing to the exclusive $d(e,e'p)n$ reaction.}
\label{edepn_diagrams}
\end{figure}

Within the one-photon exchange approximation  the  Feynman diagrams that describe the reaction  Eq.(\ref{reaction})   given in  Fig.(\ref{edepn_diagrams}). For definiteness, we consider a proton to be the knocked-out nucleon.
 
\vspace{0.15in} 

These diagrams represent different mechanisms in electro-disintegration of the deuteron and can be categorized as follows:

\begin{itemize}

\item[(a)] {\bf PWIA contribution:} In Fig.(\ref{edepn_diagrams}(a)), the situation in which the nucleon struck by the virtual photon is detected,  while the undetected one is a spectator, corresponds to a  mechanism called the direct plane-wave impulse approximation, hereafter plane-wave impulse approximation~(PWIA). The alternative possibility in which the virtual photon strikes the undetected nucleon while the detected one emerges as a spectator is called the spectator-PWIA. No final state interaction~(FSI)  is involved. Therefore, the final nucleons emerge as plane waves.

\item[({b})] {\bf Direct FSI contribution:} In this case Fig.(\ref{edepn_diagrams}({b})) the struck proton rescatters off the spectator neutron and is detected in the final state.

\item[(c)] {\bf Charge-Interchange FSI contribution:} In the case of  Fig.(\ref{edepn_diagrams}({c})) the  struck-nucleon undergoes a  charge interchange  interaction with the spectator nucleon.

\item[(d)] {\bf Intermediate State Resonance Production:} In Fig.(\ref{edepn_diagrams}(d)), the electromagnetic interaction  excites the nucleon into a resonance state  which then rescatters with the spectator nucleon resulting in the final proton and neutron.
 
\item[(e,f)] {\bf Meson Exchange Contributions:} In Fig.(\ref{edepn_diagrams} (e) and (f)), the electromagnetic interaction  takes places with the mesons which are exchanged between initial nucleons in the deuteron.

\item[(g)] {\bf Non-Nucleonic Contributions:} The last three terms contributing to the reaction Fig.(\ref{edepn_diagrams} (g), (h), and (i)) are sensitive  to  non-nucleonic components of the deuteron wave function. The first two represent an initial state with baryonic resonance(s), whereas (i) corresponds to the hidden-color component contributions.

\end{itemize}

\section{Main Features of High Energy Approximation Electro-Disintegration}
\label{HEP}

Among all the reaction mechanisms accounting for deuteron electro-disintegration, only the PWIA (Fig.\ref{edepn_diagrams} (a)) provides direct access to the NN structure of deuteron. 
However, a reliable estimate of the PWIA contribution to the reaction requires a systematic way to account for all the remaining reaction mechanisms  (diagrams Fig.\ref{edepn_diagrams}(b)-(i)) discussed in the previous section.   
As we explain below, most of the reaction mechanisms can be suppressed if we consider the reaction (\ref{reaction})  at sufficiently high energy and momentum transfer while guaranteeing that the struck nucleon  (carrying almost all the momentum of the virtual photon) is significantly more energetic than the recoiling nucleon.
Explicitly,  in the Lab frame, we must guarantee,
\begin{align}
| {\bf p}_\text{r}^\text{lab} |  \ =  | {\bf q}^{\text{lab}} - {\bf p}_{\text{f}}^\text{lab} |  \  \ll  |{\bf p}_{\text{f}}^\text{lab} |  \sim  \  {\bf q}^\text{lab} |  \  \ge \ Q   \gtrsim 2 ~ \text{GeV/c}
\label{henc}
\end{align}
where, the variables are defined in  Fig.(\ref{edeNN}). The consequences of applying the conditions (\ref{henc}) are different for each of the individual Feynman diagram in Fig.(\ref{edepn_diagrams}(b)-(i)). It is worth noting that this last observation constitutes the central paradigm of effective field theory, which is that the relevant physics does not come from a graph-by-graph analysis, but rather comes from a scale-by-scale (energy) breakdown.

In the next chapter, we will see that the scattering amplitude for the PWIA (Fig.\ref{edepn_diagrams}(a)) is proportional to the deuteron wave function. In the Lab frame, the spectator-PWIA amplitude is then proportional to $\psi_d(p_\text{f}^{\text{lab}})$, with $p_\text{f}^{\text{lab}}\simeq$ few~GeV/c. In contrast, the direct-PWIA graph is proportional to $\psi_d(p_\text{r}^{\text{lab}})$, with $ p_\text{r}^{\text{lab}}\simeq $ few hundred~MeV/c. From where it follows that for the PWIA the spectator mechanism is kinematically suppressed with respect to the direct one.

Diagrams containing meson exchange currents (Fig.\ref{edepn_diagrams}(e) and (f)) bear a dynamical suppression since, under the condition $ m_\text{meson}^2 \ll Q^2 \sim 1$GeV$^2 $,  they are suppressed compared to the PWIA term by an additional factor of  $Q^{-6}$\cite{Sargsian:2001ax,Sargsian:2002wc}. 

The dynamical suppression also occurs in the charge-exchange diagram (Fig.\ref{edepn_diagrams}(c)). The charge-exchange mechanism ($pn\rightarrow np$) has an additional factor of $s^{-1/2}$ in the amplitude and a stronger $t$ dependency compared to the direct-FSI rescattering of the nucleons
($pn\rightarrow pn$)
\cite{Sargsian:2009hf}.

The contributions of the intermediate (baryonic) resonance production mechanism (Fig\ref{edepn_diagrams}(d)), with the Delta production channel (${\color{red}\gamma^*} N \rightarrow \Delta$) being the most important, are subjected to both kinematic and dynamical suppressions.
Kinematically, it is possible to control the amount of energy transferred ($q^0_\text{lab}$) to the target such that intermediate exited states are suppressed. This corresponds to probing the side of the quasi-elastic peak that is far from the inelastic $\Delta$ electroproduction threshold. 
Dynamically, due to the spin-flip nature of the transition ${\color{red}\gamma^*} N \rightarrow \Delta$, a much steeper falloff in the transition form-factor is expected as $Q^2$ increases, compared to the elastic scattering ${\color{red}\gamma^*} N \rightarrow N$
\cite{Stoler:1993yk,Ungaro:2006df}.

Previous studies \cite{Sargsian:2009hf,Cosyn:2010ux,Boeglin:2015cha} analyzed the available deuteron electro-disintegration experimental data and were able to conclude that optimal enhancement of the direct PWIA with respect to direct FSI  contributions  occurs when, 
\begin{equation}
p_{\text{r},\text{T}} \ll p^\text{lab}_{\text{r},z}
\label{min FSI}
 \end{equation}
However, FSI contributions are always present and must be included in the calculations.

\section{High Energy Reactions and the Light-Front Framework}
\label{Intro LF}

One of the main characteristics of high energy scattering is that the process evolves along a light-like direction (e.g., \cite{Feynman, KS(1970), LB(1980), Frankfurt:1981mk, BPP(1997)}), making the light-front framework the most natural choice to describe the reaction. 
An essential feature of the LF approach, which can be exploited when describing high-energy reactions with composite systems, is the possibility of separating the relative momentum between constituents from the system's total momentum. 
This means that analogous to non-relativistic QM, the center of mass momentum will be kinematical, i.e., it will not carry any information about the dynamics since the three components of $\textbf{P}_\text{CM}$ commute with the Hamiltonian  \cite{Leutwyler1978,Bakker1979,K1980}.  This feature allows for the definition of observables such as form factors, as well as model-dependent quantities like momentum distributions, and spectral functions, all of them given in terms of LF wave functions which exclusively depend on the internal variables\footnote{More precisely, this is true for any frame collinear with $\bf{P}_\text{CM}$, which without loss of generality we can choose along the z-axis.} \cite{Brodsky2000}. 

\begin{figure}[h] 
\centering
\includegraphics[scale=0.5]{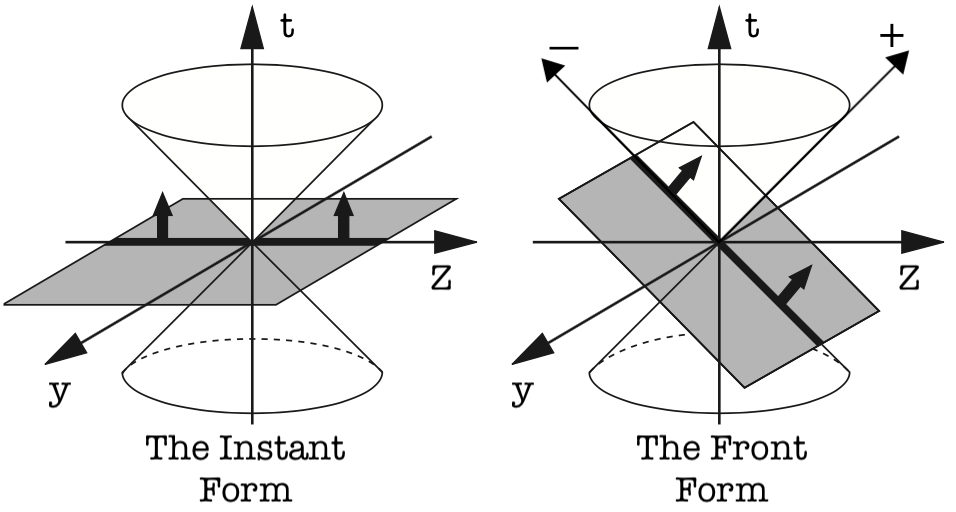}	
\caption{Dirac's Instant-  and Front-Forms  of Canonical Quantization. The equal-``time'' quantization planes are shown.}
\label{IF LF frames}
\end{figure}

The Light Front dynamics was first formulated by Dirac \cite{Dirac1949} in a pioneering study on Hamiltonian quantization. Dirac showed that the observer's time is not the only possible choice for describing the evolution of a system and introduced three consistent frameworks fulfilling this purpose, out of which two of them have been found very useful in the description of quantum mechanical systems. They are the Instant-Form {(IF)} and Light-Front-Form {(LF)} of dynamics, which differ by the respective fixed ``time'' surface used for the evaluation of (anti)commutators, Fig.(\ref{IF LF frames}).

\vspace{0.15in}

Among the advantages of a LF description are the absence of vacuum fluctuations and the possibility to  choose a good parameterization of ``time" for which the negative energy contribution to the interacting particle's propagator 
vanishes  (see Chapter \ref{PWIA}).  The latter allows for a consistent definition of relativistic wave functions for composite systems, called light-front wave functions { (LF-WF)} \cite{Frankfurt:1981mk,BPP(1997)}.

There are different conventions for the LF notation and scalar product. We use the  Lepage-Brodsky convention \cite{LB(1980)} (see  \hyperref[App. LFPT]{Appendix B}), 
\begin{align}\label{dot product}
	& x^{-} = t - z   \nonumber \\ 
	& x^+ = t+z = \tau  \nonumber \\
	& x^{\mu }=\left( x^{+}, {x}^-, x, y \right) = \left( x^{+}, {x}^-, \textbf{x}_{\textbf{T}}\right) \nonumber \\ 
	& p^{\mu }=\left( p^{+} , p^{-}, \textbf{p}_{\text{T}}\right) =(E+p_z, E-p_z, p_x, p_y) \\ 
	\ x & \cdot p={ 1 \over 2} \left( x^+ p^-  + x^- p^+\right)  -  \textbf{x}_{\text{T}} \cdot  \textbf{p}_{\text{T}} \nonumber 
	\end{align} 
The evolution is parameterized by the LF-time $x^+$, therefore, the Hamiltonian is $p^-$, since it is the operator conjugated to the time parameter. 

The quantization surface is defined by $x^+ = t+z = 0$, which represents a front of light moving along the $\hat{z}$-axis.
The states are labeled by the kinematical variables\footnote{Together with other quantum numbers like spin/helicity, etc.}: $p^+=E+p_z$, and, $\textbf{p}_\text{T} $, which are analogous to the standard labeling of states by the 3-momentum.  
For a nucleus with the atomic number $A$, instead of using the component $p^+$ for the nucleons, it is often convenient to use as a kinematical variable the light-front longitudinal momentum fraction, which is taken proportional to the atomic number ($A$),  i.e., $\alpha=Ap^+/p^+_A$. In our study of deuteron structure we will use,  $\alpha_{i,r}=2p^+_\text{i,r}/p^+_d$, for the interacting and recoil nucleons (see Fig. \ref{edepn_diagrams}).

\subsubsection{High Energy Electro-Disintegration in LF variables}

The Lab and CM reference frames of  Figs.(\ref{Lab Frame} and \ref{CM Frame}) belongs to a family of collinear frames  (chosen here along the z-axis) that can be parameterized in terms of the LF longitudinal momentum of the deuteron, $p_d^+$ \cite{Cosyn:2020kwu}. 
The LF components of deuteron and virtual photon momenta can be written in terms of the invariant kinematical variables as, 
\begin{align}\label{refframeQ}
 p_d^\mu & \equiv   (p_{d}^{+}, p_{d}^{-} ,{{\bf p}_{d}}_\text{T}) =  
\left({Q^2\over x\sqrt{s}}\left[1 + {x\over \tau} + \sqrt{1+ {x^2\over \tau}}\right], 
       {Q^2\over x\sqrt{s}}\left[1 + {x\over \tau} -  \sqrt{1+ {x^2\over \tau}}\right],  {\bf 0}_\text{T} \right) 
\\
 q^\mu & \equiv   (q^{+},q^{-}, {\bf q}_\text{T}) =  
\left({Q^2\over x\sqrt{s}}\left[1 - x - \sqrt{1+ {x^2\over \tau}}\right], 
       {Q^2\over x\sqrt{s}}\left[1 - x + \sqrt{1+ {x^2\over \tau}}\right],  {\bf 0}_\text{T} \right)  \nonumber 
\end{align} 
where, $\tau={Q^2\over m_d^2}$.  
With the  light-front components for the momentum of deuteron and the virtual photon given by Eqs.(\ref{refframeQ}), we define the the following \mbox{quantities} relevant to the considered reaction,
\begin{equation}\label{lc_fractions}
\alpha_\text{f} = {2 p_\text{f}^{+}\over p_d^{+}},  \ \ \alpha_\text{r} = {2 p_\text{r}^{+}\over p_d^{+}}, \ \  \alpha_q = {2 q^{+}\over p_d^{+}},  \ \ \mbox{and,} \ \  
\alpha_\text{i} = \alpha_\text{f}- \alpha_q = 2-\alpha_\text{r},
\end{equation}
where, $\alpha_\text{f}$ and $\alpha_\text{r}$ are the LF $+$ component of the deuteron momentum carried by the final knock-out  nucleon and recoil nucleon respectively, 
whereas $\alpha_q$ correspond to the longitudinal LF momentum fraction between the virtual photon and deuteron.
Note that they have been rescaled by a factor of 2.

The light-cone momentum fraction of the bound nucleon $\alpha_\text{i}$ is defined through the energy-momentum conservation. 
The crucial feature of these momentum fractions is that they are invariant for boosts in the $\hat z$ direction, taken here anti-parallel to the momentum transfer $\bf q$ (see Fig. \ref{Lab Frame}).  The same is true for the transverse components of the momentum, ${\bf p}_\text{T}$.
For example, in  the lab frame of the deuteron, these variables for the recoil particle are given by, 
\begin{eqnarray}
& & \alpha_\text{r} =  {\sqrt{m_N^2 + p_{\text{r},\text{lab}}^2} - p_\text{r}^\text{lab} \cdot \cos{\theta_\text{r}^\text{lab}}\over m_d/2} \\ \nonumber
& & {p_\text{r}}_{x} = {p_\text{r}^\text{lab}}\sin{\theta_\text{r}^\text{lab}}\cos{\phi_\text{r}^\text{lab}} \\ \nonumber
& &  {p_\text{r}}_{y} = -{p_\text{r}^\text{lab}}\sin{\theta_\text{r}^\text{lab}}\sin{\phi_\text{r}^\text{lab}},
\label{labkin}
\end{eqnarray}
where,   $\theta_\text{r}^\text{lab}$ and $\phi_\text{r}^\text{lab}$ are polar and azimuthal angles of the recoil nucleon in the Lab reference frame. 
The Lab coordinates were defined in Sec.(\ref{reframe}).

\section{Relevant Diagrams for High Energy and Momentum Transfer}

\begin{figure}[h]
	\centering
	\includegraphics[scale=0.5]{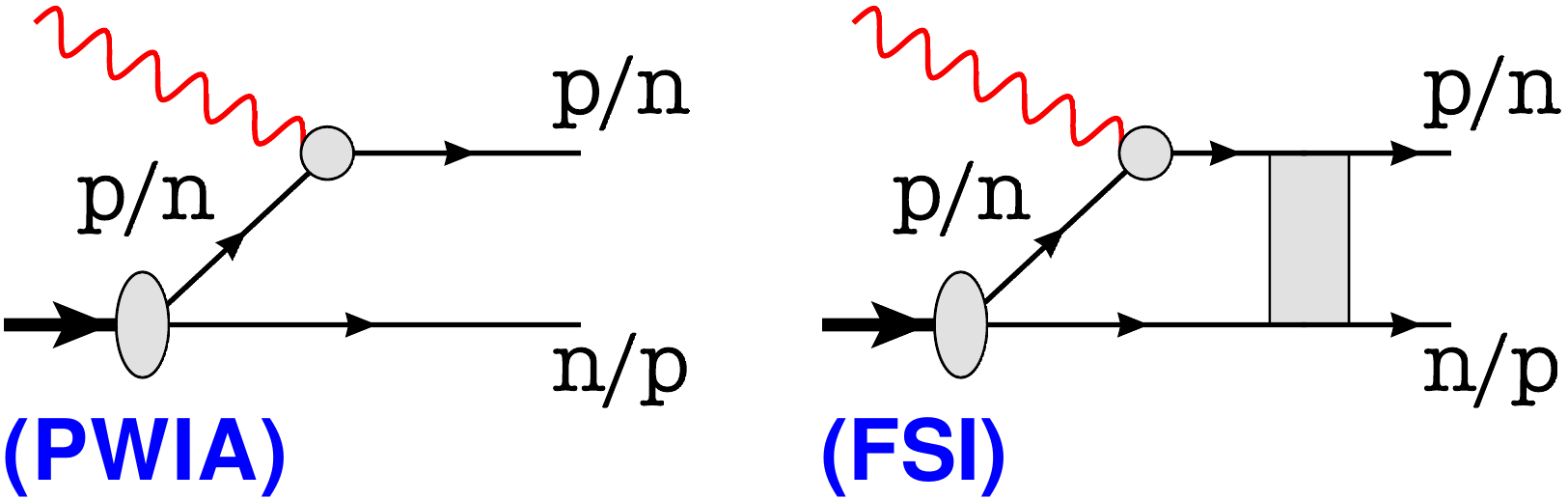}
	\caption{PWIA and FSI for deuteron electro-disintegration.}
	\label{Fig PWIA-FSI}
\end{figure}

From the diagrams involving the nucleonic content of the deuteron, the discussion of the previous section leaves us with the direct PWIA and FSI diagrams (Fig. \ref{Fig PWIA-FSI}), which are the dominant contributions in the high energy and momentum transfer kinematics considered in this work.
In addition, at very large internal momenta $700 - 750$~MeV/c a gradual increase of the contribution from the non-nucleonic components (Figs. \ref{edepn_diagrams} (g)-(i)) may appear
\cite{Frankfurt:2008zv}, which are omitted in this dissertation. 

\vspace{0.15in}

Another important advantage of 
evaluating the above diagrams at 
the high energy  and momentum transfer kinematics
is the onset of the eikonal regime. In the eikonal regime, the final state rescattering can be described through the effective $NN\rightarrow NN$, or, $NR\rightarrow NR$ amplitudes\footnote{These are the nucleon-nucleon and the nucleon-resonance scattering.}, which can be taken from high energy baryon-baryon scattering experiments.  

\begin{figure}[h]
\centering
\includegraphics[scale=0.75]{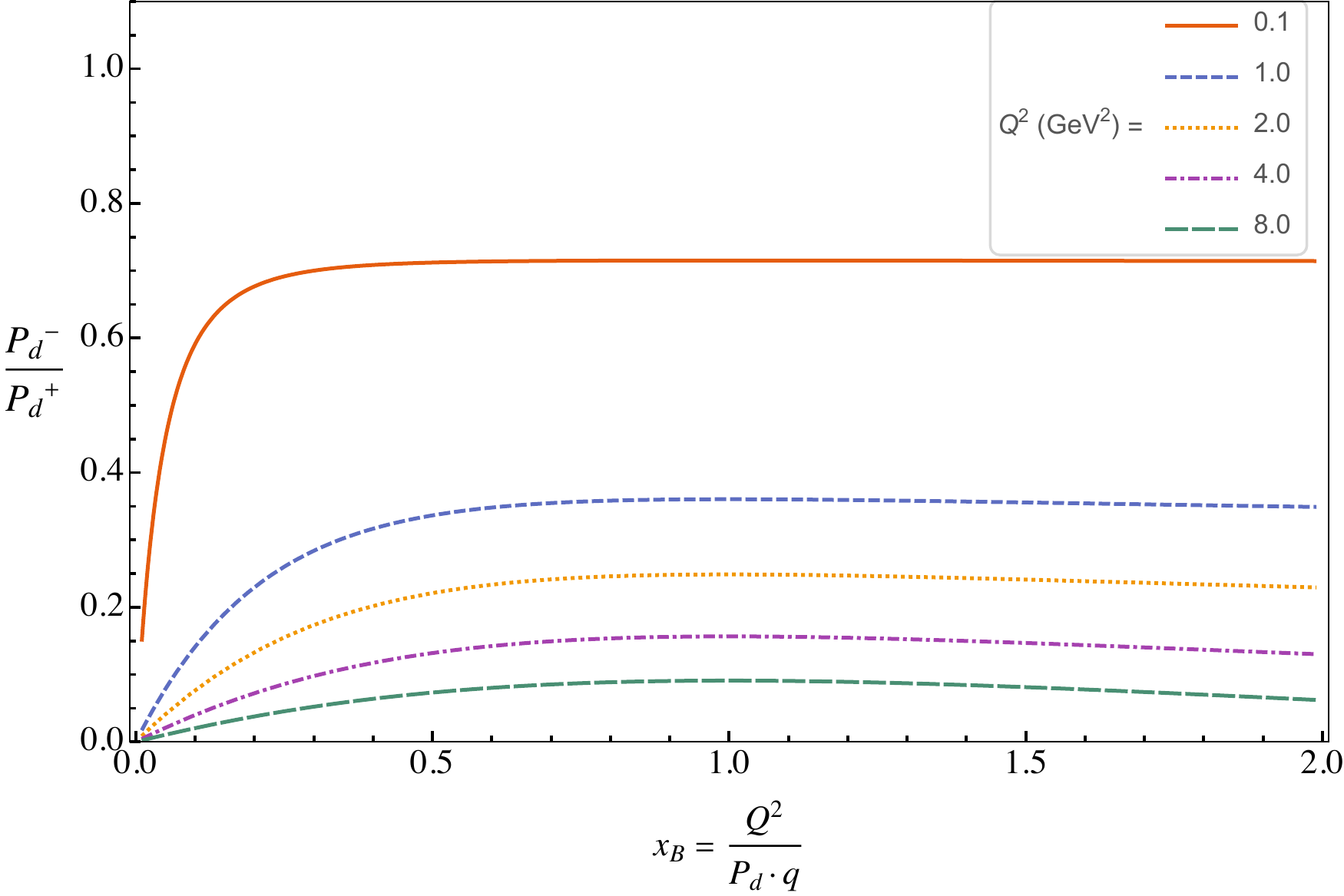}
\caption{The $x$ dependence of ${p_{d}^-\over p_{d}^+}$ ratio at different $Q^2$.}
\label{kincond}
\end{figure}

In the high energy and momentum transfer limit we have\cite{Sargsian:2001ax},
\begin{equation}
{q^0_{\text{lab}}-|{\bf q}_{\text{lab}}|\over q^0_{\text{lab}}+|{\bf q}_{\text{lab}}|}  \ll 1
\end{equation} 
which in the 
deuteron and virtual photon CM reference frame corresponds to\footnote{The components of the momenta for the deuteron and virtual photon in their CM frame are given by equations (\ref{refframeP}) and (\ref{refframeQ}) for the Lorentz and light-front coordinates respectively.},
\begin{equation}
{q^{+}\over q^{-}} \sim {p_{d}^-\over p_{d}^+} \ll 1.
\label{hlimit}
\end{equation}
In Fig.(\ref{kincond}) the above condition is checked for different $Q^2$ for the reaction of Eq.(\ref{reaction}).

As the figure shows, one expects the high energy approximation will start to be valid at $Q^2\gtrsim 2$~GeV$^2$ and improve further with an increase of $Q^2$. 
Assuming that the condition of Eq.(\ref{hlimit}) is satisfied, the diagrams involving FSI amplitudes in Fig.(\ref{edepn_diagrams}) can be systematically calculated applying effective Feynman or Light-Front  diagrammatic methods. 
In both cases, effective transition vertices and effective amplitudes for the $NN$ or $NR$ rescattering are used (see, e.g., Ref.\cite{Sargsian:2001ax}).

\section{Electro-Disintegration Cross-Section  }

The differential cross-section for the electro-disintegration reaction within the one photon-exchange approximation allows to factorize electron and hadronic parts of the interaction in the  invariant Feynman amplitude as follows,
 \begin{equation}
{ \cal M} =    \langle \lambda_f\mid j_e^\nu\mid \lambda_i\rangle {e^2g_{\nu\mu}\over q^2 }  \langle s_f,s_r \mid A^\mu\mid s_d\rangle,
\label{M}
\end{equation}
where, $q^2$ is the virtual photon's momentum squared. The leptonic current $j_e$ is defined by,
\begin{equation}
 \langle \lambda_f\mid j_e^\nu\mid \lambda_i\rangle =  \bar{u}(k_f,\lambda_f)\gamma^\nu u(k_i,\lambda_i)
 \end{equation}
and, $\langle s_f,s_r \mid A^\mu\mid s_d\rangle$ represents the (total) invariant amplitude of ${\color{red}\gamma^*} d\rightarrow NN$ scattering. 

Using Eq.(\ref{M}) for the differential cross-section of reaction (\ref{reaction}) one obtains:
\begin{equation}\label{cross-section}
{d\sigma\over d^3k_f/\epsilon_f d^3p_f/E_f} = {1\over 4 \sqrt{(p_d \cdot k_i)^2}}  {e^4\over q^4} L^{\mu\nu} H_{\mu\nu} {\delta((q+p_d - p_f)^2 - m_N^2) \over 4 (2\pi)^5}
\end{equation}
where, terms proportional to electron's mass squared ($m_e^2$) are neglected.  
The leptonic tensor is given in terms of the elementary (relativistic) spin 1/2  electro-magnetic current,
\begin{equation}\label{L-Tensor}
 L^{\mu\nu} = {1\over 2} \sum\limits_{\lambda_1 \lambda_2} \left( \bar{u}(k_f,\lambda_f)\gamma^\nu u(k_i,\lambda_i)\right)^{\dagger}  \bar{u}(k_f,\lambda_f)\gamma^\mu u(k_i,\lambda_i) 
\end{equation}
whereas the nuclear electromagnetic tensor is expressed through the scattering amplitude $A^\mu$ as follows:
\begin{equation}\label{H-Tensor}
H^{\mu\nu} = {1\over 3}   \sum\limits_{s_d s_r s_f} \langle s_d \mid A^{\mu \dagger} \mid s_f,s_r \rangle \langle s_f,s_r \mid A^\nu\mid s_d\rangle.
\end{equation}

As a consequence of the kinematics constrain discussed in the previous sections, we only need to consider in our calculation the processes corresponding to the diagrams in Figs.(\ref{edepn_diagrams} (a) and (b)), i.e.,
\begin{equation}
A^\mu = A^{\mu} _{0,\text{dir}} 
+ A^{\mu}_{1,\text{dir}} 
\label{relevant amplitudes}
\end{equation}

In the next two chapters (\ref{PWIA} and \ref{wave_function}) we elaborate on the details involved in the calculation of the PWIA amplitude ($A^{\mu} _{0,\text{dir}}$), 
first term on the right hand side of Eq.(\ref{relevant amplitudes}). The calculation of the second term, $A^{\mu}_{1,\text{dir}}$, will be the focus of \mbox{Chapter \ref{Ch FSI}.} 
Finally, the remaining term $A^\mu_{1,\text{chex}}$, is expected to be small in the kinematics of interest for the current research,
thus, it is out of the scope of the present work.

\chapter{PLANE WAVE IMPULSE APPROXIMATION} 
\label{PWIA}

The non-triviality of an accurate description for electro-scattering from deeply bound nucleon within a nucleus was evident since the 1980s, with the first intermediate energy experiments at SACLAY\cite{SACLAY1,SACLAY2} and NIKHEF\cite{NIKHEF}. 
The first theoretical calculations relied on different prescriptions to describe the nucleon's change when bound in a nucleus. 
For example, one of the earliest models\cite{Mougey}, assumed that bound and free nucleons would be equal in all respects except for their mass, with the bound nucleon's mass estimated by momentum conservation.

Currently, the most popular model is due to de Forest\cite{deForest(1983)}. de Forest considered eight different expressions for the electro-bound-nucleon cross-section $eN_\text{bound}$, 
each of them corresponding to
different assumptions about the electromagnetic (EM) interaction. 
The EM interaction is modeled by exchanging a virtual photon (${\color{red}\gamma^*}$) between the electron and bound nucleon, and the strenght of the interaction is represented by the effective vertex $\Gamma_{{\color{red}\gamma^*} N}^\mu$ (see Fig. \ref{Deuteron-PWIA}). Once again, the bound nucleon is approximated by a free nucleon. Hence, on-shell spinors are used.  
No preference is given to any of the considered eight expressions of the $eN_\text{bound}$ cross-section, and as such, these approximations allowed to check the uncertainty due to binding effects rather than calculating their actual values\footnote{In other words, these studies remained largely qualitative.}.  

Such kind of  approaches were characteristic to the intermediate scattering energy experiments\footnote{With a momentum transfer of a few hundred MeV/c.}, where the lack of a small parameter makes impractical the 
quantification caused by the strong binding effects on the nucleon electromagnetic current. 

\vspace{0.1in}

\section{Reason for Time-Ordering}	
\label{time-ordering}

\begin{figure}[h]
	\centering
	\includegraphics[scale=0.5]{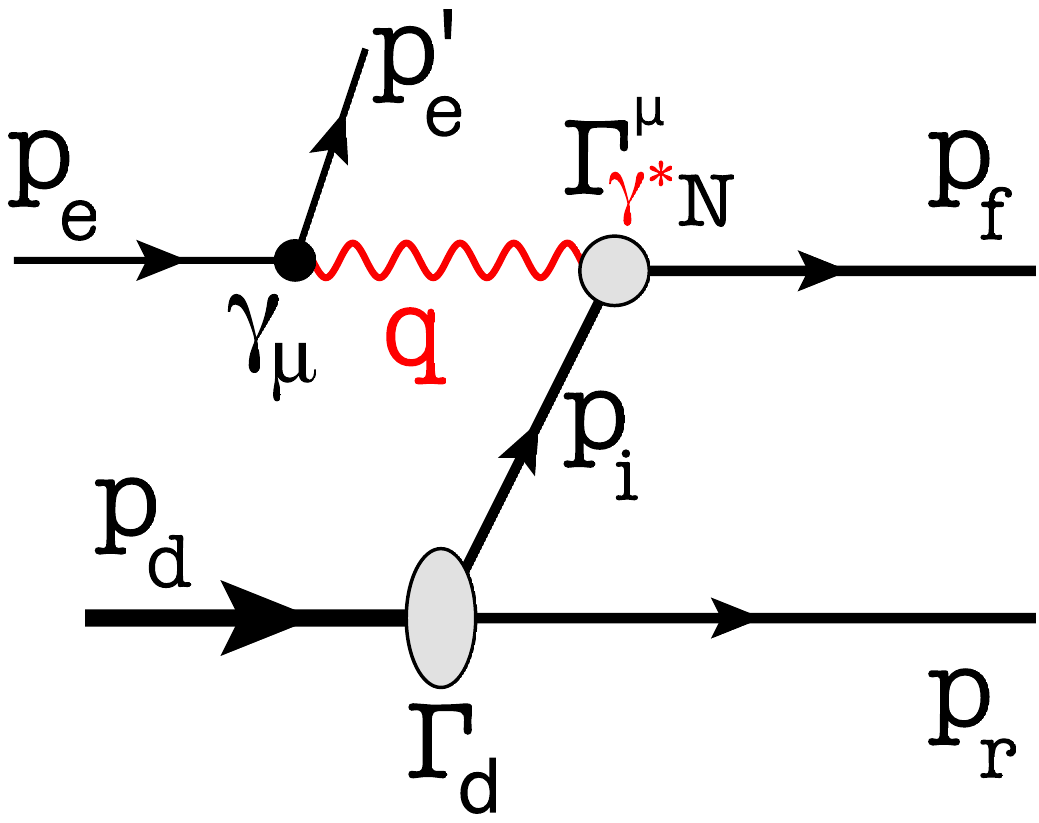}
	\caption{Exclusive electro-disinegration of the deuteron in plane wave impulse approximation (PWIA).}
	\label{Deuteron-PWIA}
\end{figure}

In this section we consider the single-photon exchange case of the reaction (\ref{reaction}) (see Fig. \ref{edeNN}) within the covariant plane wave impulse approximation~(PWIA), which is represented by the diagram of Fig.(\ref{Deuteron-PWIA}).  
The first problem we encounter in high-energy processes is the increasing probability of negative energy states.

\vspace{0.15in}

In the case of PWIA the off-shellness of the bound nucleon is completely defined by the four-momentum of  the deuteron ($p_d$) and that of the recoil nucleon ($p_r$), explicitly,
\begin{equation}
p_i = p_d - p_r
\end{equation}

The PWIA  amplitude $A_0^\mu$ is given by,  
\begin{equation}\label{A0_cov}
\langle s_f,s_r \mid A_0^\mu\mid s_d\rangle = -\bar u(p_f,s_f) \Gamma_{{\color{red}\gamma^*} N}^\mu \frac{\sh p_i + m_N}{ p_i^2-m^2_N} \bar u(p_r,s_r) \Gamma_\text{dNN}^\nu \chi^{s_d}_\nu,
\end{equation}
where, $\chi^{s_d}$ is the  spin wave function of the deuteron, and  $\Gamma_{{\color{red}\gamma^*} N}^\mu$ and  $\Gamma_\text{d}$ are covariant vertices (see Fig. \ref{Deuteron-PWIA}). 

\begin{figure}[h]
	\centering
	\includegraphics[scale=0.5]{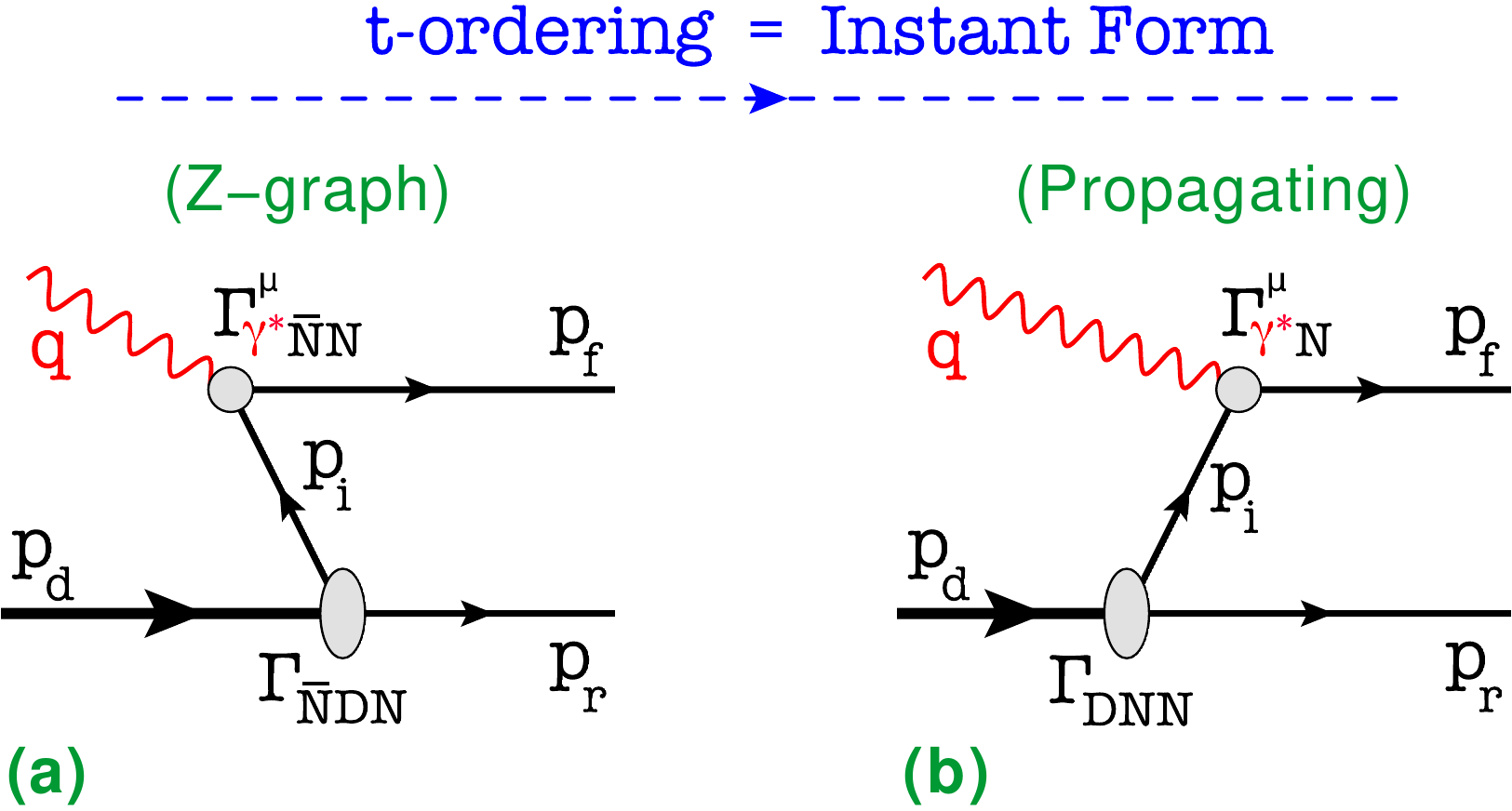}
	\caption{Representation of the covariant scattering amplitude (a) as a sum  two time-ordered 
		diagrams. (a) Production of the $\bar{\text{N}}\text{N}$ pair by the virtual photon 
		with subsequent absorption of the anti-nucleon  by the deuteron;  (b) Virtual photon scattering from the bound nucleon.}
	\label{tordered}
\end{figure}

The amplitude $A_0^\mu$ in equation (\ref{A0_cov}) contains neither the electron-bound nucleon scattering, nor the nuclear wave function in the explicit form.  The $eN_\text{bound}$ scattering and the nuclear wave function appear only when one considers the amplitude\footnote{or equivalently, the PWIA diagram (Fig. \ref{Deuteron-PWIA})} $A_0^\mu$
within time-ordered perturbation theory. In this case the covariant Feynman diagram splits into two noncovariant time-ordered processes, which are represented by the two diagrams in Fig.(\ref{tordered}).

\vspace{0.15in}

The diagram of Fig.(\ref{tordered}-b) represents a scenario in which the virtual photon interacts with the preexisting bound nucleon in the deuteron, with $\Gamma_\text{dNN}$ representing the vertex of $d\to NN$ transition and the ${\color{red}\gamma^*} N\rightarrow N$ electromagnetic interaction. 
This contribution corresponds to the noncovariant PWIA, in which case the $eA$ cross-section is expressed through the product of $eN$ cross-section and noncovariant nuclear spectral function. 

The diagram of Fig.(\ref{tordered}-a) however, represents a very different scenario. In this case, the virtual photon 
couples to an
intermediate vacuum fluctuation $\text{N}\bar{\text{N}}$ pair, which is parameterized by the transition vertex $\Gamma_{{\color{red}\gamma^*} \bar{\text{N}}\text{N}}$. 
 Subsequently, the \mbox{anti-nucleon} ($\bar {\text{N}}$) is reabsorbed into the deuteron at the $\Gamma_{\bar{\text{N}}\text{DN}}$ vertex. The latter process is not straightforward related to the 
the nucleon content in the wave function of the deuteron. 
Figure (\ref{tordered}-a) is commonly referred to as a "Z-graph" and is a purely relativistic effect. 
In the nonrelativistic limit, one deals only with the diagram of Fig.(\ref{tordered}-b), which allows to approximate the covariant scattering amplitude with the product of the nonrelativistic nuclear wave function and the amplitude for the virtual-photon scattering off the bound-nucleon (${\color{red}\gamma^*} N_\text{bound}$).  
However, the situation becomes more complicated when one is interested in bound nucleon 3-momenta comparable with its mass ($|{\bf p}_i| \sim m_N$), which can be probed at momentum transfer $|{\color{red}{\bf q}}| \gg m_N$ (see Eq. \ref{henc}).  In this case the "Z-graph" contribution (Fig.\ref{tordered}-a) becomes comparable with the one in Fig.(\ref{tordered}-b), preventing the straightforward factorization of the nuclear wave function. Thus, conventional noncovariant PWIA is inapplicable for the description of electron scattering from deeply bound (relativistic) nucleons in the nucleus.

This situation is reminiscent of the QCD processes in probing the partonic \mbox{structure} of hadrons, in which case, due to the relativistic nature of partons, the vacuum diagrams can not be neglected when the time-ordered perturbation theory is applied in the Lab reference frame of the hadron\cite{Feynman}.   
A possible solution is to consider the scattering process in the infinite momentum frame, which allows to suppress  the "Z-graphs" and consider only the diagram of Fig.(\ref{tordered}-b), for which one can introduce the wave function of the constituents.

Another solution, is to set up the problem within the framework of light-front form of dynamics, which is the one we follow in this work.
Our approach in probing deeply bound nucleon is similar to that of the partonic model. We formulate the reaction~(\ref{reaction}) in the light-front formalism, allowing us to exclude the contribution of the vacuum diagrams (Fig.\ref{tordered}-a), and consequently be able to define the  (light-front) nuclear wave function of deuteron. A remarkable characteristic of the LF wave function is that it has probabilistic interpretation similar to that of the non-relativistic nuclear theory.

\section{Light-Front Scattering Amplitude in the PWIA}
\label{LF PWIA amplitude}

\begin{figure}[h]
	\centering
	\includegraphics[scale=0.8]{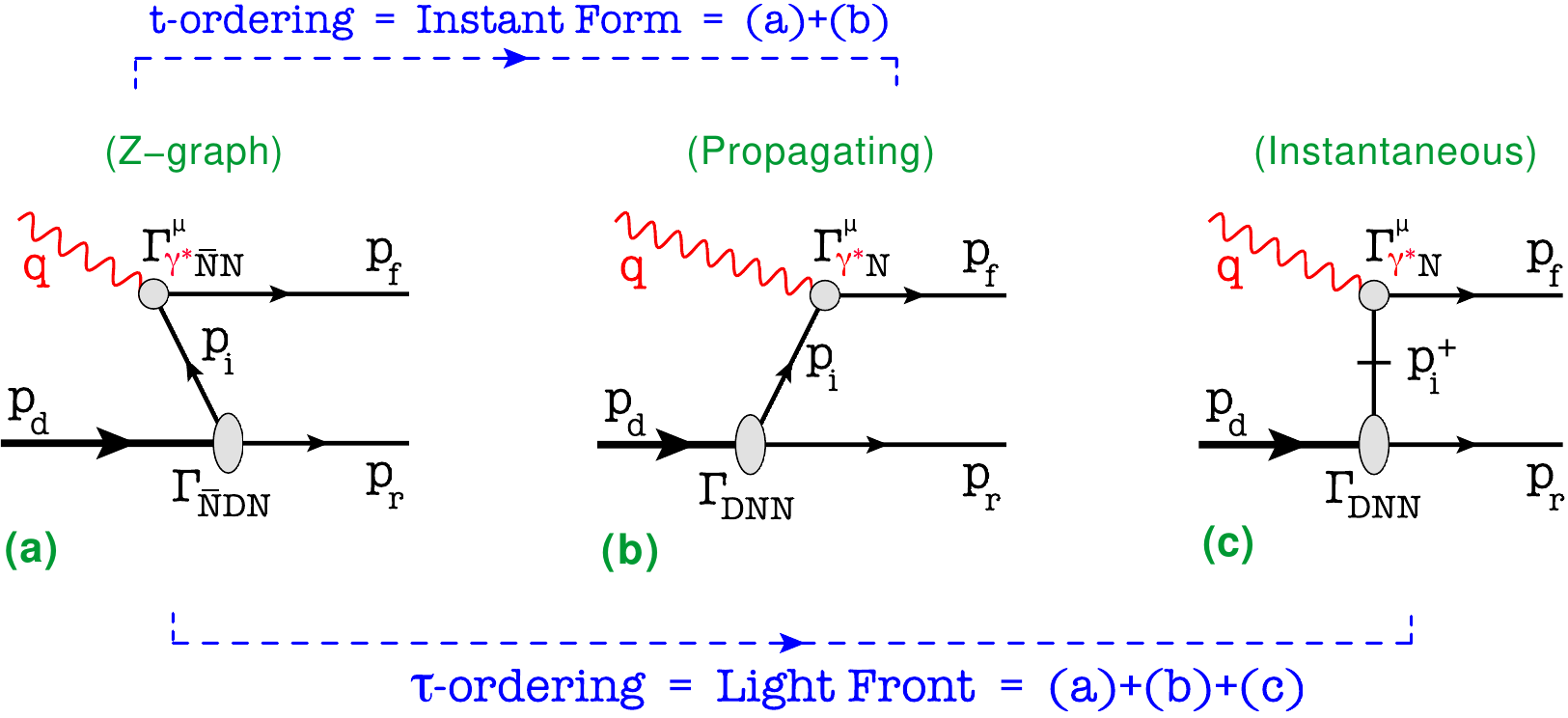}
	\caption{Representation of the covariant scattering amplitude  as a sum  of two light-cone ($\tau$)-time -ordered diagrams as well as instantaneous interaction.  (a) Production of the $\bar N N$ pair by the virtual photon with subsequent absorption of the antinucleon  by the deuteron; (b) Virtual photon scattering from the bound nucleon; (c) Instantaneous interaction of virtual photon with the bound nucleon.}
	\label{t-order}
\end{figure}

We consider now the reaction (\ref{reaction}) on the light-front, where the light-cone  time  is 
defined as $\tau\equiv t+z$. 
The calculation of the scattering amplitude corresponding to the PWIA proceeds by applying  Light-Front perturbation rules \cite{KS(1970), LB(1980)}  in an effective theory in which one identifies effective vertices for the nuclear transitions and the electron-bound nucleon scattering (see \hyperref[App. LFPT]{Appendix B}). 
 The covariant scattering amplitude Eq.(\ref{A0_cov}) is expressed as a sum of noncovariant diagrams ordered in  \mbox{$\tau$-time}, shown in Fig.(\ref{t-order}). 
 At each vertex on the $\tau$-ordered diagrams the transverse, ${\bf p}_\text{T}$, and plus, $p^+$, components of momenta are conserved.
Here, in addition to the two $\tau$ orderings analogous to the $t$-time ordering of Fig.(\ref{tordered}) there is an  additional contribution  Fig.(\ref{t-order}-c). The latter is related to the spinor nature of the bound nucleon and corresponds to the so-called instantaneous interaction.

\vspace{0.15in}

\vspace{0.15in}
\begin{figure}[h]
	\centering
	\includegraphics[scale=0.8]{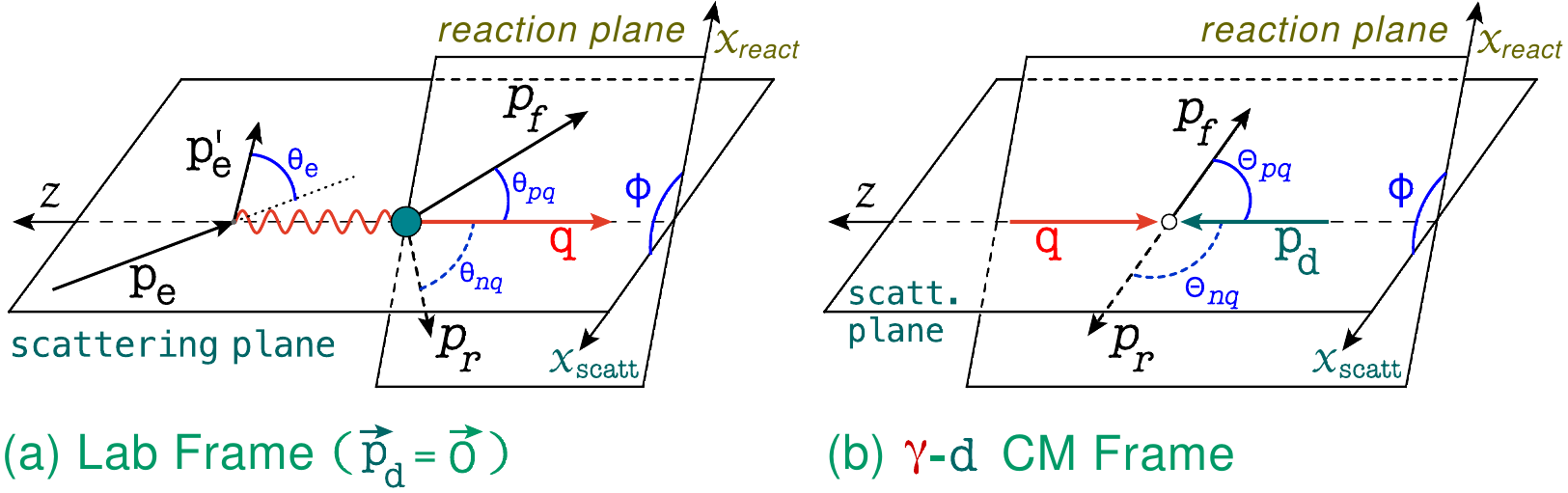}
	\caption{Scattering and Reaction Planes.}
	\label{scatt react plane}
\end{figure}

Our conventions for the scattering and reaction planes, as well as the reference frames used, are depicted in Fig.(\ref{scatt react plane}). They belong to a family of collinear frames related by boosts along the z-axis. The $\hat{z}$-axis is oriented anti-parallel to the virtual photon 3-momentum, ($\hat{z}\| -\vec{q} $).

The scattering plane corresponds to,
\begin{equation}
( \hat x_{\text{scatt}}, \hat y_{\text{scatt}},\hat z_{\text{scatt}}) = \left( { \vec q \times \hat y_{\text{scatt}} \over | \vec q \times \hat y_{\text{scatt}}|} ,  { \vec q \times  \vec p_{e} \over |\vec q \times  \vec p_{e}|} , -{\vec q \over |\vec q|} \right)
\end{equation}
where, $\vec p_e$ ($\vec p_e'$)  is the 3-momentum vector of the incoming (scattered) electron.

The reaction plane is defined by,
\begin{equation}
(\hat x_\text{react}, \hat y_\text{react},\hat z_\text{react}) = \left( { \vec q \times \hat y_{\text{react}} \over | \vec q \times \hat y_{\text{react}}|} ,  { \vec p_\text{f} \times \vec q \over |\vec p_\text{f} \times \vec q|} , -{\vec q \over |\vec q|} \right)
\end{equation} 
Notice that conventionally the $z$-axis is defined parallel to $\vec{q}$. The reason for our choice 
of reference frame, with the  $\hat z$ axis antiparallel to the transferred momentum ($\hat z  ||  -{\bf q}$) follows from the fact that in this frame we have,  $q^+ = q_0 - |\textbf{q}| < 0$.
Therefore, due to the conservation of the $+$-component of momenta at an interaction vertex (see \hyperref[App. LFPT]{Appendix B}), the diagram of  \mbox{Fig.(\ref{t-order}-a)} (Z-graph) is kinematically forbidden, since the production of the $\bar N N$ pair requires $q^+ >0$.

The 
diagram in Fig.(\ref{t-order}-b) corresponds to the amplitude $A^\mu_\text{prop}$, representing a virtual photon that knocks-out a bound nucleon, which subsequently  propagates from the 
$\Gamma_\text{dNN}$ transition vertex to the 
$\Gamma_{{\color{red}\gamma^*} \text{N}}$  interaction vertex.
Finally, in \mbox{Fig.(\ref{t-order}-c)}  we have the ``instantaneous'' amplitude  $A^\mu_\text{inst}$,   in which  the $d\rightarrow NN$ transition and ${\color{red}\gamma^*} N$ interaction take place at the same light-front time ($\tau$). 
In both diagrams, the nucleus exposes its constituents, and the scattering takes place off the bound nucleon,  allowing for a probabilistic interpretation of the scattering process with the bound nucleon.  

The propagating part of the scattering amplitude (Fig.\ref{t-order}-b) yields,
\begin{equation}
\langle s_f,s_r \mid A_\text{prop}^\mu\mid s_d\rangle =  - \bar u(p_f,s_f)\Gamma_{{\color{red}\gamma^*}  N}^{\mu} 
\  \frac{1}{p_i^+} \frac{(\sh p_i + m_N)_{\text{on}}}{ (p_d^- - p_r^- - p_{i,\text{on}}^- )} \bar u(p_r,s_r) \Gamma_\text{dNN} \chi^{s_d}
\label{A0prop}
\end{equation}
where $p_d^-$, $p_r^-$ and  $p_{i,\text{on}}^-$ are defined from the light cone energy on-shell condition,  
\begin{equation}
p^-_j = {m_j^2 + {{\bf p}_j}_\text{T}^2\over p_j^+}
\end{equation}
with, $j = d, r, (i,\text{on})$. 
The subscript "on" indicates that the light-cone components of the bound nucleon momenta are taken on-energy shell. 

Applying the rules of  \hyperref[App. LFPT]{Appendix B} to the instantaneous diagram (Fig.\ref{t-order}-c)  one obtains,
\begin{equation}
\langle s_f,s_r \mid A_\text{inst}^\mu\mid s_d\rangle =  - \bar u(p_f,s_f)\Gamma_{{\color{red}\gamma^*}  N}^{\mu}   \frac{1}{p_i^+}\left(  \frac{1}{2}\gamma^+ \right)  
\bar u(p_r,s_r) \Gamma_\text{dNN} \chi^{s_d}
\label{A0inst}
\end{equation}
Note that in both expressions (\ref{A0prop}) and (\ref{A0inst}) one has the same nuclear ($\Gamma_\text{DNN}$) and electromagnetic ($\Gamma_{{\color{red}\gamma^*}  N}$) vertices. 

For further elaborations, we introduce the LF off-energy shell of the bound nucleon ( ``-'' component), $p_i^- = p_d^- - p_r^-$, which is defined through the (LF) on-energy-shell ( ``-'' component) of the deuteron and recoil nucleon, 
\begin{equation}
\frac{1}{ p_d^- - p^-_r - p^-_{i,\text{on}} }  = \frac{1}{ p_i^- - p^-_{i,\text{on}} } = \frac{p_d^+} {m_d^2 - 4 \frac{(m_N^2 + {\bf p}_\text{T}^2)}{\alpha(2-\alpha)} }
\label{denom}
\end{equation}
where, we have used Eq.(\ref{lc_fractions}). Together with the completeness relation for on-shell spinors,
\begin{equation}
(\sh{p}_{i} + m_N )_\text{on}= \sum_{s_i} \left( u(p_i,s_i) \bar{u}(p_i,s_i)\right)_\text{on}
\label{sumrule}
\end{equation} 

Putting together the two contributions $A^\mu_\text{prop}$ and $A^\mu_\text{inst}$ the PWIA scattering amplitude is given by,
\begin{align}
A^{\mu}_{0} = A^\mu_\text{prop} + A^\mu_\text{inst} = & - \bar u(p_f,s_f) \Gamma_{{\color{red}\gamma^*}  N}^{\mu} \frac{\sum_{s_i} u(p_i,s_i) \bar{u}(p_i,s_i)}
{ \frac {\alpha}{2}\left( m^{2}_{d}-4\frac {m^{2}_N+{\bf p}_\text{T}^2}{\alpha \left( 2-\alpha \right) }\right)}  \bar u(p_r,s_r) \Gamma_\text{dNN}  \chi^{s_d}  \nonumber \\
&-  \bar u(p_f,s_f)\Gamma_{{\color{red}\gamma^*}  N}^{\mu} \frac{ \frac{1}{2} \gamma^{+} \left({p}_{i }^{-}  - {p}_{i,on}^{-} \right)}{  \frac {\alpha}{2}\left( m^{2}_{d}-4\frac {m^{2}_N+{\bf p}_\text{T}^2}{\alpha \left( 2-\alpha \right) }\right)} 
\bar u(p_r,s_r) \Gamma_\text{dNN} \chi^{s_d} 
\label{A0s}
\end{align}
Furthermore, $A^{\mu}_{0}$ can be factorized  into the product of two terms, 
\begin{equation} 
A^{\mu }_{0}=  A^{\mu}_\text{prop} + A^\mu_\text{inst}  =  \sum _{s_i}J^{\mu }_{N}  \left( p_{f}s_{f},p_{i}s_{i}\right)  \dfrac {\psi ^{s_{i} s_r s_d}_\text{LF}\left( \alpha, {\bf p}_\text{T}\right) }{\alpha }\sqrt {2\left( 2\pi \right) ^{3}}
\label{A0sb}
\end{equation}
the light-front wave function, which is defined by\cite{Frankfurt:1981mk,Artiles_Sargsian-multisrc1}:
\begin{equation}
\psi ^{s_{i}s_{r}s_{d}}_\text{LF }\left( \alpha ,{\bf p}_\text{T}\right) =-\frac {\bar {u}( p_{i}, s_i) \bar u( p_{r}, s_r) \Gamma _\text{dNN}\chi^{s_d}}{\frac {1}{2}\left( m^{2}_{d}-4\frac {m^{2}_N+{\bf p}_\text{T}^2}{\alpha \left( 2-\alpha \right) }\right) }\dfrac {1}{\sqrt {2\left( 2\pi \right) ^{3}}}
\label{wflf}
\end{equation}
and the electromagnetic current for  the bound nucleon,
\begin{equation}\label{EM current off-on}
J^{\mu}_{N} (p_f s_f,p_i s_i) = \bar{u}(p_f s_f) \Gamma_{{\color{red}\gamma^*}  N}^{\mu}  u(p_i s_i)  +  
\bar{u}(p_f s_f) \Gamma_{{\color{red}\gamma^*}  N}^{\mu} \frac{ \Delta \sh{p}_i }{2m_N}  u(p_i s_i)
\end{equation}
where, $ 2 \Delta \sh{p}_i = \gamma^{+} \left( {p}_{i }^{-}  - {p}_{i ,\text{on}}^{-} \right)$.
For more details we refer the reader to  Ref.\cite{Vera_Sargsian:2018}.

The current operator of Eq.(\ref{EM current off-on}) is half off-shell, i.e., the initial state of the nucleon is off-shell while the final state is on-shell. It differs from the free nucleon EM current in the additional last term.
The primary focus of the next section is the calculation of this half-offshell electromagnetic current.

\section{Nucleon EM Current and Form Factors}

The most general parameterization   for the    matrix element  electromagnetic (EM) current operator $(j_{EM}^\mu)_\text{free}$ of the free nucleon satisfying Lorentz and gauge invariance consists of two form factors (FF), its matrix elements are given by, 

\begin{equation}\label{nucleon free FF}
\left\langle p^{\prime}, s'\left|j_{\mu}\right| p, s\right\rangle=\overline{u}\left(p^{\prime}, s'\right)\left[F_{1}(Q^2) \gamma_{\mu}+ \frac{F_{2}(Q^2)}{2 m_N} i \sigma_{\mu \nu} q^{\nu}\right] u(p, s)
\end{equation} 
where, $m_N$ is the nucleon mass, $p$ and $s$ ($p' $ and $s' $) are the initial (final) nucleon momentum and spin respectively. The Dirac and Pauli  form factors,  $F_1$ and $F_2$, are scalar functions of the four-momentum transfer squared, $q^2 = (p' − p)^2 < 0$  $\ (Q^2=-q^2 > 0)$, and $\sigma_{\mu \nu}= {i \over 2}(\gamma_\mu \gamma_\nu - \gamma_\nu \gamma_\mu)$.

The FF are normalized at $Q^2 = 0$ as follows,
\begin{equation}
F_{1}^{p}(0)=1, \quad F_{1}^{n}(0)=0, \quad F_{2}^{p}(0)=\kappa_{p}, \quad F_{2}^{n}(0)=\kappa_{n}
\end{equation}
with, $\kappa_p = 1.79$, $(\kappa_n = −1.91)$, the anomalous magnetic moment of proton (neutron) in units of nuclear magneton. 

For a static target ($m \rightarrow \infty$) the charge and magnetization distributions are given as  Fourier transforms of the Sachs FF,
\begin{equation}
\begin{aligned} 
G_{E}(Q^2) &=F_{1}(Q^2)-\tau F_{2}(Q^2) \\ 
G_{M}(Q^2) &=F_{1}(Q^2)+F_{2}(Q^2) 
\end{aligned}
\end{equation}
where, $\tau=Q^2 /\left(4 m^{2}\right)$. 
When the recoil of the target is not small and can not be ignored, the interpretation of $G_E$ and $G_M$  as the Fourier transforms of the charge and magnetization distributions can still be valid if we describe the process in the  Breit frame (BF), which is defined by: $Q^2=\textbf{q}^2_\text{BF}  \  (q^0_\text{BF}=0)$.
This follows from the fact that in the BF, the initial three-momentum $(\textbf{p}_i)$ of the interacting particle is equal to minus its final three-momentum $(\textbf{p}_f=-\textbf{p}_i)$, henceforth during the interaction, the target appears to be (on average) as if it is at rest. Thus, the Breit frame picture is the closest to have the target at rest during the interaction. 
Furthermore, we have
\begin{align}
\left\langle p^{\prime}, s'\left|j_{\mu}\right| p, s\right\rangle \ \overrightarrow{ \ BF \ } \  \left\langle -p, s'\left|j_{\mu}\right| p, s\right\rangle=\overline{u}\left(-p, s'\right) \Gamma_\mu  u(p, s)
\end{align}
showing, that the matrix elements of the electromagnetic current receive an equivalent contribution from relativistic effects (like Lorentz contraction), which arise from the target's motion.

\subsubsection{Contributions to the Electromagnetic Current}

To identify the propagating and instantaneous parts of the electromagnetic current in Eq.(\ref{EM current off-on}), we consider first the electromagnetic vertex $\Gamma_{{\color{red}\gamma^*}  N}^{\mu}$.
Since the final state of the interacting nucleon is on mass shell, and only the  positive light-front energy projections enter in the amplitude, we are led to the half off-shell vertex function in the general form (see e.g. \cite{Bincer60,Koch90,Koch96}):
\begin{align}
\Gamma_{{\color{red}\gamma^*}  N}^{\mu} = \gamma^{\mu }F_{1} + i \sigma^{\mu \nu} q_{\nu}  {F_{2} \over 2 m_N} + q^{\mu} F_{3}
\label{eNvertex}
\end{align}
where the form-factors $F_{1,2,3}=F_{1,2,3}(m_N^2,p_i^2,q^2)$ are functions of Lorentz invariants constructed from the momenta of initial and final nucleons and momentum transfer $q$. 
In general one expects $F_{1,2}(m_N^2,p_i^2,q^2)$ not to be identical with the corresponding on-shell nucleon form-factors ($F_{1,2}(m_N^2,m_N^2,q^2)$).  
This difference is due to the modification of the internal structure of nucleons in the nuclear medium. Such modification, in principle, should originate from the dynamics similar to the one responsible for the medium modification of partonic distributions of bound nucleon, commonly referred to as EMC effect \cite{EMC}.
This, however, is out of the scope of our discussion since we are interested only in the effects related to the off-shellness of the interacting nucleon's electromagnetic current. Thus, in the numerical estimates, we will use unmodified nucleon form-factors measured for free nucleons.

Concerning $F_{3}$, it does not contribute to the cross-section of the process due to the gauge invariance of the leptonic current: $q_\mu j_e^\mu = 0$.  
However, for consistency, one can estimate the $F_{3}$ form-factor based on the fact that due to the conservation of the momentum sum rule in the light-front approach, the electromagnetic current of the bound-nucleon is conserved: 
\begin{align}
q_\mu J_N^\mu=0.
\label{qcons}
\end{align}

Using equations (\ref{EM current off-on}) and (\ref{eNvertex}) one obtains: 
\begin{align}
F_{3} = F_{1} { \shl{q}  \over Q^2}
\end{align} 
Substituting the later into Eq.(\ref{eNvertex}) one can separate the propagating and instantaneous parts of the electromagnetic vertex, which are given by,
\begin{align}
\Gamma_{{\color{red}\gamma^*}  N}^{\text{(prop)} \mu} = \gamma^{\mu }F_{1} + i \sigma^{\mu \nu} q_{\nu}  {F_{2} \over 2 m_N}
\label{Vertex-on}
\end{align} 
and, 
\begin{align}
\Gamma_{{\color{red}\gamma^*}  N}^{\text{(inst)} \mu} 
=  \left( \gamma^{\mu }F_{1} + i \sigma^{\mu \nu} q_{\nu}  {F_{2} \over 2 m_N} \right) \frac{ \Delta\sh{p}_i}{2m_N} - F_1{q^{\mu} \over q^2} \sh{q} \Big( {\bf 1} + \frac{ \Delta\sh{p}_i}{2m_N} \Big)
\label{Vertex-off}
\end{align} 
where, $\Delta{p}_i^\mu=p^\mu_{i } - p^\mu_{i, \text{on}}$.

Because $\Delta p_i^+ = { \Delta p_{i}}_\text{T} = 0$, we have $2 \Delta\sh{p}_i=\gamma^{+} \left( p^-_i - p^-_{i, \text{on} }\right)$, and we use that, 

\begin{align}
\Delta{p}_i^- = -q^- + (p_f^- - p_{i, \text{on}}^-) = \dfrac{Q^2}{q^+} -  \dfrac{m^{2}_N+{\bf p}_\text{T}^2 }{p_f^+ p_i^+} q^+ = \frac{1}{p_d^+} \left(m_d^2 - 4 \frac{(m_N^2 + {\bf p}_\text{T}^2)}{\alpha(2-\alpha)} \right)
\label{off-shell_factor}
\end{align} 
as well as,
\begin{align}
2 \Delta{p}_i \cdot p_i = \Delta{p}_i^- p_i^+ = p_i^2 - m_N^2
\end{align}
which allows to express the electromagnetic current in boost-invariant (along the $z$-axis) variables.

The separation of the electromagnetic vertex into propagating and instantaneous parts in equations (\ref{Vertex-on}) and (\ref{Vertex-off}) allows to separate the electromagnetic current in Eq.(\ref{EM current off-on}) into corresponding parts in the following form: 
\begin{align}
J_{N}^\mu (p_f s_f,p_i s_i) = J^{\mu}_\text{prop} (p_f s_f,p_i s_i) + J^{\mu }_\text{inst}(p_f s_f,p_i s_i) 
\label{Jsum}
\end{align}
where,
\begin{eqnarray}
J^{\mu}_\text{prop} (p_f s_f,p_i s_i) &= & \bar{u}(p_f s_f) \Gamma_{{\color{red}\gamma^*}  N}^{\text{(prop)} \mu} u(p_i s_i) \nonumber \\
J^{\mu }_\text{inst}(p_f s_f,p_i s_i) &= & \bar{u}(p_f s_f) \Gamma_{{\color{red}\gamma^*}  N}^{\text{(inst)} \mu}  u(p_i s_i)
\label{J_on_off} 
\end{eqnarray}

It is worth mentioning that even though the propagating vertex in Eq.(\ref{Vertex-on}) has the same form as the free on-shell nucleon vertex the corresponding electromagnetic current $J^{\mu}_\text{prop}$ does not correspond to an on-shell scattering amplitude, since \mbox{$q^\mu\ne p_f^\mu - p_{i,\text{on}}^\mu$.} Also, the current conservation (Eq. \ref{qcons}) is satisfied only for the sum of the propagating and instantaneous currents in Eq.(\ref{Jsum}).

\subsubsection{Off-Shell Parameter of  $ \ e N_{\text{ bound } } $ Scattering}

While the off-shell effects in the propagating vertex of Eq.(\ref{Vertex-on}) are kinematical, due to the fact that $q^\mu\ne p_f^\mu - p_{i, \text{on}}^\mu$, the off-shell effects in the instantaneous vertex are dynamical.  The latter interaction arises exclusively due to the binding of the nucleon.  As it follows from Eq.(\ref{Vertex-off}) the strength of the instantaneous vertex is proportional  to the magnitude of the factor $\Delta  p_i^-$  defined in Eq.(\ref{off-shell_factor}).
One can express  the $\Delta  p_i^- $ factor through a boost invariant quantities by defining the light-front reference frame such that the four-momenta of the deuteron, $p_d^\mu$ and   momentum transfer $q^\mu$ are:
\begin{align}
p_d^\mu & =   \left({Q^2\over m_N} , {\bf 0}_\text{T} , {m_d^2  m_N\over Q^2} \right) \\ 
q^\mu  & =  \left(-{Q^2 x\over m_N  \left(1 + \sqrt{1 + {4m_N^2x^2\over Q^2}} \right)} , {\bf 0}_\text{T} , {m_N\over x}  \left( 1 + \sqrt{1 + {4m_N^2x^2\over Q^2}} \right) \right)
\end{align}
Using the above definitions we  introduce the off-shell parameter $\eta$ such that, 
\begin{equation}
\Delta{p}_i^-  = -{m_N}\eta
\end{equation}
where, 
\begin{align}
\eta = {1\over Q^2} \left(4 \frac{(m_N^2 + {\bf p}_\text{T}^2)}{\alpha(2-\alpha)} - m_d^2 \right)
\label{eta}
\end{align}
which coincides with the ratio between the binding effects on the nucleon and the resolution of the probe ($Q^2$).

In the next section, we show how the half-off-shell cross-section can be expanded in powers of the parameter $\eta$, hence  providing a universal measure to off-shell effects in the quasi-elastic reaction.

\section{ Electron Bound-Nucleon Scattering Cross-Section  }
\label{IV}

In many practical applications one needs to evaluate the electron--bound-nucleon cross-section $\sigma_{eN}$ as it is defined in reference \cite{deForest(1983)}.
Such a cross-section is calculated within PWIA in which case using Eq.(\ref{A0sb})  the nuclear electromagnetic tensor of Eq.(\ref{H-Tensor}) can be expressed as follows:
\begin{align}\label{H=H_N*rho}
H^{\mu \nu}=H^{\mu \nu }_{N}(p_f,p_i) \ \rho_d \left( \alpha, {\bf p}_\text{T}\right) \ \dfrac {2-\alpha }{\alpha ^{2}} \ 2\left( 2\pi \right) ^{3}
\end{align}
where, the spin averaged light-front density matrix of the deuteron $\rho_d(\alpha, {\bf p}_\text{T})$ and bound-nucleon electromagnetic tensor $H^{\mu \nu }_{N}(p_f,p_i)$ are defined by,
\begin{align}\label{LF-density}
\rho \left( \alpha, {\bf p}_\text{T}\right) =  \dfrac {1}{2s_d+1} \  \frac {1}{2} 
\sum _{s_{d},s_i,s_r }\dfrac {\left| \psi ^{s_{i}s_{r}s_{d}}_\text{LF}\left( \alpha,{\bf p}_\text{T}\right) \right| ^{2}}{2-\alpha}
\end{align}
and,
\begin{align}\label{HN}
H^{\mu \nu }_{N}=\frac {1}{2}\sum_{s_{i}s_{f}=-1/2}^{1/2}  J_{N}^{\nu}\left( p_{f}s_{f}, p_{i}s_{i}\right)^{\dagger}  J^{\mu }_{N}\left( p_{f}s_{f}, p_{i}s_{i}\right)
\end{align}

In the  invariant cross-section for the PWIA (Eq. \ref{cross-section}), we replace the hadronic tensor $H^{\mu \nu }$ by the factorization of  Eq.(\ref{H=H_N*rho}), which yields,
\begin{align}\label{Inv-DCS}
{d\sigma\over d^3k_f/\epsilon_f d^3p_f/E_f}  =   {1\over 2 p_d \cdot k_i}  {\alpha^2_\text{EM}\over q^4} L_{\mu\nu} H^{\mu \nu }_{N}\rho \left( \alpha, {\bf p}_\text{T}\right) \dfrac {2-\alpha }{\alpha ^{2} } \delta\left( p_r^2 - m_N^2\right)
\end{align}
where, $\alpha_\text{EM}=e^2/(4\pi)$ is the electro-magnetic coupling constant.  Finally, introducing the Light-Front nuclear spectral function (Ref.\cite{Frankfurt:1981mk}),
\begin{align}\label{Spectral function}
S_d^\text{LF}(\alpha, {\bf p}_\text{T}) = \rho_d \left( \alpha, {\bf p}_\text{T}\right) \dfrac {2-\alpha }{\alpha ^{2} } \delta\left( p_r^2 - m_N^2\right)
\end{align}
one can present the differential cross-section as a product of $\sigma_{eN}$ and the spectral function as follows:
\begin{align}\label{factorized_cross-section}
{d\sigma\over  d\epsilon_f d\Omega_{k_f}  d^3p_f}  =   \sigma_\text{eN} \ S_d^\text{LF}(\alpha, {\bf p}_\text{T})
\end{align}
were, the off-shell electron--bound-nucleon cross-section assumes the form,
\begin{align}\label{sigma_eN}
\sigma_\text{eN} =  {1\over 2 m_d \epsilon_i} \ {\epsilon_f \over E_f} \  {\alpha^2_\text{EM}\over q^4} \ L_{\mu\nu} H^{\mu \nu }_{N}
\end{align}
Here, $\epsilon_i$  ($\epsilon_f$)  is the initial (final) scattered electron energy, and $E_f$  represents the energy of the knock-out nucleon in the final state.  	

It is worth mentioning that within the PWIA, the factorization showed  in Eq.(\ref{factorized_cross-section}) is universal for any nuclei, in which case one needs to replace the deuteron spectral function by  the  light-front spectral function  of the nucleus being considered.

\subsubsection{Bound-Nucleon Structure Functions }

In calculating $\sigma_\text{eN}$ in Eq.(\ref{sigma_eN}) it is convenient to present it through the four independent structure functions of the nucleon $w^N_\text{L}$, $w^N_\text{TL}$, $w^N_\text{T}$ and $w^N_\text{TT}$ in the form:
\begin{align}\label{sigma_eN-Gross}
\sigma_\text{eN} =  {1\over 2 m_d E_f}  \  {\sigma_\text{Mott}}  \left( v_\text{L} w_\text{L}^N + v_\text{TL} w_\text{TL}^N \cos\phi + v_\text{T} w_\text{T}^N + v_\text{TT} w_\text{TT}^N \cos(2\phi) \right)
\end{align}
where, 
\begin{align}
\sigma_\text{Mott} = {\alpha^2 \cos({\theta\over 2})^2\over 4 \epsilon_i^2 \sin({\theta\over 2})^4} 
\end{align}
is the Mott cross section, with  $\theta$  being scattered electron angle. In equation (\ref{sigma_eN-Gross}) the coefficient functions $v_{i}$ are given by,
\begin{align}
v_L &=   \frac{Q^4}{{\bf q}^4} \nonumber \\
v_T &=   \frac{Q^2}{2{\bf q}^2}  +  \tan^2\frac{\theta}{2}  \nonumber  \\ 
v_{TT} &=    \frac{ Q^2}{ 2 {\bf q}^2} \nonumber \\
v_{TL} &=   \frac{Q^2}{  {\bf q}^2}\left(  \frac{Q^2}{{\bf q}^2}  +  \tan^2\frac{\theta}{2}  \right)^{1/2}
\end{align}
where,  $Q^2 = 4\epsilon_i\epsilon_f\sin({\theta\over 2})^2$,  and $\bf q$ is the three momentum of the virtual photon.
The above defined  structure functions of the bound nucleon can be related to the light-front components of the nucleonic electromagnet tensor as follows (see \mbox{\hyperref[App. LFPT]{Appendix B}}):
\begin{align} \label{V-Gross}
w^N_\text{L} & = \frac{ {\bf q}^2 }{4 Q^2} \left( H^{++}{ Q^2 \over (q^+)^2 } + 2 H^{+-} + {(q^+)^2 \over  Q^2 } H^{- \  -} \right)\nonumber  \\
w^N_\text{TL} & = {|{\bf q}| \over q^+} \left( H^{+ \parallel}_N + H^{- \parallel }_N {(q^+)^2 \over Q^2} \right) \nonumber  \\ 
w^N_\text{T} & =  H^{\parallel \parallel }_N +  H^{\bot \bot} _N \nonumber  \\
w^N_\text{TT}  & = H_N^{\parallel \parallel} - H_N^{\bot \bot}   
\end{align} 
where, $\pm$ correspond to $t\pm \hat z$ directions on the light-front, with $\hat z$ defined in the negative direction of the transferred three momentum $\bf q$. The transverse components are chosen as follows: the perpendicular direction is defined by,  $ \textbf{n}_{\bot}=\frac{\textbf{p}_f\times \textbf{q} }{|\textbf{p}_f\times \textbf{q}|}$, and the parallel unit vector projection is,  $\textbf{n}_{\parallel}=\frac{ \textbf{q}  \times  \textbf{n}_{\bot} }{|\textbf{q}  \times  \textbf{n}_{\bot}|}$.  The scattering and reaction planes of the reaction are defined in Fig.(\ref{scatt react plane}).

Using now the Eq.(\ref{HN}) and the expression of the bound nucleon electromagnetic current from equations (\ref{Jsum}) and (\ref{J_on_off}) one can calculate the nucleon structure functions explicitly. 
In what follows we split the structure functions into two terms,
\begin{align}
w^N_i  =  w^N_{i, \text{prop}} + w^N_{i, \text{inst}} \ , \quad  \text{for,  i = L, TL, T, TT}
\end{align}
where, the subscript ``prop'' corresponds to the structure functions calculated using only the propagating part of the  electromagnetic current $J_\text{prop}^\mu$, whereas the terms with the subscript ``inst'' correspond to the contribution from $J_\text{inst}^\mu$, and its interference with $J_\text{prop}^\mu$.

Using the explicit expressions  for the  currents, given by  equations (\ref{Jsum}) and (\ref{J_on_off}), we calculate the above structure functions expressing them  through the \mbox{off-shell} parameter $\eta$ (Eq. \ref{eta})  as follows\footnote{In \hyperref[App Nucleonic EM Current]{App.(C)} we write the structure functions explicitly in terms of light-front kinematical variables (see Eq.(\ref{V_LF})).}: 
{\small
\begin{align}
w_{\text L ,\text{prop}}^N   = &  \  {\bf{q}}^2 \Big[ F_1^2  \tau^{-1} \left(  1+ {{\bf p}_\text{T}^2 \over m_N^2} + \tau \eta_i(\eta_i+\eta_q)   \right) -   F_1 F_2 {\kappa } \left( 2+ \eta_q   \right)   \nonumber \\
&  +  F_2^2 \kappa^2 \left(  {{\bf p}_\text{T}^2 \over m_N^2} + \tau(1 + \eta_q) \right) \Big]   \\ 
\vspace*{0.3cm}
w_{\text L  ,\text{inst}}^N  = &  \  {\bf{q}}^2  \Big[ F_1^2  \eta_i \Big(  \tau \eta_i(1 + \eta_q)  - 2 - \eta_q    \Big)  + F_1 F_2 \kappa \Big( \tau \eta_i \left( 2-2\eta_i-\eta_q \right) +\eta_q \Big)   
\nonumber \\
&  + F_2^2 \kappa^2 \tau  \Big( \tau \eta_i(\eta_i+\eta_q) - \eta_q   \Big) \Big] 
\end{align}
\begin{align}
\hspace{-1.cm}
w_{\text{TL}  ,\text{prop}}^N   = &   \   2 \ |{\bf{q}}|  \  |{\bf p}_{\text{T}}| \left(  F_1^2 +   F_2^2 \kappa^2  \tau  \right)\left[  2 +4{\alpha_N \over \alpha_q}  + 2\eta_i + \eta_q \right]  \\ 
\vspace*{0.3cm}
\hspace{-1.cm}
w_{\text{TL}  ,\text{inst}}^N   = & \   2 \ |{\bf{q}}| \ |{\bf p}_\text{T}|  \left(  F_1^2 +   F_2^2 \kappa^2  \tau  \right)  \left( 1 - \tau \eta_i   \right) \eta_q 
\hspace{3.5cm}  
\end{align}
\begin{align} 
w_{\text T  ,\text{prop} }^N  = & \ 4m_N^2 \Big[  F_1^2 \left( { {\bf p}_\text{T}^2 \over m_N^2} + 2 \tau(1 + \eta_q) \right) + 2  F_1 F_2 {\kappa } \tau \left( 2 + \eta_q   \right)  
 \nonumber  \\
&+  F_2^2 \kappa^2 \tau  \left(  2 + {{\bf p}_\text{T}^2 \over m_N^2} + 2 \tau \eta_i(\eta_i+\eta_q)   \right) \Big] \\
\vspace*{0.3cm}
w_{\text T  ,\text{inst} }^N  = & \  2 Q^2 \Big[ F_1^2  \Big( \tau \eta_i(\eta_i+\eta_q) - \eta_q   \Big)  + F_1 F_2 \kappa \Big( \tau \eta_i \left( 2\eta_i+\eta_q - 2 \right) - \eta_q \Big)  \nonumber \\
&  + F_2^2 \kappa^2 \tau  \eta_i \Big(  \tau \eta_i(1 + \eta_q)  - 2 - \eta_q    \Big)  \Big] 
\end{align}
\begin{align}
\hspace{-1.cm}
w_{\text{TT}  ,\text{prop}}^N   = & \  4 p_{\textbf{T}}^2 \left(  F_1^2 +   F_2^2 \kappa^2  \tau  \right) \\ 
\vspace*{0.3cm}
\hspace{-1.cm}
w_{\text{TT} ,\text{inst}}^N   = &  \  0 
\hspace{8.5cm}   
\label{V_LFtoRP}
\end{align}
}
where, $\tau=Q^2/(4 m_N^2)$, $\ \eta_i=\eta \ \alpha_N/2$, $\ \eta_q=\eta \ \alpha_q/2$. 
Alternatively, one can write, 

\begin{equation}
\eta_i   =   - {2 \Delta{p}_i \cdot p_i  \over Q^2}=  {(m^{2}_N+{\bf p}_\text{T}^2 ) \over Q^2 } \dfrac{\alpha_q}{\alpha_f }  -  \dfrac{\alpha_N}{\alpha_q} 
\end{equation}
\begin{equation}
\eta_q    =  -  {2 \Delta{p}_i \cdot q \over Q^2} =  {(m^{2}_N+{\bf p}_\text{T}^2 ) \over Q^2} \dfrac{\alpha_q^2}{\alpha_f \alpha_N}  -  1 
\end{equation}

\vspace{0.3cm}
The structure functions in Eq.(\ref{V_LFtoRP}) are invariant under boosts along the direction of transfer momentum, which follows directly from the fact that they are expressed through the boost invariant variables $\eta$,  $\alpha_i$, $\alpha_q$ and $\alpha_f$. 
Since many experiments in  probing high momentum bound nucleons are  performed in the fixed target experiments  it is convenient to express the above variables  through the four momenta measured in the lab frame. Considering the Lab reference frame, in which $\hat{z} || {\bf q}$, the  $\alpha_i$, $\alpha_q$ and $\alpha_f$ parameters can be expressed as follows,
\begin{eqnarray}
&& \alpha_i = 2 - \alpha_r = \alpha_f - \alpha_q  \\ 
\nonumber \\  
\alpha_r = {2 (E_r - p_r \cos\theta_r) \over m_d}  \
, 
&&\alpha_q = {2 (q_0 - {\bf q})\over m_d} \
, \quad
\alpha_f = {2 (E_f - p_f \cos\theta_f) \over m_d}
\end{eqnarray}
where,  $\ p_d^\mu = (m_d,0) $, $\ q^\mu = (q_0, {\bf q}) $, $\ p_r^\mu = (E_r, {\bf p}_r) $, and  $\ p_f^\mu = (E_f, {\bf p}_f) $,   are the target (deuteron), virtual photon, recoil and struck nucleon  four-momenta measured in the Lab frame.

\section{Numerical Estimates}
\label{V}

We present numerical estimates for kinematics which will be explored in experiments planned for 12 GeV upgraded Jefferson Lab.  In all calculations below, we take the initial energy of the electron beam $\epsilon_i = 11$~GeV.
In order to quantify the extent of the binding effects we consider the ratio,
\vspace{-0.2cm}
\begin{equation}
R = {\sigma_\text{eN}\over \sigma_\text{eN}^\text{on}}
\label{R}
\end{equation}
where, $\sigma_{eN}$ is the cross-section of electron bound-nucleon scattering that was defined in Eq.(\ref{sigma_eN}),
while  $\sigma_\text{eN}^\text{on}$ corresponds to the same cross-section for the electron scattering off the free moving nucleon with the same initial momenta. 

\begin{figure}[H]
\hspace{-0.8cm}
	\includegraphics[scale=0.335]{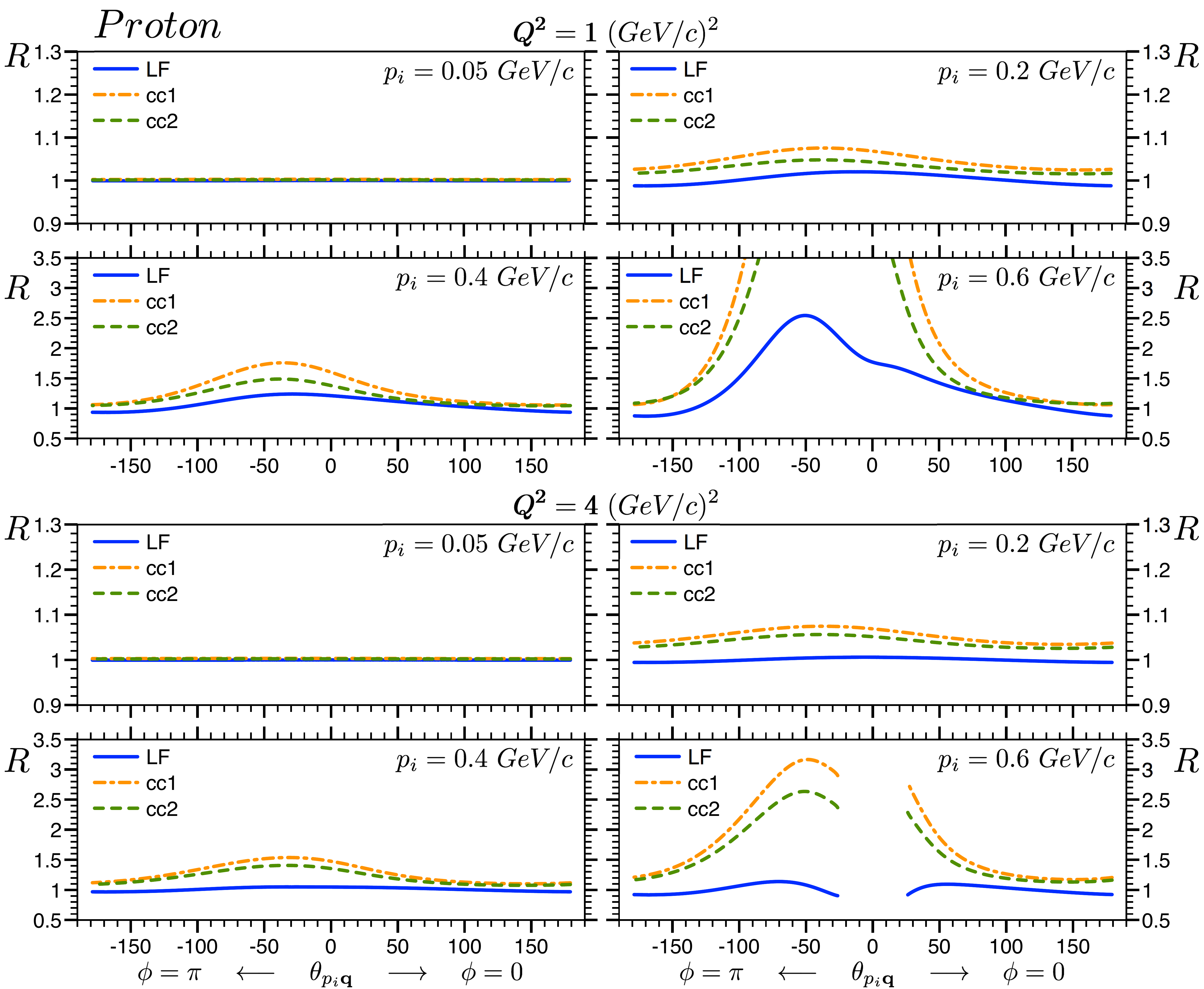}	
\caption{Angle dependence of R (Eq. \ref{R}). The solid (blue) curves are the LF calculations, dashed (green) and dash-dotted (orange) curves corresponds to cc2 and cc1 de Forest approximations\cite{deForest(1983)} for the off-shell cross section $\sigma_{eN}$. The {\em minus}  sign in $\theta_{p_{i}q}$ axis corresponds to an angle of $\phi=180^0$ between scattering and reaction planes (Fig. \ref{scatt react plane}). Calculations are done with a value of $\epsilon_i = 11$~GeV for the initial electron (beam) energy.}
	\label{theta_dep_Q2_p}
\end{figure}

\vspace{0.15in}		
\begin{figure}[h]
\hspace{-0.8cm}
	\includegraphics[scale=0.335]{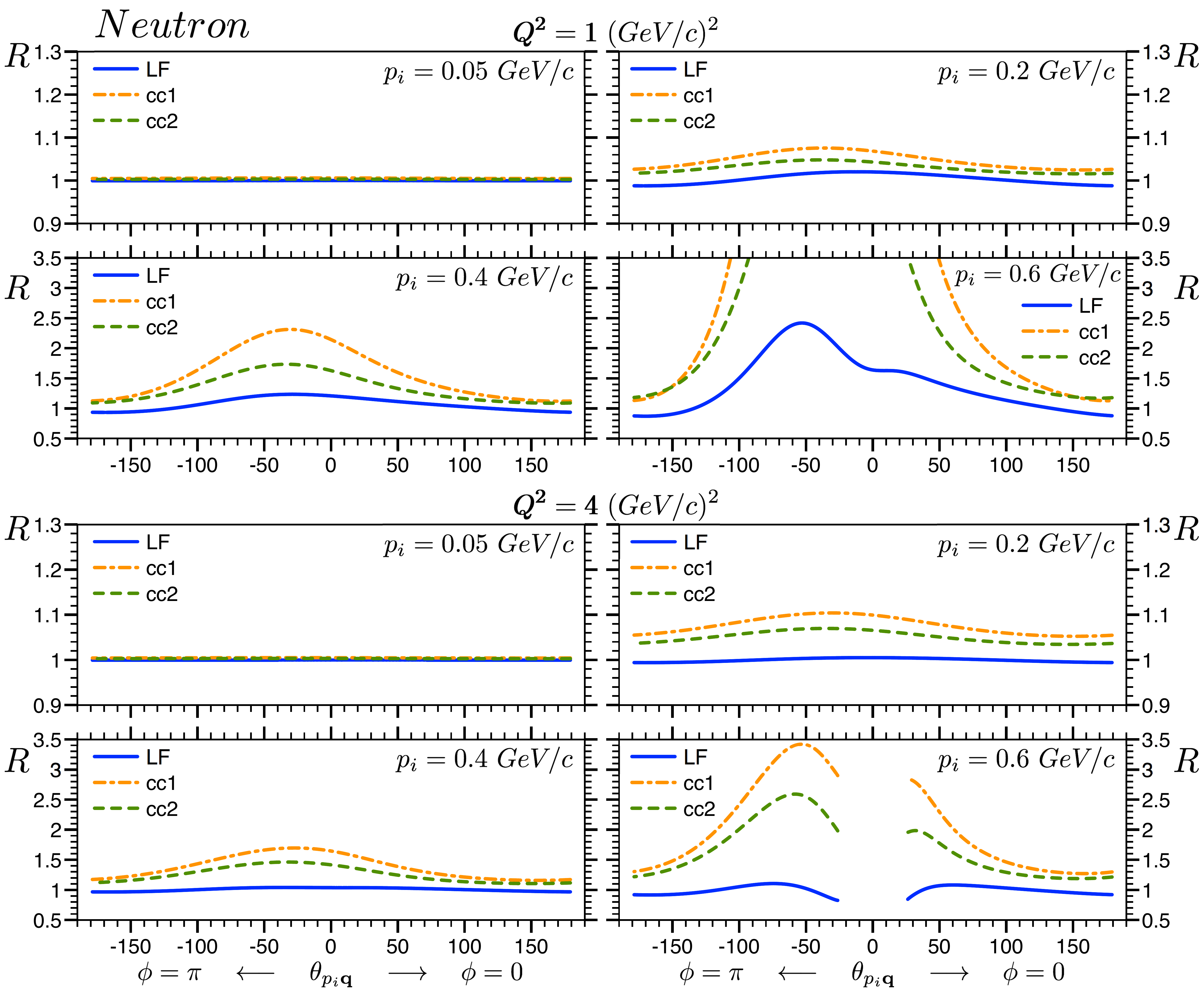}
	\caption{The same as in Fig.(\ref{theta_dep_Q2_p})  but for scattering from bound neutron.}
	\label{theta_dep_Q2_n}
\end{figure}

We consider first the dependence of $R$ on ``traditional'' kinematical parameters, which define the electronuclear processes such as initial momentum of the bound nucleon ($p_i$) its relative angle with respect to the transferred 3-momentum ($\bf q$) as well as the virtuality of the transferred momentum ($Q^2$). Additionally, we compare the predictions of LF approximation with that of the de Forest formalism\cite{deForest(1983)}, which is commonly used in the analysis of the experimental data. In all these estimates, we use the same parameterization for the electric and magnetic form-factors of the nucleons. These parameterizations are the same as for the free nucleon. Thus we do not consider the effects related to the possible modification of the charge and magnetic current distributions in the bound nucleon.

\begin{figure}[h]
\hspace{-1.2cm}
	\includegraphics[scale=0.35]{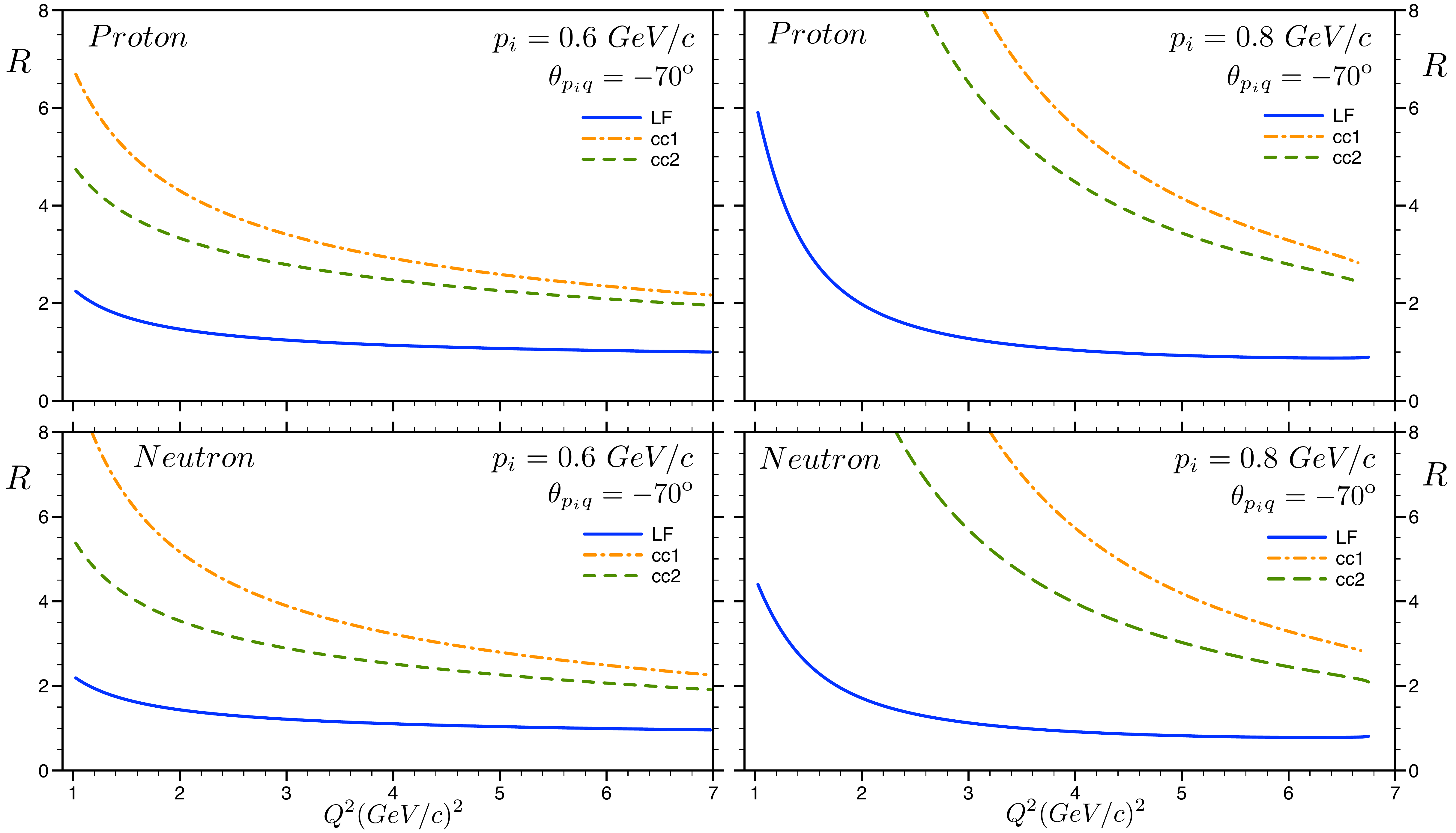}
	\caption{The $Q^2$ dependence of the off-shell effects for $\theta_{p_iq} = -70^0$ for proton and neutron targets.}
	\label{Q2_dep_p_n}
\end{figure}

In Fig.({\ref{theta_dep_Q2_p})  and Fig.(\ref{theta_dep_Q2_n}) we compare the angular dependences of ratio  $R$  at different values of missing momenta at fixed $Q^2=1$ and $4$~(GeV/c)$^2$ for bound proton and neutron respectively.  As  Fig.({\ref{theta_dep_Q2_p}) shows  LF approximation predicts off-shell effects for $Q^2=1$ (GeV/c)$^2$ as large as $40-250\%$  for bound proton momenta $\ge 400$~MeV/c. Even larger effects are expected within the de Forest approach\cite{deForest(1983)}. It can be observed from the figures, that the predictions based on the LF formalism and the de Forest approximation significantly diverges close to the kinematical limit of the scattering process, as shown in calculations for $p_i = 600$~MeV/c. Note that because of the different magnitude and signs of the proton and neutron form-factors, we expect different off-shell contributions for scattering from a bound proton or neutron. However, as can be seen from Figs.(\ref{theta_dep_Q2_p} and \ref{theta_dep_Q2_n}), qualitatively the dependence of the ratio $R$ on the kinematical parameters of the reaction is similar for both proton and neutron.

An important feature of LF calculations following from Fig.(\ref{theta_dep_Q2_p}) and Fig.(\ref{theta_dep_Q2_n}) is that the off-shell effects diminish with the increase of $Q^2$. This reflects the nature of the LF dynamics, in which case the harder the probe (larger $Q^2$), the lesser is the sensitivity to the binding effects of the target nucleon. It is worth mentioning that no such behavior exists in the de Forest approximation, since in this case, part of the off-shell effects are kinematical. In particular, the energy of the bound nucleon is taken to be equal to the on-shell energy for the given momentum of the nucleon, with the phase space of the initial nucleon being proportional to ${1\over \sqrt{m_N^2 + {\bf p}_i^2}}$.			
		
\vspace{0.15in}
\begin{figure}[H]
\hspace{-0.9cm}
	\includegraphics[scale=0.33]{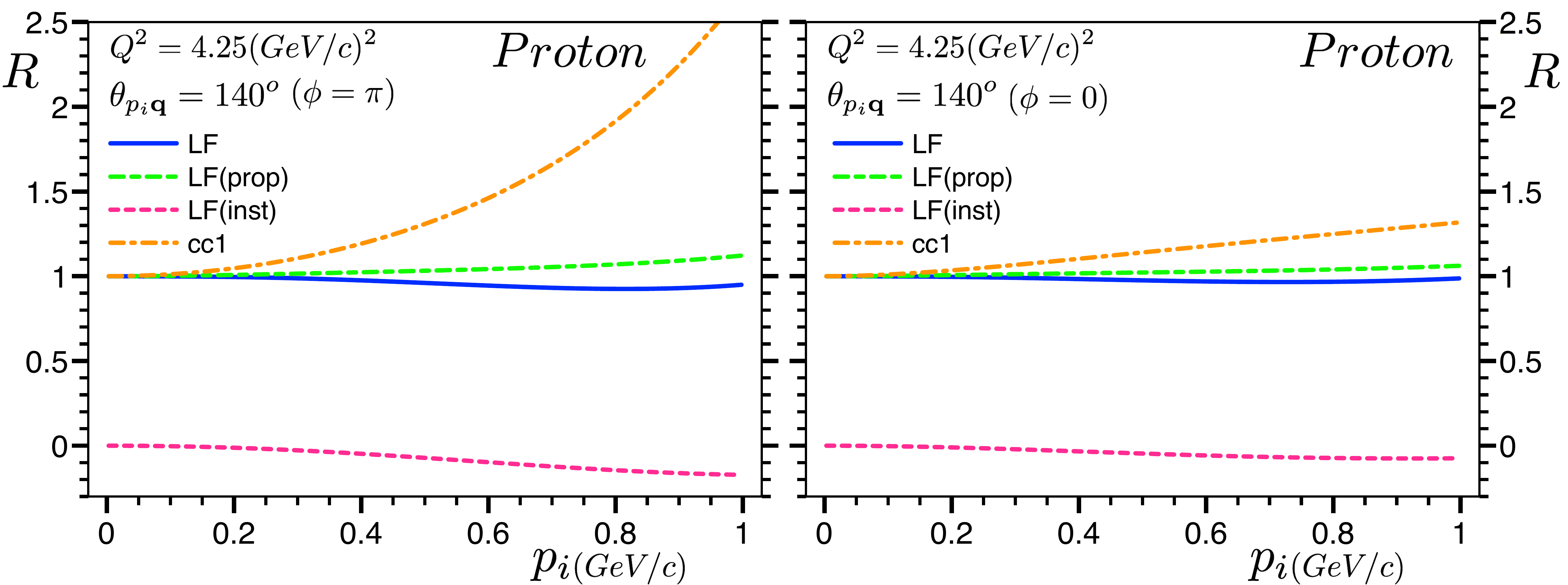}
	\caption{The left panel presents the off shell effects expected for the experiment of References\cite{Boeglin:2011mt,Yero:2020}.  The right-panel shows the off shell effects for kinematics with, $\phi = 0$. }
	\label{pi_dep_p}
\end{figure}		

The off-shell effects suppression due to the increase of transfer momentum ($Q^2$) is addressed in Fig.(\ref{Q2_dep_p_n}), where the $Q^2$ dependence of the ratio $R$ for proton and neutron with initial momenta  $p_i = 600$~ and $800$~MeV/c is shown. Here we choose \mbox{$\theta_{p_iq} = -70^0$,} for which large off-shell effects are observed in Fig.(\ref{theta_dep_Q2_p}) and Fig.(\ref{theta_dep_Q2_n}). These calculations indicate that, even at such high bound nucleon momentum, the off-shell effects predicted by the Light-Front calculation are not more than $10\%$  already at \mbox{$Q^2\sim 4$~GeV$^2$.}

For practical purposes in Fig.(\ref{pi_dep_p}), we estimate the dependence of the off-shell effects on the momentum of the bound nucleon for kinematics relevant to the JLab experiment\cite{Yero:2020},  which is aimed at probing deuteron structure at very large internal momenta. As the figure shows for both the angles between scattering and reaction planes ($\phi$), the light-front approach predicts off-shell effects to be less than $8\%$ for all kinematics with the latter value happening at   $p_i = 850$~MeV/c.

\vspace{0.15in}		
\begin{figure}[h]
\hspace{-1.3cm}	
	\includegraphics[scale=0.34]{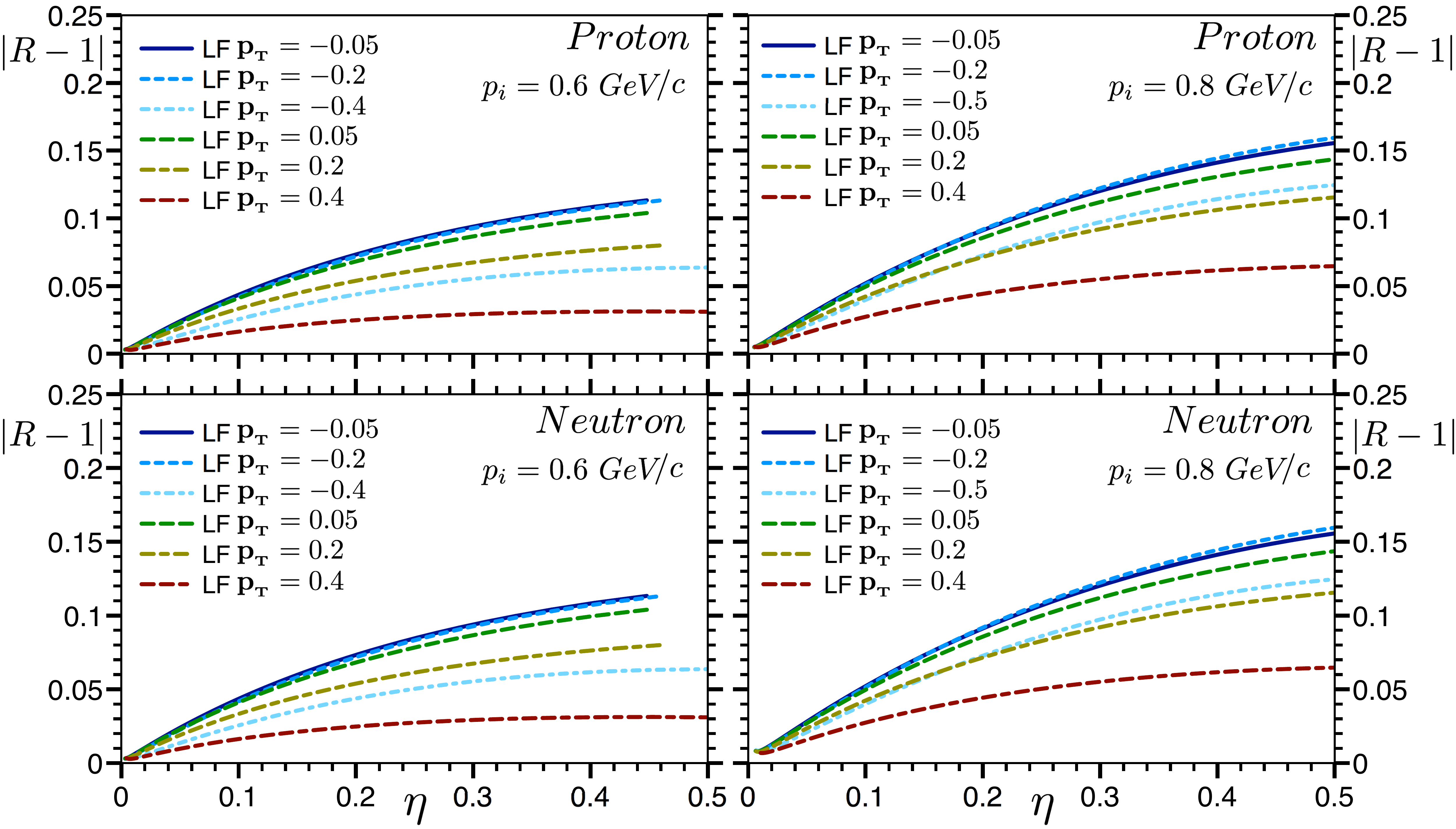}	
	\caption{ The $\eta$ parameter dependence of the off-shell effects $|R-1|$ for $p_i = 0.6$ and $0.8$~GeV/c at different values of the transverse momentum ${\bf p}_\text{T}$. }
	\label{eta-graph}
\end{figure}		
		
We discuss now why the parameter $\eta$  introduced in Eq.(\ref{eta}) can be used as a universal parameter for estimation of the off-shell effects for any kinematic conditions of the electro-production reaction. For this, in Fig.(\ref{eta-graph}) we calculate the $\eta$ dependence of $|R-1|$ for very large magnitudes  of bound nucleon momenta ($p_i = 600$ and $800$~MeV/c)  at different values of transverse momentum ${\bf p}_\text{T}$.  

Note that the expected off-shell effects will be much less for smaller values of $p_i$. As the figure shows, for any possible scenarios of kinematics, the off-shell effects can be confined below $5\%$ as soon as $\eta <0.1$. This represents a strong indication that the variable $\eta$ can be considered a universal parameter for controlling the off-shell effects in the reaction mechanism for electronuclear processes. The universality rest in the fact that if our goal is to probe a bound nucleon with very large momenta, we can use the corresponding LF longitudinal momentum fraction $\alpha$ together with the transverse momentum ${\bf p}_\text{T}$ to calculate the required $Q^2$, such that it makes $\eta < 0.1$. Satisfying this condition guarantees that the off-shell effects in the electromagnetic current can be neglected up to the bound-nucleon momentum planned to probe.

\chapter{RELATIVISTIC DEUTERON WAVE FUNCTION} 
\label{wave_function}

We dedicate this chapter to the study of the second term that makes up the deuteron electro-disintegration PWIA amplitude (Eq. \ref{A0sb}), i.e., the deuteron wave function. 
Since 1970's considerable theoretical efforts have been made towards a consistent formulation of the relativistic framework for the description of deuteron structure (see e.g. \cite{Frankfurt:1977vc,Frankfurt:1981mk,Buck:1979ff,Karmanov:1980mc,Cooke:2001kz}).
 However, the experimental verification of the various approaches was somewhat limited since the existing nuclear labs could only measure the elastic and inclusive scattering of electrons from deuteron at sufficiently large momentum transfer.  
Such measurements probe the deuteron wave function only indirectly since the measured cross-sections are sensitive to the integrated properties of the deuteron structure, which are obscured considerably by long-range phenomena (see the discussion in Ref.\cite{Boeglin:2015cha}).  
As a result, it was not possible to establish the validity of any theoretical approach unambiguously.  
The present situation is unprecedented due to recent high energy measurement of exclusive disintegration, in which case, due to the set up of the eikonal regime of FSI (Fig.\ref{edepn_diagrams} (b) and (c)), it is possible to isolate the PWIA contribution (Fig. \ref{edepn_diagrams} (a)) which directly probes the deuteron wave function. 

Our goal is to obtain a parameterization of the deuteron wave function that allows us to evaluate the $pn$ component in the limit of large internal momenta.
We use a light-front diagrammatic definition of the deuteron wave function, in which case it can be related to the light-front time-ordered amplitude describing the electro-disintegration reaction. 
Within this approach, the deuteron is first resolved into its proton-neutron constituents, and one of the nucleons then interacts with the external electromagnetic probe, Fig.(\ref{ddiagram}). 

\vspace{0.15in}
\begin{figure}[h]
\center
\includegraphics[scale=0.55]{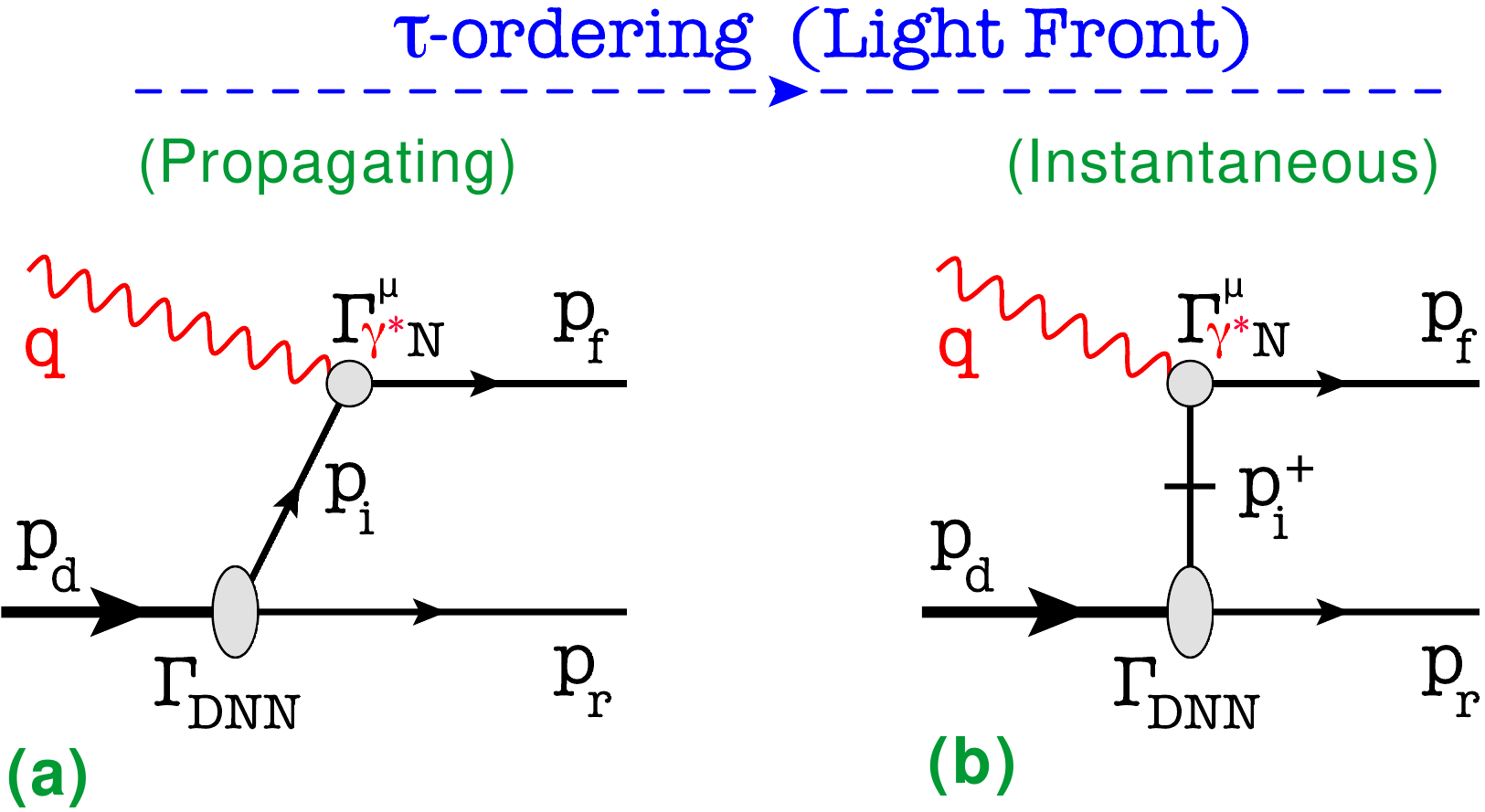}
\caption{LF-ordered diagrams contributing to the PWIA}
\label{ddiagram}
\end{figure}

The LF deuteron wave function is then defined as in Eq.(\ref{wflf}), which is rewritten here for convenience,
\begin{equation}
\psi_{s_{i}s_{r}}^{\lambda_{d} \text{(LF)}} = -\frac { \bar {u}^\text{(LF)}_{( p_{i}, s_i)} \bar u^\text{(LF)}_{( p_{r}, s_r)} \Gamma _\text{dNN}^\mu \chi^{\lambda_d \text{(LF)}}_\mu} {\frac {1}{2}\left( m^{2}_{d}-4\frac {m^{2}_N+{\bf p}_\text{T}^2}{\alpha \left( 2-\alpha \right) }\right) }\dfrac {1}{\sqrt {2\left( 2\pi \right) ^{3}}}
\label{wflf1}
\end{equation}
As a reminder, we have specified in Eq.(\ref{wflf1}) that the nucleon spinors ($\bar{u}_{i,r}$) that enter in the wave-function are light-front spinors, likewise the deuteron polarization vector ($\chi^{\lambda_d}$) is in the light-front parameterization. For the explicit form of these objects we refer the reader to the section \hyperref[App Light Front Spin]{App.(G.2)}.

\section{General Properties of the Deuteron Wave Function in the Light-Front}
\label{wave_function_gen}

To explore the general properties of the deuteron light-front wave function, 
first, we absorb the 
denominator in Eq.(\ref{wflf}) into the vertex function,
\begin{equation}
 {\Gamma_\text{dNN}\over {\frac {1}{2}\left( m^{2}_{d}-4\frac {m^{2}_N+{\bf p}_\text{T}^2}{\alpha \left( 2-\alpha \right) }\right) } \sqrt{2(2\pi)^3}} \to {\Gamma_\text{dNN} \over \sqrt{2}}
\end{equation}
The $1/\sqrt{2}$ factor in the denominator of the vertex function is kept explicitly to make a straightforward correspondence with the non-relativistic deuteron wave function in the small momentum limit.

\begin{figure}[h]
\center
\includegraphics[scale=0.55]{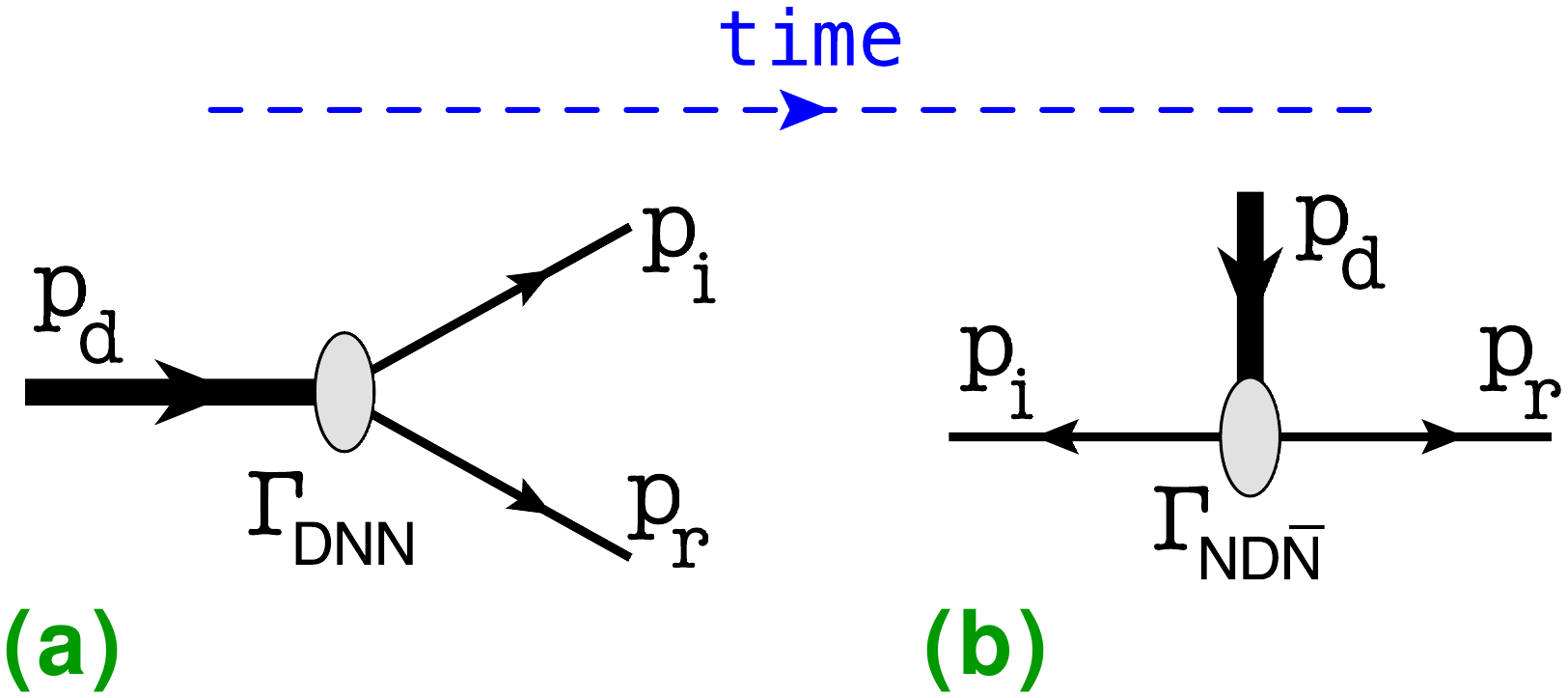}
\caption{(a) Represents the $d\to N_i N_r$ vertex. (b) Represents the $\bar{N}_i d \to N_r$ vertex}
\label{crossing diagrams}
\end{figure}

In Fig.(\ref{crossing diagrams}) the crossing symmetry between channels $d\to N_i N_r$ and $\bar N_i d\to N_r$ is shown schematically. Applying the 
charge conjugation operation allows us to express the deuteron wave function in the form,
\begin{align}
\psi^{\lambda_d}_{s_i,s_r} 
& = -\bar u(p_r,s_r) \bar u(p_i,s_i) {\Gamma^\mu_\text{dNN} \over \sqrt{2}} \chi_\mu^{\lambda_d} \nonumber \\ 
& =  -\bar u(p_r,s_r) 
{\Gamma^\mu_\text{NdN} \over \sqrt{2}}
(\rm{i} \gamma_2\gamma_0) \bar u(p_i,s_i)^T
\chi_\mu^{\lambda_d}
 \label{dwave_lf2}
\end{align}
where, 
$\bar u(p_i,s_i)^\text{T}$ means the transposition of $\bar u(p_i,s_i)= u(p_i,s_i)^\dagger \gamma^0$, and the two vertex functions are related by the charge conjugation operation, 
$ \Gamma^\mu_{\text{dNN}}= \mathcal{C} [\Gamma^\mu_{\text{NdN}}]$. 
Furthermore, making use of the identity,
\begin{equation}
(i \gamma_2\gamma_0)\bar u(p_i,s_i)^\text{T} = \sum\limits_{s_i^\prime}\gamma_5\epsilon_{s_i s_i^\prime} u(p_i,s^\prime_i)
\end{equation}
results in the following convenient expression for the deuteron LF wave-function,  
\begin{equation}
\psi_{s_i,s_r}^{\lambda_d} (p_d,p_i,p_r) = -\sum\limits_{s_i^\prime}\bar u(p_r,s_r) { \Gamma^\mu_\text{NdN}(p_d,p_i,p_r) \over \sqrt{2} } \gamma_5 \epsilon_{s_i s_i^\prime} u(p_i,s_i^\prime)\chi_\mu^{\lambda_d} 
\label{NdNwf}
\end{equation}
Here, $\epsilon_{i j}$ is the two-dimensional (totally) anti-symmetric (Levi-Civita) tensor, and the subscripts, $s_i,s_i^\prime=\pm 1/2$, label the spin states of the nucleon.

The advantage of the above representation for the deuteron wave function is that it allows to write down a general covariant expression for the vertex function, $\Gamma^\mu_\text{NdN}$.  
Since deuteron is a spin 1 particle with even parity\footnote{Symmetric under space reflection.}, it has the transformation properties of a pseudo-vector particle. Therefore, the $\gamma_5$ matrix in the wave function (Eq. \ref{NdNwf}) indicates that the vertex $\Gamma^\mu_\text{NdN}$ must transform as a Lorentz vector. 
Hence, the construction of the most general $\Gamma^\mu_\text{NdN}$ vertex function is reduced to write down all the independent Dirac bi-linear terms that can be built out of a set of independent four-vectors. 
For example, the three independent on-shell four-momenta of the particles connected to the vertex, i.e., the deuteron, the proton, and the neutron \cite{Karmanov:1995}.

In order to separate the effects due to the internal relative momentum between the nucleons (responsible for the bound state dynamics), from those related with the motion of the system as a whole (kinematic effects), a more useful choice of three independent variables are the two-body total momentum, 
\begin{equation}
P_\text{NN}^\mu=p_i^\mu + p_r^\mu
\end{equation}  
the two nucleon relative momentum, 
\begin{equation}
p_\text{rel}^\mu={p_i^\mu - p_r^\mu \over 2}
\label{p relative}
\end{equation}  
and the four-vector, 
\begin{equation}
\Delta^\mu = P_\text{NN}^\mu - p_d^\mu 
\label{delta}
\end{equation}
which accounts for the difference between the bound and on-shell nucleons. 
In our present case (two-body problem) the four-vector $\Delta^\mu$
can always be chosen to be light-like, i.e., $\Delta^2=0$. 
Then, $\Delta^\mu = (\Delta^+,{\boldsymbol \Delta}_\text{T},\Delta^-)$ can be chosen in the direction of light-front-energy, $\Delta^\mu = (0,{\bf 0}_\text{T},\Delta^-)$, which guarantees that the longitudinal light-front ($p^+$) and transverse (${\bf p}_\text{T}$) components of momentum are conserved at the deuteron to $pn$ transition vertex ($\Gamma^\mu_\text{NdN}$). 
As a consequence, the application of light-front-time ordering technic is consistent.
For more details, we refer the reader to   \hyperref[App 2body basis]{App.(H)}. 

The only non-vanishing component of the four-vector $\Delta^\mu$ is therefore the light-front off-shell energy for the $pn$ system, which is given by,
\begin{equation}
\Delta^- = P_\text{NN}^- - p_d^- = {s_\text{NN} - m_d^2 \over p_d^+}
\end{equation}
where, $s_{NN}$ is the invariant mass of the on-shell NN system, 
\begin{equation}
s_\text{NN} = 4{m_N^2 + {\bf p}_\text{T}^2 \over \alpha(2 - \alpha)} = P_\text{NN}^2 = M_\text{NN}^2
\label{sNN}
\end{equation}
Note that ${\bf p}_\text{T}^2={{\bf p}_i^2}_\text{T}={{\bf p}_r^2}_\text{T}$, and the symmetry of the denominator $\alpha(2 - \alpha)=\alpha_i \alpha_r $ in Eq.(\ref{sNN}) with respect to the labels $i,r$ allows us to omit indices referring to particular nucleon species. 
It can be checked that in the static limit, ${\bf p}_\text{T}^2 \to 0$, $\alpha \to 1$, ( ${\bf p}_{i,r}^2 \to 0$) the light-front off-shell energy approaches to the deuteron binding energy $\Delta^-\approx 2|\epsilon_B|$.
 
With the above choice of four-momenta, the general form of the $\Gamma^\mu_\text{NdN}$ vertex can be written in the following form:
 \begin{align}
 \Gamma^{\mu} = & \Gamma_{1} \gamma^{\mu} +\Gamma_{2} \frac{(p_i-p_r)^{\mu}}{2m} + \Gamma_{3} \frac{\Delta^{\mu}}{2m} + \Gamma_{4} \frac{(p_i-p_r)^{\mu} \sh{\Delta}}{ 4m^2}  
+ i \Gamma_{5} \frac{1}{4 m^{3}} \gamma_{5} \epsilon^{\mu \nu \rho \sigma}{p_{d}}_\nu \ (p_i-p_r)_{\rho} \Delta_\sigma  \nonumber \\
& + \Gamma_{6} \frac{ \Delta^{\mu} \sh{\Delta}}{4m^{2}}
\label{vertex}
\end{align}
Each of the six $\Gamma_{i}$ are scalar\footnote{In other words, they remain invariant under Lorentz transformations.} functions that depend on $p_\text{rel}$ via the bound state constrain:
\begin{equation}
p_d^2=m_d^2 = (P_\text{NN} - \Delta_p)^2 = M_\text{NN} ^2 - 2 P_\text{NN} \cdot \Delta_p
\end{equation}
where, we have taken into consideration the light-like nature of the $\Delta$ four-vector, $\Delta_p^2=0$.
It follows that,
\begin{equation}
2 P_\text{NN} \cdot \Delta_p =  M_\text{NN}^2 -m_d^2 \quad \overset{\text{LF coord} }{\longrightarrow } \quad \Delta_p^- = {M_\text{NN}^2 -m_d^2 \over P_\text{NN}^+} = {M_\text{NN}^2 -m_d^2 \over p_\text{d}^+} 
\end{equation}
which in the center of momentum frame of the $NN$ system results in 
(see the discussion in \hyperref[Two-Body basis - Momenta]{App. H.1} and \hyperref[App Dirac bilinears - Canonical]{App. I.3}),
\begin{equation}
{\Delta_p^-}_\text{cm} = {M_\text{NN}^2 -m_d^2 \over M_{12}} = { 4 E_{k}^2 - m_d^2 \over 2 E_{k}} \approx { 4 E_{k}^2 - 4m^2 \over 2E_{k}} = { 2 {\bf k}^2 \over E_{k}}
\end{equation}

The computation of the vertex functions over the entire allowed kinematic region requires a self-consistent theory of the $NN$ interaction, which is currently an impractical task. 
However, we can continue our study of the transition vertex \mbox{(Eq. \ref{vertex})} by limiting our attention to high momentum transfer kinematics, where it can be simplified\footnote{It is worth to remind ourselves that, as established in  Sec.(\ref{HEP}), this kinematics permits the unambiguous identification of events involving large internal momentum.}. 
 
As a first step, we observe that in the high energy kinematics (see Eq. \ref{hlimit}) and for small\footnote{Here we are considering $\alpha \sim 1$, and ${\bf k}_\text{T}^2 \ll m_N^2$. In the next paragraph we take care of the case when $|{\bf k}|$ may be large.} $|{\bf k}|$, the off-shell light-front-energy to mass ratio,
\begin{equation}
{\Delta^- \over 2m_N} = 2{m_N^2 + {\bf k}^2_\text{T} \over m_N p_d^+} \sim 2m_N/p_d^+
\end{equation}
can be considered a small parameter. 

In the CM frame of the deuteron and the virtual-photon, for high momentum transfer we have $m_N p_d^+ \gg m_N m_d \sim 2 m_N^2$. Therefore, we can classify each of the six vertex functions regarding its contribution to the transition vertex ($\Gamma^\mu$) by counting how many powers of the small factor $(2m_N/p_d^+)$ they contain. 
As follows from Eq.(\ref{vertex}), $\Gamma_1$, $\Gamma_2$, and $\Gamma_5 $ correspond to the leading order contribution, they have zero power of the small factor, $(2m_N/p_d^+)^0$. On the other hand, $\Gamma_{3,4}$ and $\Gamma_{6}$ are suppressed by one power, $(2m_N/p_d^+)^1$, and two powers, $(2m_N/p_d^+)^2$, respectively.

\vspace{0.15in}
 \begin{figure}[h]
\hspace{-.9cm}
\includegraphics[scale=0.9]{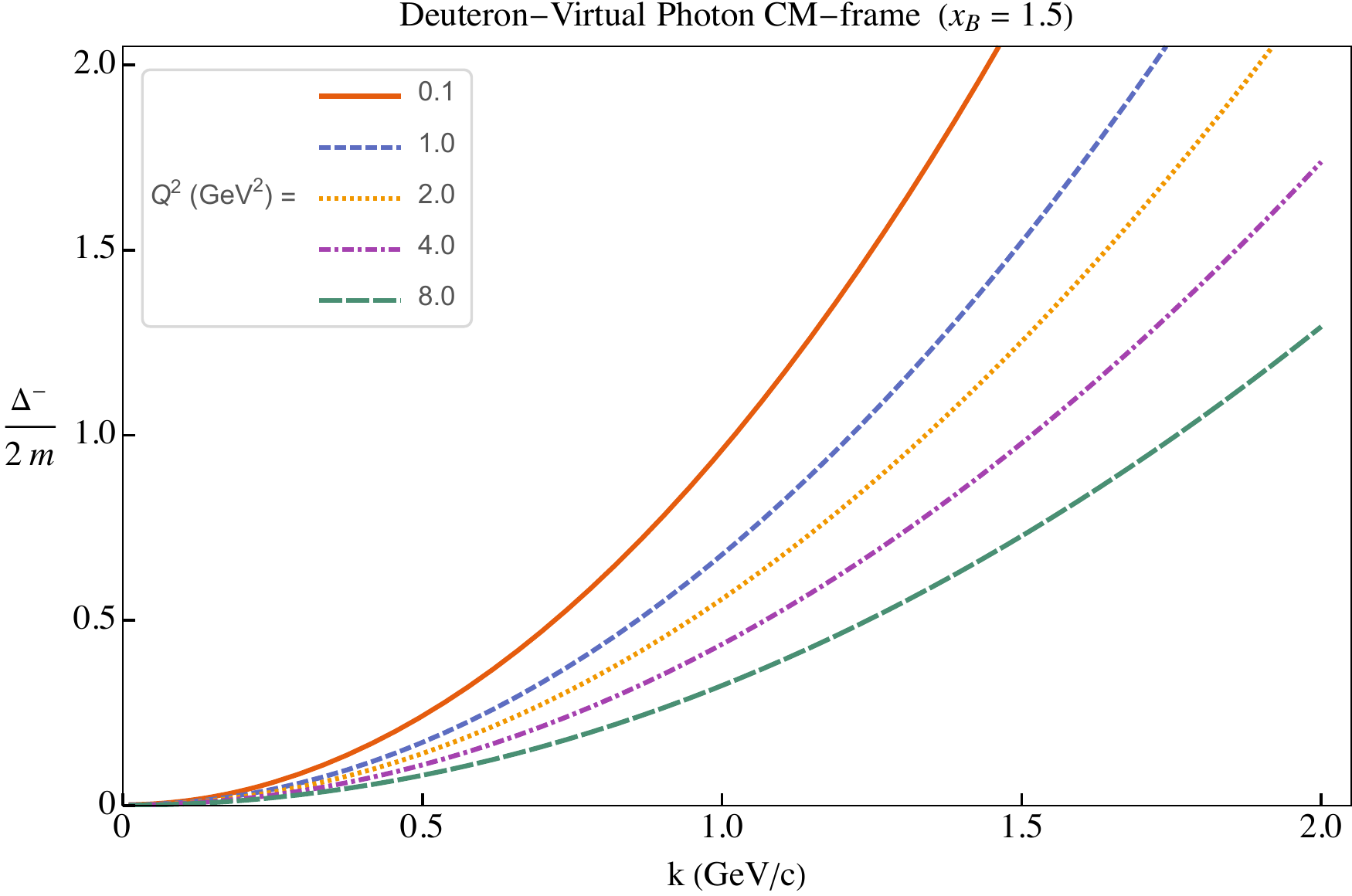}
\caption{The momentum, $k$ dependence of ${\Delta^-\over 2m}$ factor at different 
$Q^2$.}
\label{delta-over2m}
\end{figure}

The conditions under which ${\Delta^- \over 2m_N}$ can be considered a small parameter depend also on the magnitude of the internal momentum. 
In Fig.(\ref{delta-over2m}) we show the dependence of ${\Delta^-\over 2m_N}$ as a function of the internal momentum magnitude ($k=\sqrt{-k^2}=\sqrt{{\bf k}^2}$) for different values of momentum transfer ($Q^2=-q^2$). In order to take advantage of the high energy kinematics, the plot is presented in the ${\color{red}\gamma^*} d$ CM reference frame (Fig. \ref{CM Frame}). It is easy to see in the figure that for $Q^2\ge 4$ GeV/c the parameter remains small, ${\Delta^-\over 2m_N}\le 1/2$, up to high values of the internal momentum  \mbox{$k \sim 1$ GeV/c.} 
 
The next step is to use the fact that at the limit of small $NN$ relative 3-momentum (${\bf p}_\text{rel}$), the non-relativistic quantum mechanical wave function of the deuteron must be recovered. As a result, we will be able to constraint the form of the vertex function that dominates in the high energy limit.

\vspace{0.25in}

\section{Calculation of the Deuteron Wave Function in the Light Front }
\label{wave_function_calc}

Keeping the leading terms in Eq.(\ref{vertex}), containing no powers of $(2m_N/p_d^+)$, the deuteron LF-wave-function reduces to,
\begin{align}
\psi_{s_r s_i}^{\lambda_d}
=  -\sum\limits_{s_i^\prime}\bar u(p_r,s_r) 
 \left\{ \Gamma_1\gamma^\mu + 
\Gamma_2{(p_i-p_r)^\mu\over 2  m_N} + i \Gamma_5{1\over 2m^3_N}\epsilon^{\mu \nu \rho \sigma}{p^+_d}_{\nu} {(p_i - p_r)}_\rho \Delta_\sigma  \gamma_5 \right\}  &      \nonumber \\
 \times \gamma_5 {\epsilon_{s_i,s_i^\prime} \over \sqrt{2}}  u(p_i,s_i^\prime)\chi_\mu^{\lambda_d} \qquad  & 
\label{dwave_lf3}
\end{align}
where, 
${\bf p}_\text{T} = {{{\bf p}_i}_\text{T} - {{\bf p}_r}_\text{T} \over 2}$, 
and the deuteron polarization vectors are chosen as (see Appendix section \hyperref[App Separation into Angular and Radial Dependencies]{I.4} 
and Fig.\ref{Vec Diff}),
\begin{equation}
\chi_\mu^{\lambda_d} = \left(\chi_0^{\lambda_d}, {\bf \chi}_\text{T}^{\lambda_d} ,\chi_z^{\lambda_d}\right) = 
\left({ {P_\text{NN}}_z {s_d}_z \over M_\text{NN}}, {{\bf s}_d}_\text{T}, {E_\text{NN} {s_d}_z \over M_\text{NN}} \right)
\end{equation}
with, 
\begin{align}
{\bf P}_\text{NN} = ( {\bf 0}_\text{T}, {p_1}_z + {p_2}_z), \ \   E_\text{NN} = \sqrt{M_\text{NN}^2 + {\bf P}_\text{NN}^2} \ , \ \  \mbox{and}, \ \ 
M^2_\text{NN} = s_{NN} = 4{(m_N^2+ {\bf p}_\text{T} ^2)\over \alpha(2-\alpha_1)}
\end{align}

Since the wave function in Eq.(\ref{dwave_lf3})  is Lorentz boost invariant along the $z$-axis, we can perform the calculation in the more convenient reference frame obtained by boosting with the velocity ${\bf v} = { {\bf P}_{12} \over E_{12}}$. Such a transformation will result in a wave function of the form:
\begin{align}
\psi_{d}^{\lambda_d}(\alpha_i,{\bf k}_\text{T} ) =  -\sum\limits_{\lambda_2,\lambda_1,\lambda_1^\prime}\bar u(-k,\lambda_2) 
 \left\{ \Gamma_1\gamma^\mu + 
\Gamma_2{{ k}^\mu\over m_N}  +   \sum\limits_{i=1}^{2}
 i\Gamma_5{1\over 2m^3_N}\epsilon^{\mu + i  -}p^{\prime +}_{d} k_{i} \Delta^{\prime-} \gamma_5  \right\}  &      \nonumber \\
 \times \gamma_5 {\epsilon_{\lambda_1,\lambda_i^\prime}\over \sqrt{2}} u(k,\lambda^\prime_1)s_\mu^{\lambda_d}  & 
\label{dwave_lf4}
\end{align}
where, $ k^\mu = (0,{\bf k}_\text{T} ,k_z)$ with $ {\bf k}_\text{T} = {{\bf p}_1}_\text{T}$, and, 

\begin{eqnarray}
k_z & = & \Lambda(v)^z_{\mu}p_1^\mu = {E_{12}\over M_{12}}(p_{1z} - {P_{12}\over E_{12}}E_{1}) = E_{k}(\alpha_1-1) \nonumber \\
\vspace{.2cm}
-k_z & = & \Lambda(v)^z_{\mu}p_2^\mu = {E_{12}\over M_{12}}(p_{1z} - {P_{12}\over E_{12}}E_{1}) = E_{k}(\alpha_2-1)\nonumber \\
{\bf k}_\text{T} & = & {{\bf p}_1}_\text{T} = -  {{\bf p}_2}_\text{T} \nonumber \\
E_k & = & \Lambda(v)^0_{\mu}p_1^\mu  = \Lambda(v)^0_{\mu}p_2^\mu = E_{k} = \sqrt{m_N^2 + k^2}
\label{kdef}
\end{eqnarray}
with\footnote{More details can be found in  \hyperref[App Dirac bilinears - Canonical]{App.(I.3)}. }, 
\begin{equation}
{\bf k}^2 = k_z^2 + {\bf k}_\text{T}^2 \ , \quad \text{and}, \quad  E_{k} = \sqrt{m_N^2 + {\bf k}^2} = {\sqrt{S_{NN}}\over 2}
\end{equation}

Notice that, $p^{\prime +}_{d}$ and  $\Delta^{\prime-}$ correspond  to the Lorentz boosts of respective unprimed quantities  and are expressed as follows:

\begin{align}
p^{\prime +}_{d} & =   {1\over M_{12}} (E_{12}-{P_{12}}_z)p^+_d = {1\over\sqrt{s_{NN}}} 
\left[4(m_N^2 + {\bf k}_\text{T}^2)\over \alpha_1(2-\alpha_1)\right] = \sqrt{s_{NN}} = 2 E_k
\nonumber \\
 \Delta^{\prime-} & =  {1\over M_{12}} (E_{12}+{P_{12}}_z)\Delta^{-} = 
 {1\over \sqrt{s_{NN}}}\left[ {4(m_N^2 + {\bf k}_\text{T}^2)\over\alpha_1(2-\alpha_1)}-m_d^2\right]
 \nonumber \\
 p^{\prime +}_{d} \Delta^{\prime-}   &  = p^{+}_{d} \Delta^{-}  = \left[ {4(m_N^2 + {\bf k}_\text{T}^2)\over\alpha_1(2-\alpha_1)}-m_d^2\right]
 \approx 4{\bf k}^2
\end{align}
where, the last relation is correct up to the binding energy of the deuteron.
Finally, the polarization vector of the deuteron is three-dimensional $s_\mu^{\lambda_d} = (0,{\bf s^\lambda_d})$ in which

\begin{equation}
{\bf s}_d^1 = - {1\over \sqrt{2}} (1,i,0), \ \ \ {\bf s}_d^{-1} = {1\over \sqrt{2}} (1,-i,0) \ \ \ {\bf s}_d^0 = (0,0,1)
\end{equation}
where, the $z$-axis is defined along the direction of the deuteron momentum in the reference frame defined in Sec.(\ref{reframe}).

For numerical calculations of the LF deuteron wave function we adopt the following approximation. Since the term related to the vertex function $\Gamma_5$ is proportional to the virtuality factor 
$p^{+}_{d} \Delta^{-} \approx 4{\bf k}^2$,
at small internal momentum the deuteron LF wave function will approach to the traditional non-relativistic form, with its characteristic
 $S$- and $D$-waves components of the deuteron.  On this regard, one can follow the approach of Ref.\cite{Frankfurt:1981mk}, which is based on the assumption of the angular condition according to which, in the on-shell limit the $\Gamma_5$ term is absent.
Then one can relate the $\Gamma_1$ and $\Gamma_2$ vertices to the radial functions of the $S$- and $D$-wave components in the deuteron wave function, which are evaluated on the momenta $k$ defined in Eq.(\ref{kdef}).

We relate the vertices  $\Gamma_1$ and $\Gamma_2$ to the radial functions $S$- and $D$-wave components in the same way, however keeping the $\Gamma_5$ term which can be evaluated from comparison with specific observables of deuteron electro-disintegration sensitive to the transverse momenta of the bound nucleon in the deuteron.   
Note that such an approach  requires rescaling of the $S$- and $D$- contribution into the overall normalization of the deuteron wave function (see section \ref{wf norm}). 
Following the above prescription one obtains for the $\Gamma_1$ and $\Gamma_2$ vertices (see Appendix section \hyperref[App Separation into Angular and Radial Dependencies]{I.4}),
\begin{align}
\Gamma_1(k) & =   {1\over 2\sqrt{E_k}}\left(U(k) + {W(k)\over \sqrt{2}}\right){1\over \sqrt{4\pi}}\nonumber \\
\Gamma_2(k) & =  {m_N \sqrt{E_k}\over 2 k^2}\left({U(k)(E_k-m_N)\over E_k} - {W(k)\over \sqrt{2}}{(m_N+2E_k)\over E_k}\right){1\over \sqrt{4\pi}}
\end{align}

As it follows from the above relations, both vertex functions are finite and depend on the magnitude of momentum $k$. 
The two vertex functions $\Gamma_1$ and $\Gamma_2$ are shown in Fig.(\ref{vertices G12}) as a function of $k$ for the three realistic potentials, Paris, AV-18, and CD-Bonn. 
One can not make a similar correspondence for $\Gamma_5$ vertex function since as we will show below it corresponds to the effective P-wave that vanishes in the non-relativistic limit.

 \begin{figure}[h]
\hspace{-2.2cm}
\includegraphics[scale=0.73]{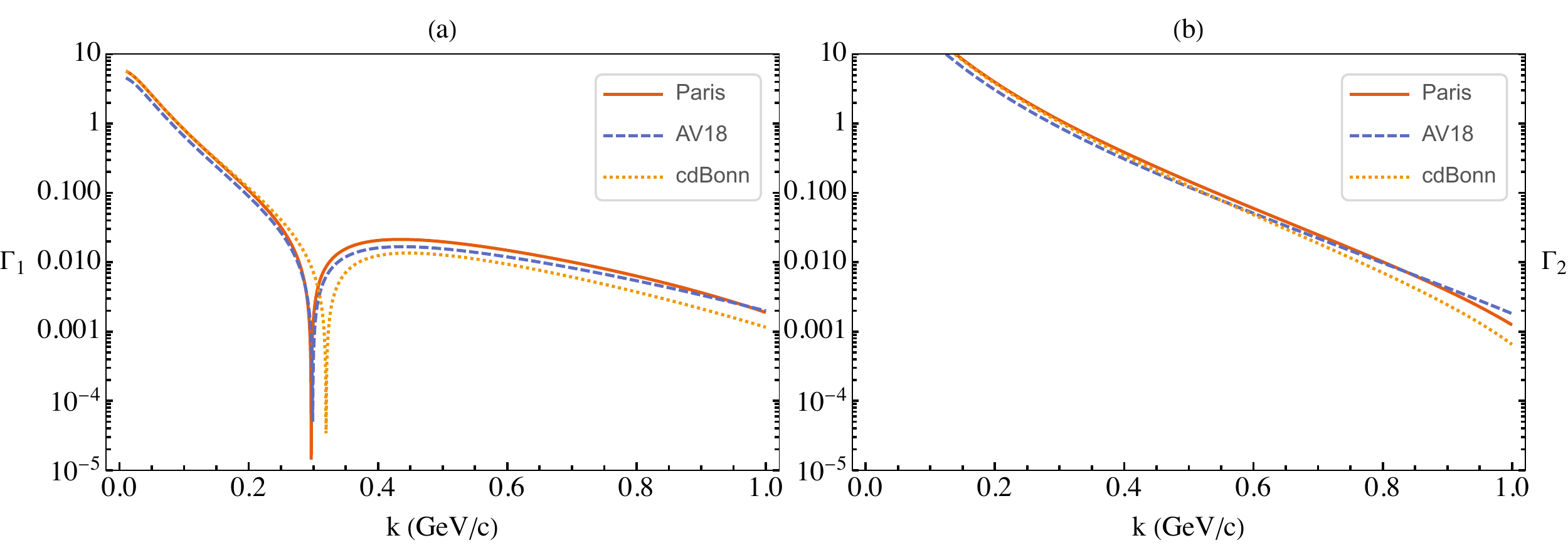}
\caption{The momentum, $k$ dependence of the vertex function $\Gamma_1$ and $\Gamma_2$ calculated 
for different $NN$ potentials.}
\label{vertices G12}
\end{figure}

To proceed, it is convenient to represent the LF deuteron wave function through to two-dimensional spinors in the form (see Appendix \hyperref[RWF]{I} for more details),
\begin{align}\label{wf 125}
\hspace{-.85cm}
\psi_d^{\lambda_d}(\alpha_1,k_t,\lambda_1,\lambda_2)   = &  {1\over \sqrt{4\pi}}\sum\limits_{\lambda_1^\prime}\phi^\dagger_{\lambda_2} \sqrt{E_k}\left[  U(k) ( \sigma \cdot {\bf s}_d^{\lambda} )
- {W(k)\over \sqrt{2}}\left( { 3 (\sigma \cdot {\bf k})({\bf k} \cdot {\bf s}_d^\lambda)\over k^2} - \sigma \cdot {\bf s}_d^\lambda \right) \right. \nonumber \\
 &  \left. \qquad\qquad \hspace{.85cm} +  {4\sqrt{E_k}\over \sqrt{3}} \left({k\over m_N}\right)^3  f_5(k) \sum\limits_{i=1}^{2}{-i\sqrt{3}( {\bf k} \times {\bf s}_d^\lambda )_z \over k}\right]
{\epsilon_{\lambda_1,\lambda_1^\prime}\over \sqrt{2}} \phi_{\lambda^\prime_1}
\end{align}
In Eq.(\ref{wf 125}) the angular dependency of the $f_5$ factor can be written as,

\begin{equation}
-i\sqrt{3}\sum\limits_{i=1}^{2}{-i\sqrt{3}( {\bf k} \times {\bf s}_d^{\pm 1})_z\over k} =  \pm \sqrt{4\pi}Y^{\pm}_1(\theta,\phi)
\end{equation}
which allows to identify the radial part of the effective ``P"-wave in the form,

\begin{equation}\label{P-wave def}
{k^2\over m_N^2} P(k) = {k^2\over m_N^2} \left[{4\sqrt{E_k}\over \sqrt{3}}{k\over m_N}  f_5(k) \right]
\end{equation}

\vspace{.2cm}
In Eq(\ref{P-wave def}), one power of ${k\over m_N}$ is included in the definition of the radial wave function to satisfy the quantum-mechanical relation that for a radial wave function corresponding to the orbital angular momentum quantum number $l$, in the limit of small momentum ($k$) it behaves as, $\lim_{k\to 0}  R_l(k) \sim k^l$. The remaining ${k^2\over m_N^2}$ factor indicates the relativistic nature of the ``P"-wave term that vanishes in the non-relativistic limit.

\begin{figure}[h]
\hspace{-1.cm}
\includegraphics[scale=0.8]{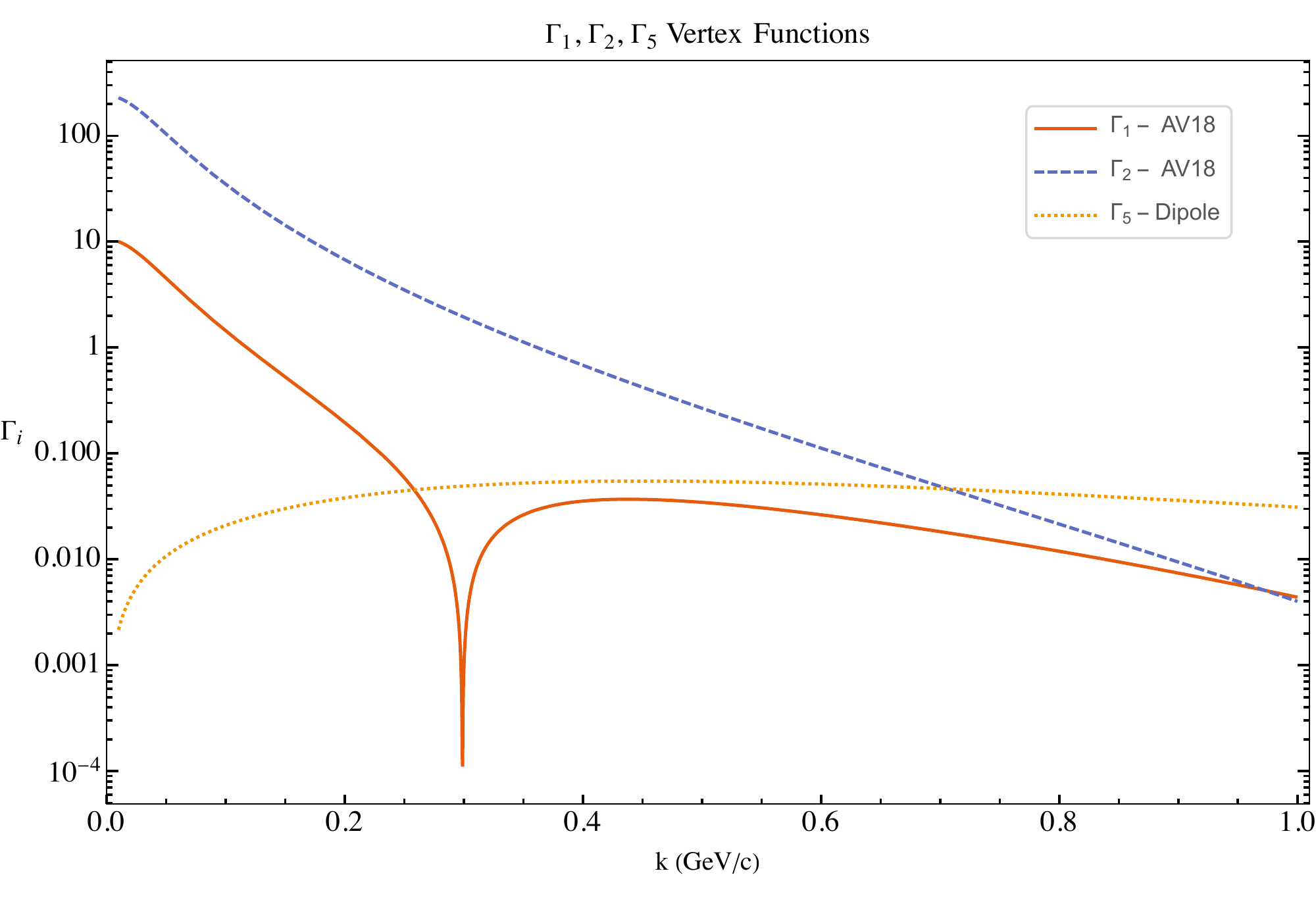}
\caption{The momentum, $k$ dependence of the vertex function $\Gamma_1$, $\Gamma_1$, and $\Gamma_5$.  $\Gamma_1$, $\Gamma_1$ have been calculated using the CD-Bonn potential, while for $\Gamma_5$ has been parameterized using the dipole model shown in the text.}
\label{vertices G125}
\end{figure}

For a  comparison among the three vertex functions we plot them in Fig.(\ref{vertices G125}).  
Here, we have employed for $\Gamma_5$ a parameterization using a simple dipole-like model,
\begin{equation}
\Gamma_5 (k)  = P (k)    = {k\over m_N} {4\sqrt{E_k}\over \sqrt{3}}   \frac{1}{\sqrt{4 \pi }} \frac{ g_{\omega }}{ \left(k^2+m_{\omega}^2\right)^2} 
\label{dipole model}
\end{equation}
with a typical value for the coupling, $\frac{ g_{\omega }^2}{4\pi}=20$ \cite{Machleidt2001}.
The dipole-like term is physically motivated, e.g., it provides the correct asymptotic behavior, for small momentum ($k\to 0$)  the P-wave behaves as $k^{L}$ with $L=1$, and for large $k$ the vertex function ($\Gamma_5 (k) $) scales as  $k^{-2}$, which guarantees that the momentum distribution will scale as $k^{-4}$, which is shown in Fig.(\ref{Mom dist scaling}). Therefore, the dipole parameterization can be seen as a simple interpolation between the two limit behavior of the momentum distribution, that is, for small and high momenta. 

\begin{figure}[H]
\hspace{-2.2cm}
\includegraphics[scale=0.73]{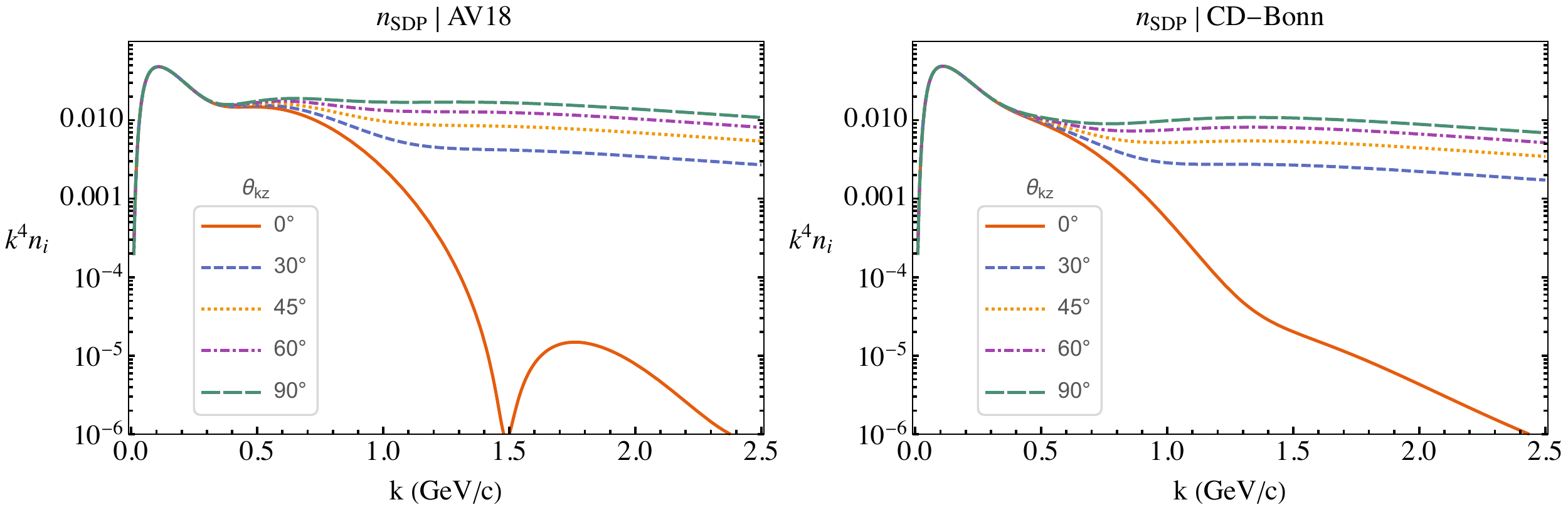}
\caption{The $k^{-4}$ scaling of the momentum distribution. The right panel only includes the S and D  radial wave functions calculated with the AV18 potential. The left panel also includes the P-wave-like term.}
\label{Mom dist scaling}
\end{figure}

\begin{figure}[h]
\hspace{-2.2cm}
\includegraphics[scale=0.74]{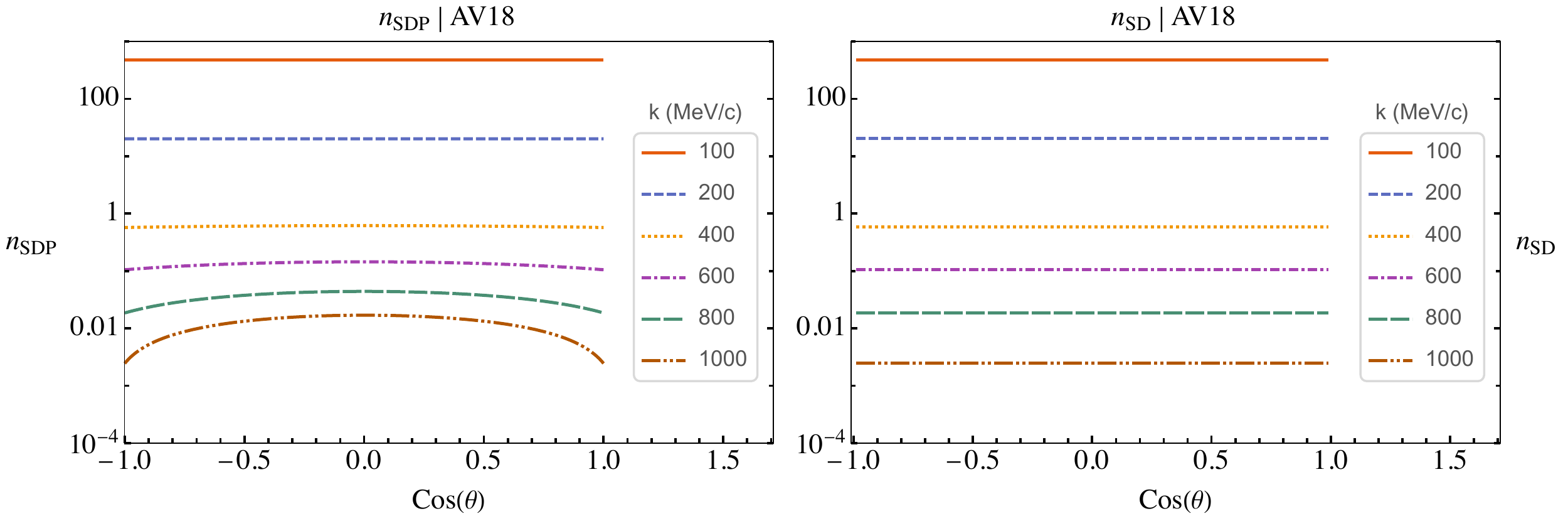}
\caption{Angular dependency of $n_{SDP}$ using the AV18 potential for the S and D radial wave functions, and the dipole model for the P-wave.}
\label{Ang dist AV18}
\end{figure}

In Fig.(\ref{Ang dist AV18}) we show the angular dependency for different values of internal momentum ($\bf{k}$) that results from the inclusion of the $\Gamma_5$ term in the deuteron's wave function.
The panel on the right only includes the radial S and D wave functions, while the left panel shows the effect of including also the  P-wave-like structure within the dipole model. As expected, at low momentum they are similar, and at high momentum there is an angular dependency coming from the P-wave term. Moreover, it is easy to see that (for each value of k) the (constant) angular distribution from $n_{SD}$ (with only S and D waves) lays at the minimum of $n_{SDP}$ (which includes also the P-wave), this minimum occurs at  $\cos{\theta}=-1,0,1$, i.e., parallel, antiparallel, and transverse to the 3-momentum transfer ($\bf q$).
  
Since the ``effective P"-wave does not contribute to the $\lambda=0$ polarization of the deuteron, 
the most important implication of this effect will be the  polar angle dependency of the unpolarized momentum distribution function extracted at large momenta.  Another effect will be the enhancement of the tensor-polarization of the deuteron, again in the large momentum limit.  

\section{Normalization}
\label{wf norm}

The normalization condition of the above wave function is defined according to the observable quantities such as deuteron baryonic number, or  the charge form factor \mbox{$G_C(Q^2=0)=1$}. Both approaches result in the normalization condition 
(see \hyperref[Deuteron EM Current]{\mbox{Appendix J}}),
\begin{equation}
{1\over 3} \sum\limits_{\lambda_d, s_r,s_i}\int \mid \psi_{d}^{s_i,s_r,\lambda_d} (\alpha, {\bf p}_\text{T})\mid^2 {d\alpha\over (2-\alpha)\alpha}d^2{\bf p}_\text{T}=1
\label{normcon}
\end{equation}

Using the dipole model for the $\Gamma_5$ vertex function, we can estimate the angular averaged one-body momentum distribution. A comparison with the equivalent momentum distribution that only takes into account the S and D waves contributions is shown in Fig.(\ref{norm G125}). In both cases the S and D-waves are calculated using the AV18 potential.

\begin{figure}[H]
\hspace{-1.8cm}
\includegraphics[scale=0.85]{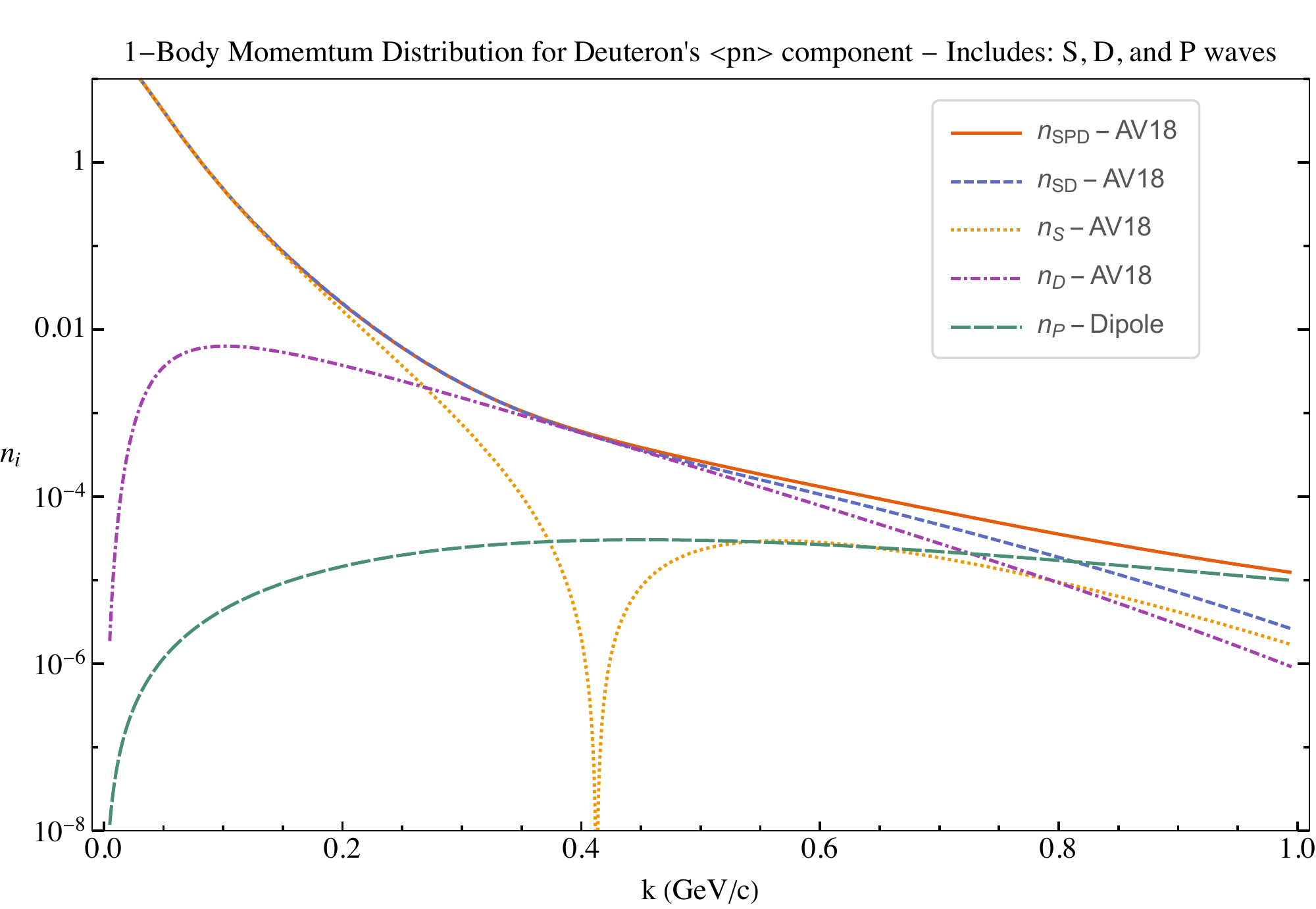}
\caption{Momentum distribution, $n(k)=n_S(k)+n_D(k)+n_P(k)$, and $n(k)=n_S(k)+n_D(k)$, together with individual contributions, $n_S(k)$, $n_D(k)$, $n_P(k)$.}
\label{norm G125}
\end{figure}

\chapter{FINAL STATE INTERACTIONS}
\label{Ch FSI}

During the last two decades significant efforts have been made  in the calculation of FSI effects in high $Q^2$  electro-nuclear processes (see e.g. Refs.\cite{Frankfurt:1996xx,Sargsian:2001ax,CiofidegliAtti:2004jg,Jeschonnek:2008zg,Laget04,eheppn1,eheppn2,Sargsian:2009hf,disfsirev}).  
In the kinematics where the momentum transferred at the rescattering vertex is significantly smaller than the momentum of the fast nucleon there is an especially reliable approach called generalized eikonal approximation (GEA)\cite{Frankfurt:1996xx,Sargsian:2009hf}.
The GEA is a self-consistent procedure  the relativistic effects associated with the large momentum of the nucleons involved in the reaction, and provideds a  theoretical framework for calculating FSI effects relevant to studies of the nuclear structure at short distances.

\section{The Generalized Eikonal Approximation}
In the kinematics in which the struck nucleon carries about the same high momentum ({$\ge 1$}~Gev/c) of the virtual photon, the final state interaction process can be described within GEA \cite{Frankfurt:1996xx,Sargsian:2001ax}.  
The GEA is especially reliable in the situations in which the momentum transfer in the rescattering vertex is significantly smaller than the momentum of the fast nucleon.
This covariant approach is based in  Feynman rules defined for effective interaction vertices.

In this chapter we carry out the calculation of the {FSI} diagrams of  \mbox{Fig.(\ref{edepn_diagrams} (b))}
within the LF, i.e., 
we rewrite the (covariant) Feynman diagram as the {\color{blue} $\tau$-ordered} non-covariant diagrams in {Fig.(\ref{graphFSI})}.
The variables used for the {\color{blue} $\tau$-ordered} FSI transition amplitudes are defined in {Fig.(\ref{graphFSI})}, where {$p_f$} and {$p'_f$} are the final and intermediate momentum of the fast (struck) nucleon.  

\begin{figure}[h]
\centering
\includegraphics[scale=0.75]{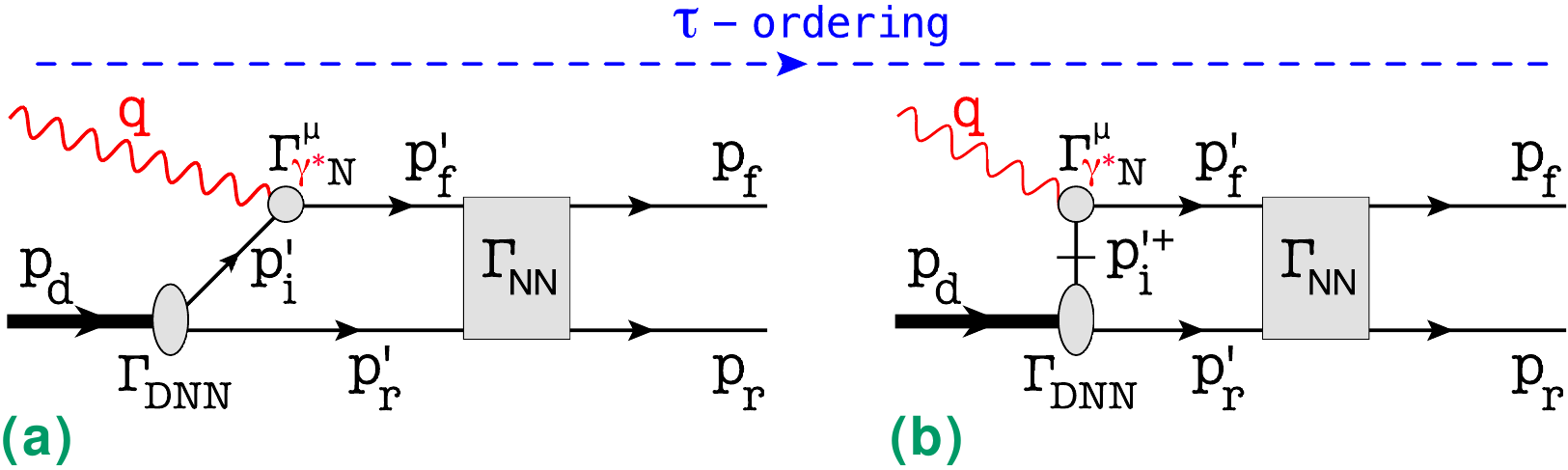}  
\caption{FSI $\tau$-ordered diagrams.} 
\label{graphFSI}
\end{figure}

In the case the virtual photon transfers a large momentum to the struck nucleon, and working in the reference frame of {Fig.(\ref{scatt react plane})}, the intermediate light-front longitudinal momentum of the nucleon will be very small, {$p'^+_f \sim 0$}.
Because the LF longitudinal component of the momentum {($p^+$)} is conserved by the interaction, and it is positive for real particles, it follows that choosing events for which {$p^+_f \sim p'^+_f \sim 0$}, {i.e. ${p_f}_z \sim |{\textbf{q}}|$}, the re-scattering between the struck and the recoiling nucleon is constrained to be mostly along the direction transverse to {$\textbf{q}$}. 
This peculiar feature of the {GEA}   leads to a strong angular anisotropy for the corresponding {FSI} and  can be used to aid the extraction of the probability distributions provided by the {PWIA} diagram, e.g. selecting kinematics that minimize the {FSI} effects.

\subsection{Calculation of Final State Interactions Amplitudes}

To calculate final state interaction  one considers the light-front diagrams of Fig.(\ref{graphFSI}). 
 Similar to the case of PWIA (Chapter \ref{PWIA}), the interacting nucleon enters with a propagating and instantaneous parts.
which results in an electromagnetic 
current as in Eq.(\ref{EM current off-on}).


The latter will result in a diagram of Fig.(\ref{graphFSI} (b)) where the vertical 
dashed line indicates the intermediate state in light-front time sequence of the scattering process.
Note that this diagram still contains instantaneous propagators of recoil (r$^\prime$) and struck (f$^\prime$) 
nucleons whose contributions will be discussed below.
Applying now effective light-front diagrammatic rules from  \hyperref[App. LFPT]{Appendix B}, for the scattering amplitude one obtains,
\begin{equation}
A_1^\mu = \int {\bar u_{h_f}(p_f)\bar u_{h_r}(p_r) \Gamma_{NN}[ \shl p_{r^\prime} + m] [ \shl p_{f^\prime} + m]\Gamma_{\gamma^* N}^\mu 
[ \shl p_{i^\prime} + m] \Gamma_{DNN} \chi^{\lambda_d}\over p^+_{f^\prime}\ D_2  \ p^+_{i^\prime} \ D_1 }
{d p^+_{r^\prime}\over p^+_{r^\prime}} {d^2 p^\perp_{r^\prime}\over 16\pi^3}
\label{A1a}
\end{equation}
where, $D_1$ and $D_2$ correspond to the light-front energy denominators, given by the expressions,
\begin{eqnarray}
D_1 &  = & p_d^- - p^-_{r^\prime} -  p^-_{i^\prime} + i\epsilon \nonumber \\
D_2 &  = & p_d^- - p^-_{r^\prime}  + q^- -  p^-_{f^\prime} + i\epsilon
\end{eqnarray}
and $\Gamma_{NN} $ represents the $NN$ rescatering amplitude.

For further derivation we introduce light-cone momentum fractions, 
\begin{equation}
\alpha_{i^\prime}  = 2{p^+_{i^\prime}\over p_d^+} \ \ \mbox{and} \ \ \alpha_{r^\prime}  = 2{p^+_{r^\prime}\over p_d^+}
\label{alphaprimes}
\end{equation}
and use the "+" component conservation to obtain, $\alpha_{f^\prime} = \alpha_{i^\prime} + \alpha_q$.

We now consider the instantaneous parts of propagators of recoil and struck nucleons. For the recoil nucleon we get,
\begin{equation}
 \shl p_{r^\prime} + m = \sum\limits_{h_r^\prime} u_{h_{r^\prime}}(p_{r^\prime}) \bar u_{h_{r^\prime}}(p_{r^\prime}) + {2(p_{r^\prime}^2- m^2)\gamma^+\over \alpha_{r^\prime}p_{d}^+}
 \label{propr}
 \end{equation}
where, the instantaneous part will be dominated in the integral of Eq.(\ref{A1a}) at $\alpha_{r^\prime}\rightarrow 0$ limit 
which corresponds to strongly virtual nucleon emerging form the deuteron vertex and instantaneously interacting with 
struck nucleon. Moreover, the magnitude of the momentum transfer in the rescattering vertex is  
$\sim \alpha_r^2p_d^{+,2}$, which provide another factor of suppression for the term contributing to the instantaneous part  of the recoil nucleon propagator. 
 For the struck nucleon propagator we have, 
\begin{equation}
 \shl p_{f^\prime} + m = \sum\limits_{h_f^\prime} u_{h_{f^\prime}}(p_{f^\prime}) 
 \bar u_{h_{f^\prime}}(p_{f^\prime}) + {2(p_{f^\prime}^2- m^2)\gamma^+\over \alpha_{f^\prime}p_{d}^+}
 \label{propf}
 \end{equation}
The 
reason leading to
the suppression of the instantaneous terms as compared to the propagating term is that, 
due to the knock-out kinematics we have $p_{f^\prime}^2\approx m^2$, hence the term is dominated by the 
$\alpha_{f^\prime}\rightarrow 0$ limit. Therefore, the momentum transfer in the rescattering amplitude is again large 
$\sim \alpha_f^2p_d^2$. 

Thus within the 
approximation
for which one expects that FSI is dominated by small momentum transfer rescattering (eikonal 
approximation) one can keep only propagating terms from Eq.(\ref{propr}) and Eq.(\ref{propf}) in the amplitude of 
Eq.(\ref{A1a}), yielding:
 \begin{align}
A_1^\mu = &   \sum\limits_{h_{r^\prime},h_{f^\prime}} 
\int {\bar u_{h_f}(p_f)\bar u_{h_r}(p_r)\Gamma_{NN} u_{h_{r^\prime}}(p_{r^\prime})  u_{h_{f^\prime}}(p_{f^\prime}) \over 
p^+_{f^\prime}\ D_2} \nonumber \\
& \times {  \bar u_{h_{f^\prime}}(p_{f^\prime})  \Gamma_{\gamma N}^\mu 
[ \sum\limits_{h_{i^\prime}} u_{h_{i^\prime}}(p_{i^\prime}) \bar u_{h_{i^\prime}}(p_{i^\prime}) + 
{\gamma^+\over 2} (p_{i^\prime}^- - p_{i^\prime,on}^-) ] 
\bar u_{h_{r^\prime}}(p_{r^\prime}) \Gamma_{DNN} \chi^{\lambda_d}\over   
\alpha_{i^\prime} {1\over 2} \left[ m_d^2 - {4(m^2+ p_{i^\prime,\perp}^2)\over \alpha_{i^\prime}(2-\alpha_{i^\prime} )}\right]}
{d \alpha_{r^\prime}\over \alpha_{r^\prime}} {d^2p^\perp_{r^\prime}\over 16\pi^3}
\label{A1b}
\end{align}
where, in the derivation we used the definitions of $\alpha_{r^\prime}$ and $\alpha_{i^\prime}$ from Eq.(\ref{alphaprimes}) as 
well as the relation:
\begin{equation}
D_1 = {1\over p_d^+}\left[m_d^2 - {4(m^2 + p_{i^\prime,\perp})\over \alpha_{i^\prime}(2-\alpha_{i^\prime})}\right]
\end{equation}

Using now the definition of Light-Front wave function from Eq.(\ref{wflf}), propagation and instantaneous component of 
nucleon's electromagnetic current from 
Eq.(\ref{EM current off-on}), 
as well as defining the $NN\to NN$ scattering amplitude as:
\begin{equation}
F_{NN}(p_3,h_3,p_4,h_4;p_1,h_1,p_2,h_2) \equiv \bar u_{h_3}(p_3)\bar u_{h_4}(p_4)\Gamma_{NN} 
u_{h_1}(p_1)u_{h_2}(p_2)
\label{FNN}
\end{equation}
for scattering amplitude one obtains:
 \begin{align}
A_1^\mu  = &    -\sum\limits_{h_{r^\prime},h_{f^\prime}h_{i^\prime}} 
\int {F_{NN}(p_{f},h_f,p_r,h_r;p_{f^\prime},h_{f^\prime},p_{f^\prime},h_{r^\prime})  \over 
p^+_{f^\prime}\ D_2} 
   J_N^\mu(p_{f^\prime},h_{f^\prime};p_{i^\prime},h_{i^\prime})  \nonumber \\
& \qquad\qquad \times {\psi_d(\alpha_{i^\prime},p_{i^\prime,\perp})\sqrt{2}\sqrt{16\pi^3} \over \alpha_{i^\prime}}
 {d \alpha_{r^\prime}\over \alpha_{r^\prime}} {d^2p^\perp_{r^\prime}\over 16\pi^3}
\label{A1b}
\end{align}

Now we consider the denominator $p^+_{f^\prime}\ D_2$ taking into account that the 
reaction (\ref{reaction}) is quasi-elastic and satisfies energy-momentum conservation according to which:
\begin{equation}
(p_d - p_r +q)^2 = m_N^2.
\label{QEkin}
\end{equation}
Using above equation and on-shellness of the recoil nucleon in the intermediate state, one obtains:
\begin{eqnarray}
p^+_{f^\prime}\ D_2 &= & (p_d - p_r + q)^2 + 2(p_r-p_{r^\prime})(p_d - p_r + q) + (p_r - p_{r^\prime})^2 - m_N^2 + i\epsilon \\ \nonumber
& = & {s+Q^2 - q^{+}p_{d}^-\over 2} \left[ \alpha_r - \alpha_{r^\prime} + \delta + i\epsilon\right]
\end{eqnarray}
with
\begin{equation}
\delta = {4\over (p_d^{+})^2} \left({p_{r,\perp}^2\over \alpha_r} - {p_{r^\prime,\perp}^2\over \alpha_{r^\prime}}\right)
\approx {4 (p_{r,\perp}^2- p_{r^\prime,\perp}^2) \over (p_d^{+})^2 \alpha_r},
\label{delta}
\end{equation}
were in the last part we used the fact that due to the peaking of the deuteron wave function at small internal momenta,
the integral in Eq.(\ref{A1b}) is dominated by $p_{r^\prime,\perp}^2\sim 0$, thus  effect due to the 
replacement of   $\alpha_{r^\prime}$ by $\alpha_{r}$  in the ${p_{r^\prime,\perp}^2\over \alpha_{r^\prime}}$ term will be 
negligible.  Note that in the above equation $p_d^{+}$ is defined according to Eq.(\ref{refframeQ}).

Substituting  Eq.(\ref{delta}) in Eq.(\ref{A1b}) one obtains:
\begin{align}
A_1^\mu   = &   \sum\limits_{h_{r^\prime},h_{f^\prime}h_{i^\prime}} 
\int {F_{NN}(p_{f},h_f,p_r,h_r;p_{f^\prime},h_{f^\prime},p_{f^\prime},h_{r^\prime})  \over 
 {s+Q^2 - q^{+}p_{d}^-\over 2} \left[ \alpha_r - \alpha_{r^\prime} + \delta + i\epsilon\right]} 
   J_N^\mu(p_{f^\prime},h_{f^\prime};p_{i^\prime},h_{i^\prime})  \nonumber \\
& \qquad\qquad \times {\psi_d(\alpha_{i^\prime},p_{i^\prime,\perp})\sqrt{2}\sqrt{16\pi^3} \over \alpha_{i^\prime}}
 {d \alpha_{r^\prime}\over \alpha_{r^\prime}} {d^2p^\perp_{r^\prime}\over 16\pi^3}.
\label{A1c}
\end{align}

Before to proceed with the calculation 
we evaluated that $\delta$ function for several  high energy kinematics,
using  the empirical observation that average transverse momentum transferred in  $NN\to NN$ scattering 
is about $250$~MeV/c.  
As it turns out 
the $\delta$ is negligible enough to be ignored 
in the calculations. The outcome of this is that the rescattering amplitude $A_1^\mu$ evaluated at the pole value of 
the denominator conserves the variable $\alpha_r$. This represents a unique feature of high energy scattering 
on the light-front. 

To evaluate the integral in Eq.(\ref{A1c}) we use the relation:
\begin{equation}
{1\over  \alpha_r - \alpha_{r^\prime} + \delta + i\epsilon} = -i\pi \delta(\alpha_{r^\prime} - (\alpha_r+\delta)) 
+ {\cal P} {1\over  \alpha_r - \alpha_{r^\prime}+\delta}
\label{delprin}
\end{equation}
where the second part of the right hand side of equation corresponds to the contributed in Eq.(\ref{A1c}) where 
integral over $\alpha_{r^\prime}$ is evaluated through the  principal value  integration.  
Substituting Eq.(\ref{delprin}) in Eq.(\ref{A1c}) and taking the delta function through the $d\alpha_{r^\prime}$ integration, we split the expression for the amplitude $A_{1}^\mu$ into two parts. 
The first  in which the rescattering is  defined by on-shell elastic $NN$ scattering amplitude and the other in which NN scattering is half-on-shell. 

Furthermore, it is convenient to redefine the $NN$ scattering amplitude in the form:
\begin{equation}
F_{NN}(p_{f},h_f,p_r,h_r;p_{f^\prime},h_{f^\prime},p_{f^\prime},h_{r^\prime}) = \sqrt{s(s-4m_N^2)} 
f_{NN}(s,t,h_f,h_r;h_{f^\prime},h_{r^\prime})
\label{fnn}
\end{equation}
In the case of small angle scattering, $f_{NN}$ corresponds to the diffractive scattering amplitude, which has a well known form as a function of the invariant momentum transfer,
\begin{equation}
t = (p_{r} - p_{r^\prime})^2.
\end{equation}
and can be extracted from  {$NN$} cross-section measurements.

Substituting Eq.(\ref{delprin}) and Eq.(\ref{fnn}) into Eq.(\ref{A1c}) one obtains:
\begin{align}
A_1^\mu  = &   i\sum\limits_{h_{r^\prime},h_{f^\prime}h_{i^\prime}} 
{\sqrt{2}\sqrt{16\pi^3}\over 2}\int {\sqrt{s(s-4m_N^2)}f^\text{on}_{NN}(s,t,h_f,h_r;h_{f^\prime},h_{r^\prime})  \over 
 s+Q^2 - q^{+}p_{d}^-} 
   J_N^\mu(p_{f^\prime},h_{f^\prime};p_{i^\prime},h_{i^\prime})  \nonumber \\
& \qquad \qquad \qquad\qquad \times {\psi_d(\tilde \alpha_{i},p_{i^\prime,\perp}) \over \tilde \alpha_{i}\tilde \alpha_{r}}
 {d^2p^\perp_{r^\prime}\over (2\pi)^2} + \nonumber \\
 &-  \sum\limits_{h_{r^\prime},h_{f^\prime}h_{i^\prime}} 
{\cal P}\int {\sqrt{s(s-4m_N^2)}f^\text{off}_{NN}(s,t,h_f,h_r;h_{f^\prime},h_{r^\prime})  \over 
 {s+Q^2 - q^{+}p_{d}^-\over 2} \left[ \alpha_r - \alpha_{r^\prime} + \delta \right]} 
   J_N^\mu(p_{f^\prime},h_{f^\prime};p_{i^\prime},h_{i^\prime})  \nonumber \\
& \qquad \qquad \times {\psi_d(\alpha_{i^\prime},p_{i^\prime,\perp})\sqrt{2}\sqrt{16\pi^3} \over \alpha_{i^\prime}}
 {d \alpha_{r^\prime}\over \alpha_{r^\prime}} {d^2p^\perp_{r^\prime}\over 16\pi^3},
\label{A2c}
\end{align}
where, $f_{NN}^\text{on}$ and $f_{NN}^\text{off}$ amplitudes correspond to one-shell and half-off-shell amplitudes of $NN$
scattering.

\subsubsection{Forward Final Sate Interaction Contribution}
In this case the lines labeled by momenta of $p_{r^\prime},p_{r}$ and $p_{f^\prime},p_{f}$ correspond to the 
same nucleons and $f_{NN}$ corresponds to a elastic forward scattering amplitude, which in high energy limit 
can be parameterized in the following form:
\begin{equation}
f^\text{on}_{NN}(s,t,h_f,h_r;h_{f^\prime},h_{r^\prime}) = \sigma_{tot}(i + \alpha)e^{{B\over 2}t}\delta_{h_r,h_{r^\prime}}\delta_{h_f,h_{f^\prime}},
\end{equation}
where $\sigma_{tot}$, $\alpha$ and $B$ correspond to the total cross section, real part and slope factor of 
small angle elastic NN scattering amplitude which can be taken from experiments on elastic NN scattering.

For the off-shell part we use similar parameterization as in Ref.\cite{Sargsian:2009hf}:
\begin{equation}
f^\text{off}_{NN}(s,t,h_f,h_r;h_{f^\prime},h_{r^\prime})  = f^\text{on}_{NN}(s,t,h_f,h_r;h_{f^\prime},h_{r^\prime}) 
e^{{B\over 2}(m_\text{off}^2-m_{N}^2)}
\label{foff}
\end{equation}
where,
\begin{equation}
m_\text{off}^2 = (m_d-E_{r^\prime} + q_0) - ({\bf q-p_{r^\prime}})^2
\end{equation}
with, $E_{r^\prime}$, $q_0$,  $\bf q$ and $\bf p_{r^\prime}$ are defined in the lab frame of the deuteron.

%
%
%
%

\section{Numerical Evaluations and Discussions of Results}

We  now proceed  to study the effects of including  the Final State Interaction  mechanism for the reaction as compared with only account for the the PWIA. To this end, we define the following ratio between the two cross sections, 
\begin{equation}
R={\sigma^\text{PWIA+FSI} \over \sigma^\text{PWIA} }
\end{equation}
The cross section $\sigma^\text{PWIA}$, is calculated by inserting in Eq.(\ref{cross-section}) the probability amplitude $A^{\mu} _{0,\text{dir}}$, given by Eq.({\ref{A0sb}}). On the other hand, $\sigma^\text{PWIA+FSI}$ is the result of using the amplitude $A^\mu = A^{\mu} _{0,\text{dir}} + A^{\mu}_{1,\text{dir}} $ (Eq. \ref{relevant amplitudes}), where $A^{\mu}_{1,\text{dir}}$ is given by Eq.(\ref{A2c}).

\begin{figure}[h]
\hspace{-2.cm}
\includegraphics[scale=0.72]{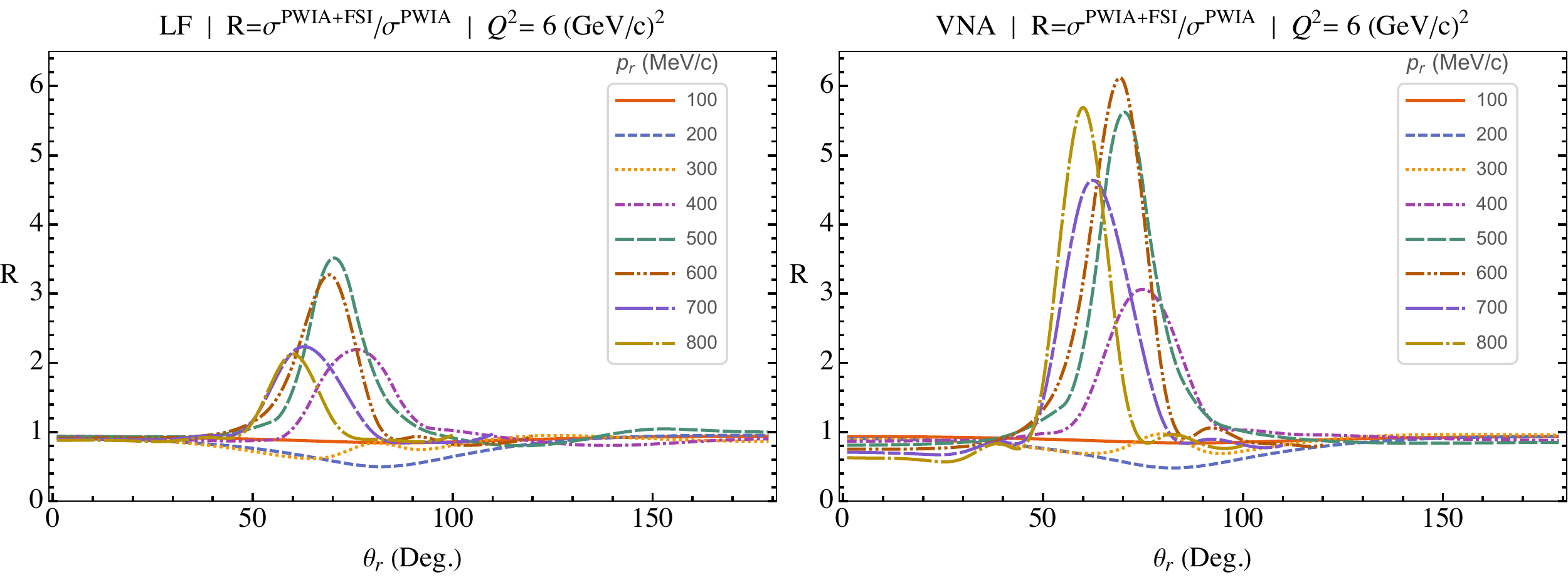}
\caption{ependence of the ratio R on the recoil angle of the neutron for a kinematics with $Q^2=6$~(GeV/c)$^2$, and different values of missing momenta ($p_r$). The left panel shows the ratio $R$ between the two cross sections calculated in this work within the Light-Front framework. The right panel shows the same ratio, but for calculations based on the Virtual Nucleon Approximation from M. Sargsian\cite{Sargsian:2009hf}.
}
\label{Ratio of PWIA-FSI to PWIA}
\end{figure}

The numerical calculations 
are presented in {Fig.(\ref{Ratio of PWIA-FSI to PWIA})}. It can be seen from the figure that both calculations account for the characteristic anisotropy of the eikonal regime, found in \cite{Sargsian:2009hf}. As expected, at small values of the momentum, the prediction for the ratio $R$ from both calculations (VNA and LF) is similar, while for high momenta their predictions diverge. The difference at high momentum is the result of relativistic effects, which are not accounted for in the VNA. The disagreement starts to be noticeable already at $p_r \sim 300$~MeV/c, and it becomes quite prominent as the value of the momentum increases. 
The result derived within the LF formalism 
shows that the effects of the FSI mechanism are much more sensitive to the increase of the missing momentum that what would  be expected from the use of VNA.
In particular, the LF result predicts that the contribution of the FSI, relative to that from the PWIA, drops rapidly as the missing momentum increases.
We expect these interesting results will emerge from experiments as more data for the high energy kinematics electro-disintegration of the deuteron become available.

\chapter{SUMMARY OF THE RESULTS}	
\label{Summary}

In this dissertation, we have developed a theoretical technique for the description of the deuteron electro-disintegration reaction in high energy kinematics within a completely relativistic approach based on Light Front dynamics. The advantage of this new procedure is twofold, it simplifies the calculations without sacrificing accuracy, and it also allows a transparent interpretation of the physical processes involved in the reaction.

Explicitly, we present a new approach for calculating the electromagnetic transition current of a nucleon from a bound state to a free state. We also introduce a new procedure to describe the relativistic nucleonic composition of the deuteron in the form of its LF wave function and identify a new term that dominates the relativistic structure of the $NN$ bound system.

The relativistic formulation of these two objects, that is,  the electromagnetic current of the bound nucleon and the wave function of the deuteron, are the main achievements of this dissertation.

\section{The Light-Front Electromagnetic Current of the Bound-Nucleon}


Based on the light-front  approach we calculated electron-deuteron scattering within PWIA which allowed us to isolate the electron-bound-nucleon scattering cross-section, $\sigma_{eN}$.  Within the LF framework the contribution from processes where the exchanged photon couples to non-nucleonic constituents can be tamed. In particular the so called Z-graphs naturally disappears while the off-shell nature of the  nucleon results in the appearance of an extra  term in the electromagnetic current of electron-bound nucleon scattering called the instantaneous term.
In deriving $\sigma_{eN}$ we separated the propagating and instantaneous contributions in the electromagnetic current which allowed explicitly to trace the effects associated with the binding of the nucleon.  
Furthermore, within the LF framework we were able to identify the parameter (defined as $\eta$) that universally characterizes the extent of the off-shellness of electromagnetic current.

The derived off-shell cross esction $\sigma_{eN}$ is used to estimate the expected off-shell effects in electro-nuclear processes in kinematics relevant to the 12 GeV energy upgraded Jefferson Lab experiments.  We compared the LF predictions with that of the de Forest approximation widely used by experimentalists to estimate the off-shell effects in the reaction mechanism of electro-nuclear processes. These comparisons indicate that practically in all kinematic cases the  LF approach predicts less off-shell effects at $Q^2\ge 1$~GeV$^2$ than the de Forest approximation does. Most importantly the LF approach predicts a significant drop of the off-shell effects with an increase of $Q^2$ which intuitively can be understood as  a decrease in the sensitivity of the hard processes on the off-shellness of the target nucleon.

We also examined our conjecture that the $\eta$-variable can be considered as a universal parameter in controlling off-shell effects. We found that for wide range of kinematics the off-shell effects can be suppressed on the level of $5\%$ as as soon as $\eta < 0.1$. The latter gives an effective method for  controlling the uncertainties in the reaction mechanism for large varieties of electro-nuclear processes probing deeply bound nucleons in the nucleus.

Finally, it is worth mentioning that even though we  considered the $eA$  scattering  within  PWIA the obtained expressions for electromagnetic current are applicable also for scattering amplitudes in which the final state interaction between outgoing nucleons is considered within eikonal approximation.  
The main contribution to the re-scattering amplitude coincides with the pole value of the struck nucleon propagator in the intermediate state. Hence,  the  electromagnetic current is the same half-off-shell  electromagnetic current  of  Eq.(\ref{Jsum}).

\section{Development of Light-Front Wave Function of the Deuteron}

There have been intensive theoretical efforts aimed to the description of the relativistic wave function of the deuteron during the past decades, however, limited experimental data \cite{Boeglin:2015cha,Gilman:2001yh} has been available.
Moreover, the vast majority of experiments were in low and intermediate energy domain and  the handful of high energy experiments involved inclusive and elastic processes neither of them being able to probe directly high momentum component of deuteron wave function.

In the present work, our main goal was to develop a calculation that allows to  probe the high momentum components  of the deuteron nucleonic ($pn$)  wave function. 
This requires a proper relativistic description of the deuteron wave function in terms of bound proton and neutron.   
We consider the deuteron wave function on the Light Front, in which case the vacuum fluctuations that obscure the probability amplitude to find the pre-existing $pn$ component with large relative momentum are removed.

Applying effective LF diagrammatic rules to the scattering amplitude for the processes depicted in Fig.(\ref{ddiagram}) 
the LF wave function of deuteron is introduce by Eq.(\ref{dwave_lf2}). The transition vertex $\Gamma_\text{dNN}$ is parameterized by 6 invariant function as in Eq.(\ref{vertex}), of which the first two ($\Gamma_1, \ \Gamma_2$) are related to the familiar S and D deuteron's radial partial waves. Within the leading contribution, we must include also the $\Gamma_5$ vertex function, which is related to a P-wave-like angular structure. 
Physical motivations pointed out to the use of a dipole-like parameterization  as a  simple model for the vertex function $\Gamma_5$ (Eq. \ref{dipole model}). 
It was shown that the dipole-like term provides the correct asymptotic behavior for small momentum ($k\to 0$), for which the P-wave behaves as $k^{L}$ with $L=1$. Moreover, it assures that for large momentum $k$ the vertex function $\Gamma_5 (k) $ scales as  $k^{-2}$, which in turns guarantees that the momentum distribution will scale as $k^{-4}$, which is shown in Fig.(\ref{Mom dist scaling}). 
Finally, the inclusion of the P-wave term modifies the relative contributions to deuteron's wave function normalization.
Within the dipole-like model for the $\Gamma_5$ term, we can estimate the angular averaged one-body momentum distribution, which is shown in  Fig.(\ref{norm G125}) together with the equivalent momentum distribution that only takes into account the S and D waves contributions. The main result is the dominance of the P-wave term in the deuteron's wave function for high internal momentum configurations.

%
%
%



%
%





\cleardoublepage
\appendix


\renewcommand{\thechapter}{} 
\phantomsection
\addcontentsline{toc}{chapter}{APPENDICES}
\chapter*{\normalfont APPENDICES}
\renewcommand{\thechapter}{\Alph{chapter}} 


\setcounter{secnumdepth}{0}


\newpage

\chapter
{Notation and Conventions} 
\phantomsection\label{App. Notation}


\renewcommand{\theequation}{A.\arabic{equation}}
\setcounter{equation}{0}

Dirac matrices are defined in four-dimension space-time by the conditions (Clifford algebra), 
\begin{equation}\phantomsection\label{App Clifford algebra}
\gamma_\mu \gamma_\nu + \gamma_\nu \gamma_\mu = 2g_{\mu\nu}
\end{equation}
 They are used to generate a basis for the linear operators in Dirac space, i.e. the (4-dimensional)  spin 1/2-spinor space with definite parity\footnote{They are eigenstates of the parity operator. For the definition of the parity operator see section \hyperref[App. Discrete Symmetries]{App.(I.4)}.}. 
 The basis is formed by the 5 (multi-)linear operators, 
 \begin{align}
 \mathbf{1}_4 \ , \quad \gamma_{5} \ , \quad \gamma_{\mu} \ , \quad \gamma_{\mu}\gamma_{5} \ , \quad \sigma_{\mu\nu}=(\rm{i}/2)[\gamma_{\mu},\gamma_{\nu}]=(\rm{i}/2)(\gamma_{\mu}\gamma_{\nu}-\gamma_{\nu}\gamma_{\mu})  
 \end{align}
where, $\ \mu,\nu=0,1,2,3$ are space-time indices, the fifth gamma matrix is $ \ \gamma_{5}=\rm{i}\gamma_{0}\gamma_{1}\gamma_{2}\gamma_{3}$, and $ \mathbf{1}_4$ is the four-by-four identity matrix.

When an explicit representation for the gamma matrices is needed, they are written in the Dirac-Pauli representation,

\phantomsection\label{App gammas Dirac-Pauli rep Sec}
\begin{align}\label{App gammas Dirac-Pauli rep}
\gamma^{0}_D =\gamma^{0}=\left(\begin{array}{cc} \boldsymbol{1}_2 & 0 \\ {0} & -\boldsymbol{1}_2 \end{array}\right), 
\quad  \gamma^{i}_D=\gamma^{i}=\left(\begin{array}{cc}{0} & {\boldsymbol{\sigma}_{i}} \\ {-\boldsymbol{\sigma_{i}}} & {0}\end{array}\right), 
\quad  \gamma^{5}_D =\gamma^{5} =\left(\begin{array}{cc}{0} & \boldsymbol{1}_2 \\ \boldsymbol{1}_2 & {0}\end{array}\right)
\end{align}
where, $ \mathbf{1}_2$ is the two-by-two identity matrix and $\boldsymbol{\sigma_{i}}=(\sigma_1, \sigma_2, \sigma_3)$ is a vector whose components are the Pauli matrices,


\phantomsection\label{App Pauli matrices Sec}
\begin{align}\label{App Pauli matrices}
\sigma_1=\left(\begin{array}{cc} {0} & 1 \\ {1} & {0} \end{array}\right), 
\qquad  \sigma_2=\left(\begin{array}{cc} {0} & {-i} \\ i & {0}\end{array}\right), 
\qquad  \sigma_3=\left(\begin{array}{cc}{1} & 0 \\ 0 & {-1}\end{array}\right)
\end{align}

Because in this representation $\gamma_{0}$ appears diagonalized, the spinors that form the basis of the representation (Dirac spinors) are eigenstates of $\gamma_{0}$.
This basis is convenient for study non-relativistic limits or approximations.

\newpage

%
%

\chapter
{Diagrammatic Rules for Nuclear Scattering {Amplitudes}} 
\phantomsection\label{App. LFPT}

\renewcommand{\theequation}{B.\arabic{equation}}
\setcounter{equation}{0}

\renewcommand{\thetable}{B.\arabic{table}}
\setcounter{table}{0}

\renewcommand{\thefigure}{B.\arabic{figure}}
\setcounter{figure}{0}

A brief summary of the rules for computation of scattering amplitudes within the Light Front (LF) formalism is presented below. 
We follow the Lepage-Brodsky convention\cite{LB(1980),BPP(1997)}.

The notation for the components of a generic 4-vector is,
\begin{align}
\label{App LF 4-vector}
&  v^{\mu }=\left( v^+ , v_x , v_y , v^- \right) 
\nonumber  \\
&  v^{\pm} = v^0 \pm v_z
\end{align} 
thus, the position and momentum 4-vectors are written as,
\begin{align}
\label{App LF 4-vector x p}
&  x^{\mu }=\left( x^+ , x , y , x^- \right)=\left( x^+ ,\textbf{x}_\text{T} , x^- \right) 
\nonumber  \\
&  p^{\mu }=\left( p^+ , p_x , p_y , p^- \right)=\left( p^+ ,\textbf{p}_\text{T} , p^- \right) 
\end{align} 
The LF scalar product takes the form, 
\phantomsection\label{App.B x dot p 2 Sec}
\begin{align}\label{App.B x dot p 2}
 x \cdot p={ 1 \over 2} \left( x^+ p^-  + x^- p^+\right)  -  \textbf{x}_\text{T} \cdot  \textbf{p}_\text{T} 
\end{align} 
The LF evolution is parameterized (as usual) by the first component of $x^\mu$, which is therefore referred as the LF time. When ambiguities may appear, the LF time is single out from the other components by using a new label, a frequent choice is  $\tau=x^+$.

\begin{figure}[h]
	\centering
	\includegraphics[scale=0.8]{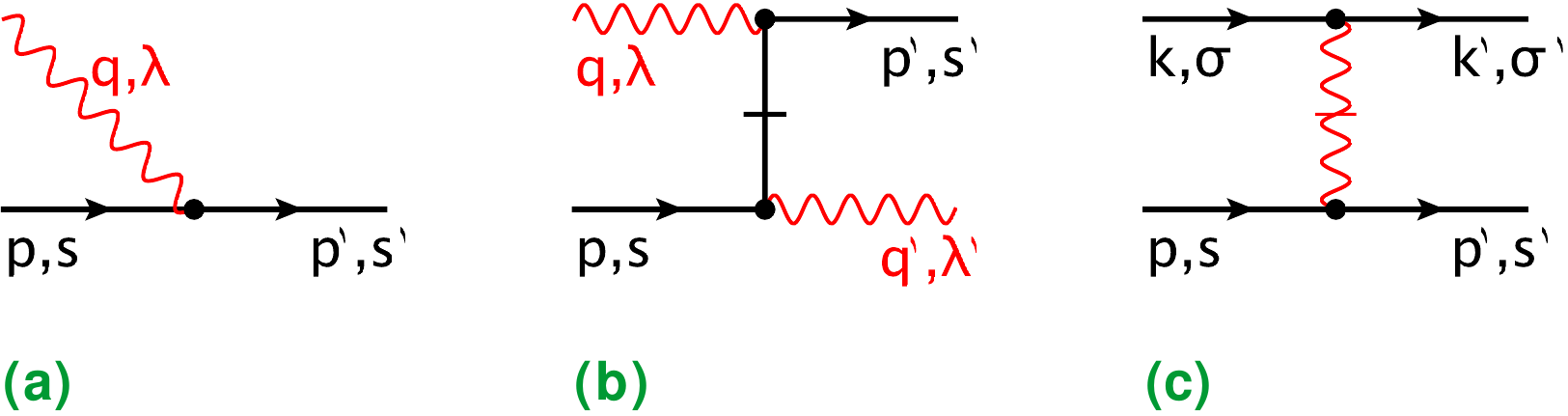}
	\caption{Example of the scattering amplitude on the light-front ($\tau=x^+$ flows from left to right).}
	\label{LCPT-Vertices-QED}
\end{figure}

Diagrammatic Rules for  effective  light-front perturbation theory can be formulated as follows:

\begin{enumerate}

	\item Draw all topologically distinct $\tau \equiv x^+$-ordered diagrams at the desired coupling power. 
In addition to  the usual advanced and retarded propagation  between two events  one needs to include  a third possibility in which the two events connected by an internal  fermion or photon interact   at the same LF $\tau$-time, commonly referred as instantaneous term.

	\item Assign to each line a four-momentum $p^\mu$ and spin $s$ (or helicity, $\lambda$) corresponding to a single on-mass-shell particle, i.e. $p^2 = m^2$. 

	\item With spin 1/2 fermions associate on-mass-shell spinors $u(p, s)$, with antifermions $v(p, s)$, with photons $\epsilon_\mu(q,\lambda)$, etc, such that, 
\begin{align}\label{App Spinors-Norm}
\bar{u}(p, s') u(p, s)  =  & - \bar{v}(p, s') v(p, s) = 2m\delta_{s  s'} \nonumber \\
\sum_{s}  u(p,s) \bar{u}(p,s)  = &  \sh{p} + m \nonumber  \\  
\sum_{s}  v(p,s) \bar{v}(p,s) = &  \sh{p} - m   \nonumber \\
(\epsilon^\mu(q, \lambda'))^* \epsilon_\mu(q, \lambda) = & - \delta_{\lambda '  \lambda} \quad , \quad  q \cdot \epsilon(q, \lambda) = 0  \nonumber \\
\sum_{\lambda} (\epsilon^\mu(q, \lambda))^* \epsilon^\nu(q, \lambda) = &  - g^{\mu \nu } + { q^\mu \eta^\nu + q^\nu \eta^\mu \over q \cdot \eta}
\end{align} 
where $\eta$ is a null vector ($\eta^2=0$), given in LC gauge by,  $ \eta = (0,0,0,2)$
	
	\item  Each intermediate state gets a factor of Light-Front energy denominator:
\begin{align}
\frac{1}{ \sum_{ini} p^- - \sum_{int} p^-  + i\epsilon} 
\end{align}
where, the sums run over LF energies of particles in the initial ($ini$) and intermediate ($int$)  states. For each particle, its LF energy is fixed by the on-mass-shell condition, $\  p^- = \frac{ m^2 + {\bf p}_\text{T}^2 }{ p^+} > 0$.
	
	\item Internal lines account for two kind of interactions:
\begin{itemize}
	\item Propagating, in which case, for a vertex like in Fig.(\ref{LCPT-Vertices-QED}-a) one has: 
\begin{align}
\bar{u}(p',s') \sh{\epsilon}(q,\lambda) u(p,s) \  \delta^2\left(\sum_{in} { {\bf p}_\text{T}}_{in} - \sum_{out} { {\bf p}_\text{T}}_{out} \right) \  \delta\left(\sum_{in} p^+_{in} - \sum_{out} p^+_{out} \right) 		
\end{align}
where, $in$  and $out$ refer to flowing into and out of the vertex. 
The $\delta$ functions at the vertex guarantee the conservation of the plus and transverse components for $in$ and $out$ momenta.
		
	\item Instantaneous. For each vertex like  in Fig.(\ref{LCPT-Vertices-QED}-b)  (fermionic), include,
\begin{align}
\bar{u}(p',s') (\sh{\epsilon}(q',\lambda'))^*  &  {\gamma^+ \over 2(q^+ - p'^+) } \sh{\epsilon}(q,\lambda) u(p,s) \nonumber \\ 
& \times \delta^2\left(\sum_{in} { {\bf p}_\text{T}}_{in} - \sum_{out} { {\bf p}_\text{T}}_{out} \right) \  \delta\left(\sum_{in} p^+_{in} - \sum_{out} p^+_{out} \right)
\end{align}

		\item And, for each vertex like in Fig.(\ref{LCPT-Vertices-QED}-c)  (vector), include,
\begin{align}
\Gamma^2 \ \bar{u}(p',s')  &  \gamma^+ u(p,s)  {1 \over (p'^+ - p^+)^2 } \bar{u}(k',\sigma') \gamma^+ u(k,\sigma) \nonumber \\
& \times \delta^2\left(\sum_{in} { {\bf p}_\text{T}}_{in} - \sum_{out} { {\bf p}_\text{T}}_{out} \right) \  \delta\left(\sum_{in} p^+_{in} - \sum_{out} p^+_{out} \right)
\end{align}
		
	\end{itemize}

	\item Each vertex in the diagram is associated with an effective transition factor 
$\Gamma$. 
For elementary interactions among bare particles the $\Gamma$ factors correspond to the fundamental vertices with coupling constants.

	\item For a composite particle $A$, represented by a state with momentum $p_A$ and spin $s_A$, the  LF wave function associated with its transition to n-constituents is defined by:
\begin{align}
\psi\left(\left\{x_{i}, {{\bf k}_i}_\text{T}, s_{i}\right\}; s_A, p_A\right)=\frac{ \left( \prod_{i=1}^{n} \chi_{i}^\dagger\left(x_{i}, {{\bf k}_i}_\text{T}, s_{i}\right) \right) \cdot \Gamma \cdot \chi_{A}\left(p_{A}, s_{A}\right)}{p_{A}^{+} \left({ p^-_A - \sum_{i=1}^n p^-_i  + i\epsilon} \right) }
\end{align}
where, $x_i$ and ${{\bf k}_i}_\text{T}$ is the LF longitudinal momentum fraction and transverse momentum of the $i$th constituent particle.
The spin wave functions of outgoing particles are described by 
$ \chi_{i}^\dagger\left(x_{i}, {{\bf k}_i}_\text{T}, s_{i}\right)$,
and $\Gamma$ is the effective vertex of the transition of particle A to n-constituents.

	\item Sum over polarizations and integrate over each internal line with the factor,
\begin{align}
\sum_{s} \int  \frac{dp_{\textbf{T}} dp^+}{2(2\pi)^3 p^+} \Theta(p^+)
\end{align}
which ensures the plus component positivity (all particles move forward in LF time).

	\item To convert incoming into outgoing lines, or particles to antiparticles, replace,
\begin{align}
u \leftrightarrow v \ , \quad \bar{u} \leftrightarrow - \bar{v} \ , \quad \epsilon \leftrightarrow \epsilon^*
\end{align}

	\item Symmetry factors must be included as usual. As well as a factor of -1 for each fermion loop, for any fermion line beginning and ending at the initial state, and for every diagram in which fermion lines are interchanged in either of the initial or final states. Also, the overall sign from Wick's theorem.

\end{enumerate}

\newpage

%
%

\chapter
{Half Off-Shell Nucleonic Electro-Magnetic Current and Tensor} 
\phantomsection\label{App Nucleonic EM Current}

\renewcommand{\theequation}{C.\arabic{equation}}
\setcounter{equation}{0}

\renewcommand{\thetable}{C.\arabic{table}}
\setcounter{table}{0}

\renewcommand{\thefigure}{C.\arabic{figure}}
\setcounter{figure}{0}


The proper requirement of charge conservation  
in 
field theory is given by the Ward-Takahashi identity, which in the problem at hand states that the equation of continuity must be satisfied at the photon-fermion vertex.
For bound particles (off the energy shell), gauge invariance requires some additions to the free EM current operator, which can make important contributions to the form factors.

The form for the off-shell EM vertex is \cite{Bincer60,Koch90,Koch96},
\phantomsection\label{App EM_vertex Sec}
\begin{equation}\label{App EM_vertex} 
\Gamma_{{\color{red}\gamma^*} N}^{\mu} = \left(\gamma^{\mu }F_{1}  + i \sigma^{\mu \nu} q_{\nu} F_{2}  {\kappa \over 2 m_N} + q^{\mu}  F_{3} \right) 
\end{equation}
This is the EM vertex one would use to study elastic scattering off constituents in a bound system like nuclei, e.g. the triangle diagram of Fig.(\ref{deuteron FF}).
%
%
%
%
The Ward-Takahashi identity can be used to impose a constrain over $\Gamma_{{\color{red}\gamma^*}  N}^{\mu}$ that allows to write the  $F_{3}$ form factor in terms of $F_1$. 
%
%
%
This is equivalent to solve, $q_\mu J_N^\mu=0$.
Using Eq.(\ref{EM current off-on}) together with 
\hyperref[App EM_vertex Sec]{Eq.(\ref*{App EM_vertex})} 
one obtains: 
\begin{equation}
F_{3}  =  F_{1} { \sh{q}   \over Q^2}
\end{equation}
Resulting in  the  off-shell EM vertex, 
\begin{equation}
\Gamma_{{\color{red}\gamma^*} N}^{\mu} =  \left(F_{1} \Big( \gamma^{\mu } +  q^\mu{ \sh{q}   \over Q^2} \Big)  + i \sigma^{\mu \nu} q_{\nu} F_{2}  {\kappa \over 2 m_N} \right)
\label{EM_vertex_full}
\end{equation}
where the form-factors ($F_{1,2,3})$ are functions of  Lorentz invariants constructed from the  initial $(p_1)$ and final $(p_2)$ nucleon's momenta  and the momentum transfer $(q)$ in Fig.(\ref{deuteron FF}).

\subsection*{C.1 Nucleonic Tensor} 
\phantomsection\label{Nucleonic Tensor}

In this section we provide explicit expressions for the components of the nucleonic tensor ($H^{\mu\nu}_N$).
Substituting Eq.(\ref{Jsum}) into Eq.(\ref{HN}), allows us to express  the nucleonic tensor as a sum of two terms:
\begin{equation}\label{HNsum} 
H^{\mu \nu}_{N } = H^{\mu \nu}_{N, \text{prop}} + H^{\mu \nu}_{N , \text{ inst}}
\end{equation} 
where, the the propagating contribution is given by,
\begin{equation}\label{HN-on_JJ}
H^{\mu \nu}_{N, \text{prop}} =  {1 \over 2} \sum_{s_i s_f} (J^{s_i s_f \ \mu}_\text{prop} )^{\dagger}  (J^{s_i s_f}_\text{prop} )^\nu  = \frac{1}{2}Tr\left[\overline{\Gamma}_{{\color{red}\gamma^*}  N}^{\text{(on)} \mu} (\sh{p}_f + m_N) \Gamma_{{\color{red}\gamma^*}  N}^{\text{(on)} \nu} (\sh{p}_{i,\text{on}} + m_N) \right]
\end{equation}
and the instantaneous by,
\begin{eqnarray}\label{HN-off_JJ}
H^{\mu \nu}_{N, \text{inst}} &=& {1 \over 2} \sum_{s_i s_f} \left( \left(J^{s_i s_f \ \nu}_{\text{off}}\right)^{\dagger}  J^{s_i s_f \ \mu}_\text{inst}  + \left(J^{s_i s_f \ \nu}_\text{prop}\right)^{\dagger}  J^{s_i s_f \ \mu}_\text{inst}  + \left(J^{s_i s_f \ \nu}_\text{inst}\right)^{\dagger}  J^{s_i s_f \ \mu}_\text{prop}  \right) \\  \nonumber
&=&   \dfrac{1}{2}  \text{Tr} \Big[ \overline{\Gamma}_{{\color{red}\gamma^*}  N}^{\text{(off)} \nu} (\sh p_f + m_N) \Gamma_{{\color{red}\gamma^*}  N}^{\text{(off)} \mu} (\sh{p}_{i,\text{on}} + m_N)  \\ 
&&  +  \overline{\Gamma}_{{\color{red}\gamma^*}  N}^{\text{(on)} \nu} (\sh p_f + m_N) \Gamma_{{\color{red}\gamma^*}  N}^{\text{(off)} \mu} (\sh{p}_{i,\text{on}} + m_N)  +   \overline{\Gamma}_{{\color{red}\gamma^*}  N}^{\text{(off)} \nu} (\sh p_f + m_N) \Gamma_{{\color{red}\gamma^*}  N}^{\text{(on)} \mu} (\sh{p}_{i,\text{on}} + m_N)  \Big]  \nonumber
\end{eqnarray}
where,  $\overline{\Gamma}_{{\color{red}\gamma^*}  N}^{\mu} = \gamma^0 \left( {\Gamma}_{{\color{red}\gamma^*}  N}^{\mu}\right)^{\dagger} \gamma^0 $. 
Notice that the initial momentum of the nucleon, $p_i$, occurring from now-on  corresponds to $p_{i,on} $, which allows to drop the on-shell label "on"   without confusion. 
With this, we can write propagating  and instantaneous  contributions of the tensor, $H^{\mu,\nu}$  as functions of the nucleon form factors $F_1$ and $F_2$ as follows:
\begin{align}\label{HN-on}
H^{\mu \nu}_{N, \text{prop}}  =& \ 2 F_1^2 \Big[ g^{\mu\nu} \left( m_N^2 - p_i \cdot p_f \right) + \left( p_i^{\mu} p_f^{\nu}  + p_i^{\nu} p_f^{\mu} \right)    \Big] 
\nonumber \\
& +    F_1 F_2 \kappa  \Big[ 2 g^{\mu\nu} q \cdot \left( p_f - p_i\right) + \left( p_i^{\mu} q^{\nu}  + p_i^{\nu} q^{\mu} \right)  -  (  p_f^{\mu} q^{\nu}   +p_f^\nu q^{\mu} ) \Big] \nonumber \\
& +   F_2^2  {\kappa^2 \over 2 m_N^2}   \Big[  g^{\mu\nu} \big[ q^2 \left( p_i \cdot p_f + m_N^2 \right)  - 2 \ q \cdot p_i \ q \cdot p_f \big] - q^2 \left( p_i^{\mu} p_f^{\nu}  + p_i^{\nu} p_f^{\mu} \right) \nonumber  \\
&   \ \ \ \   -  q^{\mu}q^{\nu}  \left( p_i \cdot p_f + m_N^2 \right)  +  q \cdot p_f \left( p_i^{\mu} q^{\nu}  + p_i^{\nu} q^{\mu} \right) +  q \cdot p_i \left( p_f^{\mu} q^{\nu}  + p_f^{\nu} q^{\mu} \right)  \Big] 
\end{align}
and the instantaneous correction as follows:
{\small
\begin{align}\label{HN-off}
\hspace{-0.95cm}
& H^{\mu \nu}_{N, \text{inst}}  =   \ 2 F_1^2 \Big[ g^{\mu\nu} \Big(  \Delta p_i \cdot\left( p_i -  p_f \right) - { \Delta p_i \cdot p_i \over  m_N^2}  \Delta p_i \cdot p_f  \Big)  + \left(\Delta p_i^\mu p_f^{\nu}  + \Delta p_i^{\nu} p_f^{\mu} \right) \Big( 1 + {\Delta p_i \cdot p_i \over  m_N^2 }   \Big)  \nonumber \\
& +   {2 \over q^2} q^{\mu}q^{\nu}  \Big( {2 \over q^2} q \cdot p_{f} \ q \cdot ( \Delta p_i + p_{i} ) -   ( p_{i} - p_f) \cdot ( \Delta p_i + p_{i} ) + { \Delta p_i \cdot p_i  \over  m_N^2}  \Big( { \Delta p_i \cdot q \over q^2}   p_f \cdot q +{\Delta p_i \cdot p_f  }   \Big) \Big)  \nonumber  \\
& -    {2 \over q^2} \left( p_i^{\mu} q^{\nu}  + p_i^{\nu} q^{\mu} \right)  q \cdot p_f   - {2 \over q^2} (  p_f^{\mu} q^{\nu}   +p_f^\nu q^{\mu} )   \Big( q \cdot ( \Delta p_i + p_{i} ) + {\Delta p_i \cdot p_i \over  m_N^2}  \Delta p_i \cdot q \Big)  \nonumber  \\
&  -   {2 \over q^2} \left(\Delta p_i^\mu q^{\nu}  + \Delta p_i^{\nu} q^{\mu} \right) q \cdot p_{f}  \Big( 1  + { \Delta p_i \cdot p_i \over  m_N^2}   \Big)  \Big]  
\nonumber \\
& +    F_1 F_2 \kappa  \Big[  g^{\mu\nu}  \Big( {  \Delta p_i \cdot p_i \over  m_N^2} q \cdot ( 2 p_f - \Delta p_i ) - 2 \Delta p_i \cdot q \Big)  + q^{\mu}q^{\nu}  \Big(  { \Delta p_i \cdot p_i  \over m_N^2 q^2} q \cdot( \Delta p_i - 2 p_{f}) -   2    \Big) \nonumber \\
&   \ \ \ \ \ \ \ \  - \left( p_i^{\mu} q^{\nu}  + p_i^{\nu} q^{\mu} \right)  + (  p_f^{\mu} q^{\nu}   +p_f^\nu q^{\mu} )  \Big] \nonumber \\
& +   F_2^2  {\kappa^2 \over 2 m_N^2}    \Big[  g^{\mu\nu} \big[ \left( q^2 \ \Delta p_i \cdot p_f  - 2 \ q \cdot \Delta p_i \ q \cdot p_f\right) \Big( 1 + {\Delta p_i \cdot p_i \over  m_N^2 }   \Big) +q^2 \ \Delta p_i \cdot p_i  \big] \nonumber  \\
&   \ \ \ \ \ \ \ -  \left( \Delta p_i^{\mu} p_f^{\nu}  + \Delta p_i^{\nu} p_f^{\mu} \right) q^2 \Big( 1 + {\Delta p_i \cdot p_i \over  m_N^2 }   \Big)  -  q^{\mu}q^{\nu}  \big[  \Delta p_i \cdot  p_f   \Big( 1 + {\Delta p_i \cdot p_i \over  m_N^2 }   \Big) -  \Delta p_i \cdot p_i  \big]  \nonumber  \\
&    \ \ \ \ \ \ \  +   \left( \Delta p_i^{\mu} q^{\nu}  + \Delta p_i^{\nu} q^{\mu} \right) q \cdot p_f  \Big( 1 + {\Delta p_i \cdot p_i \over  m_N^2 }   \Big) +   \left( p_f^{\mu} q^{\nu}  + p_f^{\nu} q^{\mu} \right) q \cdot \Delta p_i  \Big( 1 + {\Delta p_i \cdot p_i \over m_N^2 }   \Big)   \Big]
\end{align}
}

With our choice of reference frame (Fig.\ref{scatt react plane}), one can expand the $L_{\mu\nu}H^{\mu\nu}$ product  in the following form:
\begin{align}
\hspace{-0.85cm}
L_{\mu \nu} H^{\mu \nu}_N = & \left( L_{00}H^{00} - 2 L_{0z}H^{0z} + L_{zz}H^{zz} \right) + \left( -2 L_{0 \parallel }H^{0 \parallel } + 2 L_{z \parallel }H^{z \parallel} \right)  \nonumber \\
& + \frac{1}{2} \left( L_{\parallel \parallel } + L_{\bot \bot} \right) \left( H^{\parallel \parallel } +  H^{\bot \bot} \right)  + {1 \over 2 } \left( L_{\parallel \parallel } - L_{\bot \bot} \right) \left( H^{\parallel \parallel } -  H^{\bot \bot} \right)
\end{align}
Furthermore,  using the gauge-invariance  of  leptonic current, one expresses the above product in the form:
\begin{align}
L_{\mu \nu} H^{\mu \nu}_N & =  L_{00} \left( H^{00} - 2 {q^0 \over q_z } H^{0z} + \left({q^0 \over  q_z}\right)^2 H^{zz} \right) + 2 L_{0 \parallel } \left( -  H^{0 \parallel } + {q^0 \over q_z } H^{z \parallel} \right)  \nonumber \\
& +  \frac{1}{2} \left( L_{\parallel \parallel } + L_{\bot \bot} \right) \left( H^{\parallel \parallel } +  H^{\bot \bot} \right)  + {1 \over 2 } \left( L_{\parallel \parallel } - L_{\bot \bot} \right) \left( H^{\parallel \parallel } -  H^{\bot \bot} \right) \nonumber \\
& =  Q^2 (\tan(\theta/2))^2 \left( \eta_L V_{N,L} + \eta_{TL} V_{N,TL} \cos(\phi) + \eta_T V_{N,T} + \eta_{TT} V_{N,TT} \right)
\end{align}
Using the definitions of  $\eta_i$ for $ i=L,T,TL,TT$, from Eq.(\ref{eta}) for hadronic structure functions,  ($V_{N,i} $), one obtains:
\begin{align} \label{V-Gross-app}
w^N_L & = \frac{ {\bf q}^4 }{Q^4} \left( H^{00} - 2 {q^0 \over q_z } H^{0z} + ({q^0)^2 \over {\bf q}^2 } H^{zz} \right) = \frac{ {\bf q}^2 }{4 Q^2} \left( H^{++}{ Q^2 \over (q^+)^2 } + 2 H^{+-} + {(q^+)^2 \over  Q^2 } H^{- \ -} \right) \nonumber  \\
w^N_{TL} & =  2 { {\bf q}^2 \over Q^2 } \left( {q^0 \over q_z } H^{z \parallel}_N -  H^{0 \parallel }_N \right) = {|{\bf q}| \over q^+} \left( H^{+ \parallel}_N + H^{- \parallel }_N {(q^+)^2 \over Q^2} \right) \nonumber \\  
w^N_T & =  H^{\parallel \parallel }_N +  H^{\bot \bot} _N \nonumber  \\
w^N_{TT} & = H_N^{\parallel \parallel} - H_N^{\bot \bot}   
\end{align}
where we have used, $ - q_z = |{\bf q}|$, as well as the relation between components of the nucleonic tensor in light-cone and Minkowski coordinates:
\begin{align}\label{H-components-LC}
H^{00} &= { 1 \over 4 }(H^{++} + 2 H^{+-} + H^{- \ -})  \nonumber  \\
H^{0z} &= {1 \over 4} (H^{++} - H^{- \ -}) \nonumber  \\
H^{zz} &= {1 \over 4}  (H^{++} - 2 H^{+-} + H^{- \ -})  \nonumber \\   
H^{0 \parallel } &= {1 \over 2} (H^{+ \parallel} + H^{- \parallel} ) \nonumber  \\
H^{z \parallel}  &=  {1 \over 2} (H^{+ \parallel} - H^{- \parallel} )
\end{align}

From 
Eqs.(C.7, C.11)
we compute  the  explicit forms of the structure functions. 
In the reference frame of Fig.(\ref{scatt react plane}), they are given by:
{\small
\begin{align}
\hspace{-0.95cm}
w_{L \text{ prop}}^N  & =     F_1^2 {\bf{q}}^2  \dfrac{\alpha_N \alpha_f}{\alpha_q^2} \Big(  {m^{2}_N+{\bf p}_\text{T}^2  \over Q^2} \dfrac{\alpha_q^2 }
{\alpha_N \alpha_f} +  1    \Big) -   F_1 F_2 {\bf{q}}^2  {\kappa } \left(   {m^{2}_N+{\bf p}_\text{T}^2  \over Q^2} 
\dfrac{\alpha_q^2 }{\alpha_N \alpha_f} +  1    \right)   \nonumber \\
&   + F_2^2 {\bf{q}}^2  \left(  {\kappa \over 2 m_N}  \right) ^2    \left(  (m^{2}_N+{\bf p}_\text{T}^2 )\dfrac{\alpha_q^2 }{\alpha_N \alpha_f} +  4 p_{\textbf{T}}^2  \right)  \nonumber \\ 
w_{L \text{ inst}}^N   =  &   F_1^2{   \alpha_{N} \over \alpha_{q}}  {\bf{q}}^2 \left(     1 - \left( {m^{2}_N+{\bf p}_\text{T}^2  \over Q^2}  \dfrac{\alpha_q^2 }{\alpha_N \alpha_f}  \right)^2  
+  {(m^{2}_N+{\bf p}_\text{T}^2 ) \over 2m^2} \dfrac{\alpha_q }{\alpha_N }   + \left( {m^{2}_N +{\bf p}_\text{T}^2  \over Q^2} \dfrac{\alpha_q^2 }{\alpha_N \alpha_f} -  1    \right) ^2 \right)  \nonumber \\
& - 2 F_1 F_2 \kappa {\alpha_{N} \over \alpha_{q}}  {\bf{q}}^2 {   (q\cdot \Delta p_i)^2 \over m^2 Q^2} \left(    2 {\alpha_{f} \over \alpha_{q}}  + 2{  m^2 \over  q\cdot \Delta p_i   }  +1  \right) \nonumber   \\
& +   F_2^2 \Big( {\kappa \over m^2_N}  \Big)^2 {\bf{q}}^2 ~ q\cdot \Delta p_i  \left( 1 +   {q\cdot \Delta p_i  \over m^2 } { \alpha_N \alpha_f \over \alpha_q^2 }   \right) 
\end{align}
\begin{align}
\hspace{-0.95cm}
w_{TL \text{ prop}}^N  = &    |{\bf{q}}|  \dfrac{ \alpha_N   +   \alpha_f }{\alpha_q} p_{\textbf{T}} \left( 2 F_1^2 + 2 F_2^2 \Big( {\kappa \over 2 m_N}  \Big)^2 Q^2\right)  \left( 1  +  {m^{2}_N+p_{\textbf{T}}^2 \over Q^2} \dfrac{ \alpha_q^2 }{\alpha_N \alpha_f} \right)   \nonumber \\
\hspace{-0.95cm}
w_{TL \text{ inst}}^N   = &   8 |{\bf{q}}| { q \cdot \Delta p_i \over Q^2 }  {\bf p}_\text{T}  \left( 1 +{p_i \cdot \Delta p_i  \over m^2} \right) \left(   F_1^2  +  F_2^2 \Big( {\kappa \over 2 m_N}  \Big)^2  \right)
\end{align}
\begin{align}
\hspace{-0.95cm}
w_{T  \text{ prop} }^N  =& \  F_1^2\left(   2 (m^{2}_N+p_{\textbf{T}}^2 )\dfrac{\alpha_q^2}{\alpha_N \alpha_f} +  4 (p_{\textbf{T}})^2     \right)  + 2 \kappa F_1 F_2   \left(  (m^{2}_N+p_{\textbf{T}}^2 ) \dfrac{\alpha_q^2}{\alpha_N \alpha_f} +  Q^2 \right)  \nonumber  \\ 
& + F_2^2 \left(  {\kappa \over 2 m_N} \right) ^2 \left(  2 \dfrac{\alpha_N \alpha_f}{\alpha_q^2}  \left(  (m^{2}_N+p_{\textbf{T}}^2 ) \dfrac{\alpha_q^2}{\alpha_N \alpha_f} +  Q^2 \right) ^2  -  4 Q^2  p_{\textbf{T}}^2 \right)  \nonumber  \\
\hspace{-0.95cm}
w_{T  \text{ inst} }^N = &   8 F_1^2   \left( q\cdot \Delta p_i  + p_f \cdot \Delta p_i  {p_i \cdot \Delta p_i \over m^2 } \right)   +  8  F_1 F_2 \kappa \left( 1 +{p_i \cdot \Delta p_i  \over m^2} \right) \left( q\cdot \Delta p_i  - p_f \cdot q  {p_i \cdot \Delta p_i \over m^2 + p_i \cdot \Delta p_i  } \right) \nonumber  \\
& + 8 F_2^2 \left(  {\kappa \over 2 m_N}  \right) ^2  \left( 1 +{p_i \cdot \Delta p_i  \over m^2} \right) \left(  q \cdot p_f ~ q\cdot \Delta p_i  + Q^2 ~p_f \cdot \Delta p_i + Q^2  { m^2 ~p_i \cdot \Delta p_i \over m^2 + p_i \cdot \Delta p_i  } \right)    
\end{align}
\begin{align}
\hspace{-0.95cm}
w_{TT  \text{ prop}}^N = & 4 p_{\textbf{T}}^2 \left(  F_1^2 +   F_2^2  {\kappa^2 \over 4 m_N^2} Q^2   \right) \nonumber \\ 
\hspace{-0.95cm}
w_{TT \text{ off}}^N  =  &  0
\label{V_LF}
\end{align}
}

The kinematic variables, and scalar products used in the calculation are:
\begin{itemize}
\item The light-cone momentum fractions,
\begin{equation}
\alpha_N= {2 p_N^+\over p_d^+}= {2 (E_N + p_{N, z}) \over p_d^+}, \quad  \alpha_q = {2 q^+\over p_d^+}= {2(q^0 - |{\bf{q}}|) \over p_d^+}, \quad  \alpha_f = \alpha_N + \alpha_q
\end{equation}

\item The off-shell factor, 
\begin{equation} 
\Delta{p}_i^\mu=p^\mu_{i } - p^\mu_{i, on} \ , \ \text{with},  \ \  p^\mu_{i }=p^\mu_{d}-p^\mu_{r} 
\end{equation} 
Since, $\Delta p_i^+ = \Delta p_{i}^\perp = 0$, we have, $ 2 \Delta\sh{p}_i=\gamma^{+} \left( p^-_i - p^-_{i, on }\right) $, hence the minus component is given by,
\begin{equation}
\Delta{p}_i^-=p_{i }^- - p_{i \ on}^- = -q^- + (p_f^- - p_{i \ on}^-) = \dfrac{Q^2}{q^+} -   \dfrac{m^{2}_N+{\bf p}_\text{T}^2 }{p_f^+ p_i^+}  q^+ 
\end{equation}

\item The initial ($p_{i,on}^\mu$), final ($p_{f}^\mu$), and transfer ($q^\mu$)  momenta scalar product with the off-shell factor $\Delta{p}_i^\mu$ can be  written as:
\begin{eqnarray}
2 \Delta{p}_i \cdot p_i & = &   Q^2 \dfrac{\alpha_N}{\alpha_q}  -   (m^{2}_N+{\bf p}_\text{T}^2 ) \dfrac{\alpha_q}{\alpha_f }   \nonumber \\  2 \Delta{p}_i \cdot p_f& = &    Q^2 \dfrac{\alpha_f}{\alpha_q}  -   (m^{2}_N+{\bf p}_\text{T}^2 ) \dfrac{\alpha_q}{\alpha_i }\nonumber  \\  
2 \Delta{p}_i \cdot q & = &    Q^2  -   (m^{2}_N+{\bf p}_\text{T}^2 ) \dfrac{\alpha_q^2}{\alpha_f \alpha_N}
\end{eqnarray}

\end{itemize}

\newpage

\chapter{The Rotation Group} 
\phantomsection\label{App. D - Rot Group}

\setcounter{equation}{0}
\renewcommand{\theequation}{D.\arabic{equation}}

\setcounter{figure}{0}
\renewcommand{\thefigure}{D.\arabic{figure}}

The fact that fermions exist leads to the conclusion that the Rotation Group in Physics is SU(2). Relativistic Theories of Quantum Mechanics take into account this fact by using spinors for the description of fermions\footnote{ All relevant definitions are provided below.}.
However, for the sake of clarity, we  start with the more familiar case of 3-dimensional (spatial)  rotations.

\subsection*{D.1 Rotations in three dimensions} 
\phantomsection\label{Rotations 3d}

A rotation\footnote{Rotations on a 3-dimensional real vector space form a Group called SO(3). Written as matrices these are orthogonal, $R^{-1}=R^T$, and have determinant 1.} 
of a 3-dimensional vector ${ } \vec{r}= \left( x , y , z \right) \in \mathbb{R}^3$, can be seen as a transformation  that   leaves invariant the length of the vector, $\vec{r} \overset{R \ }\to \vec{r '} \implies \vec{r}^{ \ 2}={\vec{r'}}^2$.  
In matrix form, the rotations around the Cartesian axes are, 
\phantomsection\label{Rotation matrices Sec}
\begin{align}\label{Rotation matrices}
& R_x (\theta) = \left(
\begin{array}{ccc}
 1 & 0 & 0 \\
 0 & \cos (\theta ) & \sin (\theta ) \\
 0 & -\sin (\theta ) & \cos (\theta ) \\
\end{array} 
\right)
\ , \quad
R_y(\theta) = \left(
\begin{array}{ccc}
 \cos (\theta ) & 0 & -\sin (\theta ) \\
 0 & 1 & 0 \\
 \sin (\theta ) & 0 & \cos (\theta ) \\
\end{array}
\right)  \nonumber \\
&
R_z(\theta) = \left(
\begin{array}{ccc}
 \cos (\theta ) & \sin (\theta ) & 0 \\
 -\sin (\theta ) & \cos (\theta ) & 0 \\
 0 & 0 & 1 \\
\end{array}
\right)
\end{align}
Any rotation can be decompose as a sequence of rotations from 
\hyperref[Rotation matrices Sec]{Eqs.(\ref*{Rotation matrices})}
with some particular choice of angles.

\subsubsection*{Generators } 
\phantomsection\label{D.1.1 Rot Generators}

The rate of change 
of the rotation matrices 
\hyperref[Rotation matrices Sec]{Eqs.(\ref*{Rotation matrices})}
at $\theta \to 0$ \footnote{This is, $ \lim_{\theta \to 0}{1 \over \theta} R_i(\theta)$.},
contain enough information such that they can be used to parameterize arbitrary rotations. 
This limit can be readily obtained from
\hyperref[Rotation matrices Sec]{Eqs.(\ref*{Rotation matrices})}
 by differentiation,  
\phantomsection\label{SO(3) gen Sec}
\begin{align}\label{SO(3) gen}
J_m=\frac{\partial R_m(\theta )}{i \partial \theta} \Big{|}_{\theta =0}
\end{align}
hence we get,
\phantomsection\label{SO(3) gen 3-dim Sec}
\begin{align}\label{SO(3) gen 3-dim}
J_x = \left(
\begin{array}{ccc}
 0 & 0 & 0 \\
 0 & 0 & -i \\
 0 & i & 0 \\
\end{array}
\right) 
\ , \quad
J_y = \left(
\begin{array}{ccc}
 0 & 0 & i \\
 0 & 0 & 0 \\
 -i & 0 & 0 \\
\end{array}
\right)
 \ , \quad
J_z = \left(
\begin{array}{ccc}
 0 & -i & 0 \\
 i & 0 & 0 \\
 0 & 0 & 0 \\
\end{array}
\right)
\end{align}
This construction guarantee that the rotations in
\hyperref[Rotation matrices Sec]{Eq.(\ref*{Rotation matrices})}
 can be written as \mbox{$R_m(\theta)=e^{i \theta J_m}$}, with, $m=x,y,z$. 
Furthermore, an arbitrary rotation by an angle $\theta$ around an axis represented by a unit vector  $\hat{n}$,  can be parameterized in terms of 
the matrices in
\hyperref[SO(3) gen 3-dim Sec]{Eq.(\ref*{SO(3) gen 3-dim})}
as,
\phantomsection\label{Rotations 3-dim vector Sec}
\begin{align}\label{Rotations 3-dim vector}
R(\hat{n},\theta) = e^{i \theta \hat{n} \cdot \vec{J}}  = R(\vec{n}) =  e^{i  \vec{n} \cdot \vec{J}}
\end{align}
where, 
$\vec{n}=\theta \hat{n}$. 
It follows that any 3-dimensional rotation is completely characterized by a set of three 
parameters.
Because of this feature,  the matrices 
\hyperref[SO(3) gen 3-dim Sec]{(\ref*{SO(3) gen 3-dim})}
are called the generators of rotations.

Notice that any set of 
rotations around three independent axes can be used to obtain an arbitrary rotation. Different choices of the  three independent axes correspond to different parameterizations.

\subsection*{D.2 General Correspondence between SU(2) and SO(3)} 
\phantomsection\label{Sec SU(2) and SO(3)}

If instead of a 3-dimensional space,  we work in a complex 2-dimensional space, the  transformations equivalent to the 3-dimensional rotations are written as\footnote{These are unitary matrices, $U^{-1}=U^\dagger$, with determinant equal to 1.},
\phantomsection\label{Rotations 2-dim spinor Sec}
\begin{align}\label{Rotations 2-dim spinor}
U(\hat{n}, \theta) = e^{i \theta \hat{n} \cdot {\vec{\sigma } \over 2}} = \cos{\left(\frac{\theta }{2}\right)} + i
\hat{n} \cdot \vec{\sigma } \sin{\left(\frac{\theta }{2}\right)}
\end{align}
where, $\vec{\sigma}=(\sigma_1,\sigma_2,\sigma_3)$ and 
\begin{align}
\sigma_1=\left(
\begin{array}{cc}
 0 & 1 \\
 1 & 0 \\
\end{array}
\right) \ , \quad 
\sigma_2 = \left(
\begin{array}{cc}
 0 & -i \\
 i & 0 \\
\end{array}
\right)  \ , \quad 
\sigma_3 = \left(
\begin{array}{cc}
 1 & 0 \\
 0 & -1 \\
\end{array}
\right) 
\end{align}
are the standard Pauli matrices. The parameter $\theta$ represents an angle, and $\hat{n}$ a unit vector associated with a direction in 3-dimensional space. The complex 2-dimensional space of tuples $\left( \chi _1 , \chi _2 \right) \in \mathbb{C}^2$, on which the SU(2) matrices $U(\hat{n}, \theta)$ act is called spinor space.

In general, an SU(2) transformation in spinor space  
and an SO(3) transformation in 3-dimensional space ${ } \left( x , y , z \right) \in \mathbb{R}^3$, are related by the following correspondence, 
\phantomsection\label{SU(2) SO(3) correspondence Sec}
\begin{align}\label{SU(2) SO(3) correspondence}
U(\hat{n}, \theta) = e^{i \theta \hat{n} \cdot {\vec{\sigma } \over 2}}=\cos{\left(\frac{\theta }{2}\right)} + i
\hat{n} \cdot \vec{\sigma } \sin{\left(\frac{\theta }{2}\right)} { } \longleftrightarrow  { } R(\hat{n}, \theta) = e^{i \theta 
\hat{n} \cdot \vec{J} }
\end{align}
which implies that the groups must have\footnote{At least locally, i.e., the infinitesimal transformations.} a similar structure.
Indeed, this follows from the fact that their Generators obey the same Commutation Relations,
\begin{align}
\left[{J}_l,{J}_m \right]={i\epsilon }_{{lmn}}{J}_n \ , \qquad
\left[{\sigma_l \over 2},{\sigma_m \over 2} \right]={i\epsilon }_{{lmn}}{\sigma_n \over 2}
\end{align}

\subsection*{D.3 Proof of the Correspondence between SU(2) and SO(3)} 
\phantomsection\label{D.3  map SU(2)  SO(3)}

From the vector $\vec{r}=(x,y,z)$ we can form the matrix,
\phantomsection\label{embedded of r Sec}
\begin{align}\label{embedded of r}
\vec{\sigma} \cdot \vec{r}=\left(
\begin{array}{cc}
 z & x-iy \\
 x+iy & -z \\
\end{array}
\right) 
\end{align}
This is a traceless Hermitian matrix with determinant, $\det{[\vec{\sigma} \cdot \vec{r}]}=-\vec{r} \cdot \vec{r}=-r^2$.
Since the trace and the determinant are invariant under unitary transformations, it follows that SU(2) transformations acting on $\vec{\sigma} \cdot \vec{r}$ behave similar to rotations acting on $\vec{r}$ in the sense that,
 \begin{align} 
U \vec{\sigma} \cdot \vec{r} \ U^\dagger = \vec{\sigma} \cdot \vec{r}^{\ \prime} \quad \implies \quad r^2 = (r^\prime)^2
\end{align}
In other words, they preserve the length of the 3-dimensional vector. Thus, an SU(2) transformation induces a rotation on the position vector $\vec{r}$, and vice versa. 

In order to explicitly construct a map between SU(2) and SO(3) we can build a traceless Hermitian $2\times2$ complex matrix $H$, that transform under SU(2), i.e., it has SU(2) as its group of symmetry.

Noting
that, on the 2-dimensional complex space the most general form an element of SU(2) can have is\footnote{Resulting from the two conditions, $UU^\dagger=\mathbf{1}$ and $\det{U}=1$.},
\phantomsection\label{SU(2) 2-dim Sec}
\begin{align}\label{SU(2) 2-dim}
U=\left(
\begin{array}{cc}
 a & b \\
 -b^* & a^* \\
\end{array}
\right) 
\end{align}
with, $|a|^2+|b|^2=1$. 
It is straightforward to check that an arbitrary element of the space transform as,
\begin{align} 
\chi=\left(
\begin{array}{c}
 \chi_1 \\
 \chi_2 \\
\end{array}
\right) \in \mathbb{C}^2
\  \to \ 
U \chi = \left(
\begin{array}{c}
 a \chi_1 + b \chi_2 \\
 -b^* \chi_1 + a^* \chi_2 \\
\end{array}
\right) = \left(
\begin{array}{c}
 \chi_1^\prime \\
 \chi_2^\prime \\
\end{array}
\right) = \chi^\prime 
\ \ , \ \
\chi^\prime{}^\dagger=\chi^\dagger U^\dagger
\end{align}
and satisfy,
\begin{align} 
\chi^\dagger \chi \quad & \to \quad \chi^\prime{}^\dagger \chi^\prime =  |\chi_1|^2 +  |\chi_2|^2 \\
 \chi \chi^\dagger =  \left(
\begin{array}{cc}
 |\chi_1|^2 & \chi_1 \chi_2^* \\
 \chi_1^* \chi_2 & |\chi_2|^2 \\
\end{array}
\right) 
\quad & \to \quad
\chi^\prime \chi^\prime{}^\dagger = U \chi \chi{}^\dagger  U^\dagger
\end{align}
Although, $\chi \chi^\dagger$ transform under SU(2) as we want, and it is Hermitian, it is not traceless. 
However, because $\chi$ transform like the combination: $(-i \sigma_2 \chi)^*=-i \sigma_2 \chi^*$. Explicitly we have,
\begin{align} 
\exp{\left[-i \pi {\sigma_2 \over 2}\right]} \chi^* = -i \sigma_2 \chi^*  = \left(
\begin{array}{cc}
0 & -1 \\
 1 & 0 \\
\end{array}
\right) \left(
\begin{array}{c}
 \chi_1^* \\
 \chi_2^* \\
\end{array}
\right) = \left(
\begin{array}{c}
 -\chi_2^* \\
 \chi_1^* \\
\end{array}
\right) 
\nonumber \\
-i \sigma_2 \chi^* 
\ \ \to \ \ 
\left(-i \sigma_2 \chi^*\right)^\prime =  U \left(-i \sigma_2 \chi^*\right) = \left(
\begin{array}{c}
 a (-\chi_2^*) + b \chi_1^* \\
 -b^* (-\chi_2^*) + a^* \chi_1^* \\
\end{array}
\right) = -i \sigma_2 \left( \chi^\prime\right)^* 
\end{align}
Furthermore, $\chi^\dagger$ transform like the combination: $(-i \sigma_2 \chi)^T=i \chi^T \sigma_2$, i.e.,
\begin{align} 
\left( -i \sigma_2 \chi \right)^T = \left(
 -\chi_2 \quad
 \chi_1 
\right) 
 \quad \to \quad  
\left( \left( -i \sigma_2 \chi \right)^T \right)^\prime = \left( -i \sigma_2 \left( \chi^\dagger{}^\prime \right) \right)^T 
\end{align}
then\footnote{These are outer products. They are equivalent to Projection operators, which means that this  map is  a projective representation of rotations on SU(2).}, 
$(-i \sigma_2 \chi^*) (-i \sigma_2 \chi)^T$ transform like $\chi \chi^\dagger$,
{\small
\begin{align} 
 (-i \sigma_2 \chi^*) (-i \sigma_2 \chi)^T =  \left(
\begin{array}{cc}
 -\chi_1 \chi_2 & \chi_1^2 \\
 -\chi_2^2 & \chi_1 \chi_2 \\
\end{array}
\right) 
  & \to 
(i \sigma_2 \chi^*)^\prime \left( (i \sigma_2 \chi)^T \right)^\prime  = U (i \sigma_2 \chi^*) (i \sigma_2 \chi)^T  U^\dagger
\end{align}
}
and it is traceless. 
We can now identify\footnote{Note that, hermiticity is not 
necessary for the following map, however, if the matrix is not traceless the map can not be constructed.},
\begin{align} 
 (-i \sigma_2 \chi^*) (-i \sigma_2 \chi)^T  \quad  \leftrightarrow \quad - \vec{\sigma} \cdot \vec{r}  & 
 \nonumber \\
x={1 \over 2} \left( \chi_2^2 - \chi_1^2 \right) \ ,  \quad  y={1 \over 2 i} \left( \chi_2^2 + \chi_1^2 \right) \ , & \quad  z= \chi_1 \chi_2
\end{align}

It is worth to stress that because $ -i \sigma_2 $ can be represented as,
\begin{align} 
-i \sigma_2 = \exp{\left[-i \pi {\sigma_2 \over 2}\right]}
\end{align}
it follows that the equivalent spinors and transformations, 
\begin{align} 
-i \sigma_2 \chi^* = \exp{\left[-i \pi {\sigma_2 \over 2}\right]} \chi^*
\end{align} 
can be express in terms of elements of the algebra (the generators). Therefore,  the construction is guaranteed to work for any representation, not only for the  two-dimensional case explicitly shown in this section.

\newpage

\chapter{Tensorial  Representations of the Lorentz Group}
\phantomsection\label{Tensorial Reps}

\setcounter{equation}{0}
\renewcommand{\theequation}{E.\arabic{equation}}

\setcounter{figure}{0}
\renewcommand{\thefigure}{E.\arabic{figure}}

Special Relativity requires requires that two observers' descriptions of a physical system be equivalent if a Poincaré transformation relates their reference frames (RFs). Poincare transformations include rotations in space, boosts, and translations in space-time\footnote{Lorentz transformations and space-time translations.}. 
Note that the existence of fermions indicates that the representations of the Lorentz group associated with particles in Physics are spinorial representations (see \hyperref[Spinorial Reps]{App. F}).   
However, for the sake of clarity, we  start with the four-vector representation of the Lorentz group.

\subsection*{E.1 Generators} 
\phantomsection\label{Tensorial Reps Gen}

The Lorentz Transformations include space rotations (see \hyperref[Rotations 3d]{App. D.1})
together with 
(rotationless) Boosts, which are defined below.

\subsubsection*{Rotations} 
\phantomsection\label{Tensorial Reps Rot}

The  standard representation for Generators of rotations 
can be inferred directly from 
\hyperref[SO(3) gen 3-dim Sec]{Eq.(\ref*{SO(3) gen 3-dim})}
together with the fact that they only act on spatial components of 4-vectors. Explicitly we have,
{\small
\phantomsection\label{rot gen Sec}
\begin{align}\label{rot gen}
\mathbb{J}_1=\left(
\begin{array}{cccc}
 0 & 0 & 0 & 0 \\
 0 & 0 & 0 & 0 \\
 0 & 0 & 0 & -i \\
 0 & 0 & i & 0 \\
\end{array}
\right) \ , \quad 
\mathbb{J}_2=\left(
\begin{array}{cccc}
 0 & 0 & 0 & 0 \\
 0 & 0 & 0 & i \\
 0 & 0 & 0 & 0 \\
 0 & -i & 0 & 0 \\
\end{array}
\right)\ , \quad
\mathbb{J}_3= \left(
\begin{array}{cccc}
 0 & 0 & 0 & 0 \\
 0 & 0 & -i & 0 \\
 0 & i & 0 & 0 \\
 0 & 0 & 0 & 0 \\
\end{array}
\right) 
\end{align}
}
They occupy a reduced block form of the matrix, which is related to the fact that the generators of spatial rotations form a subgroup.

\subsubsection*{Boosts}
\phantomsection\label{Tensorial Reps Boosts}

The equations relating two inertial frames that differ only by a relative motion along the z-axis with  relative  speed speed $v_{12}$  are\footnote{We use the notation, $x^\mu=(x^0,x^1,x^2,x^3)=(t,x,y,z)$, and the speed of light is set to 1,  ($c=1$).},
\begin{equation}\label{x z-boost}
\tilde x^0= \gamma  \left(x^0+ \beta   x^3\right) \ , \quad
\tilde x^3= \gamma  \left(x^3+ \beta  x^0\right) \ , \quad
\tilde x^1=x^1 \ , \quad
\tilde x^2=x^2 \ , \quad
\end{equation}
where,
\begin{equation}
\gamma =(1- v_{12}^2)^{-1/2} \ , \quad
\end{equation}
Because, $\gamma^2 - \gamma^2 \beta^2=1$, the transformation can be re-parameterized ($v_{12} { } \rightarrow  { } \phi$) in terms of hyperbolic functions,  
\begin{equation}
\gamma = \cosh{\phi} \ , \quad 
\gamma \beta =\sinh{\phi}  
\end{equation}
which leads to an expression for the Lorentz boosts that resembles rotation transformations. Furthermore,
with this parameterization  the boosts  acquire a form that simplifies the study of their  infinitesimal  limit. 

 In matrix form the z-boost can be written as, 
\begin{equation}\label{Lorentz boost - IF}
\tilde x^\mu=\left(\Lambda_\text{z-boost}\right)^\mu{}_\nu x^\nu  
\end{equation}
where the matrix $\Lambda_\text{z-boost}$ has the form,
\begin{align}
\left(\Lambda_\text{z-boost}\right)^\mu{}_\nu = \left(
\begin{array}{cccc}
\cosh \phi & 0 & 0 & \sinh \phi \\
0 & 1 & 0 & 0 \\
0 & 0 & 1 & 0 \\
\sinh \phi  & 0 & 0 & \cosh \phi
\end{array}
\right)
\end{align}
and $\phi $
can be seen as a hyperbolic angle corresponding  to a hyperbolic rotation satisfying, $\cosh^2(\phi) - \sinh^2(\phi) =1$
\footnote{ 
The main difference between the hyperbolic (boosts) and the 
circular (rotations) parameterizations is found in the domain of definition of their respective parameters. For the 
circular ones, satisfying $\cos^2(\theta)+\sin^2(\theta)=1$, the domain of $\theta $ is closed, $\theta \in [0,2\pi]$. In this case, the parameter can take any value of the domain (including the boundary), a  property called compactness. 
On the other hand, the domain for the hyperbolic angle  $\phi \in (-\infty,\infty)$  does not include the boundaries, it is not closed, therefore non-compact.}. 
Equivalently to the case of rotations 
\mbox{\hyperref[SO(3) gen Sec]{(Eq. \ref*{SO(3) gen})}},
the generator of this transformation is obtained by differentiation with respect to the parameter, 
\begin{equation}
\mathbb{K}_z=\frac{\partial \Lambda_\text{z-boost} (\phi )}{i \partial \phi} {\Big |}_{\phi =0}
\end{equation}
which results in, 
\begin{equation}
\mathbb{K}_z= \left(
\begin{array}{cccc}
 0 & 0 & 0 & -i \\
 0 & 0 & 0 & 0 \\
 0 & 0 & 0 & 0 \\
 -i & 0 & 0 & 0 \\
\end{array}
\right)
\end{equation} 
thus,
\begin{align}
\left(\Lambda_\text{z-boost}\right)^\mu{}_\nu =  \exp  \left(i \mathbb{K}_z  {\phi} \right)
\end{align}
Repeating this procedure for boosts along x- and y-axis we find the three generators of rotation-less boosts to be,
\phantomsection\label{boost gen Sec}
\begin{align}\label{boost gen}
\mathbb{K}_x= \left(
\begin{array}{cccc}
 0 & -i & 0 & 0 \\
 -i & 0 & 0 & 0 \\
 0 & 0 & 0 & 0 \\
 0 & 0 & 0 & 0 \\
\end{array}
\right) \ , \quad 
\mathbb{K}_y= \left(
\begin{array}{cccc}
0 & 0 & -i & 0 \\
 0 & 0 & 0 & 0 \\
 -i & 0 & 0 & 0 \\
 0 & 0 & 0 & 0 \\
\end{array}
\right) \ , \quad
\mathbb{K}_z= \left(
\begin{array}{cccc}
 0 & 0 & 0 & -i \\
 0 & 0 & 0 & 0 \\
 0 & 0 & 0 & 0 \\
 -i & 0 & 0 & 0 \\
\end{array}
\right)
\end{align}
The matrices of 
\hyperref[boost gen Sec]{(Eq. \ref*{boost gen})},
associated with $\mathbb{K}_i$ are called the vector or fundamental representation for the generators of standard Lorentz boosts.  

\subsubsection*{Commutations Relations}
\phantomsection\label{Commutations Relations}

The defining properties of the generators are their commutations relations, among them and with the other generators of symmetry transformations, 
\phantomsection\label{App.E.1 Lorentz algebra IF Sec}
\begin{align}\label{App.E.1 Lorentz algebra IF}
\left[\mathbb{J}_l,\mathbb{J}_m \right]= i \epsilon_{lmn}\mathbb{J}_n \ , \quad 
\left[\mathbb{J}_l,\mathbb{K}_m \right]= i \epsilon_{lmn}\mathbb{K}_n \ , \quad
\left[\mathbb{K}_l,\mathbb{K}_m \right]= - i \epsilon_{lmn}\mathbb{J}_n 
\end{align}

\subsection*{E.2 Standard or Instant-Form  4-vector Representation  }
\phantomsection\label{App. IF boost rep}

The standard boost is conventionally defined as a rotationless boost in the direction of the relative velocity between  two frames of reference (observers). Let us assume that in the ``first'' reference frame 
a massive particles is at rest. The components of the particle four-momentum in the ``first'' frame are then,  
\begin{align}
\overset{\circ}p{}^\mu=(m,0,0,0)
\end{align}
A standard boost to the ``second'' frame transforms  the momentum components to, $p^\mu=(E,\vec p \hspace{.05cm})$.
In other words, this is how an observer in the ``second'' reference frame perceives the momentum of the particle. 
Note that we have assumed that the four components of the momentum are parameterized as, $p^\mu=(p^0,p^1,p^2,p^3)=(E,p_x,p_y,p_z)$, which is familiar from any textbook in special relativity.
In matrix form we can write the standard boost as,
\phantomsection\label{Lorentz boost - IF Sec}
\begin{align}\label{Lorentz boost - IF}
p^\mu & = \left( \Lambda^\text{IF} \right)^\mu{}_\nu \ \overset{\circ}p{}^\nu
\end{align}
where, IF stands for Instant Form of dynamics, to be defined in section  \hyperref[App. IF Dynamics]{App.(E.3)}   below. 

The main feature of the Instant Form boost is that they are defined to be rotationless, which  implies the following parameterization in terms of the generators $\mathbb{K}_i$ 
\hyperref[boost gen Sec]{(Eq. \ref*{boost gen})},
\phantomsection\label{Lorentz boost parameterization - IF Sec}
\begin{align}\label{Lorentz boost parameterization - IF}
\Lambda^\text{IF}  & = \exp  \left(i \vec{\mathbb{K}} \cdot \vec{\phi} \right) = \exp  \left(i \phi \vec{\mathbb{K}} \cdot \hat{\phi} \right) 
\end{align}
where\footnote{$\phi=|\vec{\phi}|$, and, $\hat\phi=\vec\phi/\phi$.}, $\vec{\phi}=\phi\hat\phi=(\phi_x,\phi_y,\phi_z)$, correspond to three parameters that completely characterizes each transformation.
Note that because the generators $\mathbb{K}_i$  do not commute 
\hyperref[App.E.1 Lorentz algebra IF Sec]{(Eqs. \ref*{App.E.1 Lorentz algebra IF})},
it follows that\footnote{Or any other order for what matters.},
\begin{align}
\Lambda^\text{IF}  & = \exp  \left(i \vec{\mathbb{K}} \cdot \vec{\phi} \right) \ne \exp  \left(i {\mathbb{K}}_x {\phi}_x \right) \exp  \left(i {\mathbb{K}}_y {\phi}_y \right) \exp  \left(i {\mathbb{K}}_z {\phi}_z \right)
\end{align}
The parameterization\footnote{Where all the generators are collected in one single exponent, which in general is a linear combination of the generators.} of 
\hyperref[Lorentz boost parameterization - IF Sec]{Eq.(\ref*{Lorentz boost parameterization - IF})},
is called the Canonical parameterization.

\subsubsection*{Explicit Representation}
\phantomsection\label{App. Explicit Representation}


In order to obtain an explicit representation 
we substitute  the vector representation for the boosts generators 
\hyperref[boost gen Sec]{(Eq. \ref*{boost gen})},
into the definition for the IF boost 
\hyperref[Lorentz boost parameterization - IF Sec]{Eq.(\ref*{Lorentz boost parameterization - IF})}\footnote{Notice that by definition the dummy index in the Lorentz transformation $\Lambda^\text{IF}$ is a lower index.},
we have,

{\small
\begin{align}
(\Lambda^\text{IF})^\mu{}_\nu  &= \left(
\begin{array}{cccc}
 \cosh (\phi ) & \hat\phi_1 \sinh (\phi ) & \hat\phi_2 \sinh (\phi ) & \hat\phi_3 \sinh (\phi ) \\
 \hat\phi_1 \sinh (\phi ) & \hat\phi_1^2 \cosh (\phi )+\hat\phi_2^2+\hat\phi_3^2 & 2 \hat\phi_1 \hat\phi_2 \sinh ^2\left(\frac{\phi }{2}\right) & 2 \hat\phi_1 \hat\phi_3 \sinh ^2\left(\frac{\phi }{2}\right) \\
 \hat\phi_2 \sinh (\phi ) & 2 \hat\phi_1 \hat\phi_2 \sinh ^2\left(\frac{\phi }{2}\right) & \hat\phi_2^2 \cosh (\phi )+\hat\phi_1^2+\hat\phi_3^2 & 2 \hat\phi_2 \hat\phi_3 \sinh ^2\left(\frac{\phi }{2}\right) \\
 \hat\phi_3 \sinh (\phi ) & 2 \hat\phi_1 \hat\phi_3 \sinh ^2\left(\frac{\phi }{2}\right) & 2 \hat\phi_2 \hat\phi_3 \sinh ^2\left(\frac{\phi }{2}\right) & \hat\phi_3^2 \cosh (\phi )+\hat\phi_1^2+\hat\phi_2^2 \\
\end{array}
\right) 
\end{align}
}

The meaning of the parameters come from the action of $\Lambda^\text{IF}$ on the 4-momentum associated with a particle at rest, $\overset{\circ}p{}^\mu=(m,0,0,0)$, which produces,
\begin{align}
 p{}^\mu = (E,\vec p) =  (\Lambda^\text{IF})^\mu{}_\nu \overset{\circ}p{}^\nu  = m \left(\cosh (\phi ),\hat\phi_1 \sinh (\phi ),\hat\phi_2 \sinh (\phi ),\hat\phi_3 \sinh (\phi ) \right)
\end{align}
thus, showing that the parameters are,

\phantomsection\label{App boost parameters IF Sec}
\begin{align}\label{App boost parameters IF}
\cosh (\phi) = {E \over m}  \ , \quad 
\hat{\phi}_j \sinh (\phi ) = {p_j \over m} 
\end{align}
from where we get,

\phantomsection\label{App boost parameters 2 IF Sec}
\begin{align}\label{App boost parameters 2 IF}
(\hat\phi)^2 (\sinh (\phi))^2 = (\hat\phi)^2 \left(1 + \cosh (\phi)^2\right) = \left({p_j \over m}\right)^2 \ \implies  \ \hat \phi = {\vec{p} \over |\vec{p}|} \ , \  \sinh (\phi ) = { |\vec{p}| \over m}
\end{align}
Using now the relations for the half ``angle'' of hyperbolic functions we find,  

\phantomsection\label{App boost parameters phi-half IF Sec}
\begin{align}\label{App boost parameters phi-half IF}
\cosh \left(\frac{\phi }{2}\right) = \sqrt{\frac{E+m}{2m}} \ , \quad \sinh \left(\frac{\phi }{2}\right) = \sqrt{\frac{E-m}{2m}} 
\end{align}
which results in,

\begin{align}
(\Lambda^\text{IF})^\mu{}_\nu  = \left(
\begin{array}{cccc}
 {E \over m} &  {p_1 \over m} &  {p_2 \over m} &  {p_3 \over m} \\
  {p_1 \over m} &  {p_1^2 \over |\vec p|^2} {E \over m}+ {p_2^2 \over |\vec p|^2}+ {p_3^2 \over |\vec p|^2} & 2  {p_1 \over |\vec p|}  {p_2 \over |\vec p|} \frac{E-m}{2m} & 2  {p_1 \over |\vec p|}  {p_3 \over |\vec p|} \frac{E-m}{2m} \\
  {p_2 \over m} & 2  {p_1 \over |\vec p|}  {p_2 \over |\vec p|} \frac{E-m}{2m} &  {p_2^2 \over |\vec p|^2} {E \over m}+ {p_1^2 \over |\vec p|^2}+ {p_3^2 \over |\vec p|^2} & 2  {p_2 \over |\vec p|}  {p_3 \over |\vec p|} \frac{E-m}{2m} \\
  {p_3 \over m} & 2  {p_1 \over |\vec p|}  {p_3 \over |\vec p|} \frac{E-m}{2m} & 2  {p_2 \over |\vec p|}  {p_3 \over |\vec p|} \frac{E-m}{2m} &  {p_3^2 \over |\vec p|^2} {E \over m}+ {p_1^2 \over |\vec p|^2}+ {p_2^2 \over |\vec p|^2} \\
\end{array}
\right)
\end{align}
Note that, 

\begin{align}
{p_1^2 \over |\vec p|^2} {E \over m}+ {p_2^2 \over |\vec p|^2}+ {p_3^2 \over |\vec p|^2} = {p_1^2 \over |\vec p|^2} {E - m \over m}+ 1 = {p_1^2 \over (E + m) m}+ 1
\end{align}
and with similar manipulations for other matrix elements, we arrive to the well known result,

\phantomsection\label{boost IF parameterization Sec}
\begin{align}\label{boost IF parameterization}
(\Lambda^\text{IF})^\mu{}_\nu  = \left(
\begin{array}{cc}
 {E \over m} &  {\vec p \over m}  \\
  {\vec p \over m} &  \delta_{ij} +  {p_i p_j \over (E + m) m}  \\
\end{array}
\right)
\end{align}

\subsection*{E.3 Instant Form Dynamics}
\phantomsection\label{App. IF Dynamics}

Note that the so called ``standard'' or Instant Form definition is actually a parameterization choice 
over the geometrical structure of space-time. 
In the next section 
(\hyperref[App. LF Dynamics]{App. E.4}) 
we will see how a different parameterization choice results in a different transformation law, i.e., a different relation between the momentum components of a particle at rest and its momentum components when is moving.   
A useful parameterization entails to relate different regions (events) in space-time in a consistent manner\footnote{Agrees with physical experiments.} while providing a reasonably ``simple'' interpretation.  
Among the possible (re)parameterizations that can be made, three has been shown to be 
most useful.
They differ on the direction along which the evolution of a system is parameterized, which is also referred as the dynamics of a physical system,  and the parameter is called time. 
The three forms of dynamics were unveiled in 1949 by P.A.M. Dirac in a seminal paper \cite{Dirac1949}. Dirac called them the Instant Form (IF), the Light Front Form, and the Point Form.

For the Instant Form, the chosen direction of the evolution time is our experiential time $t$, and evolution is understood as the study of changes occurring from one instant of time $t$ to another (later) one.
Therefore, 
Poincare transformations that leave invariant the condition $t=\text{\it constant}$, will have no effect on the description of the evolution of a system\footnote{Or better say, the way we experience the evolution of the system.}.
Since they do not interfere with evolution, the generators of this transformations are called kinematical. 
On the other hand, the generators of Poincare transformations that do not leave invariant the condition $t=\text{\it constant}$ are called dynamical.

\subsubsection*{Instant Form Kinematical Generators}
\phantomsection\label{Instant Form Kinematical Generators}

 The kinematical generators in the case of the Instant Form of dynamics are spatial-rotations and spatial-translations\footnote{Related to the canonical momentum operators},
\begin{equation}
\mathbb{J}_x \ , \quad
\mathbb{J}_y \ , \quad
\mathbb{J}_z \ , \quad 
\mathbb{P}_x \ , \quad
\mathbb{P}_y  \ , \quad
\mathbb{P}_z  
\end{equation}
They  form a subgroup of the Poincare group, {\it i.e.}, the subgroup of (space) rotations and translations.

\subsubsection*{Instant Form Dynamical Generators}
\phantomsection\label{App. IF Dynamical Generators}

The generators of the Dynamics, also called Hamiltonians, are the energy (generator of time $(t)$ translations) and standard boosts,
\begin{equation}
\mathbb{E} \ , \quad
\mathbb{K}_x \ , \quad
\mathbb{K}_y  \ , \quad
\mathbb{K}_z  
\end{equation}
They generate transformations that do not fulfill the constrain, $t_0=\text{constant}$. 
Hence, knowledge about the system at times different than $t_0$ is needed in order to perform the transformation, which implies having solved the dynamics\footnote{At least over some interval of time.}.

\subsection*{E.4 Light Front Dynamics} 
\phantomsection\label{App. LF Dynamics}

Although the standard form of dynamics is the most familiar and natural to us, it turns out that for fast-moving systems\footnote{``Close'' to the speed of light.}, there are other parameterizations of the dynamics that simplify the description and interpretation of the system's evolution. For the present work, the most relevant is the so-called Light Front form of dynamics (LF), discussed in this section.  
In the Light-Front parameterization for evolution, the parameter ``time" is denoted by $\tau$ and chosen along a light-like direction, which is  (conventionally) chosen to be  $\tau=x^+=t+z$.

We will follow the same procedure used in Sec.(\hyperref[App. IF boost rep]{App. E.2}) to find the explicit four-vector representation of the standard (IF) boosts, i.e.,  \hyperref[boost IF parameterization Sec]{Eq.(\ref*{boost IF parameterization})}.
However, now we want to find the matrix for the Light-Front boosts. 
As we shall see, in the case of the four-vector representation, it all boils down  to identify which are the parameters in the definition of the general LF boost.
This identification is provided in section \hyperref[App.E.4.1 LF boost rep]{App.(E.4.1)}.

\subsubsection*{Light Front Kinematical Generators}
\phantomsection\label{Light Front Kinematical Generators}

The LF kinematical generators are defined by linear combinations of the standard generators in such a way that its action leaves invariant the hyperplane, {}$x^+=0$, these are: 

\begin{itemize}

\item Rotation about the LF axis,
\begin{align}
\mathbb{J}_3=\mathbb{J}_z \ , \quad 
\end{align}

\item Translations within the $x^+=0$ plane,
\begin{align}
\mathbb{P}^+ = \mathbb{P}_0 + \mathbb{P}_z \ , \quad
\mathbb{P}_x \ , \quad
\mathbb{P}_y 
\end{align}

\item LF Boosts,
\begin{align}
\mathbb{G}_1=\mathbb{G}_x=\mathbb{K}_x-\mathbb{J}_y\ , \quad
\mathbb{G}_2=\mathbb{G}_y=\mathbb{K}_y+\mathbb{J}_x\ , \quad
\mathbb{K}_3=\mathbb{K}_z
\end{align}

\end{itemize}

Substituting the 4-vector representation of the IF generators 
(Eqs. \hyperref[rot gen Sec]{\ref*{rot gen}} and \hyperref[boost gen Sec]{\ref*{boost gen}}) we find the explicit representation for the kinematical LF boost generators. 

For $\mathbb{G}_1$ and $\mathbb{G}_2$, we have
\phantomsection\label{LF transverse boosts Sec}
\begin{align}\label{LF transverse boosts}
\mathbb{G}_1=\left(
\begin{array}{cccc}
 0 & -i & 0 & 0 \\
 -i & 0 & 0 & -i \\
 0 & 0 & 0 & 0 \\
 0 & i & 0 & 0 \\
\end{array}
\right) \ , \quad
\mathbb{G}_2=\left(
\begin{array}{cccc}
 0 & 0 & -i & 0 \\
 0 & 0 & 0 & 0 \\
 -i & 0 & 0 & -i \\
 0 & 0 & i & 0 \\
\end{array}
\right)
\end{align}

The LF boosts commutation relations follows directly from using the IF algebra 
\hyperref[App.E.1 Lorentz algebra IF Sec]{Eq.(\ref*{App.E.1 Lorentz algebra IF})}, 
\phantomsection\label{LF boost gen CR Sec}
\begin{align}\label{LF boost gen CR}
[\mathbb{K}_3 , \mathbb{G}_1] = i  \mathbb{G}_1 \ , \quad
[\mathbb{K}_3 , \mathbb{G}_2] = i \mathbb{G}_2  \ , \quad
[\mathbb{G}_i , \mathbb{G}_j ] = 0
\end{align}


\subsubsection*{Light Front Dynamical Generators}
\phantomsection\label{App. LF Dynamical Generator}

As expected, the LF Dynamical Generators relate regions of space-time outside the plane, {}$x^+=0$, 
\begin{equation}
\mathbb{P}^- = \mathbb{P}_0 - \mathbb{P}_z \ , \quad
\mathbb{D}_x=\mathbb{K}_x+\mathbb{J}_y \ , \quad
\mathbb{D}_y=\mathbb{K}_y-\mathbb{J}_x 
\end{equation}
where, $\mathbb{P}^- $ is called the LF energy and generates the LF time evolution, whereas  $\mathbb{D}_x$, $\mathbb{D}_y$, are called  LF rotations and they change the light-like direction chosen to parameterize the evolution. 

\subsubsection*{E.4.1 Vector Representation of the Light Front Boost }
\phantomsection\label{App.E.4.1 LF boost rep}

A general LF boost is conventionally defined\footnote{For its action on a particle.} as a two step process, first a boost in the direction of LF propagation ($z-$axis) from rest, 
\phantomsection\label{momentum at rest IF Sec}
\begin{align}\label{momentum at rest IF}
\overset{\circ}p{\hspace{0.05cm}}^\mu=(m,0,0,0) 
\end{align}
to a frame where it achieve its final component of momentum along the $z-$axis ($p_z$)\footnote{Chosen here as the spatial direction for the light-like parameterization of evolution. The intermediate result would be, $p^\mu_\text{int}=(E_\text{int}, 0, 0, p_z)$.}, followed by a linear combination of generators\footnote{Note that these generators do not change the null plane, $x^+=0$.} such that attains its final transverse momentum $\vec{p}_\text{T}$, explicitly
\phantomsection\label{Lorentz boost - LF Sec}
\begin{align}\label{Lorentz boost - LF}
\Lambda^\text{LF}_\text{def}  = \exp  \left[i \vec{\mathbb{G}} \cdot \vec{\text{v}}_\text{T} \right] \cdot  \exp \left[i \mathbb{K}_3\eta \right] 
\end{align}
where, $\vec{\text{v}}_\text{T}=(\text{v}_1,\text{v}_2)$ and $\eta$ correspond to three parameters that completely characterizes the LF boost. 

It is worth to mention that in the definition of the LF boost  the order between the action of $\mathbb{K}_3$ and $\mathbb{G}_i$ matters, since they do not commute. 
On the other hand, for $\mathbb{G}_1$ and $\mathbb{G}_2$ we have
\begin{align}
[\mathbb{G}_1,\mathbb{G}_2]=0
\end{align}
hence, they can be trivially added into one exponent like in 
\hyperref[Lorentz boost - LF Sec]{Eq.(\ref*{Lorentz boost - LF})}. 
In other words, we can write,
\phantomsection\label{LF boost alternative def Sec}
\begin{align}\label{LF boost alternative def}
\Lambda^\text{LF}_\text{def} & = \exp  \left[i \vec{\mathbb{G}} \cdot \vec{\text{v}}_\text{T} \right] \cdot  \exp \left[i \mathbb{K}_3\eta \right] \nonumber \\
&= \exp  \left[i \mathbb{G}_1 \text{v}_1 \right] \cdot \exp  \left[i \mathbb{G}_2 \text{v}_2 \right] \cdot  \exp \left[i \mathbb{K}_3\eta \right] \nonumber \\
&= \exp  \left[i \mathbb{G}_2 \text{v}_2 \right] \cdot \exp  \left[i \mathbb{G}_1 \text{v}_1 \right] \cdot  \exp \left[i \mathbb{K}_3\eta \right] 
\end{align}
where, the parameters $\vec{\text{v}}_\text{T}=(v_1,v_2)$ and $\eta$ are the same in the three statements.
Although, the definition of LF boosts 
\hyperref[Lorentz boost - LF Sec]{Eq.(\ref*{Lorentz boost - LF})}
is more elegant, the three statements in 
\hyperref[LF boost alternative def Sec]{Eq.(\ref*{LF boost alternative def})}
are equivalent.

Substituting the LF parameterization of the generators 
\hyperref[LF transverse boosts Sec]{(Eq. \ref*{LF transverse boosts})}
in the definition of LF boosts 
\hyperref[Lorentz boost - LF Sec]{(Eq. \ref*{Lorentz boost - LF})},
we arrive at,

{\small
\begin{align}
\hspace{-0.5cm}
(\Lambda^\text{LF}_\text{def})^\mu{}_\nu & = \left(
\begin{array}{cccc}
 \frac{1}{2} e^{-\eta } \left(1+e^{2 \eta }+e^{2 \eta } \text{v}_1^2+e^{2 \eta } \text{v}_2^2\right) & \text{v}_1 & \text{v}_2 & \frac{1}{2}
e^{-\eta } \left(-1+e^{2 \eta }+e^{2 \eta } \text{v}_1^2+e^{2 \eta } \text{v}_2^2\right) \\
 e^{\eta } \text{v}_1 & 1 & 0 & e^{\eta } \text{v}_1 \\
 e^{\eta } \text{v}_2 & 0 & 1 & e^{\eta } \text{v}_2 \\
 -\frac{1}{2} e^{-\eta } \left(1-e^{2 \eta }+e^{2 \eta } \text{v}_1^2+e^{2 \eta } \text{v}_2^2\right) & -\text{v}_1 & -\text{v}_2 & \frac{1}{2}
e^{-\eta } \left(1+e^{2 \eta }-e^{2 \eta } \text{v}_1^2-e^{2 \eta } \text{v}_2^2\right) \\
\end{array}
\right) 
\end{align}
}
The meaning of the parameters comes from acting with $\Lambda^\text{LF}_\text{def}$ on the 4-momentum associated with a particle at rest 
\hyperref[momentum at rest IF Sec]{(Eq. \ref*{momentum at rest IF})},
which results in,
\begin{align}
p^\mu & =(E,p_x,p_y,p_z) =  (\Lambda^\text{LF}_\text{def})^\mu{}_\nu  \overset{\circ}p{}^\nu  \nonumber \\
& = \frac{m}{2} \left( e^{-\eta } + e^{\eta } + e^{\eta }  \vec{\text{v}}_T^2 \ , \ e^{\eta }  \text{v}_1 \ , \ e^{\eta }  \text{v}_2 \ , \  e^{\eta } - e^{-\eta } - e^{\eta } \vec{\text{v}}_T^2 \right)
\end{align}
 using now the LF coordiantes,  $p^+=E+p_z$ and $p^-= E-p_z$, we get
\begin{align}
(p^+ , p_x , p_y , p^-) = m \left( e^{\eta } \ , \ e^{\eta }  \text{v}_1 \ , \ e^{\eta }  \text{v}_2  \ , \  e^{-\eta } + e^{\eta } \vec{\text{v}}_T^2 \right)
\end{align}
showing that the parameters must  fulfill the conditions, 
\phantomsection\label{LF parameters Sec}
\begin{align}\label{LF parameters}
p^+ = m \ e^{\eta }  \ , \quad 
\vec{p}_T = m \ e^{\eta }  \vec{\text{v}}_T \ , \quad 
p^- = m \ e^{- \eta } + m \ e^{\eta } \vec{\text{v}}_T^2  
\end{align}
Thus, we arrive to the following simple form for the LF boost parameters,
\phantomsection\label{LF boost parameters Sec}
\begin{align}\label{LF boost parameters}
e^{\eta } = {p^+ \over m}   \ , \quad 
 \vec{\text{v}}_T = {\vec{p}_T \over p^+}   \ , \quad 
p^- = {m^2 +\vec{p}_T^2 \over p^+}  
\end{align}

\subsubsection*{E.4.2 Light Front Boost in Light Front Coordinates}
\phantomsection\label{App.E.4.2 LF boost rep in Light Front Coord}

A question may rise when following the procedure in the previous section. 
Why are we using
\hyperref[Lorentz boost - LF Sec]{Eq.(\ref*{Lorentz boost - LF})}\footnote{LF boost definition.} to transform the 4-vector  $\overset{\circ}p{\hspace{0.05cm}}^\mu=(m,0,0,0)$\footnote{Momentum at rest in standard coordinates.}? 
The answer is that $(\Lambda^\text{LF}_\text{def})^\mu{}_\nu $ is written  in  standard coordinates as well. 
To write down $\Lambda^\text{LF}_\text{def}$ in light-front coordinates we must perform a change of coordinates\footnote{This change of coordinates do not correspond to any Lorentz transformation. Note that it does not leave the Lorentzian metric ($g^{\mu\nu}=\text{diag}(-1,1,1,1)$) invariant.}. 
The representation of this transformation in space-time is,
\begin{equation}
\Omega_{\text{IF} \to \text{LF}} = \left(
\begin{array}{cccc}
 1 & 0 & 0 & 1 \\
 0 & 1 & 0 & 0 \\
 0 & 0 & 1 & 0 \\
 1 & 0 & 0 & -1 \\
\end{array}
\right)
\end{equation}
Its action on the momentum of a particle at rest results in, 
\phantomsection\label{momentum at rest LF Sec}
\begin{align}\label{momentum at rest LF}
\Omega_{\text{IF} \to \text{LF}} \ \overset{\circ}p{}^\mu_\text{IF} = (m,0,0,m) = \overset{\circ}p{}^\mu_\text{LF}
\end{align}
which applied to the LF boost produces,
\phantomsection\label{LF boost def in LF coord Sec}
\begin{align}\label{LF boost def in LF coord}
\hspace{-0.95cm}
\left(\Omega_{\text{IF} \to \text{LF}} \cdot \Lambda^\text{LF}_\text{def} \cdot \Omega_{\text{IF} \to \text{LF}} ^{-1} \right)^\mu{}_\nu  
= (\Lambda^\text{LF})^\mu{}_\nu  &=
\left(
\begin{array}{cccc}
 e^{\eta } & 0 & 0 & 0 \\
 e^{\eta } \text{v}_1 & 1 & 0 & 0 \\
 e^{\eta } \text{v}_2 & 0 & 1 & 0 \\
 e^{\eta } \vec{\text{v}}_\text{T}^2 & 2 \text{v}_1 & 2 \text{v}_2 & e^{-\eta } \\
\end{array}
\right) 
=\left(
\begin{array}{cccc}
 {p^+ \over m } & 0 & 0 & 0 \\
 { p_1\over m } & 1 & 0 & 0 \\
 { p_2\over m } & 0 & 1 & 0 \\
  { \vec{p}_\text{T}^2 \over m p^+ }  & { 2 p_1 \over p^+ }  &  { 2 p_2 \over p^+ } &  { m \over p^+ }  \\
\end{array}
\right)
\end{align}
Hence, by adopting LF coordinates we obtain more simple expressions. 
In particular, the action of $ \Lambda^\text{LF} $ on the momentum of a particle at rest, also written in LF coordinates 
\hyperref[momentum at rest LF Sec]{(Eq. \ref*{momentum at rest LF})},
reproduces  the results in 
\hyperref[LF parameters Sec]{Eq.(\ref*{LF parameters})},
which is the meaning of the last equality in 
\hyperref[LF boost def in LF coord Sec]{Eq.(\ref*{LF boost def in LF coord})}.

From 
\hyperref[LF boost def in LF coord Sec]{Eq.(\ref*{LF boost def in LF coord})},
it can be seen explicitly that, contrary to the IF case, the LF ``space'' components of momentum, $p^+,\vec{p}_\text{T}$, do not receive any contribution from the $p^-$ component when is boosted from rest. Therefore, LF boosts can be performed over states labeled by  $p^+$ and $\vec{p}_\text{T}$  independently of any knowledge about the LF energy (dynamics), i.e., the LF boosts are kinematic.

\subsection*{E.5 ``Canonical'' Parameterization}
\phantomsection\label{E.5 ``Canonical'' Parameterization}

In order to use the conventional method for deriving the  irreducible representations of the Lorentz group,
corresponding to particles with non zero spin, we must transform the boost definition into the exponential map form (Canonical form), 
\phantomsection\label{canonical boost Sec}
\begin{align}\label{canonical boost}
\Lambda_\text{Definition}  \quad \to \quad \Lambda_\text{Canonical} = \exp \left[ i ( \ldots ) \right]
\end{align}
where all the generators appear in a single exponential.
The definition of standard boost is already in this form. However, the LF definition must be transformed.

\subsubsection*{E.5.1 LF Boost in ``Canonical'' Parameterization}
\phantomsection\label{E.5.1 LF Boost in ``Canonical'' Parameterization}

We have identified what are the parameters that characterize the LF boosts. 
To be able to use the general methodology for deriving the irreducible representations of the Lorentz group we must recast the LF boost definition into the canonical form 
\hyperref[canonical boost Sec]{(Eq. \ref*{canonical boost})} \footnote{Note that the generators of LF boost form a closed algebra 
\hyperref[LF boost gen CR Sec]{Eq.(\ref*{LF boost gen CR})},
hence no others generators appear in the exponent.},
\phantomsection\label{LF boost canonical provisional Sec}
\begin{align}\label{LF boost canonical provisional}
\Lambda^\text{LF}_\text{Can} = \exp \left[ i \vec{\mathbb{G}} \cdot \vec{\tilde{v}}_\text{T} + i \mathbb{K}_3 \tilde\eta \right]
\end{align}
where, the parameters $\vec{\tilde{v}}_\text{T}=(\tilde{v}_1,\tilde{v}_2)$ and $\tilde{\eta}$, are in principle different from those in
\hyperref[LF boost parameters Sec]{Eq.(\ref*{LF boost parameters})}.
Indeed, the matrix form of 
\hyperref[LF boost canonical provisional Sec]{Eq.(\ref*{LF boost canonical provisional})}
is,

{\footnotesize 
\begin{align}
\hspace{-0.85cm}
\left(\Lambda^\text{LF}_\text{Can} \right)^\mu{}_\nu = \left(
\begin{array}{cccc}
 \frac{ \left(e^{\tilde{\eta}}+e^{- \tilde{\eta} }\right) \tilde{\eta}{}^2+\left(e^{\tilde{\eta}\over2}-e^{\tilde{\eta}\over2 }\right)^2\left( \tilde{v}_1^2+ \tilde{v}_2^2\right)} {2\tilde{\eta} ^2} & \frac{\left(1-e^{-\tilde{\eta} }\right) \tilde{v}_1}{\tilde{\eta} } & \frac{ \left(1-e^{-\tilde{\eta} }\right) \tilde{v}_2}{\tilde{\eta} } & \frac{ \left(e^{\tilde{\eta}}-e^{- \tilde{\eta} }\right) \tilde{\eta}{}^2+\left(e^{\tilde{\eta}\over2}-e^{\tilde{\eta}\over2 }\right)^2\left( \tilde{v}_1^2+ \tilde{v}_2^2\right)} {2\tilde{\eta} ^2}  \\
 \frac{-\left(1-e^{\tilde{\eta} }\right) \tilde{v}_1}{\tilde{\eta} } & 1 & 0 & \frac{-\left(1-e^{\tilde{\eta} }\right) \tilde{v}_1}{\tilde{\eta} } \\
 \frac{-\left(1-e^{\tilde{\eta} }\right) \tilde{v}_2}{\tilde{\eta} } & 0 & 1 & \frac{-\left(1-e^{\tilde{\eta} }\right) \tilde{v}_2}{\tilde{\eta} } \\
 \frac{ \left(e^{\tilde{\eta}}-e^{- \tilde{\eta} }\right) \tilde{\eta}{}^2-\left(e^{\tilde{\eta}\over2}-e^{\tilde{\eta}\over2 }\right)^2\left( \tilde{v}_1^2+ \tilde{v}_2^2\right)} {2\tilde{\eta} ^2}  & \frac{ \left(e^{-\tilde{\eta} }-1\right) \tilde{v}_1}{\tilde{\eta} } & \frac{\left(e^{-\tilde{\eta} } -1 \right) \tilde{v}_2}{\tilde{\eta} } & \frac{ \left(e^{\tilde{\eta}}+e^{- \tilde{\eta} }\right) \tilde{\eta}{}^2-\left(e^{\tilde{\eta}\over2}-e^{\tilde{\eta}\over2 }\right)^2\left( \tilde{v}_1^2+ \tilde{v}_2^2\right)} {2\tilde{\eta} ^2}  \\
\end{array}
\right)
\end{align}
}
which in LF coordinates acquires the form, 
\phantomsection\label{LF boost - Canonical form - Lorentz Coord - vector Rep Sec}
\begin{align}\label{LF boost - Canonical form - Lorentz Coord - vector Rep}
\left(\Lambda^\text{LF}_\text{Can} \right)^\mu{}_\nu  = \left(
\begin{array}{cccc}
 e^{\tilde{\eta} } & 0 & 0 & 0 \\
 \frac{\left(e^{\tilde{\eta} }-1\right) {\tilde{v}}_1}{\tilde{\eta} } & 1 & 0 & 0 \\
 \frac{\left(e^{\tilde{\eta} }-1\right) {\tilde{v}}_2}{\tilde{\eta} } & 0 & 1 & 0 \\
 \frac{e^{-\tilde{\eta} } \left(e^{\tilde{\eta} }-1\right)^2 \left({\tilde{v}}_1^2+{\tilde{v}}_2^2\right)}{\tilde{\eta} ^2} & \frac{2 e^{-\tilde{\eta} } \left(e^{\tilde{\eta} }-1\right) {\tilde{v}}_1}{\tilde{\eta} } & \frac{2 e^{-\tilde{\eta} } \left(e^{\tilde{\eta} }-1\right) {\tilde{v}}_2}{\tilde{\eta} } & e^{-\tilde{\eta} } \\
\end{array}
\right)
\end{align}

To identify the meaning of the parameters in the matrix $\Lambda^\text{LF}_\text{Can}$ we make the following observation. 
The matrices 
\hyperref[LF boost - Canonical form - Lorentz Coord - vector Rep Sec]{Eq.(\ref*{LF boost - Canonical form - Lorentz Coord - vector Rep})}
and 
\hyperref[LF boost def in LF coord Sec]{Eq.(\ref*{LF boost def in LF coord})} 
must transform the LF momentum components of a particle at rest 
\hyperref[momentum at rest LF Sec]{Eq.(\ref*{momentum at rest LF})}
into the very same transformed momentum.
In other words, the two matrices must be identical.
Comparing them we find the relations between the two set of parameters,
\phantomsection\label{LF boost parameters - canonical Sec}
\begin{align}\label{LF boost parameters - canonical}
\tilde{\eta}=\eta  = \log{\left( p^+ \over m \right)}   \ , \quad 
 \vec{\tilde{v}}_\text{T}  = {\eta \over 1 - e^{ - \eta }} \vec{\text{v}}_\text{T} = {\eta \over 1 - e^{ - \eta }} {\vec{p}_\text{T}  \over p^+}     = {\eta \over p^+ - m} {\vec{p}_\text{T} }     
\end{align}

Then the LF boost in ``canonical'' form and in terms of ``good'' parameters is,

\phantomsection\label{LF boost canonical Sec}
\begin{align}\label{LF boost canonical}
\Lambda^\text{LF} = \exp \left[ i \vec{\mathbb{G}} \cdot \vec{\text{v}}_\text{T}  \frac{\eta }{1-e^{-\eta }} + i \mathbb{K}_3 \eta \right] = \exp \left[ i {\eta \over p^+ - m}  {\vec{p}_\text{T}  } \cdot  \vec{\mathbb{G}}  + i \eta  \mathbb{K}_3 \right]
\end{align}

Note that compared with the canonical re-parameterization  
\hyperref[LF boost canonical provisional Sec]{(Eq. \ref*{LF boost canonical provisional})},
the definition of LF boosts 
\hyperref[Lorentz boost - LF Sec]{(Eq. \ref*{Lorentz boost - LF})}
results in more simple
relations between the boost parameters and the LF variables $p^+$ and  $\vec{p}_\text{T}$\footnote{Compare 
\hyperref[LF boost parameters Sec]{Eq.(\ref*{LF boost parameters})}
with 
\hyperref[LF boost parameters - canonical Sec]{Eq.(\ref*{LF boost parameters - canonical})}.}.
On this regard, it is preferred as the definition for the LF boosts. However, the canonical form is more useful for generalizations to spinorial representations\footnote{See \hyperref[Spinorial Reps]{Appendix F}.}. 

\subsection*{E.6 Higher Rank Tensor Representations}

A rank $k$ tensor $T^{\mu_1 \cdots \mu_k}$, transform on each index as a four-vector representation, i.e., each index transform under the action of a matrix $\Lambda^\text{LF}$.
All the tensorial representations can be reduced to more simple ones.
The more simple ones, called irreducible, are preferred because they reflect the correct number of degrees of freedom for particles that transform under such a rule.

\newpage

\chapter{Spinorial Representations of the Lorentz Group} 
\phantomsection\label{Spinorial Reps}

\setcounter{equation}{0}
\renewcommand{\theequation}{F.\arabic{equation}}

\setcounter{figure}{0}
\renewcommand{\thefigure}{F.\arabic{figure}}


In  Appendix section \hyperref[Sec SU(2) and SO(3)]{(D.2)}, we  found that 
3-dimensional  (Euclidean)  vectors\footnote{Meaning three real numbers forming an ordered array (3-component tuples), and with the property  that they transform under rotations by matrix multiplication with  
\hyperref[Rotations 3-dim vector Sec]{(\ref*{Rotations 3-dim vector})}.}, 
and 2-dimensional (Pauli) spinors\footnote{Complex 2-component tuples transforming under rotations with the half angle law of
\hyperref[Rotations 2-dim spinor Sec]{Eq.(\ref*{Rotations 2-dim spinor})}.}  
can be related  via  
\hyperref[SU(2) SO(3) correspondence Sec]{Eq.(\ref*{SU(2) SO(3) correspondence})}.
It follows that each rotation in 3-dimensional space corresponds to two different rotations in spinor space, i.e.,  
for a unit vector $\hat{n}$ the map of 
\hyperref[SU(2) SO(3) correspondence Sec]{Eq.(\ref*{SU(2) SO(3) correspondence})}
means that when the two-component spinor rotates a full cycle, as $\theta$ runs from $0$ to $4\pi$, 
the corresponding  Euclidean vector cover two cycles 
in 3-dimensional space\footnote{This is the statement that SU(2) is the double covering group of SO(3)}.

This result is so general that can be used to define what we mean by a spinor. 
An $m$-dimensional spinor is a complex $m$-tuple, which is an element of a complex vector space\footnote{It has the structure of a vector space under the operations of vector addition and multiplication by scalars.} transforming under rotations with the corresponding law,
\phantomsection\label{Rotations spinor Sec}
\begin{align}\label{Rotations spinor}
U(s,\hat{n}, \theta) = e^{i \theta \hat{n} \cdot {\vec{\mathbb{J} } }{}^{(s)}} 
\end{align}
where, $\vec{\mathbb{J}}^{(s)}=(\mathbb{J}_1^{(s)},\mathbb{J}_2^{(s)},\mathbb{J}_3^{(s)})$, are the generators for the spin ($s=1/2,\ 1,\ 3/2,\ldots$) representation of SU(2), and  $\hat{n}$ is a unit vector in the direction of the rotation axis, $\hat{n}^2=1$. The dimension $m$ and the spin are related by, $m=2s+1$. Note that for half integer spin representations 
\hyperref[Rotations spinor Sec]{Eq.(\ref*{Rotations spinor})}
results in the half angle transformation law, in particular, for the spin 1/2 representation we have, $\vec{\mathbb{J}}{}^{(1/2)} \to \vec{\sigma}/2$.  

The  correspondence between rotations in 3-dimensional space and spinor space\footnote{For the sake of simplifying the notation, whenever it does not lead to confusion, sometimes we will omit the label for the spin.}, i.e., the map,
\begin{align}
\vec{r} \in \mathbb{R}^3 \to \vec{r} \cdot \vec{\sigma} = \left(
\begin{array}{cc}
  z &  x-i y \\
  x+i y & - z \\
\end{array}
\right) \in \mathbb{C}^2
\end{align}
induces for every spatial rotation in $\mathbb{R}^3$ an SU(2) transformation in $\mathbb{C}^2$, such that,
\phantomsection\label{SU(2) SO(3) map Sec}
\begin{align}\label{SU(2) SO(3) map}
U(s,\hat{n}, \theta) = e^{i \theta \hat{n} \cdot {\vec{\mathbb{J} } }{}^{(s)}} { } \longleftrightarrow  { } R(\hat{n}, \theta) = e^{i \theta 
\hat{n} \cdot \vec{J} }
\end{align}
This relation shows that all spinors with definite spin $s$ transform under  spatial rotations with the very same law. 
This is to be contrasted with Lorentz boosts transformations. We will see that for each spin there are two types of 
spinors, which can be identified from each other by their distinct transformation rules under boosts.
We say that for spinor representations there are two inequivalent Irreducible Representations (IRReps) for each spin. 
In general, the two IRReps are related by ``parity conjugation''. 
There is however a caveat, since  parity conjugation must act only on ``space'' components, the {\it appropriate} ``parity conjugation'' operation 
depends on 
what particular parameterization
is used to describe the geometry of  space-time. 

\subsection*{F.1 Correspondence between so(1,3) and sl(2,$\mathbb{C}$)}
\phantomsection\label{F.1 Correspondence between so(1,3) and sl(2,C)}

In the previous section,
we have established the explicit correspondence between the rotation a spinor undergoes in spinor space, and 
the space rotation experienced by a 
vector in space-time 
\hyperref[SU(2) SO(3) map Sec]{Eq.(\ref*{SU(2) SO(3) map})}\footnote{Note that the difference between the original and rotated vector is space-like, which means there is a frame where the rotation takes place on a fixed time hyper-surface.}.
 This allows to compare spinors on frames of reference that differ by some rotation of the space coordinates.
We want now to obtain the analogous correspondence for the case of Lorentz boosts, i.e., we want to know how the components of a spinor transform between inertial frames that only differ by a relative velocity\footnote{ Note that, like in the case of rotations, as seen by a third observer (which could be on one of the two frames related by $v_{12}$) a 4-vector on the frames of reference related by $v_{12}$ differ by a space-like 4-vector, $\Delta p^2=(p^\mu-\overset{\circ}{p}{}^\mu)^2=2m(m-E)<0$. Hence, there is a frame where the change is of the form, $\Delta \tilde{p}{}^\mu=(0,\vec{p})$, this is the Breit frame.}.

There is a correspondence between Lorentz transformations in 4-dimensional space-time and spinor space, i.e., the map,
\begin{align}
x^\mu \in \mathbb{M}^4 \to x^\mu \sigma_\mu = \left(
\begin{array}{cc}
  t+z &  x-i y \\
  x+i y & t - z \\
\end{array}
\right) \in \mathbb{C}^2
\end{align}
induces for every Lorentz transformations in $\mathbb{M}^4$ an SL(2,$\mathbb{C}$) transformation in $\mathbb{C}^2$, such that,
\phantomsection\label{SL(2,C) SO(1,3) map Sec}
\begin{align}\label{SL(2,C) SO(1,3) map}
D(\Lambda) = e^{i \vec{\theta}  \cdot {\vec{\mathbb{J} }}{}^{(s)} \pm \vec{\phi}  \cdot {\vec{\mathbb{K} } }{}^{(s)}} 
{ } \longleftrightarrow  { } 
\Lambda_{(\vec{\theta},\vec{\phi})} 
\end{align}

The procedure followed here 
is similar to that one used in relating SO(3) with SU(2), i.e. finding a map between the real group and the complex covering group\cite{Weinberg1964,Ryder1996}.
For the 4-vector representation, we found that the standard Generators
satisfy the algebra,
\begin{align}
\left[\mathbb{J}_l,\mathbb{J}_m \right]= i \epsilon_{lmn}\mathbb{J}_n \ , \quad 
\left[\mathbb{J}_l,\mathbb{K}_m \right]= i \epsilon_{lmn}\mathbb{K}_n \ , \quad
\left[\mathbb{K}_l,\mathbb{K}_m \right]= - i \epsilon_{lmn}\mathbb{J}_n 
\end{align}
therefore, the combinations\footnote{ Complexification of the algebra.},
\begin{align}
\mathbb{A}_m=\frac{1}{2}\left(\mathbb{J}_m+i \mathbb{K}_m\right)  \ , \quad 
\mathbb{B}_m =\frac{1}{2}\left(\mathbb{J}_m-i \mathbb{K}_m\right)
\end{align}
satisfy the commutation relations of the SU(2) algebra\footnote{Thus, each set ($\mathbb{A}_m$ and $\mathbb{B}_m$) generates the SU(2) group.},
\phantomsection\label{A and B algebra Sec}
\begin{align}\label{A and B algebra}
\left[\mathbb{A}_l,\mathbb{A}_m\right] = i \epsilon_{lmn} \mathbb{A}_n  \ , \quad 
\left[\mathbb{B}_l,\mathbb{B}_m\right] = i \epsilon_{lmn} \mathbb{B}_n  \ , \quad 
\left[\mathbb{A}_l , \mathbb{B}_m \right] =0 
\end{align}
where, indices $l$, $m$, and $n$ run over $1,2,3$. Since the new generators  $\vec{\mathbb{A}}$ and $\vec{\mathbb{B}}$ commute with each other, 
it follows that they generate independent transformations\footnote{It follows that the Lorentz Group can be mapped (locally) to: 
$\text{SU}(2) \otimes \text{SU}(2)$.}.

Complex vectors transforming only under $\mathbb{A}$ correspond to constrain the generators to satisfy, 
\phantomsection\label{Right handed transformations Sec}
\begin{align}\label{Right handed transformations}
\mathbb{B}_m=0 \quad \equiv \quad  \mathbb{K}_m \to \mathbb{K}_m^R = -i  \mathbb{J}_m 
\end{align} 
which are classified by an angular-momentum-like label\footnote{Since they generate the SU(2) group, the representations can be classified in the same way as the Rotation group.}, $j_R$, and they are referred as the right-handed chiral representations. 
Analogously, complex vectors transforming only under  $\mathbb{B}$  correspond to constrain the generators to satisfy, 
\phantomsection\label{Left handed transformations Sec}
\begin{align}\label{Left handed transformations}
\mathbb{A}_m=0 \quad \equiv \quad  \mathbb{K}_m \to \mathbb{K}_m^L =  i \mathbb{J}_m 
\end{align} 
which are classified by another (independent) angular-momentum-like label $j_L$, and are called  left-handed chiral representations\footnote{Note that both 
\hyperref[Right handed transformations Sec]{Eq.(\ref*{Right handed transformations})}
and 
\hyperref[Left handed transformations Sec]{Eq.(\ref*{Left handed transformations})} 
are consistent with the commutation relations of
\hyperref[App.E.1 Lorentz algebra IF Sec]{Eqs.(\ref*{App.E.1 Lorentz algebra IF})}.}.

Therefore, choices 
\hyperref[Right handed transformations Sec]{Eq.(\ref*{Right handed transformations})}
and 
\hyperref[Left handed transformations Sec]{Eq.(\ref*{Left handed transformations})}
correspond to  two kind of complex vectors each transforming differently. 
As can be seen from 
\hyperref[A and B algebra Sec]{Eq.(\ref*{A and B algebra})}
these are the only two independent possibilities.
This result is summarized by the statement that any arbitrary representation of the Lorentz group\footnote{ Here we always refer to the simple connected proper orthochronous subgroup of the Lorentz group. } can be written in terms of $\vec{\mathbb{A}}$ and $\vec{\mathbb{B}}$. Henceforth, the representations of the Lorentz group are classified by  two angular momentum labels $(j_R,j_L)$  corresponding to $\mathbb{A}_r$ and $\mathbb{B}_l$ respectively. 

The Lorentz transformations for right an left handed (chiral) spinors then become,
\phantomsection\label{A and B transformations Sec}
\begin{align}\label{A and B transformations}
  (j_R,0) &: \quad  D^R_{(\Lambda)}{}^{(s)} = D_{(\Lambda)}^{(s)} = \exp\left[ i \vec{\mathbb{J}}{}^{(s)} \cdot \left(\vec{\theta } - i \vec{\phi }\right) \right] \nonumber  \\  
 (0,j_L) &: \quad  D^L_{(\Lambda)}{}^{(s)} = \bar{D}_{(\Lambda)}^{(s)} = \exp\left[ i \vec{\mathbb{J}}{}^{(s)} \cdot \left(\vec{\theta } + i \vec{\phi }\right) \right]
\end{align} 
where, $\Lambda$ is an arbitrary Lorentz transformation with parameters $\vec{\theta}$, and, { }$\vec{\phi}$, 
characterizing (space) rotations and (pure) standard boosts respectively\footnote{See 
\hyperref[Rotations 3-dim vector Sec]{Eq.(\ref*{Rotations 3-dim vector})}
and 
\hyperref[Lorentz boost - IF Sec]{Eq.(\ref*{Lorentz boost - IF})}.}. 
 The transformation laws 
 \hyperref[A and B transformations Sec]{Eq.(\ref*{A and B transformations})}
 show that both type of spinors transform in the same way under rotations but not under boosts. Thus, they only differ from each other when have been boosted, i.e. there is no meaningful way to distinguish them when they are at rest.

It is worth to stress the generality of this construction.
Notice that nowhere in the derivation we need to make reference to what spin are we dealing with, i.e. a knowledge of the algebra\footnote{Which must be satisfied by every particular representation.} was enough. 
This construction is used to represent chiral spinors with the spin labeled by  $j_{R, L}$, which translates into a representation of the generators $ \vec{\mathbb{J}} $ of SU(2).
The advantage of these construction over a tensorial one, is that it does not introduces superfluous degrees of freedom which need to be removed later.
For example, the spin 1 representation can be obtained from the tensor product of two spin $1/2$ representations, $1/2 \otimes 1/2 = 1 \oplus 0$, from where we must remove the trivial spin 0 representation. 



\subsection*{F.2 Discrete Space-Time Symmetries and Bi-spinors}
\phantomsection\label{App. Discrete Symmetries}
\subsubsection*{Parity and Time reversals}

Since the algebra of the generators is the same for every representation, we can use the scalar representation to explicitly show these statement. For the scalar representation the indices on the matrices are space-time indices, thus the Parity and Time reversal operations are represented as,
\phantomsection\label{Parity and Time reversal Sec}
\begin{align}\label{Parity and Time reversal}
\mathcal{P}^{(0)}= \left(
\begin{array}{cccc}
 1 & 0 & 0 & 0 \\
 0 & -1 & 0 & 0 \\
 0 & 0 & -1 & 0 \\
 0 & 0 & 0 & -1 \\
\end{array}
\right)  \ , \quad
\mathcal{T}^{(0)}= \left(
\begin{array}{cccc}
 -1 & 0 & 0 & 0 \\
 0 & 1 & 0 & 0 \\
 0 & 0 & 1 & 0 \\
 0 & 0 & 0 & 1 \\
\end{array}
\right)
\end{align}
which transform the generators of rotations 
\hyperref[rot gen Sec]{Eq.(\ref*{rot gen})}
and boosts 
\hyperref[boost gen Sec]{Eq.(\ref*{boost gen})}
by,
\phantomsection\label{Parity transf of gen Sec}
\begin{align}\label{Parity transf of gen}
 \mathcal{P}^{(0)} \mathbb{J}_m^{(0)} \left(\mathcal{P}^{(0)}\right)^{-1}  =\mathcal{P}^{(0)} \mathbb{J}_m^{(0)} \mathcal{P}^{(0)}  = + \mathbb{J}_m^{(0)}  \quad & \implies \quad \vec{\mathbb{J}}  \quad \overset{\mathcal{P}}{\longleftrightarrow } \quad \vec{\mathbb{J}} \nonumber \\
 \mathcal{T}^{(0)} \mathbb{J}_m^{(0)} \left(\mathcal{T}^{(0)}\right)^{-1}  =\mathcal{T}^{(0)} \mathbb{J}_m^{(0)} \mathcal{T}^{(0)}  = + \mathbb{J}_m^{(0)}  \quad & \implies \quad \vec{\mathbb{J}}  \quad \overset{\mathcal{T}}{\longleftrightarrow } \quad \vec{\mathbb{J}} \nonumber \\
 \mathcal{P}^{(0)} \mathbb{K}_m^{(0)} \left(\mathcal{P}^{(0)}\right)^{-1}  =\mathcal{P}^{(0)} \mathbb{K}_m^{(0)} \mathcal{P}^{(0)}  = - \mathbb{K}_m^{(0)}   \quad & \implies \quad \vec{\mathbb{K}} \quad \overset{\mathcal{P}}{\longleftrightarrow } \quad - \vec{\mathbb{K}}  \nonumber \\
 \mathcal{T}^{(0)} \mathbb{K}_m^{(0)} \left(\mathcal{T}^{(0)}\right)^{-1}  = \mathcal{T}^{(0)} \mathbb{K}_m^{(0)} \mathcal{T}^{(0)}  = - \mathbb{K}_m^{(0)}   \quad & \implies \quad \vec{\mathbb{K}}  \quad \overset{\mathcal{T}}{\longleftrightarrow } \quad - \vec{\mathbb{K}}
\end{align}

It follows from 
\hyperref[Right handed transformations Sec]{Eq.(\ref*{Right handed transformations})}
and 
\hyperref[Left handed transformations Sec]{Eq.(\ref*{Left handed transformations})}
that, the right $(j_R,0)$ and left $(0,j_L)$  handed representations (corresponding to the same spin)  transform into one another under Parity or Time reversal,
\phantomsection\label{Parity-Time chiral reps Sec}
\begin{align}\label{Parity-Time chiral reps}
(j,0)  \underset{\text{Time}}{\overset{\text{Parity}}{\longleftrightarrow } } (0,j) 
\quad , \quad 
\phi_R  \underset{\text{Time}}{\overset{\text{Parity}}{\longleftrightarrow } } \phi_L
\end{align}

%
%

\subsubsection*{Bi-Spinors}
\phantomsection\label{App. Bi-Spinors}


For some theoretical models Parity is consider a symmetry\footnote{This means that the interactions are invariant under Parity, like it is the case for QED and QCD.}, then it follows that the two type of spinors\footnote{These spinors describe particles with spin that are subjected  to interactions ruled by such models.}, are not independent anymore. 
In other words, 
if Parity is not taken into account the two type of spinors are independent. However, introducing Parity as a symmetry leads to a constrain which must be satisfied by both spinors at the same time. 
It is no longer convenient to treat them separately, and we shall consider the (Weyl) bi-spinor $\chi$ \footnote{The situation is similar to consider 3-vectors instead of individual components for the study of spatial rotations.},
\phantomsection\label{Weyl bi-spinors Sec}
\begin{align}\label{Weyl bi-spinors}
\chi =\left(
\begin{array}{c}
\phi_R \\
\phi_L \\
\end{array}
\right)
\end{align}
As it follows from the relations 
\hyperref[Parity-Time chiral reps Sec]{(\ref*{Parity-Time chiral reps})}, 
under parity the Weyl bi-spinor transform as,
\begin{align}
\chi &= \left(
\begin{array}{c}
\phi_R \\
\phi_L \\
\end{array}
\right) \quad  \overset{\text{Parity}}{\longrightarrow} \quad \mathcal{P}[ \chi] = \left(
\begin{array}{c}
\phi_L \\
\phi_R \\
\end{array}
\right) 
\end{align}
Hence, the representation for Parity in Weyl bi-spinor space is, 
\begin{align}
\mathcal{P} [\chi] =  {\bf P }_W  \chi = \left(
\begin{array}{cc}
 0 & 1 \\
 1 & 0 \\
\end{array}
\right)\left(
\begin{array}{c}
\phi_R \\
\phi_L \\
\end{array}
\right)  =  \gamma^0_W  \chi
\end{align}
i.e., ${\bf P}_W=\gamma^0_W$ is the Weyl representation of the $\gamma^0$ matrix\footnote{Note that different conventions may differ from this by a phase factor.}.

Under Lorentz Transformations ($\Lambda $), the Weyl bi-spinor $\chi$ is transformed by\footnote{$M_{(\Lambda)}^W$ is the representation of the Lorentz Transformation $\Lambda$ in Weyl's bi-spinor space.},
\begin{align}
 \chi=\left(
\begin{array}{c}
\phi_R \\
\phi_L \\
\end{array}
\right) 
\to 
 M_{(\Lambda)}^W \chi & = \left(
\begin{array}{cc}
 \exp\left[ i \vec{\mathbb{J}}{}^{(s)} \cdot \vec{b} \right] & 0 \\
 0 & \exp\left[ i \vec{\mathbb{J}}{}^{(s)} \cdot \vec{b}^* \right] \\
\end{array}
\right)\left(
\begin{array}{c}
\phi_R \\
\phi_L \\
\end{array}
\right)
\nonumber \\
&= \left(
\begin{array}{cc}
 D_{(\Lambda)} & 0 \\
 0 & \bar{D}_{(\Lambda)} \\
\end{array}
\right)\left(
\begin{array}{c}
\phi_R \\
\phi_L \\
\end{array}
\right) 
\end{align}
where, 
\phantomsection\label{App parameter b Sec}
\begin{align}\label{App parameter b}
\vec{b}=\vec{\theta } - i \vec{\phi }
\end{align}
On the other hand, the parity transformed bi-spinor changes as,
\begin{align}
 \mathcal{P} [\chi] =\left(
\begin{array}{c}
\phi_L \\
\phi_R \\
\end{array}
\right)  \to  M_{(\Lambda)}^W \chi & = \left(
\begin{array}{cc}
 \exp\left[ i \vec{\mathbb{J}}{}^{(s)} \cdot \vec{b}^* \right] & 0 \\
 0 & \exp\left[ i \vec{\mathbb{J}}{}^{(s)} \cdot \vec{b} \right] \\
\end{array}
\right)\left(
\begin{array}{c}
\phi_L \\
\phi_R \\
\end{array}
\right)
\nonumber \\
&=\left(
\begin{array}{cc}
 \bar{D}_{(\Lambda)} & 0 \\
 0 & {D}_{(\Lambda)} \\
\end{array}
\right)\left(
\begin{array}{c}
\phi_L \\
\phi_R \\
\end{array}
\right) 
\end{align}
which shows that the Weyl spinor and its parity related partner transform differently under Lorentz transformations\footnote{The three $\mathbb{J}{}^{(s)}_l$ are the $(2s+1)$ dimensional representation for the generators of the rotation group,  and the three ${b}_l$ correspond to the chosen parameterization.}.

Since right- and left-chiral spinors transform differently  under boosts, whereas under rotations they transform with the same unitary matrix, it follows that they differ from each other only when their momentum is not zero.
In other words, there is no Chiral classification for spinors at rest {(Isotropy of space)}. 
Hence, for massive spinors at rest we must have, $\overset{\circ}{\varphi} _R=\overset{\circ}{\varphi} _L$. 
Since at rest there is no distinction between them,  we must consider the  two linear independent  combinations, 
 \begin{align} 
\overset{\circ}{\psi} =\frac{1}{\sqrt{2}} \left(
\begin{array}{c}
 \overset{\circ}{\varphi} _R+\overset{\circ}{\varphi} _L \\
\overset{\circ}{ \varphi} _R-\overset{\circ}{\varphi }_L \\
\end{array}
\right)
\end{align}
These are called Dirac bi-spinors. 
Dirac bi-spinors $\psi$ are obtained from Weyl bi-spinors $\chi$ by the transformation,
\begin{align} 
\psi =  T_{W-D} \chi =  T_{W-D} \left(
\begin{array}{c}
 \varphi _R \\
 \varphi _L \\
\end{array} 
\right) = \frac{1}{\sqrt{2}}\left(
\begin{array}{cc}
 \mathbf{1}_{2s+1} & \mathbf{1}_{2s+1} \\
 \mathbf{1}_{2s+1} & -\mathbf{1}_{2s+1} \\
\end{array}
\right)
\left(
\begin{array}{c}
 \varphi _R \\
 \varphi _L \\
\end{array}
\right)
\end{align}
The matrix $T_{W-D}$ perform a change from Weyl representation to Dirac representation.
For the Lorentz transformations we have\footnote{$M_{(\Lambda )}^{D}$ is the representation of the Lorentz Transformation $\Lambda$ in Dirac's  bi-spinor space.},
\begin{align} 
\hspace{-.3cm}
M_{(\Lambda )}^{D}=
T_{W-D} M_{(\Lambda )}^{W} T_{W-D}^{-1} = {1 \over 2} \left(
\begin{array}{cc}
 \exp\left[i \vec{\mathbb{J}} \cdot\vec{b} \right]+\exp\left[i \vec{\mathbb{J}} \cdot\vec{b}^* \right] & \exp\left[i \vec{\mathbb{J}} \cdot\vec{b} \right]-\exp\left[i \vec{\mathbb{J}} \cdot\vec{b}^* \right] \\
\exp\left[i \vec{\mathbb{J}} \cdot\vec{b} \right]-\exp\left[i \vec{\mathbb{J}} \cdot\vec{b}^* \right] & \exp\left[i \vec{\mathbb{J}} \cdot\vec{b} \right]+\exp\left[i \vec{\mathbb{J}} \cdot\vec{b} \right] \\
\end{array}
\right)
\end{align}


An arbitrary Dirac bi-spinor is obtained  by  applying the representation of boosts in Dirac bi-spinor space to an at-rest bi-spinor,
\phantomsection\label{Dirac bi-spinor Sec}
\begin{align}\label{Dirac bi-spinor}
\hspace{-.3cm}
\psi _{(p)} & = \frac{1}{\sqrt{2}} \left(
\begin{array}{c}
 \varphi^{R}_{(p)} + \varphi^{L}_{(p)} \\
 \varphi^{R}_{(p)} - \varphi^{L}_{(p)} \\
\end{array}
\right)
\nonumber \\
& = \frac{1}{2} \left(
\begin{array}{cc}
 \exp\left[i \vec{\mathbb{J}} \cdot\vec{b} \right]+\exp\left[i \vec{\mathbb{J}} \cdot\vec{b}^* \right] & \exp\left[i \vec{\mathbb{J}} \cdot\vec{b} \right]-\exp\left[i \vec{\mathbb{J}} \cdot\vec{b}^* \right] \\
\exp\left[i \vec{\mathbb{J}} \cdot\vec{b} \right]-\exp\left[i \vec{\mathbb{J}} \cdot\vec{b}^* \right] & \exp\left[i \vec{\mathbb{J}} \cdot\vec{b} \right]+\exp\left[i \vec{\mathbb{J}} \cdot\vec{b} \right] \\
\end{array}
\right) 
{1 \over \sqrt{2}} \left(
\begin{array}{c}
 \varphi^{R}_{(\overset{\circ}{p})} + \varphi^{L}_{(\overset{\circ}{p})} \\
 \varphi^{R}_{(\overset{\circ}{p})} - \varphi^{L}_{(\overset{\circ}{p})} \\
\end{array}
\right)
\end{align}
Similarly, the Dirac representation of Parity is obtained by,
\begin{align}
 {\bf P}_{D} = T_{W-D} \gamma^0_{W} T_{W-D}^{-1} =\gamma^0=\left(
\begin{array}{cc}
 \mathbf{1}_{2s+1} & 0 \\
 0 & -\mathbf{1}_{2s+1} \\
\end{array}
\right)
\end{align}
Under parity the Dirac bi-spinor transform as,
\begin{align}
\psi _{(p)}=\frac{1}{\sqrt{2}} \left(
\begin{array}{c}
 \varphi^{R}_{(p)} + \varphi^{L}_{(p)} \\
 \varphi^{R}_{(p)} - \varphi^{L}_{(p)} \\
\end{array}
\right) \quad \to \quad \mathcal{P} [\psi] & = {\bf P}_{D} \frac{1}{\sqrt{2}} \left(
\begin{array}{c}
 \varphi^{L}_{(p)} + \varphi^{R}_{(p)} \\
 \varphi^{L}_{(p)} - \varphi^{R}_{(p)} \\
\end{array}
\right) 
\nonumber \\
& =\frac{1}{\sqrt{2}} \gamma^0 \left(
\begin{array}{c}
 \varphi^{R}_{(p)} + \varphi^{L}_{(p)} \\
 -(\varphi^{R}_{(p)} -  \varphi^{L}_{(p)}) \\
\end{array}
\right) = \psi _{(p)}
\end{align}
which means that Dirac spinors are eigenvectors of Parity. The Dirac bi-spinors ($\psi$), shown in 
\hyperref[Dirac bi-spinor Sec]{Eq.(\ref*{Dirac bi-spinor})},
are  eigenvectors of the (space) Parity operator with eigenvalue $+1$ and they correspond to particles. The eigenvectors with eigenvalues $-1$ are associated with anti-particles.
Thus, when parity is a good symmetry\footnote{Which is to say, that during the evolution of the system the interaction between particles do not change their Parity.}, it is convenient to work with Dirac bi-spinors.

\newpage

\chapter{Spinors and Polarization Vectors for Single \mbox{Particle} States}
\phantomsection\label{App spinors}

\setcounter{equation}{0}
\renewcommand{\theequation}{G.\arabic{equation}}

\setcounter{figure}{0}
\renewcommand{\thefigure}{G.\arabic{figure}}

\subsection*{G.1 Canonical Spin}
\phantomsection\label{Canonical Spin}

\subsubsection*{Spin  1/2 particle - Canonical  }
\phantomsection\label{App Spin 1/2 particle - Canonical}

The canonical spinors for a (free) spin 1/2 particle are obtained from at-rest spinors by applying a rotationless boost\footnote{They are solution of Dirac's (homogeneous) equation.}. 
This is equivalent to substituting in  
\hyperref[Dirac bi-spinor Sec]{Eq.(\ref*{Dirac bi-spinor})}
the spin $1/2$ representation of the rotation group
\hyperref[App Pauli matrices Sec]{(Eq. \ref*{App Pauli matrices})}, 
while we set the rotation parameter to zero ($\vec\theta=\vec0$)  in
\hyperref[App parameter b Sec]{Eq.(\ref*{App parameter b})}.
We first observe that,
\begin{align} 
\exp\left[\vec{\mathbb{J}}^{(s=1/2)} \cdot\vec{b} \right] \to \exp\left[{\vec{{\sigma}} \over 2} \cdot\vec{\phi} \right] = \cosh{\left({|\vec{\phi}| \over 2}\right)} \ \mathbf{1}_2 + \vec{{\sigma}} \cdot {\vec{\phi} \over |\vec{\phi}|} \sinh{\left({|\vec{\phi}| \over 2}\right)}
\end{align}
Using now the identification of the boost parameters in equations 
\hyperref[App boost parameters IF Sec]{(\ref*{App boost parameters 2 IF})} 
and
\hyperref[App boost parameters phi-half IF Sec]{(\ref*{App boost parameters phi-half IF})} 
we find the following representation of the boosts for canonical Dirac bi-spinors,
\begin{align} 
\hspace{-.3cm}
\mathcal{B}_{(s=1/2 )}^\text{IF}  =  \left(
\begin{array}{cc}
 \cosh{\left({|\vec{\phi}| \over 2}\right)} \ \mathbf{1}_2 & \vec{{\sigma}} \cdot {\vec{\phi}  \over |\vec{\phi}|} \sinh{\left({|\vec{\phi}| \over 2}\right)} \\
 \vec{{\sigma}} \cdot {\vec{\phi}  \over |\vec{\phi}|} \sinh{\left({|\vec{\phi}| \over 2}\right)} &\cosh{\left({|\vec{\phi}| \over 2}\right)} \ \mathbf{1}_2 \\
\end{array}
\right)
 =  \left(
\begin{array}{cc}
 \sqrt{\frac{E+m}{2m}} & \vec{{\sigma}} \cdot {\vec p \over \sqrt{2m(E+m)}}  \\
  \vec{{\sigma}} \cdot {\vec p \over \sqrt{2m(E+m)}}  & \vec{{\sigma}} \cdot {\vec p \over m}  \\
\end{array}
\right)
\end{align}
%
The spinors themselves follow straightforwardly from
\hyperref[Dirac bi-spinor Sec]{Eq.(\ref*{Dirac bi-spinor})}, explicitly we get,
\begin{equation}
u(\vec{p}, s)=\left(\frac{E+m}{2m}\right)^{\frac{1}{2}}\left(\begin{array}{c}{\overset{\circ}{\chi}(s)} \\ {\frac{\vec{\sigma} \cdot \vec{p}}{E+m} \overset{\circ}{\chi}(s)}\end{array}\right)
\end{equation}
where, $m$ is the mass of the spin 1/2 particle, and the two-component spinors $\phi(s)$ are given by,
\begin{equation}
\overset{\circ}{\chi}\left(+{1/2}\right)=\left(\begin{array}{c}{1} \\ {0}\end{array}\right), \quad \overset{\circ}{\chi}\left(-{1/2}\right)=\left(\begin{array}{c}{0} \\ {1}\end{array}\right)
\end{equation}

 The canonical states transform properly under rotations, 
\begin{equation}
\mathcal{D}[R]^{(1/2 )}_\text{IF} \  u(\vec{p}, s)=\sum_{s^{\prime}}   \ u\left(R( \vec{p} \hspace{.05cm} ), s^{\prime}\right) D_{s^{\prime} s}^{(1 / 2)}[R]
\end{equation}
where, the representation of rotations $\mathcal{D}[R]$ follows from setting now the boost parameter to zero ($\vec\phi=\vec0$)  in
\hyperref[App parameter b Sec]{Eq.(\ref*{App parameter b})}.
Hence, we have
\begin{align} 
\exp\left[\vec{\mathbb{J}}^{(s=1/2)} \cdot\vec{b} \right] \to \exp\left[i {\vec{{\sigma}} \over 2} \cdot\vec{\theta} \right] = \cos{\left({|\vec{\theta}| \over 2}\right)} \ \mathbf{1}_2 + i \vec{{\sigma}} \cdot {\vec{\theta} \over |\vec{\theta}|} \sin{\left({|\vec{\theta}| \over 2}\right)}
\end{align}
and,
\begin{align} 
\hspace{-.3cm}
\mathcal{D}[R]^{(1/2 )}_\text{IF}  
=  \left(
\begin{array}{cc}
 \exp\left[i {\vec{{\sigma}} \over 2} \cdot\vec{\theta} \right] & 0 \\
 0 & \exp\left[i {\vec{{\sigma}} \over 2} \cdot\vec{\theta} \right] \\
\end{array}
\right)
\end{align}

%


\subsubsection*{Spin 1 particle - Canonical}
\phantomsection\label{Spin 1 particle - Canonical}

The general set of  polarization vectors for deuteron (with ${\vec{p}_{d}}{}_\text{T} = \vec 0$)
are obtained by boosting\footnote{Applying a rotationless Lorentz boost transformation relating the frame where the deuteron is at rest with that one in which it has momentum, $ p_d=(E_d,0,0,p_{dz})$.} 
the at-rest polarization vectors, we get,
\begin{align}
\begin{array}{l}
\chi_{\mathrm{C} }^0 = \left({p_{dz} \over M_d},0,0, {E_{d} \over M_d}\right) \quad \overrightarrow{ \text {rest frame}} \quad  \left(0,0,0, 1 \right)  \\
\chi_{\mathrm{C}  }^{+1}=\left(0, \frac{-1}{\sqrt{2}}, \frac{-\mathrm{i}}{\sqrt{2}}, 0\right) \\
\chi_{\mathrm{C} }^{-1}=\left(0, \frac{1}{\sqrt{2}}, \frac{-\mathrm{i}}{\sqrt{2}}, 0\right)
\end{array}
\end{align}


\subsection*{G.2 Light Front Spin and Melosh Rotation} 
\phantomsection\label{App Light Front Spin}

\subsubsection*{Spin 1/2 particle - LF}
\phantomsection\label{App Spin 1/2 particle - LF}

Light-front states of a massive particle are obtained from its rest frame state by first boosting in the z-direction to obtain the desired $p^+$, followed by a light-front transverse boost from the $z$-axis to obtain the desired transverse momentum $\vec p_\text{T}$. The light-front spinors at any momentum are thus given by
 
\begin{align}
u_{(p,1/2)}=\frac{1}{\sqrt{2 p^{+}}}\left(\begin{array}{c}{p^{+}+m} \\ {p_{r}} \\ {p^{+}-m} \\ {p_{r}}\end{array}\right)
\quad u_{(p,-1/2)}=\frac{1}{\sqrt{2 p^{+}}}\left(\begin{array}{c}{-p_{\ell}} \\ {p^{+}+m} \\ {p_{\ell}} \\ {-p^{+}+m}\end{array}\right)
\end{align}
\begin{align}
v_{(p,1/2)}=\frac{1}{\sqrt{2 p^{+}}}\left(\begin{array}{c}{-p_{\ell}} \\ {p^{+}-m} \\ {p_{\ell}} \\ {-p^{+}-m}\end{array}\right) \quad v_{(p,-1/2)}=\frac{1}{\sqrt{2 p^{+}}}\left(\begin{array}{c}{p^{+}-m} \\ {p_{r}} \\ {p^{+}+m} \\ {p_{r}}\end{array}\right)
\end{align}

The relation between light front  and canonical spinors (Melosh rotation) is given by,
\begin{align}
\Omega_{1 / 2}(p)=\left(\begin{array}{cc}{\beta_{1 }(p)} & {0} \\ {0} & {\beta_{1 }(p)}\end{array}\right)
\end{align}
where the $2\times 2$ block matrix $\beta_{1 / 2} $ is given by,
\phantomsection\label{Melosh 2-component Sec}
\begin{align}\label{Melosh 2-component}
\beta_{1 / 2}(p)=\frac{1}{N_{1 / 2}}\left(\begin{array}{cc}{p^{+}+m} & {-p_{\ell}} \\ {p_{r}} & {p^{+}+m}\end{array}\right)
\end{align}
Like any rotation the determinant of this matrix must be 1, from where it follows  the value of the normalization constant, 
\begin{align}
1=\text{Det}[\beta_{1 / 2}]  = \frac{1}{N_{1 / 2}^2} \left( (p^+ + m)^2 + \vec{p}_{\text{T}}^{\hspace{.05cm} 2} \right) = \frac{1}{N_{1 / 2}^2} \left( 2(E+m) p^{+} \right) 
\end{align}
this is,
\begin{align}
N_{1 / 2}=  \sqrt{ (p^+ + m)^2 + \vec{p}_{\text{T}}^{\hspace{.05cm} 2} } =  \sqrt{ 2(E+m) p^{+} }
\end{align}
Furthermore, the angle of rotation can be easily identified by comparing $\beta_{1/2 }(p)$ with the spin 1/2 representation of rotations (Wigner matrices) parameterized by Euler angles\footnote{This is, $R^{M}_{1/2} = d^{1/2}_{\lambda^\prime,\lambda} $, where $d^j_{\lambda^\prime,\lambda}$ are the Wigner d-functions for spin  $ j$.},
\begin{align}
R^{M}_{1/2}=\left(\begin{array}{cc}
{ \cos {\theta_{M} \over 2} } & {- \sin{\theta_{M} \over 2} }  \\
{\sin{\theta_{M} \over 2}  } & {\cos {\theta_{M} \over 2} } \end{array}\right)
\end{align}
Notice that for the correct angle $\theta_M$, both matrices $R^{M}_{1/2}$ and $\beta_{1 / 2}$ correspond to the same rotation. There most be a coordinate transformation relating them. Since the trace of a matrix is invariant under a change of coordinates, we have,
\begin{align}
\Tr{\beta_{1 / 2}} = \Tr{R^{M}_{1/2}}
\end{align}
from where we get the relation,
\phantomsection\label{Melosh angle Sec}
\begin{align}\label{Melosh angle}
\frac{p^{+}+m}{\sqrt{2(E+m) p^{+}}}=\cos {\theta_{M} \over 2} \quad \to \quad \cos \theta_{M}=1-\frac{\vec{p}_{\text{T}}^{\hspace{.05cm} 2}}{(E+m) p^{+}} \quad \underrightarrow{\vec{p}_{\text{T}}^{\hspace{.05cm} 2}=0} \quad  \theta_{M}= 0
\end{align}
hence, if the spin 1/2 particle moves along the LF there is no rotation to correct for.

The inverse Melosh rotation is the complex conjugate transposed (adjoint) of 
\hyperref[Melosh 2-component Sec]{Eq.(\ref*{Melosh 2-component})}.
\begin{align} 
 \beta^{-1}_{1 / 2}(p)=\frac{1}{N_{1 / 2}}\left(\begin{array}{cc}{p^{+}+m} & {p_{\ell}} \\ {-p_{r}} & {p^{+}+m}\end{array}\right) \nonumber \\
\end{align}
which corresponds to the change, $\vec{p}_{\text{T}} \to - \vec{p}_{\text{T}}$.

\subsubsection*{Spin 1 particle - LF}
\phantomsection\label{App Spin 1 particle - LF}
 
For this case we use the at-rest reference four-vector,
\begin{align}
\begin{array}{llll}{\qquad \quad + \quad \ \  1 \quad \ \ 2 \quad \ \ -} \\ {\stackrel{\circ}{p}^{\mu}=\left(\begin{array}{llll}{ m} & {\ \ 0} & {\ \ 0} & {\ \ m}\end{array}\right)}
\end{array}
\end{align}
which is time-like, $({\stackrel{\circ}{p})^2}/m^2=1$. 
The polarization vectors
(as usual) correspond to eigenvectors of the little group 
with different eigenvalues, we will use the labels, $\lambda = \pm 1,0$. 

%

Explicitly, the at-rest light-front polarization vectors for a massive spin 1 particle are,
\begin{align}
\varepsilon_{+}^{\mu}(\stackrel{\circ}{p})= \left(\begin{array}{c}{0}  \\ {\frac{-1}{\sqrt{2}}} \\ {\frac{-i}{\sqrt{2}}} \\ {0} \end{array}\right) = -\varepsilon^\mu_r ,
\quad
\varepsilon_{-}^{\mu}(\stackrel{\circ}{p})=  \left(\begin{array}{c} {0} \\ {\frac{1}{\sqrt{2}}} \\ {\frac{-i}{\sqrt{2}}} \\ {0} \end{array}\right) =\varepsilon^\mu_\ell ,
\quad
\varepsilon_{0}^{\mu}(\stackrel{\circ}{p})=  \left(\begin{array}{c}  {1} \\ {0} \\ {0} \\ {-1} \end{array}\right)
\end{align}
and, the corresponding polarization vectors in a frame where the particle has momentum $p$ are given by,
\phantomsection\label{lf polarization vectors spin 1 Sec}
\begin{align}\label{lf polarization vectors spin 1}
\varepsilon_{+}^{\mu}(p)= \left(\begin{array}{c}{0}  \\ {\frac{-1}{\sqrt{2}}} \\ {\frac{-i}{\sqrt{2}}} \\ {\frac{-\sqrt{2} p_{r}}{p^{+}}} \end{array}\right) ,
\quad
\varepsilon_{-}^{\mu}(p)=  \left(\begin{array}{c} {0}  \\ {\frac{1}{\sqrt{2}}} \\ {\frac{-i}{\sqrt{2}}} \\ {\frac{\sqrt{2} p_{\ell}}{p^{+}}} \end{array}\right) ,
\quad 
\varepsilon_{0}^{\mu}(p)= \left(\begin{array}{c} {{\frac{p^{+}}{m}} }  \\ {\frac{p_{1}}{m}} \\ {\frac{p_{2}}{m}} \\ -{\frac{\vec{p}_{\text{T}}^{\hspace{.05cm} 2}+m^{2}}{m p^+ }} \end{array}\right)
\end{align}
where, $p_{r}=p_1 + i p_2$ and $p_{\ell}=p_1 - i p_2$.

Notice that the transformation law guarantee the constrain, $\varepsilon^*\varepsilon=-1$. This polarization vectors are a 4-dimensional representation of spinors, not 4-vectors. The fact that in the case of spin 1 there are three of them, which together with a time-like vector can be used as a basis for the 4-dimensional Minkowski space is just a coincidence that may end up being very confusing.

For matrix elements of spin 1 operators, e.g., the deuteron electromagnetic current 
(see \hyperref[deuteron EM current Sec]{Eq. \ref*{deuteron EM current}}), 
which has been written on a basis of light-front polarization vectors 
\hyperref[lf polarization vectors spin 1 Sec]{(Eq. \ref*{lf polarization vectors spin 1})}
we have,
\begin{align}
\left\langle p_{d}^{\prime} \lambda_{d}^{\prime}\left|J^{\mu}_d\right| p_{d}, \lambda_{d}\right\rangle={\varepsilon_{\lambda_d^{\prime}}^{\alpha}}^* J_{\alpha \beta}^{\mu} \varepsilon_{\lambda_d}^{\beta}=\varepsilon_{\lambda_d^{\prime}}^\dagger J_d^{\mu} \varepsilon_{\lambda_d} = {J_\text{LF}^{\mu}}_{(\lambda_d^{\prime} ,{\lambda_d})}
\end{align}
Then, the relation with the corresponding matrix elements written on a basis  of canonical polarization vectors (in the same reference frame),  is given by a Melosh rotation\footnote{Here, we use $\varepsilon$ to denote LF polarization vectors, and $\epsilon$ to denote canonical ones.},
\begin{align}
{J_\text{LF}^{\mu}} = { \beta_{1 }}^{\dagger} {J_\text{C}^{\mu}}{ \beta_{1 }} \to {J_\text{LF}^{\mu}}_{(\lambda_\text{lf}^{\prime} ,{\lambda_\text{lf}})}
= { \beta_{1 }}^{*}_{(\lambda_\text{lf}^{\prime} ,{\lambda_\text{c}^{\prime}})} {J_\text{C}^{\mu}}_{(\lambda_\text{c}^{\prime} ,{\lambda_\text{c}})}{ \beta_{1 }}_{(\lambda_\text{c} ,{\lambda_\text{lf}})} 
= { \beta_{1 }}^{*}_{(\lambda_\text{lf}^{\prime} ,{\lambda_\text{c}^{\prime}})} {\epsilon_{\lambda_\text{c}^{\prime}}^{\alpha}}^{*} J_{\alpha \beta}^{\mu} \epsilon_{\lambda_\text{c}}^{\beta} { \beta_{1 }}_{(\lambda_\text{c} ,{\lambda_\text{lf}})}
\end{align}
where the Melosh rotation matrix $\beta_{1}$ is given by,
\phantomsection\label{Melosh spin 1 Sec}
\begin{align}\label{Melosh spin 1}
\beta_{1 }(p)=\frac{1}{N_{1}}\left(\begin{array}{ccc} {(p^{+}+m)^2} & {-\sqrt{2} (p^{+}+m) p_{\ell}} & {p_{\ell}^2} \\ {\sqrt{2} (p^{+}+m) p_{r}} & { (p^{+}+m)^2 - \vec{p}_{\text{T}}^{\hspace{.07cm} 2}} & {-\sqrt{2} (p^{+}+m) p_{\ell}} \\ {p_{r}^2} & {\sqrt{2} (p^{+}+m) p_{r}} & {(p^{+}+m)^2} \end{array}\right)
\end{align}
Like any rotation the determinant of this matrix must be 1, from where it follows  the value of the normalization constant, 
\begin{align}
N_{1 }= (p^+ + m)^2 + \vec{p}_{\text{T}}^{\hspace{.07cm} 2} =  2(E+m) p^{+} 
\end{align}
Furthermore, the angle of rotation can be easily identified by comparing $\beta_{1 }(p)$ with the spin 1 representation of rotations (Wigner matrices) parameterized by Euler angles\footnote{This is, $R^{M}_{1} = d^1_{\lambda^\prime,\lambda} $, where $d^j_{\lambda^\prime,\lambda}$ are the Wigner d-functions for spin  $j$.},
\begin{align}
R^{M}_{1}=\left(\begin{array}{ccc}
{\frac{1}{2} \left(1+\cos \theta_{M}\right)} & {-\frac{1}{\sqrt{2}} \sin \theta_{M} } &{\frac{1}{2} \left(1-\cos \theta_{M}\right)}  \\
{\frac{1}{\sqrt{2}} \sin \theta_{M} } & \cos \theta_{M} & {-\frac{1}{\sqrt{2}} \sin \theta_{M} } \\
{\frac{1}{2} \left(1-\cos \theta_{M}\right)} & {\frac{1}{\sqrt{2}} \sin \theta_{M} } & {\frac{1}{2} \left(1+\cos \theta_{M}\right)}
\end{array}\right)
\end{align}
Notice that for the correct angle $\theta_M$, both matrices $R^{M}_{1}$ and $\beta_{1}$ correspond to the same rotation. There must be a coordinate transformation relating them. Since the trace of a matrix is invariant under a change of coordinates, we have,
\begin{align}
\Tr{\beta_{1}} = \Tr{R^{M}_{1}}
\end{align}
from where we get the relation,
\begin{align}
\frac{3\left(p^{+}+m\right)^{2} - \vec{p}_{\text{T}}^{\hspace{.07cm} 2}}{\left(p^{+}+m\right)^{2} + \vec{p}_{\text{T}}^{\hspace{.07cm} 2}}=1+ 2\cos \theta_M \quad \to \quad \cos \theta_{M}=1-\frac{\vec{p}_{\text{T}}^{\hspace{.07cm} 2}}{(E+m) p^{+}} \quad \underrightarrow{\vec{p}_{\text{T}}^{\hspace{.07cm} 2}=0} \quad  \theta_{M}= 0
\end{align}
which, as expected coincides with the rotation angle for a spin 1/2 particle 
\hyperref[Melosh angle Sec]{\mbox{(Eq. \ref*{Melosh angle})}}.
Thus, analogous to the case of spin 1/2, for a spin 1 particle moving along the LF there is no rotation. 

The inverse Melosh rotation is the complex conjugate transposed (adjoint) of 
\hyperref[Melosh spin 1 Sec]{Eq.(\ref*{Melosh spin 1})},
\begin{align} 
 \beta^{-1}_{1}(p)=\frac{1}{N_{1}}\left(\begin{array}{ccc} {(p^{+}+m)^2} & {\sqrt{2} (p^{+}+m) p_{\ell}} & {p_{\ell}^2} \\ {-\sqrt{2} (p^{+}+m) p_{r}} & { (p^{+}+m)^2 - \vec{p}_{\text{T}}^{\hspace{.07cm} 2}} & {\sqrt{2} (p^{+}+m) p_{\ell}} \\ {p_{r}^2} & {-\sqrt{2} (p^{+}+m) p_{r}} & {(p^{+}+m)^2} \end{array}\right) 
\end{align}
which corresponds to the change, $\vec{p}_{\text{T}} \to - \vec{p}_{\text{T}}$.

\newpage

\chapter{Basis for the Relativistic Two-Body Problem } 
\phantomsection\label{App 2body basis}

\setcounter{equation}{0}
\renewcommand{\theequation}{H.\arabic{equation}}

\setcounter{figure}{0}
\renewcommand{\thefigure}{H.\arabic{figure}}

\subsection*{H.1 Two-Body basis - Momenta} 
\phantomsection\label{Two-Body basis - Momenta}

Our task is to provide a consistent decomposition of deuteron momentum.
Let us start with the description of the two-body system.
Some variables necessary to characterize the state of the system are kinematical (also called external), like the total momentum,
\begin{align}
P_\text{cm} = p_1 +p_2 
\end{align}
others are  dynamical (internal) variables, like the relative momentum ($p_\text{rel}$), which are defined 
by decomposing  the  momenta of each constituent particle into its projection onto the total momentum of the 2-body system and the rest, explicitly 
\begin{align}
p_1^\mu & =  {  p_1  \cdot P_\text{cm} \over  P_\text{cm}^2}P_\text{cm}^\mu  + p_\text{rel}^\mu  \\
p_2^\mu & = {  p_2 \cdot P_\text{cm} \over  P_\text{cm}^2}P_\text{cm}^\mu  - p_\text{rel} ^\mu
\end{align}
therefore, the relative momentum is,
\phantomsection\label{prel Sec}
\begin{align}\label{prel}
p_\text{rel}^\mu= { \left( p_1 - p_2 \right)^\mu \over 2} - { \left( p_1 - p_2 \right) \cdot P_\text{cm} \over 2 P_\text{cm}^2}P_\text{cm}^\mu 
\end{align}
The main reason for the above definition of $p_\text{rel}$ is the following orthogonality condition
\begin{align}
 p_\text{rel} \cdot P_\text{cm} = 0
\end{align}
which guarantee a simple separation between kinematical and dynamical  variables.

\subsubsection*{Useful Intermediate Results}

The invariant mass of the system ($M_{12}=M_{pn}$) is an important quantity that will appear frequently,
\phantomsection\label{M12 Sec}
\begin{align}\label{M12}
M_{12}^2 = P_\text{cm}^2 = s_\text{NN} = m_1^2 + m_2^2 + 2 p_1 \cdot p_2
= m_1^2 + m_2^2 + 2 E_1 \cdot E_2 - 2 \vec{p}_1 \cdot \vec{p}_2
\end{align}

The relative momentum can be written in term of the invariant mass,
\begin{align}
p_\text{rel}^\mu= { p_1^\mu \over 2} \left( 1- {m_1^2 - m_2^2 \over M_{12}^2} \right) -  { p_2^\mu \over 2} \left( 1+ {m_1^2 - m_2^2 \over M_{12}^2} \right)
\end{align}
For every practical purpose the masses of proton and neutron are equal. Hence, in our particular case ($m_1= m_2$) the relative momentum acquires a very simple form,  
\phantomsection\label{prel (M12) Sec}
\begin{align}\label{prel (M12)}
p_\text{rel}^\mu= { p_1^\mu  - p_2^\mu \over 2} 
\end{align}

\subsubsection*{H.1.1 Momenta - IF}

In the center of momentum of the two-body system the expressions on our previous section achieve their simplest form.
In the case of IF this is,
\phantomsection\label{kc Sec}
\begin{align}\label{kc}
k_\text{c}^\mu= {  E_2  \over \sqrt{P_\text{cm}^2}} p_1^\mu - {  E_1  \over \sqrt{P_\text{cm}^2}} p_2^\mu
\end{align}
where,
\begin{align}
E_{1,2}  =  {  p_{1,2}   \cdot P_\text{cm} \over  \sqrt{P_\text{cm}^2}} = {P_\text{cm}^2  + p_{1,2}^2 - p_{2,1}^2  \over  2 \sqrt{P_\text{cm}^2}} = {P_\text{cm}^2  + m_{1,2}^2 - m_{2,1}^2  \over  2 \sqrt{P_\text{cm}^2}}={P_\text{cm}^2    \over  2 \sqrt{P_\text{cm}^2}} = {M_{12} \over 2}
\end{align}
with, $M_{12}^2 = P_\text{cm}^2 $.

This is a crucial result, it shows that in this frame both particles have the same energy. Therefore, the relative momentum decouples from the ``evolution" time\footnote{This is the time that defines the hyper-plane  $t=0$  of Instant Form.}. For numerical calculations it leads to two important simplifications. On one hand, the description of the internal dynamics happens in a three-dimensional space, reducing the required computational power. On the other hand, the reduction is such that the three-dimensional space is orthogonal to the evolution time direction, hence the signature of the metric is Euclidean which simplifies convergence.

The energies end up being the same only  for two particles with equal masses.
However, because the difference in energies do not depend on the internal momentum, only on the  total momentum, there is also a kinematical definition of the two-body frame that decouples internal an external motion. 

In particular, we have
\phantomsection\label{k Sec}
\begin{align}\label{k}
k_\text{c}^\mu & = k^\mu= { p_1^\mu - p_2^\mu \over 2} \\
E_{12} = E_{1} +E_{2}  & =  M_{12} 
= {P_\text{cm}^2    \over  \sqrt{P_\text{cm}^2}} 
= {  4m^2 - 4 k^2 \over M_{12} } =4 {  m^2 + \vec k^2 \over M_{12} } - E_k 
\end{align}

In this frame\footnote{As stated above, this is the rest frame for the two-body system, $\vec P_\text{cm}=0$.}  the components of the vector $\Delta_{p}$ become,
\phantomsection\label{delta IF Sec}
\begin{align}\label{delta IF}
\Delta_{p}^0 & = p_d^0-(p_1^0 + p_2^0)=p_d^0-E_k  \\
\vec{\Delta}_{p} & = \vec{p}_d-(\vec{p}_1 + \vec{p}_2) =\vec{p}_d={\vec{p}_d \over \sqrt{\vec{p}_d^2}} {\vec{k}_c^2 \over E_k} =\vec{n} {\vec{k}_c^2 \over E_k} 
\end{align}
which shows that in the two-body rest frame the deuteron momentum is not zero.

\subsubsection*{H.1.2 Momenta - LF}

 In  the Light Front Form of dynamics we have,
\begin{align}
p^+=p^0+\vec p.\vec{n} \ , \quad p^-=p^0-\vec p.\vec{n} \ , \quad p^-={m^2+p_\text{T}^2 \over p^+} \ , \quad  \vec p_\text{T}\cdot \vec{n}=0 
\end{align}
we shall choose, $\vec{n}$ along deuteron's momentum, which we identify with the positive $z$ direction\footnote{Therefore, we have assumed that, $\vec P^\text{cm}_\text{T}=0$, and, in agreement with the previous section, $p_1^2 = p_2^2=m^2$.}.

The equivalent relations to 
Eqs.(\hyperref[kc Sec]{\ref*{kc}}, 
\hyperref[delta IF Sec]{\ref*{delta IF}})
in the LF are\footnote{Notice that the term, $\alpha(2-\alpha)$, is the same independently of which particle, $1, \text{or}, \ 2$, it refers to. Whenever a situation like this appears the labels will be omitted in order to keep the expressions uncluttered.},
\phantomsection\label{delta minus LF Sec}
\begin{align}\label{delta minus LF}
\Delta_p^- = p_d^- -( p^-_1 + p^-_{2})   = \frac{1}{p_d^+} \left(M_d^2 - 4 \frac{m_N^2 + k_\text{T}^2}{\alpha(2-\alpha)}\right) 
\end{align}
with all other components zero. It is worth to stress that within the LF framework, the kinematical picture is not only more simple, e.g., $\Delta_p^\mu$ has only one non-vanishing component, but also more general since the component of $P^\mu_\text{cm}$ along the $z$ direction does not have to be zero.

In this case the relative momentum is written as,
\phantomsection\label{k- Sec}
\begin{align}\label{k-}
k_\text{LF}^- = { \left( p_1 - p_2 \right)^- \over 2} - { \left( p_1 - p_2 \right) \cdot P_\text{cm} \over 2 P_\text{cm}^2}P_\text{cm}^-  
= {   1 - \alpha_1 \over  P^{+}_{\text{cm}} }  2 {m^2 + k_{\text{T}}^2 \over \alpha(2-\alpha) }   
\end{align}
	\hyperref[k- Sec]{Eq.(\ref*{k-})}
	clearly establish that the evolution time decouples when, $\alpha=1$ \footnote{This case reduces to the previous section when $\vec P_\text{cm}=0$.}, or,
	$P^+_\text{cm}\alpha(2-\alpha) \gg 4m^2$, since the numerator is (approximately) bounded by the term \mbox{$ (1-\alpha_1)4m^2$}.
This argument shall become clear when we use the WF in the scattering amplitude, because then $\ k_\text{LF}^-\ $ is the conjugate variable to time\footnote{Dynamical time.} in the transition 
amplitude. 

From 
\hyperref[prel Sec]{Eq.(\ref*{prel})}
we also have,
\begin{align}
k_\text{LF}^+ = { P_\text{cm}^+ \over 4 P_\text{cm}^2 } (\alpha_1 -1)  \left( P_\text{cm}^2 + 4 {m^2 + k_{\text{T}}^2 \over \alpha(2-\alpha) } \right) =   { P_\text{cm}^+ \over 2} (\alpha_1 -1) 
\end{align}
whereas the transverse components of the relative momentum are the same to those of the Instant frame. Notice that,
\begin{align}
k_\text{LF}^+ k_\text{LF}^- = - (\alpha_1 -1)^2 {m^2 + k_{\text{T}}^2 \over \alpha(2-\alpha) } 
\end{align}
Therefore, after using, 
\begin{align}
k_z = (1-\alpha_1){M_{12} \over 2}
\end{align}
we get,
\begin{align}
k^2 = k_\text{LF}^2 = - k_{\text{T}}^2 - k_z^2 
\end{align}
which explicitly shows that both formulations (IF and LF) agree with each other.

\subsection*{H.2 Two-Body basis - Spin}
\phantomsection\label{Two-Body basis - Spin}


The Pauli-Lubanski pseudo-vector, 
\phantomsection\label{P-L spin Sec}
\begin{align}\label{P-L spin}
W^\mu = {1\over 2} \epsilon^{\mu \alpha \beta \delta} P_\alpha J_{\beta\delta}
\end{align}
is the closest to a relativistic (covariant) spin axial vector. 
Because the spin is an axial vector it is frame dependent, only at rest it can be defined unambiguously. Moreover, it depends on the origin of coordinates, which for a multi-particle relativistic theory does not have unambiguous definition\footnote{For composite states, binding via strong interactions (and also Gravity at large scale), the ambiguity on the definition of a relativistic spin operator remains an unsolved problem. Here, we overstep the problematic issues by studying a region of phase space where ambiguities can be suppressed, i.e. they can be kept under control under certain conditions, which are presented below.}.

For massive particles, which is the only case considered here, the spin operator is defined by,
\phantomsection\label{spin Sec}
\begin{align}\label{spin}
(0,{\mathbf J}_\text{cm})=\Lambda_\text{cm}^{-1}({\bf P}_\text{cm})^\mu_{\ \nu} W^\nu/M
\end{align}
where, {$\Lambda_\text{cm}^{-1}({\bf P}_\text{cm})^\mu_{\ \nu} \ $}  represents a  Lorentz transformation from  a reference frame where the system have  momentum {${\bf P}_\text{cm}=0$} \footnote{If there is only one particle, this frame coincides with its rest frame. If there are more than one particle, it refers to the center of momentum frame of the system.}.

The Pauli-Lubanski operator 
\hyperref[P-L spin Sec]{(Eq. \ref*{P-L spin})}
for a Dirac particle is\footnote{Which is just the result of substituting in 
\hyperref[P-L spin Sec]{Eq.(\ref*{P-L spin})}
the bi-spinor representation of the Angular Momentum tensor (Spin 1/2 representation extended by Parity), 
{${\bf J}_{\mu\nu}= \sigma_{\mu\nu}/2 = i [\gamma_\mu , \gamma_\nu]/4 $}. 
Remind that ${\bf J}_{\mu\nu}= \sigma_{\mu\nu}/2$ correspond to the bi-spinor representation for the Generators of the Lorentz Group (proper and orthochronous).},  
\phantomsection\label{P-L spin 1/2 Sec}
\begin{align}\label{P-L spin 1/2}
W_{1/2}^\mu (P) = {1 \over 2} (P^\mu + m \gamma^\mu) \gamma_5
\end{align}
Notice that\footnote{Strictly speaking, the spinors are only the components of the state, and the operator should be evaluated on the state (ket). However, this notation is less cluttered and easier to read.},
\begin{align}
 W_{1/2}^\mu (P)  {u}{(p,s)} & =   {1 \over 2} \left(P^\mu + m \gamma^{\mu} \right) \gamma_{5} {u}{(p,s)} 
 = {1 \over 2} \left(p^\mu + m \gamma^{\mu} \right) \gamma_{5} {u}{(p,s)}  \nonumber  \\ 
& =  W_{1/2}^\mu (p)  {u}{(p,s)} 
\end{align} 
\begin{align}
\overline{W}_{1/2}^\mu (P) {u}{(p,s)} & =  \gamma^0 \left( W_{1/2}^\mu (P) \right)^\dagger \gamma^0  {u}{(p,s)}   
=  {1 \over 2}  \gamma^0 \left(P^\mu - m (\gamma^{\mu})^\dagger \right) \gamma_{5}  \gamma^0  {u}{(p,s)} \nonumber \\
& =  {1 \over 2}  \left(-P^\mu + m \gamma^{\mu} \right) \gamma_{5}  {u}{(p,s)} =  {1 \over 2}  \left(-p^\mu + m \gamma^{\mu} \right) \gamma_{5}  {u}{(p,s)} \nonumber  \\  
& =  W_{1/2}^\mu (-p)  {u}{(p,s)} 
\end{align} 
and, 
\begin{align} 
\overline{u}{(p,s)} W_{1/2}^\mu (P)  & =   {u}^\dagger{(p,s)} \gamma^0 W_{1/2}^\mu (P)  = {1 \over 2}  {u}^\dagger{(p,s)} \left(-P^\mu - m (\gamma^{\mu})^\dagger \right) \gamma_{5}  \gamma^0 \nonumber  \\
& = {1 \over 2}  {u}^\dagger{(p,s)} \left( \left(-P^\mu + m \gamma^{\mu} \right) \gamma_{5} \right)^\dagger  \gamma^0 = {1 \over 2}  \left( \left(-P^\mu + m \gamma^{\mu} \right) \gamma_{5} {u}{(p,s)} \right)^\dagger  \gamma^0  \nonumber  \\
& =  (W_{1/2}^\mu (-p) u(p,s) )^\dagger \gamma^0  
= \overline{u}{(p,s)}   {1 \over 2} (-p^\mu + m \gamma^\mu) \gamma_5 \nonumber \\
& = \overline{u}{(p,s)} W_{1/2}^\mu (-p) 
\end{align}
\begin{align}
\overline{u}{(p,s)} \overline{ W}_{1/2}^\mu (P)   & =   {u}^\dagger{(p,s)}  \left( W_{1/2}^\mu (P) \right)^\dagger \gamma^0 = {1 \over 2}  {u}^\dagger{(p,s)}  \left(P^\mu - m (\gamma^{\mu})^\dagger \right) \gamma_{5} \gamma^0 \nonumber \\
& = {1 \over 2}  {u}^\dagger{(p,s)}  \left( - (-P)^\mu - m (\gamma^{\mu})^\dagger \right) \gamma_{5} \gamma^0 \nonumber  \\
& = \overline{u}{(p,s)} W_{1/2}^\mu (p) 
\end{align}

Thus, the Pauli-Lubanski operator is not Dirac self-adjoint\footnote{This is a necessary condition for any operator with observable eigenvalues.}. However, we can use it to construct a self-adjoint (observable) spin 1/2  operator, 
\phantomsection\label{spin op 1/2 self adj Sec}
\begin{align}\label{spin op 1/2 self adj}
 S_{1/2} (P)  =  {1 \over 2m} \left(W_{1/2}(P) + \overline{W}_{1/2}(P) \right) 
\end{align}
where, we have divided by the mass to make sure that at rest it coincides with the standard non-relativistic notion of spin.
This definition is consistent ,
i.e. it produces the same result no matter to which spinor side  of the matrix element it is applied.

In 
bi-spinor space, we have,
\phantomsection\label{spin op 1/2 self adj matrix elements Sec}
\begin{align}\label{spin op 1/2 self adj matrix elements}
[S_{1/2}(p)]_{s_2,s_1} =  {1 \over 2m} \overline{u}{(p,s_2)} \left(W_{1/2}(P) + \overline{W}_{1/2}(P) \right) {u}{(p,s_1)}
\end{align}
which results in\footnote{This is the simplest non-trivial case of a spin density matrix: $\rho_{1/2}= {1\over 2}(a \mathbf{1} + \textbf{s} \cdot \bm{\sigma})$.},
\phantomsection\label{spin op 1/2 Sec}
\begin{align}\label{spin op 1/2}
 [S^\mu (p)]_{s_2,s_1} = {1 \over 4m} \overline{u}{(p,s_2)} \left( (p-p)^\mu + 2m\gamma^{\mu} \right) \gamma_{5}  u(p,s_1) = {1 \over 2} \overline{u}{(p,s_2)}  \gamma^{\mu} \gamma_{5}  u(p,s_1) 
\end{align}

\subsubsection*{H.2.1 Two-Body Spin 1/2 - Canonical}
\phantomsection\label{Two-Body Spin 1/2 - Canonical}

If we use in 
\hyperref[spin op 1/2 Sec]{Eq.(\ref*{spin op 1/2})}
 the canonical spinors we get the Instant Form of spin,
\phantomsection\label{spin op 1/2 Canonical Sec}
\begin{align}\label{spin op 1/2 Canonical}
 [S^\mu_\text{IF} (p)]_{s_2,s_1} =  {1 \over 2} \bar{u}_\text{C}{(p,s_2)}  \gamma^{\mu} \gamma_{5}  u_\text{C}(p,s_1)
\end{align}
we obtain,
\begin{align}
S_\text{IF} = \left( {\bm{p} \cdot \bm{\sigma} \over m}  \ \textbf{,} \ \bm{p}_x \frac{\bm{p} \cdot \bm{\sigma}}{m(m + E_p)} + \sigma_x \ \textbf{,} \ \bm{p}_y \frac{\bm{p} \cdot \bm{\sigma}}{m(m + E_p)} + \sigma _2 \ \textbf{,} \ {p}_z \frac{\bm{p} \cdot \bm{\sigma}}{m(m + E_p)} + \sigma _3 \right) 
\end{align}
which defines a spin-polarization vector, 
\phantomsection\label{2B polarization vector Sec}
\begin{align}\label{2B polarization vector}
\chi^\mu(p,\textbf{s}) = {1\over 2} \sum_{s_2,s_1}\left[(\textbf{s}\cdot \bm{\sigma})_{s_1,s_2}  (S^\mu_\text{IF})_{s_2,s_1}(p) \right] = {1\over 2} \text{Tr}\left[(\textbf{s}\cdot \bm{\sigma}) S^\mu_\text{IF}(p) \right]
\end{align}
where, $\textbf{s}$ is the unit  3-vector in the direction of the spin when the particle is at rest.

Repeating the steps leading to 
\hyperref[spin op 1/2 Sec]{Eq.(\ref*{spin op 1/2})},
but now for two different particles with respective momenta $\bf p$ and $\bf{p}'$,  we will arrive to the spin operator for the two body system\footnote{The Spin operator for two spin 1/2 particles is constructed out of the tensor product of the individual spin 1/2 operators, and can be conveniently separate into rotation invariant structures, scalar (singlet), vector (triplet), and the symmetric and traceless rank-2 spin tensor: $(\rho_{1})_{ij} = {1\over 3} \left( a \hspace{.05cm} \delta_{ij} + {3 \over 2} ({\bf s} \cdot  {\bf J})_{ij} + 3 \text{T}_{i l} {\bf J}_{l j}  \right)$.},
{\small
\begin{align}
\hspace{-1.1cm}
S_\text{IF }^\text{2B}=  \left( {\textbf{P}_{12} \cdot \bm{\sigma} \over M_{12}} \ \textbf{,} \ {P_{12}}_x \frac{{\textbf{P}_{12}}\cdot \bm{\sigma}}{M_{12}(M_{12} + P_{12}^0)} + \sigma_x \ \textbf{,} \ {P_{12}}_y \frac{{\textbf{P}_{12}} \cdot \bm{\sigma}}{M_{12}(M_{12} + P_\text{cm}^0)} + \sigma _y \ \textbf{,} \ {P_{12}}_z \frac{{\textbf{P}_{12}} \cdot \bm{\sigma}}{M_{12}(M_{12} + P_{12}^0)} + \sigma _z \right) 
\end{align}
}
where, the total momentum and the invariant mass of the two-body system are respectively, 
\begin{align}
P_{12}=p_1+p_2 \ , \  \text{and}, \ M_{12}^2=P_{12}^2
\end{align}  
The spin $S^\text{2B}$ defines a spin-polarization vector $\chi^\mu(p,\textbf{s})$ which reassembles that of a spin 1 particle 
(see section 
\hyperref[Spin 1 particle - Canonical Sec]{App.(\ref*{Spin 1 particle - Canonical})}). 
In the CM frame ($\textbf{P}_{12}^\text{cm}=0$), it reduces to,
\begin{align}
S_\text{IF,CM}^\text{2B} 
=  \left( 0 \ {,} \  \sigma_x \ {,} \  \sigma _y \ {,} \  \sigma _z \right) 
\end{align}

\subsubsection*{H.2.2 Two-Body Spin 1/2 - LF}
\phantomsection\label{Two-Body Spin 1/2 - LF}

Accounting for the contribution from each particle to the two body system, and restricting ourselves to collinear frames ($\textbf{P}_\text{12}^\text{T}=0$), we have, 
\phantomsection\label{spin op 1/2 LF Sec}
\begin{align}\label{spin op 1/2 LF}
S_\text{LF}^\text{2B} =  \left( {\text{P}_\text{cm}^+  \over M_{12}}  {\sigma}_z \ \textbf{,} \  \bm{\sigma}_\text{T} \ \textbf{,} \ {P_\text{cm}^2 -2M_{12}^2 \over 2 M_{12} P_\text{cm}^+} \sigma_z  \right) =  \left( {\text{p}_\text{d}^+  \over M_{12}}  {\sigma}_z \ \textbf{,} \  \bm{\sigma}_\text{T} \ \textbf{,} \ {-M_{12} \over 2  p_\text{d}^+} \sigma_z   \right) 
\end{align}

\newpage

\chapter{Relativistic Two-Body Wave Function}
\phantomsection\label{RWF}

\setcounter{equation}{0}
\renewcommand{\theequation}{I.\arabic{equation}}

\setcounter{figure}{0}
\renewcommand{\thefigure}{I.\arabic{figure}}

 
The  wave function (WF) of a bound state is dictated by the symmetries it must satisfy and is constructed out of all the degrees of freedom (dof) necessary to achieve a satisfactory description of the system.
Deuteron is a spin 1, iso-singlet (isotopic-spin zero) state that is invariant under parity transformation (space reflection).
We work on the two nucleon approximation, which is the most important component in  Fock space expansion. 
In other words we assume is made out of (mainly) two massive spin 1/2  constituents.  

It is convenient to write the WF Eq.(\ref{wflf}) in a form that permits a parameterization in terms of Dirac structures. We have,
\begin{align}
 \bar{u}_{s_i} \bar{u}_{s_r} \Gamma ^{\mu }_{\text{dNN}} \chi ^{\lambda _d}_{\mu }\to   \bar{u}_{s_i} \Gamma ^{\mu }_{\text{dNN}} \chi ^{\lambda _d}_{\mu } \left(\text{i$\gamma $}_2 \gamma _0 \right) \bar{u}_{s_r}^T= \bar{u}_{s_i} \Gamma ^{\mu }_{\text{dNN}}   \chi ^{\lambda _d}_{\mu } \gamma _5  \epsilon ^{\text{rr}'} u_{s_{r'}}
\end{align}
The first step makes use of Charge Conjugation, i.e. crossing between the exchange and annihilation channels, thus relating the  charge conjugated fermion with its antiparticle. 
The second uses CPT symmetry, i.e. the relation  between the fermion  and its antiparticle, $v_{s_r}=\gamma _5 \epsilon ^{\text{rr}'} u_{s_{r'}}$. 
Thus, we have
\phantomsection\label{wfparam Sec}
\begin{align}\label{wfparam}
\Psi ^{\lambda _d}_{s_i s_r}=\bar{u}_{s_i} \Gamma ^{\mu }_{\text{dNN}}   \chi ^{\lambda _d}_{\mu } \gamma _5  \epsilon ^{\text{rr}'} u_{s_{r'}}
\end{align}

\subsection*{I.1 General Structure of the Vertex} 
\phantomsection\label{App DNN vertex}

The most general form that satisfy all the requirements is, 
\phantomsection\label{WF Sec}
\begin{align}\label{WF}
\Phi_{s_2 s_1}^{\lambda}=\chi_{\mu (P_d)}^{\lambda} \bar{u}_{(k_2,s_2)} \Gamma^{\mu}  \left(u_{(k_1,s_1)} \right)^{\mathcal{C}}
\end{align}
where,
\phantomsection\label{vertex-App Sec}
\begin{align}\label{vertex-App}
\Gamma^{\mu} = & \Gamma_{1}  \gamma^{\mu} +\Gamma_{2} \frac{k^{\mu}}{m} + \Gamma_{3} \frac{\Delta_{p}^{\mu}}{2m}+\Gamma_{4} \frac{k^{\mu} \sh{\Delta}_p}{ 2m^2} \nonumber \\ 
& +\Gamma_{5} \frac{1}{2 m^{3}} \gamma_{5} \epsilon^{\mu \nu \rho \gamma}(p_\text{d})_\nu k_{\rho} (\Delta_{p})_\gamma+\Gamma_{6} \frac{ \Delta_{p}^{\mu} \sh{\Delta}_p}{4m^{2}} 
\end{align}
Here, $\chi^{\lambda}_\mu$ is the polarization vector with $\lambda$ labeling the 3 possible polarizations of a spin 1 particle, {\it e.g.}, $\lambda=1,0,-1$, are, right handed, longitudinal and left handed, respectively.
The covariant functions, $F_i$,  convey the dynamics for  deuteron decomposition into proton-neutron system, they are functions of the invariant mass of the two-nucleon system, 
\phantomsection\label{inv mass Sec}
\begin{align}\label{inv mass}
M_{NN}^2=s_{NN}=(E_p+E_n)^2=4(m^2+\textbf{k}^2)=4{m^2+\textbf{k}^2_\perp \over \alpha(2-\alpha)} = M_d^2 - \Delta_p \cdot P_d
\end{align}  
here, $E_{p,n}=\sqrt{m^2_{p,n}+\textbf{k}^2}$  are the respective energies of proton and neutron in the two-body center of momentum frame,
the (reduced) mass $m$ is approximated by {\small$m=(m_p+m_n)/2$}), the 3-vector  $\textbf{k}=(k_\perp, k_3)=(\textbf{k}_r -\textbf{k}_i)/2$ is half the relative momentum, where $k_r$ refers to the recoil nucleon and  $k_3$ is defined by 
\hyperref[inv mass Sec]{Eq.(\ref*{inv mass})} 
as, 
\phantomsection\label{k3 LF Sec}
\begin{align}\label{k3 LF}
k_3=(\alpha_r -1)M_{NN}/2 
\end{align}
where the longitudinal momentum fraction is, 
\begin{align}
\alpha=2{E_r + k_{r z} \over E_d + p_{dz} }=2{p^+_r \over P_d^+} 
\end{align}

Here we particularize our study to the unpolarized scattering cross-section. Therefore, in our  calculation within the Impulse Approximation (also for the case with Final State Interactions restricted to the GEA) the relevant object is the spin averaged LF density matrix of the deuteron 
(see 
\hyperref[density matrix LF Sec]{Eq.(\ref*{density matrix LF})}),

\phantomsection\label{density matrix LF spin averaged Sec}
\begin{align}\label{density matrix LF spin averaged}
\rho_d(\alpha,k_{\perp}) = {1 \over 3} \cdot {1 \over 2} \sum_{\lambda_{d},s_r,s_i}{|\Psi^{\lambda_{d}}_{s_{i} s_{r}}(\alpha,k_{\perp}) |^2 \over   (2-\alpha) }
\end{align}
which after using 
\hyperref[k3 LF Sec]{Eq.(\ref*{k3 LF})}, 
we get,

\phantomsection\label{density matrix LF in k - spin averaged Sec}
\begin{align}\label{density matrix LF in k - spin averaged}
\rho_d(\alpha,k_{\perp}) = {1 \over 3} \cdot {1 \over 2} \sum_{\lambda_{d},s_r,s_i}{ {E_k - k_3} \over  E_k} |\Psi^{\lambda_{d}}_{s_{i} s_{r}}(\alpha,k_{\perp}) |^2
\end{align}
where, 
$E_{k}=\sqrt{m^2+{\vec k}^2}$.

\subsection*{I.2 Comparison with Partial Waves Analysis}
\phantomsection\label{App Comparison with Partial Waves Analysis}

For sufficiently low energies it is reasonable to expect that the contribution to the scattering amplitude other than the lowest partial waves must be kinematically suppressed.
This means that at low relative momentum, and in the CM of the on-shell nucleons, valuable insight about the functional form of the covariant vertex functions ($\Gamma_i$) can be obtained by matching the covariant WF decomposition\footnote{Which results from the simplification of the electro-disintegration cross-section provided by the PWIA.}  with the Partial Wave decomposition\footnote{Which results from the simplification that occurs at low energy (transferred) allowing us to treat the reaction as scattering by a Potential.}.
Although, this identification provide expressions for the relativistic WF which rely on non-relativistic QM\footnote{Note that the Partial Waves depend uniquely on a relative 3-momentum, which in potential scattering theory would be the momentum transferred in the NN scattering.}.

To proceed with the matching of the two WF decompositions both of them must be expressed on a spherical basis, thus we shall use canonical spinors since these are manifestly rotational invariant\footnote{As it is well known the canonical spinors are invariant under SO(3) spatial rotations.}.
To this end we apply a change of basis to our light-front wave function Eq.(\ref{NdNwf}), which has the effect of a spatial rotation, 
known as the Melosh rotation. In general, a spin 1/2 particle with momentum $\bf p$ can be represented equivalently with light-front or canonical bi-spinors, which are related by a rotation as follows,
\begin{align}
u_s^{\text{LF}}(p^+,{\bf p}_\text{T}) = u_{s'}^C(\vec{p}) D\left[R_M(p)\right]_{s s'}
\end{align}
\begin{align}
D\left[R_M\right]=\left(
\begin{array}{cc}
 R_M & 0 \\
 0 & R_M^T \\
\end{array}
\right)
\end{align}
where, $R_M^T$ is the transpose of  $R_M$, which is the representation of the Melosh rotation in two-dimensional spinor space,
\phantomsection\label{Melosh rotation Sec}
\begin{align}\label{Melosh rotation}
R_M(p)= {1 \over \sqrt{2 p^+(E_p+m)}} \left(
 \begin{matrix}{}
 p^+ + m & \ -(p_x - {i} p_y) \\
 p_x + {i} p_y & \ p^+ + m
 \end{matrix}   \right)  
\end{align}
%
%
and  $E_p=\sqrt{m^2+ \vec{p}^2}$.

\subsection*{I.3 Dirac Bilinears - Canonical} \phantomsection\label{App Dirac bilinears - Canonical}

We use Canonical spinors to evaluate the following Dirac bilinears. The reference frame is such that, $\vec{P}_{NN}^T=0$. 

We find, for the bilinear, $\bar{u}_{p_i} \mathbf{1} u_{p_r}$,
\begin{align}
\bar{u}_{p_i} u_{p_r}=\frac{1}{\sqrt{p_{i}^0+m} \sqrt{m+p_{r}^0}} \left(\vec p_i \cdot \vec p_r+m \left(p_{\text{i}}^0+m+p_{\text{r}}^0\right)\mathbf{1}-i \sigma _j \epsilon _{jln} {p_{i}}_l   {p_{r}}_n\right)
\end{align}
for the bilinear, $\bar{u}_{p_i} \left( \gamma _5\right)u_{p_r}$,
\begin{align}
\bar{u}_{p_i}\left( \gamma _5\right)u_{p_r}=\sqrt{p_{\text{i}}^0+m} \sqrt{m+p_{\text{r}}^0} \left( \frac{ \vec{p}_r \cdot \vec{\sigma } }{p_{\text{r}}^0 + m } - \frac{ \vec{p}_i \cdot \vec{\sigma }   }{p_{\text{i}}^0+m} \right)
\end{align}
for, $\bar{u}_{p_i} \left(\gamma _0 \gamma _5\right) u_{p_r}$,
\begin{align}
\bar{u}_{p_i}\left(\gamma _0 \gamma _5\right)u_{p_r}=\sqrt{p_{\text{i}}^0+m} \sqrt{m+p_{\text{r}}^0} \left(\frac{ \vec{p}_i \cdot \vec{\sigma }   }{p_{\text{i}}^0+m}+\frac{ \vec{p}_r \cdot \vec{\sigma } }{p_{\text{r}}^0 + m }\right)
\end{align}
and for, $\bar{u}_{p_i}\left( \gamma _j \gamma _5 \right)u_{p_r}$, we find,
\begin{align}
\bar{u}_{p_i}\left( \gamma _j \gamma _5 \right) u_{p_r} =2  \sqrt{\frac{m+p_r^0}{p_i^0+m}} \sigma _j \left(\frac{p_i^0+p_r^0}{2} \mathbf{1} - \left(  {p_r}_j \sigma_j - \frac{ {p_i}_z+{p_r}_z}{2} \sigma _3 \right) \frac{\vec{p}_r \cdot \vec{\sigma } }{m+p_r^0} \right)
\end{align}

More useful for our purpose is to work  in the basis of total ($P_{NN}$) and relative ($p_\text{rel}$) momenta, here we have,
\begin{align}
\bar{u}_{p_i} u_{p_r}=\frac{2}{\sqrt{\left(2 m+E_{\text{NN}}\right){}^2-4 E_{\text{rel}}^2}} \left(\left( m E_{\text{NN}}+\frac{M_{12}^2}{2}\right) \mathbf{1}+i \sigma_j \epsilon_{jnl} {P_{\text{NN}}}_l  {p_{\text{rel}}}_n\right)
\end{align}

\begin{align}
\bar{u}_{p_i} \left( \gamma _5\right) u_{p_r}={2}\frac{\left(2 m+E_{\text{NN}}\right) \vec{p}_{\text{rel}} \cdot \vec{\sigma }-E_{\text{rel}} \vec{P}_{\text{NN}}\cdot\vec{\sigma }}{\sqrt{\left(2 m+E_{\text{NN}}\right){}^2-4 E_{\text{rel}}^2}}
\end{align}

\begin{align}
\bar{u}_{p_i}\left(\gamma _0 \gamma _5\right)u_{p_r}={2}\frac{\left(m+\frac{E_{\text{NN}}}{2}\right)  \left( \vec{P}_{\text{NN}} \cdot \vec{\sigma }\right) - 2 E_{\text{rel}} \left(\vec{p}_{\text{rel}} \cdot \vec{\sigma }\right)}{\sqrt{\left(2 m+E_{\text{NN}}\right){}^2-4 E_{\text{rel}}^2}}
\end{align}

\begin{align}
\bar{u}_{p_i}\left( \gamma _j \gamma _5\right)u_{p_r} = {2} \frac{ \sigma _j \left(m E_{\text{NN}}+\frac{M_\text{NN}^2}{2}\right) -2 p_{\text{rel}_j} \left( \vec{p}_\text{rel} \cdot \vec{\sigma}\right) +\frac{1}{2} P_{\text{NN}_j}  \left( \vec{P}_{\text{NN}} \cdot \vec{\sigma }\right) + i  \epsilon _{\text{klj}} p_{\text{rel}_k} P_{\text{NN}_l} \mathbf{1} }{\sqrt{\left(2 m+E_{\text{NN}}\right){}^2-4 E_{\text{rel}}^2}}
\end{align}

From equation 
(\ref{p relative}) 
we note that, 
\phantomsection\label{prel_is_space_like Sec}
\begin{align}\label{prel_is_space_like}
p_{\text{rel}}^2=m^2-{M_\text{NN}^2 \over 4} <0
\end{align}
i.e., $p_\text{rel}$ is a space-like four-vector. Thus, it is always possible to find a frame where the relative energy vanishes,
\begin{align}
p_\text{rel}^0=E_\text{rel}=0
\end{align}
This frame is given by the condition, 
\begin{align}
0=E_{\text{rel}}^2=p_{\text{rel}}^2+\vec{p}_{\text{rel}}^2=m^2-\frac{M_\text{NN}^2}{4}+\vec{p}_{\text{rel}}^2\Longrightarrow P_{\text{NN}}^2=M_\text{NN}^2=4 \left(m^2+\vec{p}_{\text{rel}}^2\right)
\end{align}
For  two particles of equal mass, like the nucleons, 
the condition, $E_\text{rel}=0$, 
occurs in the two-body (NN) center of momentum (CM), i.e.  $\vec{P}_\text{NN}=0$.
In this frame, it is useful to define the four-momentum of the two 
nucleons through 
$k^\mu =(E_k, {\bf k}_\text{T} , k_z)$, which has components, 
\phantomsection\label{def_k Sec}
\begin{align}\label{def_k}
E_k= {\sqrt{s_\text{NN}}\over 2} = {M_\text{NN}\over 2} = \sqrt{m_\text{N}^2 + k^2}, \quad  {\bf k}_\text{T} = {{\bf p}_{r}}_\text{T} \ , \quad \text{ and,} \quad k_z = E_k(\alpha_r -1),
\end{align}
%
then we have, 
$\vec{k}=\vec{p}_\text{rel}^\text{\ cm}/2=\vec{p}_r^\text{\ cm}=-\vec{p}_i^\text{\ cm}$, 
and the bilinears 
are given by the following expressions,
\begin{align}
\bar{u}_{-\vec{k}} u_{\vec{k}}=M_\text{NN} \mathbf{1}
\end{align}
\begin{align}
\bar{u}_{-\vec{k}} \left( \gamma _5\right)  u_{\vec{k}}= {2} \vec{k} \cdot \vec{\sigma }
\end{align}
\begin{align}
\bar{u}_{-\vec{k}}\left(\gamma _0 \gamma _5\right)u_{\vec{k}}=0
\end{align}
\begin{align}
\bar{u}_{-\vec{k}}\left(\gamma _5 \gamma _j\right)u_{\vec{k}}= M_\text{NN} \sigma_j - {2} \frac{2 k_j \left(\vec{k}\cdot\vec{\sigma }\right)}{2 m+M_\text{NN}}
\end{align}

The wave function (Eq. \ref{NdNwf}) in the two-body CM 
is then given by,
\begin{align}
{\psi}_\text{CM}^{\lambda_d} = & -2\Gamma_{1} \left( E_k (\vec{\sigma} \cdot \vec\chi_{\lambda_d}) -  \frac{1}{ m+E_{k}} ( \vec{k} \cdot \vec\chi_{\lambda_d})(\vec{k} \cdot \vec{\sigma})\right) 
-2\Gamma_{2} \left(\frac{1}{m} ( \vec{k} \cdot \vec\chi_{\lambda_d})(\vec{k} \cdot \vec{\sigma}) \right) 
\nonumber \\
&
+ 2\Gamma_{3} \left( { {\Delta}_p^\mu  {\chi}^{\lambda_d}_\mu \over 2m}(\vec{k} \cdot \vec{\sigma}) \right) 
+2 \Gamma_{4}  {\vec{k} \cdot \vec\chi_{\lambda_d}  \over m}  \left( { E_k } \frac{\vec{\Delta}_p \cdot \vec{\sigma}}{2 m} -  \frac{\vec{\Delta}_p \cdot \vec{k}}{ 2 m+2 E_{k}} {\vec{k} \cdot \vec{\sigma} \over m} \right) 
\nonumber \\
&-2\Gamma_{5} \frac{M_{NN}^2}{2 m}\left(\frac{ \vec\chi_{\lambda_d} \times \vec{k} }{m}\right) \cdot  { \vec{\Delta}_p \over 2m} -2\Gamma_{6} { {\Delta}_p^\mu  {\chi}^{\lambda_d}_\mu \over 2m} \left( { E_k }  \frac{\vec{\Delta}_p \cdot \vec{\sigma}}{2 m} -  \frac{\vec{\Delta}_p \cdot \vec{k}}{ 2 m+2 E_{k}} {\vec{k} \cdot \vec{\sigma} \over m} \right)
\end{align}

\subsection*{I.4 Separation into Angular and Radial Dependencies} 
\phantomsection\label{App Separation into Angular and Radial Dependencies}

In order to compare with the partial wave decomposition of the NN interaction we need to separate  the angular and radial variables. In other words, the vertex function must be expressed in terms of irreducible space tensors (spherical harmonics). 

We first pull out the spin part, after which the components of vectors $k$, and $\Delta$ are separated into symmetric and anti-symmetric terms. Both of this terms are invariant under rotations. The last step is to extract from the symmetric structure the trace, which is invariant by it self. 

For the structure associated with $\Gamma_1$ we have,
\begin{align}
 \Gamma_{1}  \rightarrow & - 2\Gamma_{1} \left( E_k (\vec{\sigma} \cdot \vec\chi_{\lambda_d}) -  \frac{{\vec{k}}^2}{ m+E_{k}} ( \hat{k} \cdot \vec\chi_{\lambda})(\hat{k} \cdot \vec{\sigma})\right)
  \nonumber \\
 &
 = - 2\Gamma_{1} \left( m \delta_{ij} -  (m-E_{k}) \hat k_i \hat k_j \right) \sigma_i \chi^{\lambda_d}_j 
 \nonumber \\
 & = - {2 \over 3} \Gamma_{1} \left( \left(2E_{k}+m \right) \left[ \delta_{ij} \right] + \left( E_{k} - m \right) \left[3 \hat k_i \hat k_j -\delta_{ij} \right] \right)  \sigma_i  \chi^{\lambda_d}_j 
\end{align}
where, we use the notation, $\hat k = \frac{ {\bf k} }{ |{\bf k}| } $. 
Note that in the NN-CM,  $ {\vec \chi }_d^{\lambda }=\hat{\chi }^{\lambda } - \frac{ \vec \Delta}{2m} {E_k - m  \over E_k+m}$, where 
$ \ \hat{\chi }^{\lambda } \ $  is the polarization vector for a massive spin 1 particle (like deuteron) at rest, i.e. 
\begin{equation}
\hat{\chi }^{\lambda =0}=(0,0,1)\text{,  } \quad
\hat{\chi }^{\lambda =+1}=-\frac{1}{\sqrt{2}}(1,i,0)\text{,    } \quad
\hat{\chi }^{\lambda =-1}=\frac{1}{\sqrt{2}}(1,-i,0)
\end{equation}
 The difference only appears for the longitudinal  polarization ($\lambda=0$), and it reaches about 7\% at $k \sim 1$ GeV/c. In Fig.(\ref{Vec Diff}) we show that this  correction  is much  smaller\footnote{Notice that it is compared with the square of the small parameter.} than  the small parameter ${\Delta^- \over 2m }$. Hence, the use of the more simple polarization vector 
$ \ \hat{\chi }^{\lambda } \ $  instead of the actual $ \ \vec{\chi }^{\lambda }_d$  is justified. 

For the $\Gamma_2$ term,
\begin{align}
\Gamma_{2}  \rightarrow & -  2 \Gamma_{2} {\vec k ^2 \over m}  ( \hat{k} \cdot \vec\chi^{\lambda_d})(\hat{k} \cdot \vec{\sigma})  = -  2 \Gamma_{2} {\vec k ^2 \over m}  ( \hat{k}_i \hat{k}_j )   \sigma_i \chi^{\lambda_d}_j  
\nonumber \\
&
= -  {2 \over 3}  \Gamma_{2} {\vec k ^2 \over m} \left( \left[3 \hat{k}_i \hat{k}_j -\delta_{ij}\right] + \left[ \delta_{ij} \right] \right) \sigma_i \chi^{\lambda_d}_j 
\end{align}

For the $\Gamma_3$ term,
\begin{align}
\hspace{-.6cm}
\Gamma_{3}  \rightarrow & \ 2 \Gamma_{3} \left( { {\Delta}^\mu  {\chi}^{\lambda_d}_\mu \over 2m}(\vec{k} \cdot \vec{\sigma}) \right)
= -2 \Gamma_{3} \frac{\vec{k}^{3}}{2 m^{2}}\left(\hat{k}_{i} \hat\Delta_{j}\right) \sigma_{i} \hat{\chi}_{j}^{\lambda_d}
=-2\Gamma_{3} \frac{k^{3}}{2 m^{2}}\left[\frac{1}{2}\left(\hat{k}_{i} \hat{\Delta}_{j} \pm \hat{k}_{j} \hat{\Delta}_{i}\right)\right] \sigma_{i} \hat{\chi}_{j}^{\lambda_d}
 \nonumber \\ 
&=-2 \Gamma_{3} \frac{\vec{k}^{3}}{2 m^{2}}\left[\left(\frac{1}{2} \left\{\hat{k} \hat{\Delta}\right\}_{ij} - {1\over 3}(\hat{k} \cdot \hat{\Delta}) \delta_{i j}\right)+{1\over 3}(\hat{k} \cdot \hat{\Delta}) \delta_{i j}+\frac{1}{2} \epsilon_{i j l} \epsilon_{m n l}\hat{k}_{m} \hat{\Delta}_{n}\right] \sigma_{i} \hat{\chi}_{j}^{\lambda_d}
 \nonumber \\ 
&=- \Gamma_{3} \frac{\vec{k}^{3}}{2m^{2}}\left\{  {2 \over 3} \left[ {3 \over 2} \left\{\hat{k} \hat{\Delta}\right\}_{ij} - (\hat{k} \cdot \hat{\Delta}) \delta_{i j} \right] + {2\over 3}(\hat{k} \cdot \hat{\Delta}) \left[ \delta_{i j} \right] + \left[ \epsilon_{i j l} \left( \hat{k} \times \hat{\Delta}\right)_{l} \right] \right\} \sigma_{i} \hat{\chi}_{j}^{\lambda_d}
\end{align}
where, $ \left\{\hat{k} \hat{\Delta}\right\}_{ij}=\left(\hat{k}_{i} \hat{\Delta}_{j} + \hat{k}_{j} \hat{\Delta}_{i}\right)$ indicates the symmetrization on indices $ij$,  and we use the notation,  $\hat \Delta={\vec P_\text{NN} - \vec p_d \over | \vec P_\text{NN} - \vec p_d |}$.

For the $\Gamma_4$ term,
\begin{align} 
\Gamma_{4} \ \rightarrow & \  2\Gamma_{4}  {\vec{k} \cdot \vec\chi^{\lambda_d}  \over 2m}  \left( { E_k } \frac{\vec{\Delta} \cdot \vec{\sigma}}{2 m} -  \frac{\vec{\Delta} \cdot \vec{k}}{ 2 m+2 E_{k}} {\vec{k} \cdot \vec{\sigma} \over m} \right)  
\nonumber \\ 
& 
= \Gamma_{4}  { {\vec k}^3 \over 2 m^2 E_k}   \left( \hat \Delta_i \hat k_j  \right) \sigma_i \chi^{\lambda_d}_j  -  \Gamma_{4}  { {\vec k}^3 \over 2 m^2 E_k}  (E_k - m)  \left( \hat k_i \hat k_j \hat k_l \right)  \sigma_i \chi^{\lambda_d}_j   \hat\Delta_l 
\nonumber \\ 
&
=\Gamma_{4} \frac{\vec{k}^{3}}{ 4 m^{2} }\left[ {2m \over 3 E_k}(\hat{k} \cdot \hat{\Delta}) \delta_{i j}  -  \epsilon_{i j l} \left( \hat{k} \times \hat{\Delta}\right)_{l} \right] \sigma_i \hat\chi^{\lambda_d}_j 
\nonumber \\ 
& \quad
+ \Gamma_{4} \frac{\vec{k}^{3}  (3E_k +2 m)}{ 20 m^{2} E_k}  \left( \left\{\hat{k} \hat{\Delta}\right\}_{ij} - \frac{2}{3} (\hat{k} \cdot \hat{\Delta}) \delta_{i j}\right)   \sigma_i \hat\chi^{\lambda_d}_j 
\nonumber \\ 
& \quad
- \Gamma_{4}  { {\vec k}^3 (E_k - m) \over  10 m^2 E_k}   \left[ 5 \left( \hat k_i \hat k_j \hat k_l \right) -  \delta_{ij} \hat k_l  -  \delta_{jl} \hat k_i  -  \delta_{li} \hat k_j  \right] \sigma_i \hat\chi^{\lambda_d}_j   \hat\Delta_l  
\end{align}
Notice that in principle $\hat \Delta_p$ is arbitrary. In practice however, when the deuteron to proton-neutron transition vertex is used in a break-up reaction,  $\hat \Delta_p$ is specify by the direction of the transfer momentum. In this case we get,
\begin{align}
\Gamma_{4} \ \rightarrow & \  \Gamma_{4} \frac{\vec{k}^{3}}{ 4 m^{2} }\left[ {2m \over 3 E_k}(\hat{k} \cdot \hat{\Delta}) \delta_{i j}  -  \epsilon_{i j l} \left( \hat{k} \times \hat{\Delta}\right)_{l} \right] \sigma_i \hat\chi^{\lambda_d}_j 
\nonumber \\ 
& \quad
+ \Gamma_{4} \frac{\vec{k}^{3}  (3E_k +2 m)}{ 20 m^{2} E_k}  \left( \left\{\hat{k} \hat{\Delta}\right\}_{ij} - \frac{2}{3} (\hat{k} \cdot \hat{\Delta}) \delta_{i j}\right)   \sigma_i \hat\chi^{\lambda_d}_j 
\nonumber \\ 
& \quad
- \Gamma_{4}  { {\vec k}^3 (E_k - m) \over  10 m^2 E_k}   \left[ 5 \left( \hat k_i \hat k_j  \right) \left(\hat k \cdot  \hat\Delta \right) -  \delta_{ij} \left(\hat k \cdot  \hat\Delta \right)  -   \hat\Delta_j \hat k_i  -  \hat\Delta_i \hat k_j  \right] \sigma_i \chi^{\lambda_d}_j   
\nonumber \\ 
&  
=  \Gamma_{4} \frac{\vec{k}^{3}}{ 6 m^{2}} \left[ {3 \over 2} \left\{\hat{k} \hat{\Delta}\right\}_{ij} - (\hat{k} \cdot \hat{\Delta}) \delta_{i j} \right]   \sigma_{i}  \hat{\chi}_{j}^{\lambda_d}  
- \Gamma_{4} \frac{\vec{k}^{3}}{ 4m^{2}}\left[ \epsilon_{i j l} \left( \hat{k} \times \hat{\Delta}\right)_{l} \right]   \sigma_{i}  \hat{\chi}_{j}^{\lambda_d}  
\nonumber \\
&
\quad - \Gamma_{4}  { {\vec k}^3 (E_k - m) \over  6 m^2 E_k} \left(\hat\Delta \cdot \hat k \right)  \left[3 \hat{k}_i \hat{k}_j -\delta_{ij} \right] \sigma_i \chi^{\lambda_d}_j  
+ \Gamma_{4} \frac{\vec{k}^{3}}{ 6 m E_k } \left(\hat{k} \cdot \hat{\Delta}\right) \left[  \delta_{i j} \right]  \sigma_{i}  \hat{\chi}_{j}^{\lambda_d} 
\end{align}

For the $\Gamma_5$ term,
\begin{align}
\Gamma_{5}  \rightarrow & - i \Gamma_{5} \frac{M_\text{NN}^2}{ m}\left( { \vec{\Delta} \over 2m}  \times  \frac{ \vec{k} }{m}\right) \cdot  \vec\chi^{\lambda_d}  \mathbf{1}_2 = - i \Gamma_{5} \frac{ 2 E_k \vec k^3}{ m^3}\left( \hat{\Delta} \times \hat{k} \right) \cdot  \hat\chi^{\lambda_d}   \mathbf{1}_2 
\nonumber \\
&
= - i \Gamma_{5} \frac{E_k}{ m} {\vec k^3 \over 2m^2}  \hat{k}_\text{T}  \cdot  \hat\chi_\text{T}^\lambda  \mathbf{1}_2 = - i \Gamma_{5} \frac{E_k}{ m} {\vec k^3 \over 2m^2} \left( \hat{k} \cdot  \hat\chi^\lambda  - (\hat \Delta \cdot \hat k) (\hat \Delta \cdot \hat \chi^\lambda )\right)  \mathbf{1}_2
\nonumber \\
&
= i \Gamma_{5} \frac{E_k}{ m} {\vec k^3 \over 6 m^2}  \left(  \hat k_i \left[ 3 \hat \Delta_i  \hat \Delta_j -  \delta_{ij}  \right]  - 2 \hat k_i \left[ \delta_{ij} \right]  \right)   \hat\chi^\lambda_j   \mathbf{1}_2
\end{align}
where, $ \mathbf{1}_2$ is the $2\times2$ identity matrix.

For the $\Gamma_6$ term,
{\small
\begin{align}
\Gamma_{6}  \rightarrow & - 2\Gamma_{6} { {\Delta}^\mu  {\chi}^{\lambda_d}_\mu \over 2m} \left( { E_k }  \frac{\vec{\Delta} \cdot \vec{\sigma}}{2 m} -  \frac{\vec{\Delta} \cdot \vec{k}}{ 2 m+2 E_{k}} {\vec{k} \cdot \vec{\sigma} \over m} \right) 
\nonumber \\
&
= \Gamma_{6}  { {\vec k}^4 \over 2 m^3 E_k}   \left( { E_k } \hat \Delta_i \hat \Delta_j - (E_k - m) (\hat\Delta \cdot \hat k) \hat k_i \hat \Delta_j  \right) \sigma_i  \chi^{\lambda_d}_j 
\nonumber \\
&
= \Gamma_{6}  { {\vec k}^4 \over 6 m^3 }     \left( (3 \hat{\Delta}_i \hat{\Delta}_j -\delta_{ij}) + \delta_{ij}  \right) \sigma_i  \chi^{\lambda_d}_j 
\nonumber \\
&
\quad - \Gamma_{6}  { {\vec k}^4 (E_k - m) \over 2 m^3 E_k}    (\hat\Delta \cdot \hat k) \left[ \frac{1}{6} \left( 3 \left\{\hat{k} \hat{\Delta}\right\}_{ij} - 2(\hat{k} \cdot \hat{\Delta}) \delta_{i j}\right)+{1\over 3}(\hat{k} \cdot \hat{\Delta}) \delta_{i j}+\frac{1}{2} \epsilon_{i j l} \left( \hat{k} \times \hat{\Delta}\right)_{l} \right] \sigma_{i} \hat{\chi}_{j}^{\lambda_d}
\nonumber \\
&
= \Gamma_{6}  { {\vec k}^4 \over 6 m^3 }     \left[ 3 \hat{\Delta}_i \hat{\Delta}_j -\delta_{ij}   \right] \sigma_i  \hat\chi^{\lambda_d}_j  - \Gamma_{6}  { {\vec k}^4 (E_k - m) \over 4 m^3 E_k}    (\hat\Delta \cdot \hat k) \left[ \epsilon_{i j l} \left( \hat{k} \times \hat{\Delta}\right)_{l} \right] \sigma_{i} \hat{\chi}_{j}^{\lambda_d}
\nonumber \\
&
\quad + \Gamma_{6}  { {\vec k}^4 \over 6 m^3 E_k }   \left( E_k - \left(\hat\Delta \cdot \hat k \right)^2  \left(E_k - m \right) \right)  \left[ \delta_{ij}  \right] \sigma_i  \hat\chi^{\lambda_d}_j 
\nonumber \\
&
\quad - \Gamma_{6}  { {\vec k}^4 (E_k - m) \over 3 m^3 E_k}    (\hat\Delta \cdot \hat k) \left[  \frac{3}{2} \left\{\hat{k} \hat{\Delta}\right\}_{ij} - (\hat{k} \cdot \hat{\Delta}) \delta_{i j} \right] \sigma_{i} \hat{\chi}_{j}^{\lambda_d}
\end{align}
}

We arrive then to the separation, 
\begin{align}\label{wf cm pw}
{\psi}_\text{CM}^{\lambda_d} = &  \left\{\sigma_i  \hat\chi^{\lambda_d}_j \right\}   \left\{   
 \left[ \delta_{i j} \right] 
 \left[  - {2 \over 3} \Gamma_{1} \left( 2E_{k}+m \right) 
-  {2 \over 3}  \Gamma_{2} {\vec k ^2 \over m} 
- \Gamma_{3} \frac{\vec{k}^{3}}{3 m^{2}} (\hat{k} \cdot \hat{\Delta}) 
+ \Gamma_{4} \frac{\vec{k}^{3}}{ 6 m E_k } \left(\hat{k} \cdot \hat{\Delta}\right)  \right. \right. 
\nonumber \\
& \left.
\hspace{3cm} + \Gamma_{6}  { {\vec k}^4 \over 6 m^3 E_k }  \left( E_k -  \left(\hat\Delta \cdot \hat k \right)^2  \left(E_k - m \right) \right) \right]
\nonumber \\
&
+ \left[  3 \hat k_i \hat k_j -\delta_{ij}   \right]
\left[  {2 \over 3} \Gamma_{1} \left( m - E_{k} \right) 
-  {2 \over 3}  \Gamma_{2} {\vec k ^2 \over m}  
- \Gamma_{4}  { {\vec k}^3 (E_k - m) \over  6 m^2 E_k} \left(\hat\Delta \cdot \hat k \right)  \right]
\nonumber \\
&
+ \left[ \frac{3}{2} \left\{\hat{k} \hat{\Delta}\right\}_{ij} - (\hat{k} \cdot \hat{\Delta}) \delta_{i j} \right] 
\left[ - \Gamma_{3} \frac{\vec{k}^{3}}{ 3 m^{2}} 
+ \Gamma_{4} \frac{\vec{k}^{3}}{ 6 m^{2}} 
- \Gamma_{6}  { {\vec k}^4 (E_k - m) \over 3 m^3 E_k}  (\hat\Delta \cdot \hat k) \right] 
\nonumber \\
&
+ \left[ \epsilon_{i j l} \left( \hat{k} \times \hat{\Delta}\right)_{l} \right]  
\left[ - \Gamma_{3} \frac{\vec{k}^{3}}{ 2 m^{2}}
- \Gamma_{4} \frac{\vec{k}^{3}}{ 4m^{2}}  
- \Gamma_{6}  { {\vec k}^4 (E_k - m) \over 4 m^3 E_k}    (\hat\Delta \cdot \hat k)  \right] 
\nonumber \\
& \left. 
+ \left[ 3 \hat{\Delta}_i \hat{\Delta}_j -\delta_{ij}   \right]  \left[  \Gamma_{6}  { {\vec k}^4 \over 6 m^3 }   \right]  \ \right\}
 \nonumber \\
&
+ \left\{ \mathbf{1}_2   \hat\chi^{\lambda_d}_j  \right\}  \left\{  \left[ 3 \hat \Delta_i  \hat \Delta_j -  \delta_{ij}  \right]  - 2  \left[ \delta_{ij} \right]   \right\}   \left( i \Gamma_{5} \frac{E_k}{ m} {\vec k^3 \over 6 m^2} \hat k_i \right)
\end{align} 
In this form we can now make the following associations. 
The terms proportional to $\left\{\sigma_i  \hat\chi^{\lambda_d}_j \right\}\left[ \delta_{i j} \right]$ and $\left\{\sigma_i  \hat\chi^{\lambda_d}_j \right\}  \left[  3 \hat k_i \hat k_j -\delta_{ij}   \right]$ correspond to an S-wave and D-wave transition, respectively. 
These terms must converge to the nonrelativistic deuteron S- and D-wave radial functions\footnote{From the deuteron wave function which is a solution to the  Shrodinger equation with an $NN$ interaction potential, e.g., CD-Bonn or AV18.}, when the contribution from the vertex functions $\Gamma_3$, $\Gamma_4$, and $\Gamma_6 $. The conditions for which we expect this to be a good approximation are stablished at the end of Sec.(\ref{wave_function_gen}).
Since, the terms associated with $\Gamma_{3,4,6}$ are expected to be small we focus on the angular structures associated with $\Gamma_5$, which contributes as P-wave and has a relativistic origin with no analogous in the nonrelativistic framework.

 \begin{figure}[h]
\hspace{-0.5cm}
	\includegraphics[scale=0.75]{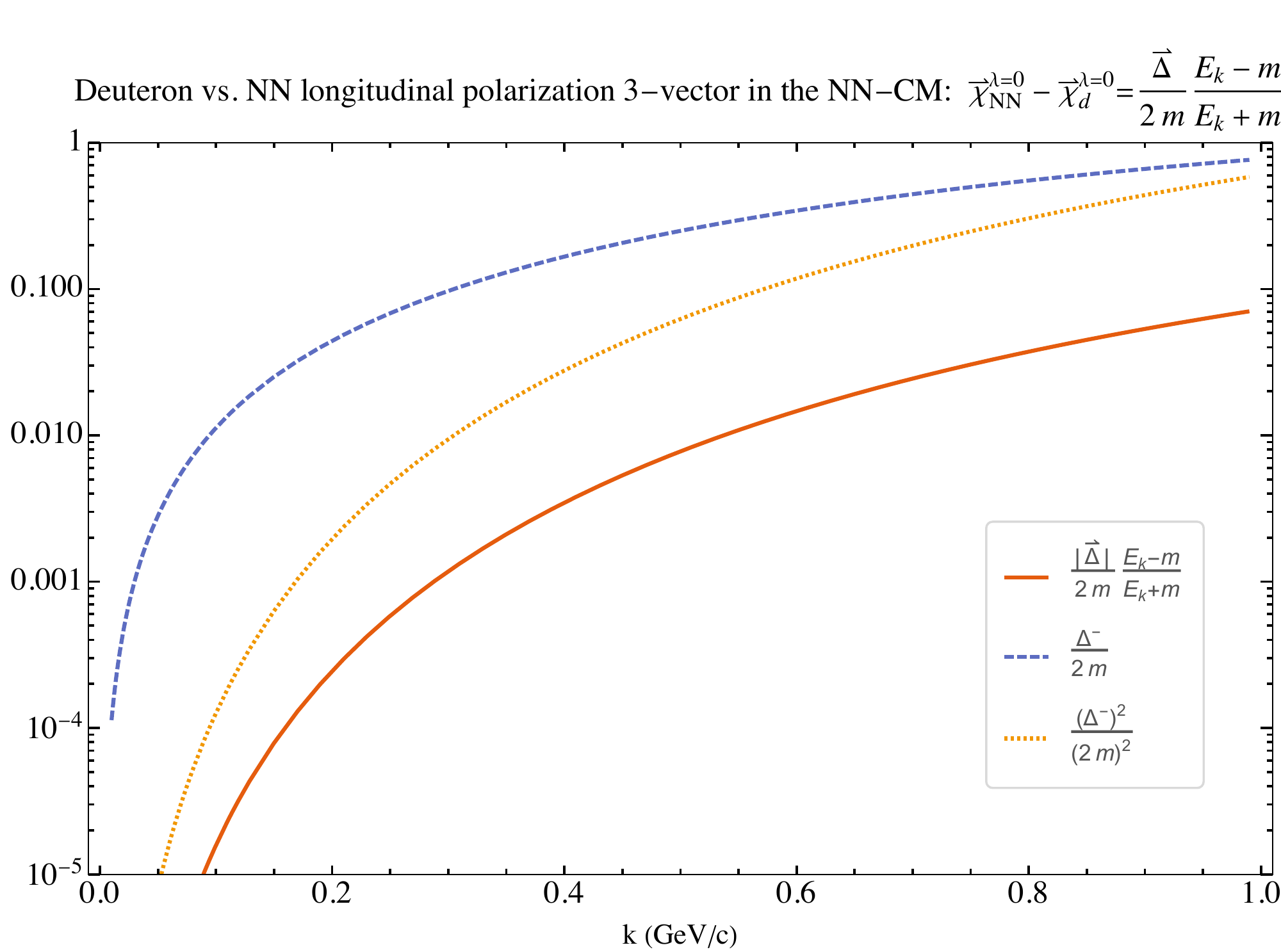}
	\caption{Deviation of deuteron longitudinal polarization 3-vector $\vec{\chi }_d^{\lambda =0}$ from that of the two-body in the NN-CM  $ \vec{\chi }_\text{NN}^{\lambda =0}$  }
	\label{Vec Diff}
\end{figure}

%

\subsection*{I.5 Dirac Bilinears - LF}

To recover the light-front wave function we must apply the Melosh rotation 
\hyperref[Melosh rotation Sec]{Eq.(\ref*{Melosh rotation})}
to the Canonical bilinear structures. 
The LF bilinears in the two-body CM and two dimensional spin space are,
\begin{align}
\bar{u}_{- \vec{k}}^\text{LF} u_{\vec{k}}^\text{LF} = \frac{M_\text{NN}}{\sqrt{k_T^2+m^2}} \left( m \mathbf{1} + \left( \vec{k}_\text{T} \cdot \vec{\sigma}_\text{T} \right) \sigma _3  \right)
\end{align}
\begin{align}
\bar{u}_{-\vec{k}}^\text{LF} \gamma _5 u_{\vec{k}}^\text{LF} = & \frac{1 }{\sqrt{k_T^2+m^2}} \left( M_\text{NN} \vec{k}_\text{T} \cdot \vec{\sigma}_\text{T} + 2m  k_3 \sigma _3 \right) \nonumber \\
& =  \frac{M_\text{NN} }{\sqrt{k_T^2+m^2}} \left(  \vec{k}_\text{T} \cdot \vec{\sigma}_\text{T} + m (\alpha -1){\sigma}_3 \right)
\end{align}
\begin{align}
\bar{u}_{-\vec{k}}^\text{LF} \left(\gamma^+ \gamma _5\right)u_{\vec{k}}^\text{LF} & = \bar{u}_{-\vec{k}}^\text{LF} \left(\gamma _3 \gamma _5\right)u_{\vec{k}}^\text{LF}  = 2 \sqrt{k_T^2+m^2} \left(  \vec{k} \cdot \vec{\sigma} -  \vec{k}_\text{T} \cdot \vec{\sigma}_\text{T}  \right) = M_\text{NN} \sqrt{\alpha(2-\alpha)} \sigma _3   
\end{align}
\begin{align}
\bar{u}_{-\vec{k}}^\text{LF} \left(\gamma^- \gamma _5\right)u_{\vec{k}}^\text{LF} & = - \bar{u}_{-\vec{k}}^\text{LF} \left(\gamma _3 \gamma _5\right)u_{\vec{k}}^\text{LF}  =  -  M_\text{NN} \sqrt{\alpha(2-\alpha)} \sigma _3 
\end{align}
\begin{align}
\bar{u}_{-\vec{k}}^\text{LF} \left( {\gamma}_1 \gamma _5\right)u_{\vec{k}}^\text{LF}  &= \frac{1 }{\sqrt{k_T^2+m^2}} \left(  M_\text{NN} m  \sigma_1  - 2 k_3 \sigma _3 k_x + \rm{i} M_\text{NN}  k_y \mathbf{1} \right) \nonumber \\
 &= \frac{M_\text{NN} }{\sqrt{k_T^2+m^2}} \left(   m  \sigma_1  - (\alpha -1) \sigma _3 k_x + \rm{i}  k_y \mathbf{1} \right)
\end{align}
\begin{align}
\bar{u}_{-\vec{k}}^\text{LF} \left( {\gamma}_2 \gamma _5\right)u_{\vec{k}}^\text{LF} 
& = \frac{M_\text{NN} }{\sqrt{k_T^2+m^2}} \left(  m \sigma_2  - (\alpha -1)  \sigma _3 k_y - \rm{i}   k_x \mathbf{1} \right)
\end{align}

\newpage

\setcounter{equation}{0}
\renewcommand{\theequation}{J.\arabic{equation}}

\setcounter{figure}{0}
\renewcommand{\thefigure}{J.\arabic{figure}}

\chapter{Deuteron Electro-Magnetic Current and Wave Function Normalization } 
\phantomsection\label{Deuteron EM Current}

\begin{figure}[h]
	\centering
	\includegraphics[scale=0.45]{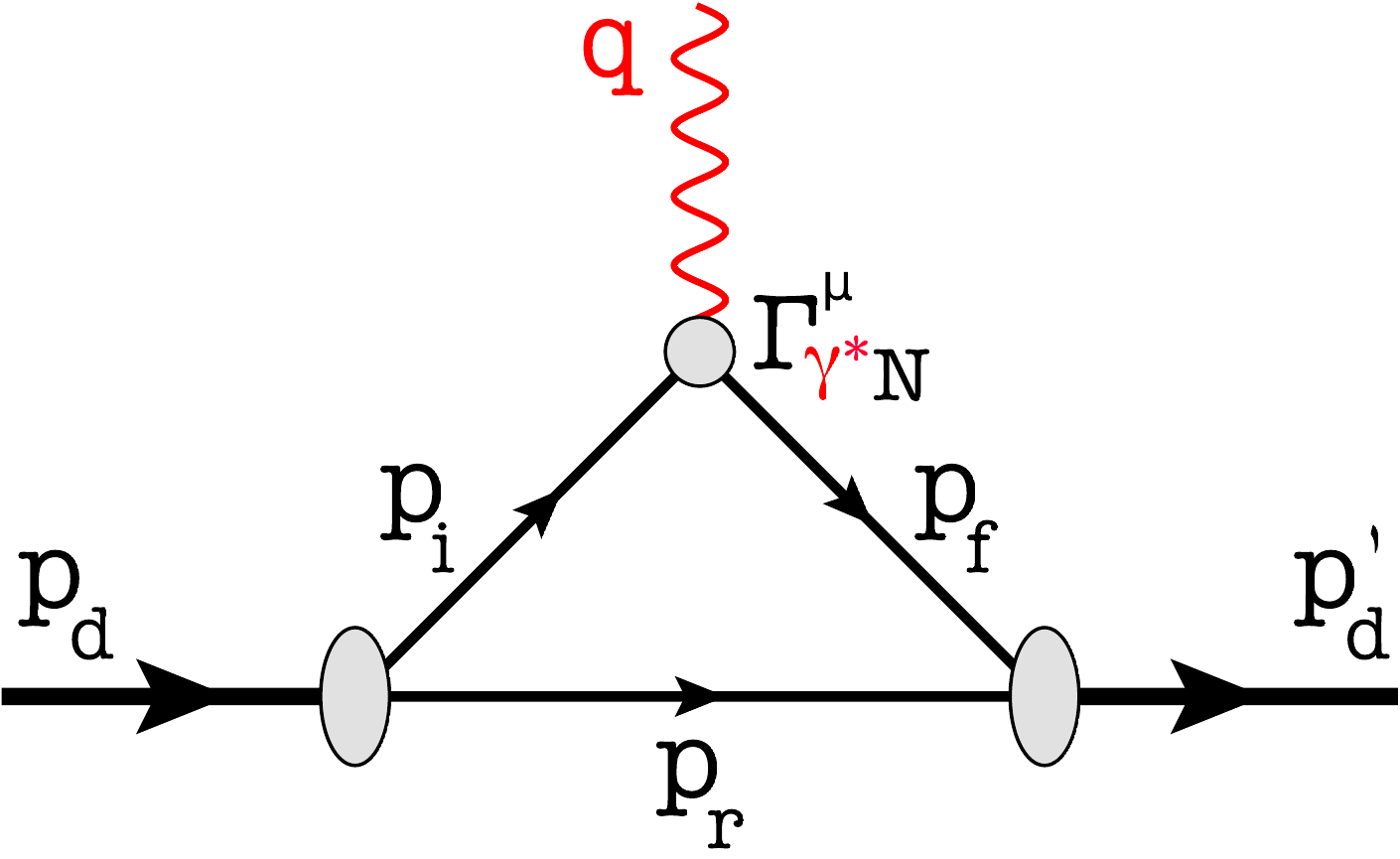}
	\caption{Deuteron elastic scattering: $d+{\color{red}\gamma^*} \rightarrow d'$}
	\label{deuteron FF}
\end{figure}

The normalization of deuteron's wave function  is fixed relative to its elastic charge form-factor:  $G_C(Q^2=0)=1$.

Equivalently to the EM nucleon current parameterization in terms of form factors, in the one-photon exchange approximation, Lorentz invariance constrains  the deuteron 
EM current to be completely described by three form factors 
\cite{Gilman:2001yh}: 
\phantomsection\label{deuteron EM current Sec}
\begin{align}\label{deuteron EM current}
\left\langle p_d^\prime s_d^{\prime}\left|J^{\mu}_d\right|p_d s_d\right\rangle = & -\Big( G_{1} (Q^{2}) (\chi^{'*} \cdot \chi) - G_{3}(Q^{2})\frac{(\chi^{'*} \cdot q)(\chi \cdot q)}{2 m_{d}^{2}}\Big)\left(p_d^{\mu}+p_d^{\prime \mu}\right) \nonumber \\
& 
+G_{M}(Q^{2})\Big(\chi^{\mu}\left(\chi^{'*} \cdot q\right)-(\chi^{\prime \mu})^{*}(\chi \cdot q)\Big)
\end{align}
where, $\chi=\chi_{s_d}$ ($\chi'=\chi_{s'_d}$) are deuteron polarization vectors.
The charge ($G_C$) and  quadrupole ($G_Q$) FF are related with $G_1$, $G_3$  and 
($G_M$)   by,
\begin{align} 
G_{C} &=G_{1}+\frac{2}{3} \tau G_{Q} \\ 
G_{Q} &=G_{1}-G_{M}+(1+\tau ) G_{3} 
\end{align}
At  $Q^2 = 0$, they are conventionally normalized to,
\begin{equation}
\begin{array}{ll}{G_{C}(0) =1} & {\text { (in units of } e )} \\ {G_{Q}(0)=Q_{d}} & {\left(\text { in units of } e / m_{d}^{2}\right)} \\ {G_{M}(0)=\mu_{d}} & {\left(\text { in units of } e / 2 m_{d}\right)}
\end{array}
\end{equation}  

The FF can be extracted from combinations of the EM current matrix elements. 
On the Briet frame and in the limit, $Q^2\rightarrow 0$, the unpolarized elastic scattering  is proportional to the charge form factor. 
For the ($+$) component of the current on the LF we obtain:
\begin{align}
{1\over 3} \sum_{s_d, s'_d} \left\langle p_d^\prime s_d^{\prime}\left|J^{+}\right|p_d s_d\right\rangle=  \left(p_d+p_d^{\prime}\right)^+ G_1(0) =  \left(p_d+p_d^{\prime}\right)^+ G_C(0)=  \left(p_d+p_d^{\prime}\right)^+  
\end{align}
In this frame we have, $p_d^\mu=(m_d\sqrt{1+\tau},0,0,-Q/2)$,  and, $(p'_d)^{\mu}=(m_d\sqrt{1+\tau},0,0,Q/2)$, which yields,
\begin{align}
{1\over 3} \sum_{s_d, s'_d} \left\langle p_d^\prime s_d^{\prime}\left|J^{+}\right|p_d s_d\right\rangle=  \left(p_d+p_d^{\prime}\right)^+ G_C(0)= 2 m_d
\end{align}
where, $\tau={Q^2 \over 4m_d^2}$.

The relation with deuteron's WF is provided by the elastic scattering diagram Fig.(\ref{deuteron FF}), 
which has the transition amplitude,

\begin{align}
\hspace{-.6cm}
	\left\langle p_{d}^{\prime} s_{d}^{\prime}\left|A^{\mu}\right| p_{d} s_{d}\right\rangle =  - \int \frac{d^{4} p_{r}}{i(2 \pi)^{4}}  \chi_{s_{d}^{\prime}}^{ \dagger} \Gamma_\text{dNN}^{\dagger} \frac{\sh{p}_{\color{red}f}+m}{p_{\color{red}f}^{2}-m^{2}+i \epsilon} \Gamma_{{\color{red}\gamma^*} N}^{\mu} \frac{\sh{p}_{\color{red}i}+m}{p_{\color{red}i}^{2}-m^{2}+i \epsilon} \Gamma_\text{dNN}   \chi_{s_{d}}  \frac{\sh{p}_{r}+m}{p_{r}^{2}-m^{2}+i \epsilon}
\end{align}
a sum over $r$ is implicit, where $ r=p,n$, accounts for the two nucleons. 

Changing 
to LF variables ($d^4p_r \rightarrow 1/2dp^- dp^+ d^2p_\perp$),
and integrating over the light-cone energy, $dp_r^-$, puts the recoil particle on shell, we obtain

\begin{align}
\left\langle p_{d}^{\prime} s_{d}^{\prime}\left|A^{\mu}\right| p_{d} s_{d}\right\rangle =   & \int  \frac{d^{2} p_{r \perp} dp_r^+}{2p_{r}^+(2 \pi)^{3}} \chi_{s_{d}^{\prime}}^{ \dagger} \Gamma_\text{dNN}^{\dagger} \frac{\sh{p}_{\color{red}f}+m}{p_{{\color{red}f}}^{2}-m^{2}+i \epsilon} \nonumber \\
& \Gamma_{{\color{red}\gamma^*} N}^{\mu} \frac{\sh{p}_{{\color{red}i}}+m}{p_{{\color{red}i}}^{2}-m^{2}+i \epsilon} \Gamma_\text{dNN} \chi_{s_{d}}  \sum_{s_r}  u(p_r,s_r) \bar{u}(p_r,s_r)
\end{align}

The evaluation of  result is that the active nucleon, 
used to construct the EM current, is off-shell for the initial ($p_{\color{red}i}$) and final ($p_{\color{red}f}$)  states. Therefore, we write the respective propagators 
as follows:

\begin{align}
\frac{\sh{p}_{{\color{red}i} ,\text{off}} + m_N}{p_{{\color{red}i},\text{off}}^2 - m_N^2 } = & \frac{ \sum_{s_{\color{red}i}} \left( u(p_{\color{red}i},s_{\color{red}i}) \bar{u}(p_{\color{red}i},s_{\color{red}i})\right) + \sh \Delta_{p_{\color{red}i} }}{ \frac{\alpha_{\color{red}i}}{2} \left(m_d^2 - 4 \frac{m_N^2 + p_{{\color{red}i} {\bf T}}^2}{\alpha_{\color{red}i}(2-\alpha_{\color{red}i})}\right)} =  \frac{ \sum_{s_{\color{red}i}} \left( u(p_{\color{red}i},s_{\color{red}i}) \bar{u}(p_{\color{red}i},s_{\color{red}i})\right) + \sh \Delta_{p_{\color{red}i} } }{\frac{\alpha_{\color{red}i}}{2} \left(m_d^2 - s_\text{NN} \right) }  
\end{align}
\vspace{-0.2cm}
\begin{align}
\frac{\sh{p}_{{\color{red}f},\text{off}} + m_N}{p_{{\color{red}f},\text{off}}^2 - m_N^2 } = & \frac{\sum_{s_{\color{red}f}} \left( u(p_{\color{red}f},s_{\color{red}f}) \bar{u}(p_{\color{red}f},s_{\color{red}f})\right) + \sh \Delta_{p_{\color{red}f} }  }{ \frac{\alpha_{\color{red}f}}{2} \left(m_d^2 - 4 \frac{m_N^2 + p_{{\color{red}f} {\bf T}}^2}{\alpha_{\color{red}f}(2-\alpha_{\color{red}f})}\right)}  = \frac{\sum_{s_{\color{red}f}} \left( u(p_{\color{red}f},s_{\color{red}f}) \bar{u}(p_{\color{red}f},s_{\color{red}f})\right) + \sh \Delta_{p_{\color{red}f} }  }{\frac{\alpha_{\color{red}f}}{2} \left(m_d^2 - s'_\text{NN} \right) }
\end{align}
Notice that in general, for the diagram of Fig.(\ref{deuteron FF}),  
\begin{align}
\alpha_{\color{red}f}= 2 \frac{p_{\color{red}f}^+}{(p'_d)^{+}} \ne \alpha_{\color{red}i}= 2 \frac{p_{\color{red}i}^+}{p_d^{+}} = 2 - \alpha_r = 2 - 2 \frac{p_r^+}{p_d^{+}} 
\end{align}

The  substitution of the propagators into the amplitude produce,

\begin{align}
\left\langle p_{d}^{\prime} s_{d}^{\prime}\left|A^{\mu}\right| p_{d} s_{d}\right\rangle =  & 
\sum_{s_{r}} \int \frac{d^{2} p_{r \perp} dp_r^+}{\alpha_{\color{red}i} \alpha_{\color{red}f} p_{r}^+} \Psi_{s_{\color{red}f} s_r}^{s_{d}^{\prime} \dagger }\left(p_{r}, p_{d}^\prime \right) \tilde{J}_{N}^{\mu} \  \Psi_{s_{\color{red}i} s_r}^{s_{d} }\left(p_{r}, p_{d} \right) \nonumber \\
= & \sum_{s_{r}} \int \frac{d^{2} p_{r \perp} d\alpha}{\alpha  (2-\alpha) (2 - \alpha + \alpha_q) } \Psi_{s_{\color{red}f} s_r}^{s_{d}^{\prime} \dagger} \tilde{J}_{N}^{\mu} \Psi_{s_{\color{red}i} s_r}^{s_{d}}
\end{align}
where, $\alpha=\alpha_r$. The WF of deuteron is given by (Eq.(\ref{wflf})), 
\begin{align}
 \Psi_{s_{i} s_r}^{s_{d}} = \frac {\bar {u}_i \bar u_r \Gamma _\text{dNN}  \chi^{s_d}}{\frac {1}{2}\left(s_\text{NN} - m^{2}_{d} \right) }\dfrac {1}{\sqrt {2}\left( 2\pi \right) ^{3/2}}
\end{align}
and the off-shell nucleon current has the following  four contributions, 
\begin{align}
\tilde{J}_{N}^{\mu}\left(s_{\color{red}f}, p_{\color{red}f} , s_{\color{red}i}, p_{\color{red}i}\right)= & \ J_{N, \text{on}}^{\mu}\left(s_{\color{red}f}, p_{\color{red}f} , s_{\color{red}i}, p_{\color{red}i}\right)  + J_{N, \text{off}}^{\mu}\left(s_{\color{red}f}, p_{{\color{red}f}, \text{off}} , s_{\color{red}i}, p_{{\color{red}i}, \text{off}} \right)  \nonumber \\ 
&+J_{N,\text{off}}^{\mu}\left(s_{\color{red}f}, p_{{\color{red}f}, \text{off}} , s_{\color{red}i}, p_{\color{red}i}\right)+J_{N, \text{off}}^{\mu}\left(s_{\color{red}f}, p_{\color{red}f} , s_{\color{red}i}, p_{{\color{red}i}, \text{off}} \right) \nonumber \\
\end{align}
where,
\begin{align}
J_{N, \text{on}}^{\mu}\left(s_{\color{red}f}, p_{\color{red}f} , s_{\color{red}i}, p_{\color{red}i}\right) = & \ \bar{u}(p_{\color{red}f} s_{\color{red}f}) \Gamma_{{\color{red}\gamma^*}  N}^{\mu}  u(p_{\color{red}i} s_{\color{red}i})  \\  
J_{N, \text{off}}^{\mu}\left(s_{\color{red}f}, p_{\color{red}f} , s_{\color{red}i}, p_{{\color{red}i}, \text{off}} \right) = & \ \bar{u}(p_{\color{red}f} s_{\color{red}f}) \Gamma_{{\color{red}\gamma^*}  N}^{\mu} \frac{ \sh \Delta_{p_{\color{red}i} }  }{2m_N}  u(p_{\color{red}i} s_{\color{red}i})   \\
J_{N, \text{off}}^{\mu}\left(s_{\color{red}f}, p_{{\color{red}f}, \text{off}} , s_{\color{red}i}, p_{\color{red}i} \right)  = & \ \bar{u}(p_{\color{red}f} s_{\color{red}f}) \frac{ \sh \Delta_{p_{\color{red}f} }  }{2m_N} \Gamma_{{\color{red}\gamma^*}  N}^{\mu}  u(p_{\color{red}i} s_{\color{red}i})   \\ 
J_{N, \text{off}}^{\mu}\left(s_{\color{red}f}, p_{{\color{red}f}, \text{off}} , s_{\color{red}i}, p_{{\color{red}i}, \text{off}} \right)= & \ \bar{u}(p_{\color{red}f} s_{\color{red}f}) \frac{ \sh \Delta_{p_{\color{red}f} } }{2m_N} \Gamma_{{\color{red}\gamma^*}  N}^{\mu} \frac{ \sh \Delta_{p_{\color{red}i} }  }{2m_N}  u(p_{\color{red}i} s_{\color{red}i})
\end{align}
Hence,  it can be written as the succinct expression,
\phantomsection\label{EM current Sec}
\begin{align}\label{EM current}
\tilde{J}_{N}^{\mu}\left(s_{\color{red}f}, p_{\color{red}f} ; s_{\color{red}i}, p_{\color{red}i}\right) =  \ \bar{u}(p_{\color{red}f} s_{\color{red}f}) \tilde{\Gamma}_{{\color{red}\gamma^*}  N}^{\mu}  u(p_{\color{red}i} s_{\color{red}i})  
\end{align}
with the vertex function given by,
\phantomsection\label{vertex off Sec}
\begin{align}\label{vertex off}
\tilde{\Gamma}_{{\color{red}\gamma^*}  N}^{\mu}   =  \ \left(1 +   \frac{ \sh \Delta_{p_{\color{red}f} }  }{2m_N}  \right)  \Gamma_{{\color{red}\gamma^*}  N}^{\mu}  \left( 1 +   \frac{ \sh \Delta_{p_{\color{red}i} }  }{2m_N}  \right)  
\end{align}

\subsection*{J.1 Wave Function Normalization from EM Charge}
\phantomsection\label{WF Normalization EM Charge}

In the reference frame of Fig.(\ref{scatt react plane}), we have $ q^2=q^+ q^- $, and  $ q^+ < 0 $. 
Then, the limit of real photon scattering ($Q^2 \rightarrow 0$) is given by,  $ q^+\rightarrow 0$. As a consequence, $p'_d \rightarrow p_d$, and $ \ p^+_{\color{red}f} \rightarrow p^+_{\color{red}i}$.
\hyperref[vertex off Sec]{Eq.(\ref*{vertex off})}, is  
\begin{align}
\tilde{\Gamma}_{{\color{red}\gamma^*}  N}^{+}   =  \ \left(1 +   \frac{\sh \Delta_{p_{\color{red}f} }  }{2m_N}  \right)    \left(F_{1} \Big( \gamma^{+ } -  q^+{ \gamma^+ {q}^-   \over 2 q^+ q^-} \Big)  + i \sigma^{+ \nu} q_{\nu} F_{2}  {\kappa \over 2 m_N} \right)   \left( 1 +   \frac{ \sh \Delta_{p_{\color{red}i} }  }{2m_N}  \right)  
\end{align}
In the real photon limit we have, 
\begin{align}
\tilde{\Gamma}^{+}_{{\color{red}\gamma^*}  N \ ({Q^2, q^+\rightarrow 0})} \rightarrow & \ \left(1 +   \frac{  \gamma^+ \Delta_{p_{\color{red}i} }^- }{4m_N}  \right)    \left(F_{1} { \gamma^{+ } \over 2} + i \sigma^{+ +} q^{-} F_{2}  {\kappa \over 4 m_N} \right)   \left( 1 +   \frac{  \gamma^+ \Delta_{p_{\color{red}i} }^- }{4m_N}  \right) \nonumber \\
& = \ \left(1 +   \frac{  \gamma^+ \Delta_{p_1 }^- }{4m_N}  \right)    \left(F_{1}  { \gamma^{+ } \over 2}  \right)   \left( 1 +   \frac{  \gamma^+ \Delta_{p_{\color{red}i} }^- }{4m_N}  \right) \nonumber
\end{align}
which reduces to,
\begin{align}
\tilde{\Gamma}^{+}_{{\color{red}\gamma^*}  N \ ({Q^2, q^+\rightarrow 0})} \rightarrow  F_{1}(0)  { \gamma^{+ } \over 2} 
\end{align}
Hence, the matrix elements of the EM current read as,
\begin{align}
\tilde{J}^{+}_{ N \ ({Q^2, q^+\rightarrow 0})} \rightarrow &  \  \bar{u}(p_{\color{red}f} s_{\color{red}f})  F_{1}(0)  { \gamma^{+ } \over 2}   u(p_{\color{red}i} s_{\color{red}i})  = F_1(0) \sqrt{p_{\color{red}f}^+ p_{\color{red}i}^+} \delta_{s_{\color{red}f} s_{\color{red}i}}  \nonumber  \\
\tilde{J}^{+}_{ N \ ({Q^2, q^+\rightarrow 0})} \rightarrow &  \  F_1(0){p_{\color{red}i}^+} \delta_{s_{\color{red}f}s_{\color{red}i}}
\end{align}

The normalization condition acquires  the form,
\begin{align}
{1\over 3} \sum_{s_d, s'_d}  \left\langle p_d^\prime s_d^{\prime}\left|J^{+} \right| p_d s_d\right\rangle & =  \left(p_d+p_d^{\prime}\right)^+ G_C(0) 
\nonumber \\
\longrightarrow \quad & {1\over 3} \sum_{s_d, s'_d} {\color{blue} 1\over 2} \sum_{{\color{blue} s_i}, s_{{\color{red}f}}, s_{r}} \int \frac{d^{2} p_{r \perp} d\alpha}{\alpha  (2-\alpha) (2 - \alpha) } \Psi_{s_{\color{red}f} s_{r}}^{s_{d}^{\prime} \dagger} \left(  F_{1}(0) p_{\color{red}i}^{+} \delta_{s_{\color{red}f} s_{\color{red}i}} \right)  \Psi_{ {\color{blue}s_i} s_r}^{s_{d}}  
\end{align}
where we have averaged over initial states 
(${
s_d}, {\color{blue}s_i}$) 
and sum over final ($s_d',s_f$)\footnote{Note that the sum over the internal index $s_r$ is shown explicitly.}. 
Taking one of the sums over deuteron polarization indices fixes, $s'_d=s_d$ (orthogonal states), and substituding $\left(p_d+p_d^{\prime}\right)^+ \rightarrow 2  p_d^+$, it becomes
\begin{align}
2 p_d^+ G_C(0) = {1\over 3} \sum_{s_d, s_{r},s_1} \int \frac{d^{2} p_{r \perp} d\alpha}{\alpha  (2-\alpha) (2 - \alpha) } \Psi_{d, s_1,s_r}^{s_{d} \dagger} ( F_{1,p}(0) +  F_{1,n}(0) ){p_1^+}  \Psi_{d, s_1,s_r}^{s_{d}}
\end{align}
Since  the charge FF of neutron vanishes at $Q^2=0$, it follows that, $ F_{1,p}(0) +  F_{1,n}(0) =  F_{1,p}(0) = 1 = G_C(0)$. 
Thus, in terms of LF variables we arrive to the following result,
\phantomsection\label{LF WF norm LF fraction Sec}
\begin{align}\label{LF WF norm LF fraction}
{G_C(0) \over F_{1,p}(0) } = {1\over 3\cdot 2} \sum_{ s_d, s_r,s_1} \int \frac{d^{2} p_{r \perp} d\alpha}{2\alpha  (2-\alpha)} \Psi_{d, s_1,s_r}^{s_{d} \dagger}  \Psi_{d, s_1,s_r}^{s_{d}}
\end{align}
where we have used the equality, $\ 2-\alpha={ 2 p_1^+ \over p_d}$.

If in equation 
\hyperref[LF WF norm LF fraction Sec]{Eq.(\ref*{LF WF norm LF fraction})} 
we change the LF variables for  $\textbf{k}=(\textbf{p}_{r\perp},k_3)$, then the integral acquires a more familiar form, explicitly, 
\begin{align}
& \alpha={2(e_r+ k_3) \over (e_r + e_1)}, \quad k_3=(\alpha-1) \left(\frac{m^2 + p_{r \perp}^2}{\alpha(2-\alpha)} \right)^{1/2}
\end{align}
In the center of mass frame, $\textbf{k}=\textbf{p}_r=-\textbf{p}_1$, the integration measure transform to, 
\begin{align}
\frac{d^{2} p_{r \perp} d\alpha}{2\alpha  (2-\alpha)} =\frac{d^{3} k}{E_k}
\end{align}
where, $E_k=\sqrt{m^2 + \textbf{k}^2}$, and we find,

\phantomsection\label{LF WF norm relative momentum Sec}
\begin{align}\label{LF WF norm relative momentum}
{1\over 3} \sum_{ s_d, s_r,s_1} \int \frac{d^{3} k}{2E_k} |\Psi_{d}^{s_{d}}(k, s_1,s_r) |^2 ={G_C(0) \over F_{1,p}(0) } = 1
\end{align}

\subsection*{J.2 Wave Function Normalization from Baryonic Number}
\phantomsection\label{WF Normalization Baryonic Charge}

	The current operator we have constructed is such that coupled to an EM probe, 
it is an EM (transition) current. Nevertheless, the baryonic nature of the nucleon allows for a normalization procedure based on baryonic charge conservation which is  more simple than the one using  EM charge conservation. 

We start by noting that the main interaction between the bound nucleons is the residual strong force. Hence, 
\hyperref[EM current Sec]{Eq.(\ref*{EM current})}
which describes the transition of deuteron to constituent nucleons can be seen as a hadronic current. Within the approximation in Fock space decomposition of deuteron as two nucleon bound state, the  baryonic charge is conserve during the interaction.

Explicitly, using the number density operator\footnote{This is the zero component of the matrix element of the current.} \cite{Frankfurt:1981mk},

\phantomsection\label{density matrix LF Sec}
\begin{align}\label{density matrix LF}
\rho_d^{s_d}(\alpha,p_\perp) =   \sum_{s_r,s_1}{|\Psi_{d}^{s_{d},s_r,s_1}(\alpha,p_{\perp}) |^2 \over   (2-\alpha) }
\end{align}
Evaluating the integral of $\rho_d^{s_d}$  results in the baryonic number 
for each deuteron polarization, which is equal to $2$ in this case, since each polarized state have 2 baryons (in this $NN$ Fock component), explicitly,

\begin{equation}
\begin{aligned}
\mathcal{B}_{s_d} &= \int \frac{d^{2} p_{\perp} d\alpha}{2\alpha } \rho_d^{s_d}(\alpha,p_\perp) = \sum_{s_r,s_1} \int \frac{d^{2} p_{\perp} d\alpha}{2\alpha} {|\Psi_{d}^{s_{d},s_r,s_1}(\alpha,p_{\perp}) |^2 \over   (2-\alpha) }  = 2
\end{aligned}
\end{equation}
For the unpolarized case we have again 2, 
\begin{align}
{1 \over 3} \sum_{s_d} \mathcal{B}_{s_d} = 2
\end{align}

Which is consistent with 
\hyperref[LF WF norm LF fraction Sec]{Eq.(\ref*{LF WF norm LF fraction})}, 
\begin{align}
\hspace{-.6cm}
 {1\over 3}\sum_{s_d,s_r,s_1} \int \frac{d^{2} p_{r \perp} d\alpha}{2\alpha  (2-\alpha)} \Psi_{d, s_1,s_r}^{s_{d} \dagger}  \Psi_{d, s_1,s_r}^{s_{d}} = {1\over 3} \sum_{s_d, s_r,s_1} \int \frac{d^{2} p_{r \perp} d\alpha}{2\alpha} {|\Psi_{d}^{s_{d}}(k, s_1,s_r) |^2 \over   (2-\alpha) } = {G_C(0) \over F_{1,p}(0) } = 1
\end{align} 
and also with 
\hyperref[LF WF norm relative momentum Sec]{Eq.(\ref*{LF WF norm relative momentum})},
from where we get,

\phantomsection\label{LF WF normalization Sec}
\begin{align}\label{LF WF normalization}
\sum_{ s_d, s_r,s_1} \int \frac{d^{3} k}{2E_k} |\Psi_{d}^{s_{d}}(k, s_1,s_r) |^2  = 3
\end{align}


\begin{vita}
\begin{center}FRANK VERA \\[4ex]
\end{center}

\noindent
\begin{tabular}{ll}
Born, Ciudad de La Habana, Cuba &  \\[2ex]

2005  \hspace{2in} & B.Sc., Physics \\
	        & Universidad de Carabobo \\
		& Valencia, Venezuela \\[2ex]
              
2006-2008            & Graduate Research Assistant \\
                & Universidad Central de Venezuela \\
		& Caracas, Venezuela \\[2ex]

2008-2009       & Ph.D. Candidate in Physics \\
		& Universidad Central de Venezuela \\
		& Caracas, Venezuela \\[2ex]

2014-2021        & Graduate Teaching Assistant \\
		& Florida International University \\
		& Miami, Florida \\[2ex]

2019        & M.Sc., Physics \\
		& Florida International University \\
		& Miami, Florida \\[2ex]

2018-2021        & Ph.D. Candidate in Physics \\
		& Florida International University \\
		& Miami, Florida \\[2ex]

\end{tabular}

\vspace{0.25in}
\noindent
PUBLICATIONS AND PRESENTATIONS \\[1ex]  
\noindent 
Reed, T., Leon, C., Vera, F., Guo, L. and Raue, B. {\em The Constituent Counting Rule and Omega Photoproduction}. Phys. Rev. C {103}, 065203 (2021).  \\[1ex]  
\noindent
Vera, F. and Sargsian, M. {\em Electron scattering from a deeply bound nucleon on the light-front}. Phys. Rev. C {98}, 035202 (2018).  \\[1ex]  
\noindent
Gaitan, R. and Vera, F. {\em GL(3,R) Gauge Theory of Gravity coupled with an Electromagnetic Field}. Ciencia 14(3), 305-308 (2006).  \\[1ex]  
\noindent
Vera, F. (2020, October). {\em Short-Range Structure of the Deuteron on the Light-Front}. Division of Nuclear Physics (DNP) Conference, Remote (Zoom).  \\[1ex]  
\noindent
Vera, F. (2019, April). {\em Probing the Deuteron at Short Distances on the Light-Front}. American Physical Society (APS) Conference, Denver, Colorado.  \\[1ex]  
\noindent
Vera, F. (2018, August). {\em Electron Scattering from Deeply Bound Nucleon on the Light-Front}. Workshop on Short Range Correlations (MIT), Boston, Massachusetts.  \\[1ex]  

\end{vita}


\begin{thebibliography}{16}


	\bibitem{Njm} 
	V.~G.~J.~Stoks, R.~A.~M.~Klomp, C.~P.~F.~Terheggen and J.~J.~de Swart,
	Phys.\ Rev.\ C {\bf 49}, 2950 (1994)


	\bibitem{Feynman} 
	R.~P.~Feynman, ``Photon-hadron interactions,''
	Reading 1972, 282p.   
	
	

	\bibitem{V18} 
	R.~B.~Wiringa, V.~G.~J.~Stoks and R.~Schiavilla,
	Phys.\ Rev.\ C {\bf 51}, 38 (1995)
	
	
	\bibitem{CDBonn} 
	R.~Machleidt,
	Phys.\ Rev.\ C {\bf 63}, 024001 (2001)
	

	
	\bibitem{Weinberg1966} 
	S.~Weinberg,
	Phys.\ Rev.\  {\bf 150}, 1313 (1966).
	
	
		  
\bibitem{Frankfurt:2008zv}
L.~Frankfurt, M.~Sargsian and M.~Strikman,
Int.\ J.\ Mod.\ Phys.\ A \textbf{23}, 2991-3055 (2008)


\bibitem{Sargsian:2009hf}
M.~M.~Sargsian,
Phys.\ Rev.\ C \textbf{82}, 014612 (2010).

	
	
\bibitem{Cosyn:2010ux}
W.~Cosyn and M.~Sargsian,
Phys.\ Rev.\ C \textbf{84}, 014601 (2011)

	
\bibitem{Cosyn:2020kwu}
W.~Cosyn and C.~Weiss,
Phys. Rev. C \textbf{102}, 065204 (2020)



\bibitem{Arrington:2011xs}
J.~Arrington, D.~Higinbotham, G.~Rosner and M.~Sargsian,
Prog.\ Part.\ Nucl.\ Phys.\  \textbf{67}, 898-938 (2012)


\bibitem{Frankfurt:1988nt}
L.~Frankfurt and M.~Strikman,
Phys.\ Rept.\  \textbf{160}, 235-427 (1988)


\bibitem{arnps} 
N.~Fomin, D.~Higinbotham, M.~Sargsian and P.~Solvignon,
Ann.\ Rev.\ Nucl.\ Part.\ Sci.\  {\bf 67}, 129 (2017). 
	
	
	\bibitem{Hen:2016kwk} 
	O.~Hen, G.~A.~Miller, E.~Piasetzky and L.~B.~Weinstein,
	Rev.\ Mod.\ Phys.\  {\bf 89}, no. 4, 045002 (2017).
	
	
	

\bibitem{Boeglin:2015cha} 
 W.~Boeglin and M.~Sargsian,
 Int.\ J.\ Mod.\ Phys.\ E {\bf 24}, no. 03, 1530003 (2015).
 


\bibitem{Harvey:1980rva}
M.~Harvey,
Nucl.\ Phys.\ A \textbf{352}, 326-342 (1981).

\bibitem{Ji:1985ky}
C.~Ji and S.~J.~Brodsky,
Phys.\ Rev.\ D \textbf{34}, 1460 (1986).


\bibitem{Hockert:1974qt}
J.~Hockert, D.~Riska, M.~Gari and A.~Huffman,
Nucl.\ Phys.\ A \textbf{217}, 14-28 (1973).

\bibitem{Fabian:1976ne}
W.~Fabian and H.~Arenhovel,
Nucl.\ Phys.\ A \textbf{258}, 461-479 (1976).

\bibitem{Ulmer:2002jn}
P.~Ulmer, \textit{et al.} ,
Phys.\ Rev.\ Lett.\  \textbf{89}, 062301 (2002).

\bibitem{Egiyan:2007qj}
K.~Egiyan \textit{et al.} [CLAS],
Phys.\ Rev.\ Lett.\  \textbf{98}, 262502 (2007).

\bibitem{Boeglin:2011mt}
W.~Boeglin~\textit{et al.} [Hall A],
Phys.\ Rev. \ Lett. \ \textbf{107}, 262501 (2011).

\bibitem{Yero:2020}
C.~Yero~{\em et al.}, [Hall C Collaboration]
Phys.\ Rev.\ Lett.\ \textbf{125}, 262501 (2020)



 
\bibitem{Frankfurt:1977vc}
L.~Frankfurt and M.~Strikman,
Nucl. Phys. B \textbf{148}, 107-140 (1979).


\bibitem{Frankfurt:1981mk}
L.~Frankfurt and M.~Strikman,
Phys. Rept. \textbf{76}, 215-347 (1981).
	

\bibitem{Buck:1979ff}
W.~Buck and F.~Gross,
Phys. Rev. D \textbf{20}, 2361 (1979)

 
\bibitem{Karmanov:1980mc}
V.~Karmanov,
Nucl. Phys. A \textbf{362}, 331 (1981).

\bibitem{Karmanov:1995}
J.~Carbonell and V. A.~Karmanov, 
Nucl. Phys. \textbf{A581}, 625 (1995)


\bibitem{Cooke:2001kz}
J.~R.~Cooke and G.~A.~Miller,
Phys. Rev. C \textbf{66}, 034002 (2002).


\bibitem{CiofidegliAtti:2000xj}
C.~Ciofi degli Atti, L.~Kaptari and D.~Treleani,
Phys. Rev. C \textbf{63}, 044601 (2001).
 

\bibitem{Laget:2004sm}
 J.~Laget,
Phys. Lett. B \textbf{609}, 49-56 (2005).
 
 	
	\bibitem{Atti:2015eda} 
	C.~Ciofi degli Atti,
	Phys.\ Rept.\  {\bf 590}, 1 (2015).
	
	
	\bibitem{srcprog} 
	J.~Arrington, D.~W.~Higinbotham, G.~Rosner and M.~Sargsian,
	Prog.\ Part.\ Nucl.\ Phys.\  {\bf 67}, 898 (2012).
	
		
	
	\bibitem{FSDS93} 
	L.~L.~Frankfurt, M.~I.~Strikman, D.~B.~Day and M.~Sargsian,
	Phys.\ Rev.\ C {\bf 48}, 2451 (1993).
	
	
	\bibitem{Kim1} 
	K.~S.~Egiyan {\it et al.} [CLAS Collaboration],
	Phys.\ Rev.\ C {\bf 68}, 014313 (2003).
	
	
	\bibitem{Kim2} 
	K.~S.~Egiyan {\it et al.} [CLAS Collaboration],
	Phys.\ Rev.\ Lett.\  {\bf 96}, 082501 (2006).
	
	
	\bibitem{Fomin:2011ng} 
	N.~Fomin {\it et al.},
	Phys.\ Rev.\ Lett.\  {\bf 108}, 092502 (2012).
	
	
	\bibitem{EIP2} 
	A.~Tang {\it et al.},
	Phys.\ Rev.\ Lett.\  {\bf 90}, 042301 (2003).
	
	
	
	\bibitem{EIP3}   R.~Shneor {\it et al.} [Jefferson Lab Hall A Collaboration],
	Phys.\ Rev.\ Lett.\  {\bf 99}, 072501 (2007).
	
	
	\bibitem{isosrc} 
	E.~Piasetzky, M.~Sargsian, L.~Frankfurt, M.~Strikman and J.~W.~Watson,
	Phys.\ Rev.\ Lett.\  {\bf 97}, 162504 (2006).  
	
	
	\bibitem{EIP4}  R.~Subedi {\it et al.},
	Science {\bf 320}, 1476 (2008).
	
	
	\bibitem{newprops} 
	M.~M.~Sargsian,
	Phys.\ Rev.\ C {\bf 89}, no. 3, 034305 (2014).
	
	
	\bibitem{Hen:2014nza} 
	O.~Hen {\it et al.},
	Science {\bf 346}, 614 (2014).   
	
	
	\bibitem{Artiles-Sargsian2016}   O.~Artiles and M.~M.~Sargsian,
	Phys. Rev.  {\bf C94}, 064318 (2016).
		
	
\bibitem{Frankfurt:1996xx}
L.~Frankfurt, M.~Sargsian and M.~Strikman,
Phys. Rev. C \textbf{56}, 1124-1137 (1997).


\bibitem{Sargsian:2001ax}
M.~M.~Sargsian,
Int. J. Mod. Phys. E \textbf{10}, 405-458 (2001).


\bibitem{Sargsian:2002wc}
M.~Sargsian {\em et al}.,
J. Phys. G \textbf{29}, no.3, R1-R45 (2003).


\bibitem{Stoler:1993yk}
P.~Stoler,
Phys. Rept. \textbf{226}, 103-171 (1993).

\bibitem{Ungaro:2006df}
M.~Ungaro \textit{et al.} [CLAS],
Phys. Rev. Lett. \textbf{97}, 112003 (2006).



\bibitem{Gilman:2001yh}
R.~A.~Gilman and F.~Gross,
J. Phys. G \textbf{28}, R37-R116 (2002)




\bibitem{Vera_Sargsian:2018}
F.~Vera and M.~M.~Sargsian,
Phys. Rev. C \textbf{98}, 035202 (2018).

	
	
	\bibitem{CiofidegliAtti:2004jg} 
	C.~Ciofi degli Atti and L.~P.~Kaptari,
	Phys.\ Rev.\ C {\bf 71}, 024005 (2005).
	
	
	
	\bibitem{Jeschonnek:2008zg} 
	S.~Jeschonnek and J.~W.~Van Orden,
	Phys.\ Rev.\ C {\bf 78}, 014007 (2008).
	
	
	\bibitem{Laget04} 
	J.~M.~Laget,
	Phys.\ Lett.\ B {\bf 609}, 49 (2005).
	
	
	\bibitem{eheppn1} 
	M.~M.~Sargsian, T.~V.~Abrahamyan, M.~I.~Strikman and L.~L.~Frankfurt,
	Phys.\ Rev.\ C {\bf 71}, 044614 (2005).
	
	
	\bibitem{eheppn2} 
	M.~M.~Sargsian, T.~V.~Abrahamyan, M.~I.~Strikman and L.~L.~Frankfurt,
	Phys.\ Rev.\ C {\bf 71}, 044615 (2005).
	
	
	\bibitem{disfsirev} 
	W.~Cosyn and M.~Sargsian,
	Int.\ J.\ Mod.\ Phys.\ E {\bf 26}, no. 09, 1730004 (2017).
	
	\bibitem{Machleidt2001}
	R. Machleidt
 Phys.\ Rev.\ C {\bf 63}, 024001 (2001).


	\bibitem{SACLAY1} 
	A.~Bussiere {\it et al.},
	Nucl.\ Phys.\ A {\bf 365}, 349 (1981).
	
	
	\bibitem{SACLAY2} 
	S.~Turck-Chieze {\it et al.},
	Phys.\ Lett.\  {\bf 142B}, 145 (1984).
	
	
	\bibitem{NIKHEF} 
	J.~F.~J.~Van Den Brand {\it et al.},
	Phys.\ Rev.\ Lett.\  {\bf 60}, 2006 (1988).
	
	
	\bibitem{Mougey}S. Frullani and J. Mougey,  Adv. \ Nucl. \ Phys. {\bf 14}, 1 (1984).
	
	
	\bibitem{deForest(1983)}
	T.~de Forest, Nuclear Physics {\bf A392}  232-248 (1983)   
	
	
	
	\bibitem{KS(1970)}
	J.~Kogut and D. Soper,
	Phys.\ Rev.\  {\bf D1}, 2901 (1970).
	
	
	\bibitem{LB(1980)}
	G.~P.~Lepage and S.~J.~Brodsky,
	Phys.\ Rev.\ D {\bf 22}, 2157 (1980).
	
	
	\bibitem{Dirac1949} 
	P.~A.~M.~Dirac,
	Rev.\ Mod.\ Phys.\  {\bf 21}, 392 (1949).
	
	

\bibitem{Weinberg1964}
S.~Weinberg,
Phys. Rev. {\bf 133}, B1318 (1964)


\bibitem{Ryder1996}
L.~H.~Ryder,   Quantum Field Theory. 2nd Edition (1996), Cambridge University Press, Cambridge.




	
	\bibitem{BPP(1997)}
	S.~J.~Brodsky, H.~C.~Pauli and S.~S.~Pinsky,
	Phys.\ Rept.\  {\bf 301}, 299 (1998).
	
	
	
\bibitem{Artiles_Sargsian-multisrc1}   O.~Artiles and M.~M.~Sargsian,
Phys. Rev.  {\bf C94}, 064318.
	
	
	
	\bibitem{KP91} 
	B.~D.~Keister and W.~N.~Polyzou.
	Adv.\ Nucl.\ Phys.\  {\bf 20}, 225 (1991).
	
	
	\bibitem{Miller00} 
	G.~A.~Miller,
	Prog.\ Part.\ Nucl.\ Phys.\  {\bf 45}, 83 (2000).
	
	%
	%
	
	\bibitem{Leutwyler1978} 
	H.~Leutwyler and J.~Stern,
	Annals Phys.\  {\bf 112}, 94 (1978).
	
	
	\bibitem{Bakker1979} 
	B.~L.~G.~Bakker, L.~A.~Kondratyuk and M.~V.~Terentev,
	Nucl.\ Phys.\ B {\bf 158}, 497 (1979).
	
	
	%
	%
	%
	
	\bibitem{K1980}
	V.~A.~Karmanov,
	Nucl.\ Phys.\ A {\bf 362}, 331 (1981).
	
	
	\bibitem{Foldy1960}
	L.~L.~Foldy,
	Phys.\ Rev.\  {\bf 122}, 275 (1961).
	
	
	\bibitem{Fuda1990} 
	M.~G.~Fuda,
	Phys.\ Rev.\ D {\bf 42}, 2898 (1990).
	
	
	\bibitem{Brodsky2000} 
	S.~J.~Brodsky, D.~S.~Hwang, B.~Q.~Ma and I.~Schmidt,
	Nucl.\ Phys.\ B {\bf 593}, 311 (2001)
	
	
	\bibitem{DFSS2018}
	D.~B.~Day, L.~L.~Frankfurt, M.~M.~Sargsian and M.~I.~Strikman,
	
	
	\bibitem{FS1991}
	L.~Frankfurt and M.~Strikman,
	In *Frois, B. (ed.), Sick, I. (ed.): Modern topics in electron scattering* 645-694
	
	
	\bibitem{Gross79} 
	W.~W.~Buck and F.~Gross,
	Phys.\ Rev.\ D {\bf 20}, 2361 (1979).
	
	
	\bibitem{Glazek84} 
	S.~D.~Glazek,
	Acta Phys.\ Polon.\ B {\bf 14}, 893 (1983).
	
	
	\bibitem{Miller13} 
	G.~A.~Miller,
	Phys.\ Rev.\ C {\bf 89}, no. 4, 045203 (2014).
	
	
	\bibitem{Bincer60}
	A. M. Bincer, Phys. Rev. {\bf 118}, 855 (1960).
	
	
	\bibitem{Koch90} 
	H.~W.~L.~Naus, S.~J.~Pollock, J.~H.~Koch and U.~Oelfke, Nucl. Phys. {\bf A509}  717-735 (1990)
	
	
	\bibitem{Koch96}
	H.~W.~L.~Naus, S.~J.~Pollock, J.~H.~Koch, Phys. Rev.  {\bf C53}, 2304 (1996)
	
	
	
	
	\bibitem{EMC}  J.~J.~Aubert {\it et al.} [European Muon Collaboration],
	Phys.\ Lett.\  {\bf 123B}, 275 (1983).
	
	
	
		

\end{thebibliography}
\end{document}